\documentclass[11pt,letterpaper]{report}
\usepackage{yalethesis}
\usepackage{amsmath}
\usepackage{amssymb}
\usepackage{epsfig}
\usepackage{graphicx}
\usepackage{subfigure}
\usepackage[T1]{fontenc}
\usepackage[ansinew]{inputenc}
\usepackage{pifont}
\usepackage[spanish,english]{babel}
\usepackage[dvips]{rotating}
\usepackage{cite}
\newlength{\originalbase}
\newcommand{\spacing}[1]{\setlength{\baselineskip}{#1\originalbase}}

\newcommand{\micron}{\mbox{$\mu\mathrm{m}$}}
\newcommand{\ie}{\textit{i.e.}}
\newcommand{\eg}{\textit{e.g.}}
\newcommand{\vs}{\textit{vs.}}
\newcommand{\sqrts}{\mbox{$\sqrt{s}$}}
\newcommand{\Tc}{\mbox{$T_c$}}
\newcommand{\sqrtsNN}{\mbox{$\sqrt{s_{_{\mathrm{NN}}}}$}}
\newcommand{\AuAu}{\mbox{$\mathrm{Au}+\mathrm{Au}$}}
\newcommand{\PbPb}{\mbox{$\mathrm{Pb}+\mathrm{Pb}$}}
\newcommand{\hminus}{\mbox{$h^-$}}
\newcommand{\hplus}{\mbox{$h^+$}}
\newcommand{\Nhminus}{\mbox{$N_{h^-}$}}

\newcommand{\ppbar}{\mbox{$p\bar{p}$}}
\newcommand{\pp}{\mbox{$pp$}}
\newcommand{\pA}{\mbox{$pA$}}
\newcommand{\NN}{\mbox{$AA$}}
\newcommand{\kzeros}{\mbox{$K^0_S$}}
\newcommand{\piminus}{\mbox{$\pi^-$}}
\newcommand{\piplus}{\mbox{$\pi^+$}}
\newcommand{\kminus}{\mbox{$K^-$}}
\newcommand{\pbar}{\mbox{$\bar{p}$}}
\newcommand{\meanpt}{\mbox{$\langle p_\perp \rangle$}}
\newcommand{\meanmt}{\mbox{$\langle m_\perp \rangle$}}
\newcommand{\pt}{\mbox{$p_\perp$}}
\newcommand{\mt}{\mbox{$m_\perp$}}
\newcommand{\gev}{\mbox{$\mathrm{GeV}$}}
\newcommand{\gevc}{\mbox{$\mathrm{GeV/}c$}}
\newcommand{\gevcc}{\mbox{$\mathrm{GeV/}c^2$}}
\newcommand{\mev}{\mbox{$\mathrm{MeV}$}}
\newcommand{\mevc}{\mbox{$\mathrm{MeV/}c$}}
\newcommand{\mevcc}{\mbox{$\mathrm{MeV/}c^2$}}
\newcommand{\fm}{\mbox{$\mathrm{fm}$}}
\newcommand{\TAA}{\mbox{$\mathrm{T}_{\mathrm{AA}}$}}
\newcommand{\siginel}{\mbox{$\sigma_{\mathrm{inel}}$}}
\newcommand{\ncoll}{\mbox{$N_{\mathrm{coll}}$}}
\newcommand{\npart}{\mbox{$N_{\mathrm{part}}$}}
\newcommand{\spp}{\mbox{$\sigma_{\mathrm{pp}}$}}
\newcommand{\sppbar}{\mbox{$\sigma_{\mathrm{p\bar{p}}}$}}
\newcommand{\sigmahad}{\mbox{$\sigma_{\mathrm{AuAu}}$}}

\newcommand{\Et}{\mbox{$E_\perp$}}

\newcommand{\ep}{\mbox{$\mathrm{e}^+\mathrm{e}^-$}}

\newcommand{\dedx}{\mbox{$\frac{dE}{dx}$}}
\newcommand{\betat}{\mbox{$\mathrm{\beta_\perp^{flow}}$}}
\newcommand{\mpi}{\mbox{$\mathrm{m_\pi}$}}
\newcommand{\dndy}{\mbox{$dN/dy$}}
\newcommand{\detdy}{\mbox{$d\Et/dy$}}
\newcommand{\effvtx}{\mbox{$\epsilon_{\mathrm{vtx}}$}}
\newcommand{\Nglobal}{\mbox{$N_{\mathrm{global}}$}}
\newcommand{\Hijing}{\mbox{\textsc{hijing}}}
\newcommand{\zvertex}{\mbox{$z_{\mathrm{vertex}}$}}
\newcommand{\Teff}{\mbox{$T_{\mathrm{eff}}$}}
\newcommand{\Tfo}{\mbox{$T_{\mathrm{f.o.}}$}}
\begin{document}
\spacing{1.2}  

\title{Charged Hadron Spectra in Au+Au Collisions at $\sqrtsNN = 130 \ \gev$}
\author{Manuel Calderón de la Barca Sánchez}
\awardate{December 2001}
\advisor{Prof. John W. Harris}
\coadvisor{Thomas S. Ullrich}


\begin{abstract}



The collision of high energy heavy ions is the most promising
laboratory for the study of nuclear matter at high energy density
and for creation of the Quark-Gluon Plasma.  A new era in this
field began with the operation and first collisions of Au nuclei
in the Relativistic Heavy Ion Collider (RHIC) at Brookhaven
National Laboratory during 2000.  This work concentrates on
measurement of global hadronic observables in Au+Au interactions
at a centre-of-mass energy of $\sqrtsNN = 130\ \gev$, which mainly
address conditions in the final state of the collision. The
minimum bias multiplicity distribution, the transverse momentum
(\pt), and pseudorapidity ($\eta$) distributions for charged
hadrons (\hminus, \hplus) are presented. Results on identified
\piminus\ transverse mass (\mt) and rapidity ($y$) distributions
are also discussed. The data were taken with the STAR detector
with emphasis on particles near mid-rapidity.

We find that the multiplicity density at mid-rapidity for the 5\%
most central interactions is $dN_{h^-}/d\eta|_{\eta=0} = 280 \pm
1_{\textrm{stat}}\pm 20_{\textrm{syst}}$, an increase per
participant of 38\% relative to \ppbar\ collisions at similar
energy. The mean transverse momentum is $0.508 \pm 0.012$ GeV/c
and is larger than in \PbPb\ collisions at lower energies.  The
scaling of the \hminus\ yield per participant nucleon pair,
obtained via a ratio of \AuAu\ to \ppbar\ \pt\ distributions, is a
strong function of \pt. The pseudorapidity distribution is almost
constant within $|\eta|<1$. The \piminus\ rapidity distribution is
also flat around mid-rapidity in the region $|y|<0.8$, with a
yield of pions for central collisions of $dN_{\pi^-}/dy|_{y=0} =
287 \pm 1_{\textrm{stat}} \pm 21_{\textrm{syst}}$. However, the
slope of the \mt\ distributions is not the same for different
rapidity bins, suggesting that boost-invariance is not fully
achieved in the collisions.



\end{abstract}


\beforepreface
%
%

\prefacesection{Acknowledgements}
\thispagestyle{empty}

I am happy to extend a heartfelt thank you to the many people who
made the completion of this work a reality. John Harris gave me
his continued support since before my arrival in New Haven and
found the time out of his ever-busy schedule to follow my
progress: from making possible my coming to graduate school at an
unusual time, to making sure my work was helpful both to me and to
the STAR collective effort. Thomas Ullrich offered continuous
guidance in this project.  He always had insightful suggestions
and infinite patience with my questions, be it physics, C$^{++}$
or any other subject. Vielen Dank f\"ur deine ganze Hilfe! John
and Thomas were not just my mentors, but also my friends.

My gratitude goes to Brian Lasiuk for helping me many times since
I began the heavy ion path at CERN and to the grad students, Matt,
Mike, Jon, Betya and Sevil for the shared experiences during this
exciting and crazy time. I gratefully acknowledge all the STAR
collaboration, in particular the \hminus\ gang; Peter Jacobs,
Spiros Margetis, Zhangbu Xu, and the ``Godfathers''; Herbert
Str\"obele, Tom Trainor and Fuquiang Wang, who all contributed to
make the analysis the best it could be; Mike Lisa, an enthusiastic
one-man think tank who helped shape several details of the
analysis, and was always helpful when I came with questions; and
all the Spectra conveners with whom I had the pleasure of working,
Bill Llope, Craig Ogilvie and Raimond Snellings.

Saving the best for last, I want to thank {\Pifont{psy}Alex\'ia},
who made our time together the most wonderful ever. {\Pifont{psy}
S\'{} agap\'{w} para pol\'{u}, kai e\'imai p\'{a}nta trel\'{o}V
gia esena.} \selectlanguage{spanish} Y más que nada, a mi papá, a
mi madre, a Laura y a Cathy, quienes son el núcleo de donde
provengo. \selectlanguage{english}

\pagenumbering{roman}
\tableofcontents
\figurespagetrue
\tablespagetrue
\listoffigures
\listoftables


\chapter{Introduction}
\label{ch:Introduction}
\pagenumbering{arabic}

Based on statistical Quantum Chromodynamics, our current
expectation is that strongly interacting matter at extreme energy
density ($\epsilon \gtrsim 1\ \gev/\fm^3$) and temperature ($T
\gtrsim 170\ \mev$) is found in a state where hadrons no longer
exist as discrete entities \cite{collins:75,Polyakov:1978vu}. The
relevant degrees of freedom for such a system are those of the
underlying partons, and we label this state the Quark-Gluon Plasma
(QGP).\cite{muller:85,Mclerran:1986zb,hwa:90,hwa:95}

Such a state of matter is believed to be the one in which the
early universe existed in a time scale $\sim 10^{-6}$ s after the
Big Bang \cite{olive:91,kolb:90} and is also predicted to exist in
the interior of neutron stars
\cite{Ellis:1991qq,Glendenning:1989cf}.

Ultra-relativistic heavy ion collisions are the most promising
tool for the creation of a QGP in the laboratory. The appearance
and study of predicted signatures in such collisions has been the
subject of intense theoretical and experimental work for more than
two decades (for recent reviews one can refer to the proceedings
of the `Quark Matter' conferences \cite{qm:81, qm:82, qm:83,
qm:84, qm:86, qm:87, qm:88,
  qm:90, qm:91, qm:93, qm:95, qm:96, qm:97, qm:99, qm:01}).
In the Alternating Gradient Synchrotron (AGS) at the Brookhaven
National Laboratory (BNL) collisions have been produced using
beams from Si to Au with energies of 11-14 \gev\ per nucleon.  The
corresponding centre-of-mass energy for \AuAu\ collisions at the
AGS is $\sqrtsNN = 5\ \gev$. We use $\sqrtsNN$ to denote the
centre-of-mass energy per nucleon pair.  The Super Proton
Synchrotron (SPS) at the Conseil Europ\'{e}en de la Recherche
Nucl\'{e}aire (CERN) has yielded results from O, S and Pb beams
with energies of 60-200 \gev\ per nucleon, this is a
centre-of-mass energy of $\sqrtsNN = 17.2\ \gev$ for \PbPb\
collisions.

There has been considerable excitement in the field during the last
year with the commissioning of the Relativistic Heavy Ion Collider
(RHIC) at BNL \cite{Baym:2001in}. Dedicated experiments began taking
data in the summer of 2000 with the highest colliding energy heavy ion
beams available.  RHIC ran during this period with Au beams at a
center of mass energy of $\sqrtsNN = 130 \ \gev$.
Studying collisions at different energies helps to map the phase
diagram of nuclear matter, where RHIC is expected to probe the
region of high temperature and near-zero net-baryon density, a
regime accessible by QCD lattice simulations.


For all heavy ion experiments, global event observables have played an
important role. It remains true that to understand specific plasma
signatures one must first understand the global character of the
reaction dynamics.
The momentum distribution of the bulk of the particles measured in
the detectors provide evidence mainly of the final phase of the
system formed in the collision, also called the freeze-out phase.
In order to obtain information about the early stages where plasma
formation is expected to occur, one must use indirect methods such
as hard probes which must be gauged to reference observables
related to the global character of the reaction. The information
obtained in the final state in the form of particle spectra is one
of the main sources of global event information.  In addition,
global final state observables help provide limits on the possible
evolution of the system at earlier times. All experiments at RHIC
and elsewhere have some capability to measure global observables
in order to characterize and classify events, compare results with
other experiments, and perform systematic studies. Measuring the
final state particle spectra is therefore a basic requirement for
the study of the collision dynamics.

The measurements of global observables for the first year of RHIC
collisions are an exciting first topic of study in an energy
regime where perturbative phenomena are expected to dominate.  The
contribution to the total charged particle multiplicity coming
from hard processes (jets and mini-jets) at RHIC is estimated to
be between $30 - 50 \%$
\cite{Kajantie:1987pd,Eskola:1989yh,Calucci:1990hb,Calucci:1991sz}.
Since particle production is a dominant feature of the collision,
one of the first observables to study is the multiplicity of
charged particles for each event.  This quantity and related
global observables such as the transverse energy (\Et) and the
energy deposition in the very forward region as measured typically
by zero degree calorimeters (ZDCs) is related to the impact
parameter of the collision. The more central the collision is, the
more nucleons from both projectile nuclei will participate in the
interaction, and hence the more secondary hadrons will be produced
(large multiplicity and total \Et).  A geometrical correlation
between the impact parameter $b$, \Et\ and the energy reaching the
zero degree calorimeter can be given in terms of the number of
``wounded'' nucleons \cite{Bialas:1976ed}. Central (high
multiplicity) events have a special interest as it is for these
events that the largest fraction of the incoming energy will be
redistributed to new degrees of freedom. The next key observable
is the distribution of the particles in momentum space. Rapidity
and transverse momentum distributions allow one to address
properties of the reaction dynamics such as the extent to which
the nucleons are slowed down in the collision (\textit{stopping}),
the approach of the system to thermal equilibrium, and the shape
of the emitting source of particles.

The Solenoidal Tracker At RHIC (STAR) is one of the 4 experiments that
partake in the RHIC program.  The analyses presented here are based on
data taken by STAR during the 2000 summer run. The pith of the
experiment is a large acceptance cylindrical Time Projection Chamber
(TPC) placed in a uniform solenoidal magnetic field for momentum
determination. The TPC provides charged particle tracking in the
mid-rapidity region with full azimuthal coverage and particle
identification for low momentum particles.  The STAR detector is thus
ideally suited for the study of hadronic observables and will
therefore focus on such measurements early on, although the STAR
physics program includes other topics in addition to QGP physics.
Since hadrons are the most copiously produced particles in the
collision and $\pi$ mesons comprise $\sim 80\%$ of the total hadron
population, this work concentrates on the study of charged hadron
(\hminus\ and \hplus) and identified \piminus\ meson production and
momentum spectra for \AuAu\ collisions at $\sqrtsNN = 130 \ \gev$.
The minimum bias \hminus\ multiplicity distribution is presented.
For different selections of event centralities, we present the
pseudorapidity ($\eta$) and transverse momentum (\pt) distributions
for \hminus.  We compare the \pt\ distributions to those expected in
similar energy \ppbar\ collisions as reference.  We also discuss
rapidity ($y$) and transverse mass (\mt) distributions for identified
\piminus.

This work is organized in the following manner.  It starts with a
discussion on general aspects of Quark-Gluon Plasma physics and a
review of hadronic particle production at the AGS and SPS
(chapters \ref{ch:theoreticalBackground}, \ref{ch:expSearchQgp}
and \ref{ch:spectraReview}), followed by a description of the STAR
detector (chapter \ref{ch:StarExperiment}). An overview of the
tracking strategy and detector calibration procedures is then
given (chapter \ref{ch:ReconstructionCalibration}). After a
discussion of the detector simulation (chapter
\ref{ch:DetectorSimulation}) and a layout of the analysis
technique and corrections (chapter \ref{ch:nchAnalysis}), we
present the results on hadron production and identified pion
spectra and discuss their implications regarding the dynamics of
the collision features at RHIC (chapters \ref{ch:NchResults} and
\ref{ch:PiResults}). Finally, we present our conclusions (chapter
\ref{ch:Conclusions}).

%
%

\chapter{The Physics of the Quark-Gluon Plasma}
\label{ch:theoreticalBackground}

We give now an overview of the high temperature phase of QCD and
the Quark-Gluon Plasma.  For recent reviews on the subject, see
\eg\ \cite{Muller:1995rb,harris:96,bass:99}, review articles in
the proceedings of the `Quark Matter' conferences
~\cite{blaizot:99,digiacomo:99} and the growing collection of
books ~\cite{muller:85,hwa:90,hwa:95,csernai:94,wong:94}.

\section{Deconfinement in QCD}
At the fundamental level, strongly interacting matter is described
by interactions of quarks through the exchange of gluons.  The
theory that describes these interactions, quantum chromodynamics
(QCD), has the remarkable properties that at large distances or
small momenta $q$, the effective coupling constant $\alpha_s(q^2)$
is large, and it decreases logarithmically at short distances or
large momenta.  This behaviour can be seen from perturbative QCD.

The QCD Lagrangian is given by
\begin{equation}
\mathcal{L}_{QCD} = i\bar{\psi}\gamma^\mu(\partial_\mu - i g \hat{A}_\mu)\psi - m\bar{\psi}\psi
- \frac{1}{4}F^a_{\mu\nu}F_a^{\mu\nu}
\label{eq:qcdLagrangian}
\end{equation}
where the colour potential \(\hat{A}_\mu \) is a $3\times3$ matrix
(indicated by the circumflex symbol) and can be represented by a linear combination
of the 8 Gell-Mann matrices:

\begin{equation}
\hat{A}_\mu = \frac{1}{2} \sum_{a=1}^8 A^a_\mu(x)\lambda_a \ .
\label{eq:qcdColourPotential}
\end{equation}
This potential is introduced to make the Lagrangian invariant
under local (space-dependent) rotations of the 3 colour components
of the quark wavefunction $\psi$. The eight-component field
strength tensor expressed as
\begin{equation}
F^a_{\mu\nu} = \partial_\mu A_\nu^a - \partial_\nu A_\mu^a + gf_{abc}A_\mu^b A_\nu^c \ ,
\label{eq:qcdFieldTensor}
\end{equation}
where $f_{abc}$ are the antisymmetric structure constants for the
Lie group SU(3).  The product $F^a_{\mu\nu}F_a^{\mu\nu}$ also
remains invariant under a local colour gauge transformation.

The perturbative quantization proceeds in QCD in a similar manner as in Quantum
Electrodynamics.  The quadratic terms in the lagrangian
define free quark and gluon fields, described by propagators with the same
form as those in QED for electrons and photons.  The free propagators
$D_0(q^2)$ are
proportional to $1/q^2$, and if this were the only ingredient would lead
to a colour force which would fall off like $1/r$.
The main difference comes from the coupling of the gluon field to itself.
This modifies the true gluon propagator, because one has to evaluate
contributions arising from terms in the perturbation series corresponding
to the virtual creation of a pair of coloured particles from the vacuum.
We call this the `vacuum polarization' function $\Pi(q^2)$, and it turns
out to be proportional to $-1/q^2\ln(-q^2)$.

The higher order diagrams, in which the gluon interacts consecutively
once, twice, three times with the vacuum polarization, etc. can be summed
into a geometric series yielding the full propagator

\begin{eqnarray}
D(q^2) & = & D_0(q^2) + i^2 D_0(q^2)\Pi(q^2)D_0(q^2) + i^4 D_0\Pi D_0\Pi D_0 + \ldots \nonumber \\
       & = & \frac{D_0(q^2)}{1-\Pi(q^2)D_0(q^2)} \nonumber \\
       & = & \left(1/q^2\right) \frac{1}{1 + \frac{(33-2N_F)\alpha_s}{12\pi\ln(-q^2/\mu^2)}}
\label{eq:qcdPropagator}
\end{eqnarray}
where $\mu$ is a reference point introduced by renormalization, and $N_F$ counts the
number of flavours with mass below $|q^2|^\frac{1}{2}$.  So after renormalization,
the second factor in Eq.~\ref{eq:qcdPropagator} acts as a momentum-dependent
modification of the strong
coupling constant.  Combining it with $\alpha_s$ we obtain the ``running''
coupling constant

\begin{equation}
\alpha_s(q^2) = \frac{4\pi}{(11-2N_F/3)\ln(-q^2/\Lambda^2)} \propto \frac{1}{\ln(q^2/\Lambda^2)}
\label{eq:qcdAlphaS}
\end{equation}
where $\Lambda$ is a (dimensional) parameter introduced also by the renormalization
process.  We see that the running coupling exhibits a pole; in this approximation it is
at $q^2=-\Lambda^2$.  More sophisticated expressions for the gluon propagator
indicate that the pole is really at $q^2=0$, and that $\alpha_s(q^2)$
should behave as $1/q^2$ in the limit $q^2 \rightarrow 0$.

Converted into coordinate space, this means that $\alpha_s(r)$ grows like
$r^2$ for large distances, corresponding to a linearly rising potential.
The logarithm in Eq.~\ref{eq:qcdAlphaS} causes a gradual
decrease of the coupling strength between colour charges at large momenta
or small distances.  It is this property that is known as `\textit{asymptotic freedom}'.
This behaviour of the running coupling at large distances results
in the confinement of quarks (\ie\ isolated quarks are not
observed in nature).

It is important to
note that the behaviour of the strong coupling constant outlined above is
derived for interactions \textit{in vacuo}.
The usual point of
comparison for measurements and calculations of the strong
coupling constant is at the mass of the $Z^0$ boson, $M_Z = 91\
\gevcc$, where the world average is $\alpha_s(M_Z) = 0.1172 \pm
0.0045$ \cite{pdg:00}.  The typical initial momentum transfers at
even the highest RHIC energies are significantly lower than $M_Z$,
so we are really talking about a running of the coupling with
temperature.  One must instead focus on
obtaining an expression for the coupling constant using expressions for the
propagators which have corrections due to the presence of a coloured medium \cite{muller:85}.
Using this modified propagator we again sum all diagrams with successive
\textit{medium} polarization functions $\Pi(q,T)$ to obtain an effective running coupling constant.
Leaving aside the contribution of quark loop diagrams, this yields \cite{muller:85}
\begin{equation}
\label{eq:qcdAlphaSofQT}
\alpha_s(q,T) = \frac{g^2}{4\pi(1-\Pi(q,T)/q^2)} = \frac{4\pi}{11\ln(-q^2/\Lambda^2) - 48 G(q/T)}
\end{equation}
where
\begin{eqnarray}
 G(\zeta)  &=&  \int_0^\infty z dz \frac{f(z)}{e^{\zeta z}-1}\ \ \mathrm{and} \nonumber  \\
 f(z)  &=&  (z-1/2z+1/8z^3)\ln\left(\frac{|1-2z|}{|1+2z|}\right)-1+1/2z^2 .
\label{eq:gzAuxAlphaS}
\end{eqnarray}
In the $q \rightarrow 0$ limit, the polarization function
remains finite, which means that the propagator effectively
contains a mass term, $D(q,T) = 1/(q^2-\Pi(q,T))$, in a manner analogous to Debye screening in
an electrolytic medium.  This leads to the property that
a test colour charge will cause a polarization of the charges in
the coloured medium in the same way as electric charges in an
electrolyte.  In addition, an important property that follows
 from the temperature dependence of Eq. \ref{eq:qcdAlphaSofQT} is
that as the temperature increases in QCD,  the coupling becomes
weak, falling logarithmically with increasing temperature.
As a consequence, nuclear matter at very high
temperature should not exhibit confinement.

Another important property that arises from the study of the QCD
Lagrangian is that of chiral symmetry breaking (\ie\ the quarks
confined in hadrons do not appear as nearly massless constituents,
but instead possess a mass of a few hundred \mev\ that is
generated dynamically).  The expectation value $\langle\bar\psi
\psi\rangle \approx - (235\ \mev)^3$, usually called the
\textit{quark condensate}, describes the density of $q\bar{q}$
pairs found in the QCD vacuum, and the fact that it is
non-vanishing is directly related to chiral-symmetry breaking.  In
the limit of zero current quark mass, the quark condensate
vanishes at high temperature, \ie\ chiral symmetry is
restored. It is this phase of QCD which exhibits
neither confinement nor chiral-symmetry breaking that we entitle the Quark-Gluon
Plasma.

Since the quark condensate is zero in the high temperature phase and non-zero
at low temperature, it therefore acts as an order
parameter.  This behaviour leads to the expectation that the
change between the low-temperature and high-temperature phases
should exhibit a discontinuity.  For 2 (3) massless quark flavours,
universality arguments predict a second- (first-) order phase transition
\cite{pisarski:84,rajagopal:93}.  The question of the order of
the transition, or if there is a phase transition as opposed to a rapid
cross-over, for QCD with the real values for the $u$, $d$ and $s$ quark masses is still
the subject of current investigations \cite{Karsch:2000kv,karsch:01}.

\subsection{Phase Diagram}

\begin{figure}[htb]
\begin{center}
\includegraphics[width=.6\textwidth]{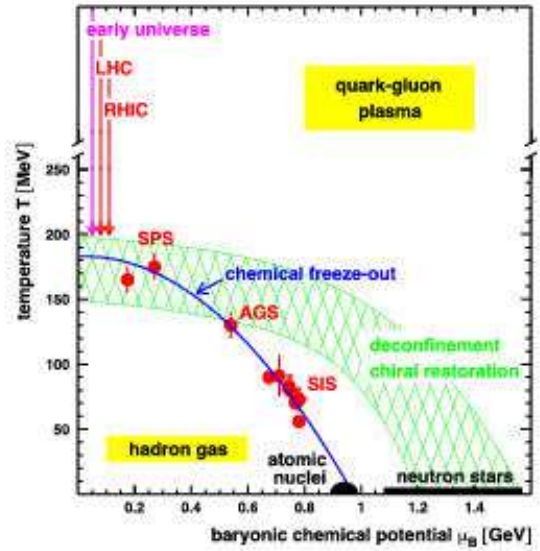}
\caption[Phase diagram of hadronic matter.]{Phase diagram of hadronic matter, showing hadron
gas and quark-gluon plasma regions. The temperature $T$
and the baryochemical potential $\mu_B$ data are derived from
particle yield ratios.  The solid curve through the data
points represents the chemical freeze-out of hadronic-matter.}
\label{fig:qcdPhaseDiagram}
\end{center}
\end{figure}

One of the first pictures that can serve as a guide to the
behaviour of QCD can be obtained from the simplified `MIT bag
model' \cite{cleymans:86}.  We can identify two regions where we
expect deconfinement to occur: when compression causes the hadrons
to overlap significantly, reaching densities 3 to 5 times higher
than those of ordinary nuclear matter (0.17 $\fm^{-3}$), or when
the temperature of the medium exceeds some critical threshold.
Figure \ref{fig:qcdPhaseDiagram} shows the usual representation of
the `$T - \mu_B$' phase diagram, where $\mu_B$ is the Baryon
chemical potential.  It is normally represented with a continuous
curve connecting the high temperature transition region at
$\mu_B=0$ with the high baryon density region at $T=0$.  The
figure shows the regions probed by the different beam energies
from the SIS to RHIC.  At high densities and near-zero temperature
we expect a deconfined phase.  This deconfined high-density phase is predicted to exist
for example in the interior of
neutron stars.  The near-zero Baryon chemical potential and high
temperature region is the one probed by the highest energy RHIC
collisions. The region where $\mu_B=0$ is
believed to be the one in which the early universe existed, and is also
accessible to numerical simulations of QCD on a lattice. The
data points in the figure are from an analysis of particle ratios
which is customarily employed to evaluate the degree of chemical
equilibration observed in the final state.  To roughly illustrate the region that is probed
by RHIC in the diagram, preliminary analysis from mid-rapidity particle ratios at RHIC
\cite{harris:01} indicate a
value of $\mu_B \simeq 0.050\ \gev$ and of the chemical freeze-out temperature of
$T_\mathrm{ch} \simeq 170\ \mev$. We discuss chemical
and kinetic freeze-out in Chapter~\ref{ch:expSearchQgp}. It should
be noted however, that we do not really know where the transition
curve really lies, or even if there is a transition as opposed to a
rapid cross-over at finite baryon number density.  First principle
numerical calculations of QCD, discussed in the next section, have
provided us with guidance, but most results so far pertain to the
$\mu_B=0$ region.

\section{Lattice QCD Results}
From sophisticated numerical simulations of QCD on the lattice, we
have gained much insight into the structure of the QCD at high
temperature.

The approach is based on the Feynman path integral: the aim is to
calculate the action on the lattice and use it to evaluate
expectation values of different observables.  The first results
used pure SU(3) gauge theory, sometimes called \textit{pure glue}
theory.  A problem arises when one introduces the (discrete)
fermion fields in a straightforward way, one obtains a `doubling'
of the flavour spectrum in the quark sector.  We get $2^d = 16$
copies of each quark species (in $d=4$ dimensions).  Different
approaches were then introduced to incorporate the fermion degrees
of freedom in a way that reduces the doubling problem.  The
approach proposed by Wilson \cite{wilson:74} solves the flavour
doubling essentially by giving the doublers a mass proportional to
1/$a$, where $a$ is the lattice spacing, so they go away in the
continuum limit. The disadvantage is that the mass introduces a
chiral-symmetry breaking piece into the action. Another action
known as the Kogut-Susskind\cite{Kogut:1975ag}, or also as the
`staggered' fermion action\cite{Susskind:1977jm,bernard:97}, has
the advantage of preserving a part of chiral symmetry. However, it
does not completely eliminate the flavour problem. The number of
doublers is reduced to 4. This leads for example to having
$4^2-1=15$ pions instead of the usual $2^2-1=3$.  In addition, the
Wilson action yields results that in general need corrections of
$\mathcal{O}(a)$, where $a$ is the lattice spacing, whereas the
staggered fermion action is fine up to $\mathcal{O}(a^2)$.

In the
last decade, there have been significant developments that provide
a solution to the doubling problem. The ``domain wall''
\cite{Furman:1995ky} approach relies on introducing a fifth
(fictitious) dimension such that the chiral zero modes live on 4D
surfaces.  If $N_5$ is the number of lattice spacings along the
fifth dimension, in the limit $N_5 \rightarrow \infty$ the left-
and right-handed fields live in surfaces in opposite ends of the
5D lattice and do not mix, so we have exact chiral symmetry with
no doublers. An alternative approach has been presented where
chiral symmetry is preserved in the lattice if the lattice Dirac
operator is of a certain form (see \eg\
Ref.~\cite{Narayanan:1993ss}). Although several constructions for
the Dirac ``overlap'' operator exist, they all satisfy an
identity, originally due to Ginsparg and Wilson
\cite{Ginsparg:1982bj}, which guarantees having exact global
chiral symmetries directly on the lattice. For a recent review of
the overlap approach, see \eg\
Refs.~\cite{Luscher:1998pq,Neuberger:1999ry}.

\subsection{Critical Temperature and Energy Density}\label{sec:CritTempEnergyDensity}

Over recent years, thermodynamic calculations on the lattice have
steadily been improved.  This is partly due to the much improved
computer resources, however equally important has been (and will
continue to be) the development of improved discretization
schemes, \ie\ improved actions.

Early calculations of the QCD transition temperature performed
with standard Wilson fermions \cite{wilson:74,bitar:91} and
staggered fermion actions \cite{bernard:97} led to significant
discrepancies of the results. These differences were greatly
reduced based on improved Wilson fermions (Clover action)
\cite{AliKhan:2000iz,bernard:97,edwards:99}, as well as improved
staggered fermions \cite{Orginos:1999cr,Karsch:2000kv}, and domain
wall \cite{Furman:1995ky} approaches.  We shall not
discuss these here, but only take the current results from
improved staggered fermion actions for the following discussion.

Typically, one can look at the dependence of the
energy density $\epsilon$ and pressure versus temperature where one
expects a change due to the increase in the number of degrees of freedom (\textit{d.o.f.}).
In order to illustrate this point, we can obtain some semi-quantitative
insight into the number of degrees of freedom and the energy density using
the following simplified scenario.
A massless non-interacting hadron gas is made up basically of pions, of which
we have 3 types ($\pi^+$, $\pi^-$, $\pi^0$) neglecting the resonances.  From
an ideal relativistic Bose gas at Temperature $T$ we obtain the energy density

\begin{equation}
\epsilon_{g} = \int\frac{d^3p}{(2\pi)^3}\frac{p}{e^{p/T}-1} 
             = \frac{4\pi T^4}{(2\pi)^3}\int_0^\infty \frac{x^3 \ dx}{e^x - 1} dx = \frac{\pi^2}{30} T^4 \ ,
\label{eq:energyDensityBose}
\end{equation}
where we have rescaled the momentum as $x = p/T$. This is the
usual Stefan-Boltzmann relation ($\epsilon = a T^4$). The energy
density for the hadron gas is therefore simply
\begin{equation}
\epsilon_{HG} = 3 \ \epsilon_{g} = 3 \ \frac{\pi^2}{30} T^4 \ .
\label{eq:energyDensityHg}
\end{equation}

For a QGP consisting of 2 massless quark flavours ($u$ and $d$) at
vanishing net baryon density ($\mu = 0$) we must sum the quark
and the gluon contributions to the energy density.  The gluon
contribution is given by Eq.~\ref{eq:energyDensityBose} times
the 16 gluonic \textit{d.o.f.} ($2_{\mathrm{spin}} \times (3^2 -1)_{\mathrm{colour}}$).
The quark contribution is obtained from a similar integral for a Fermi gas

\begin{equation}
\epsilon_{q} = \int\frac{d^3p}{(2\pi)^3}\frac{p}{e^{p/T}+1} 
               = \frac{4\pi T^4}{(2\pi)^3}\int_0^\infty \frac{x^3 \ dx}{e^x + 1} dx = \frac{7}{8}\frac{\pi^2}{30} T^4 \ .
\label{eq:energyDensityFermi}
\end{equation}
Multiplying by the number of (anti)quark \textit{d.o.f.}
($2_{\mathrm{spin}} \times 3_{\mathrm{colour}} \times N_F$) we
obtain the energy density

\begin{equation}
\epsilon_{QGP}= 16 \epsilon_{g} + 6 N_F(\epsilon_{q}+\epsilon_{\bar{q}})
              = 16 \frac{\pi^2}{30} T^4 + 12 N_F \frac{7}{8}\frac{\pi^2}{30} T^4 = (16 + \frac{21}{2}N_F) \frac{\pi^2}{30} T^4 \ .
\label{eq:energyDensityQGP}
\end{equation}

\begin{figure}[htb]
\begin{center}
\includegraphics[width=.6\textwidth]{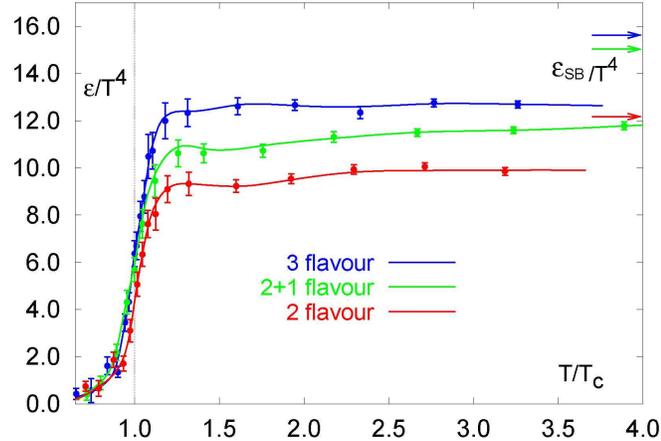}
\caption[Energy density from Lattice QCD]{Energy density $\epsilon$ obtained from a numerical
evaluation of QCD on the lattice \cite{Karsch:2000kv}.  $\epsilon$
is divided by $T^4$ to exhibit the sudden rise of the number of
thermally excited degrees of freedom near the critical temperature
$\Tc$. Arrows show the ideal gas values as given by
Eq.~\ref{eq:energyDensityQGP}.} \label{fig:qcdEnergyDensity}
\end{center}
\end{figure}

\begin{figure}[htb]
\includegraphics[width=1.\textwidth]{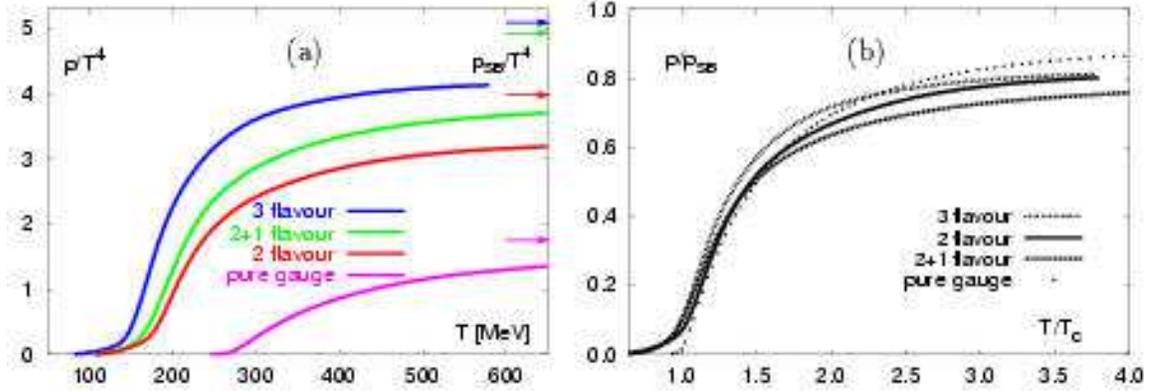}
\caption[Pressure in Lattice QCD \vs\ Temperature]{The Pressure $p$ in lattice QCD \cite{Karsch:2000kv}.
Left panel: Curves with $N_F = 0, 2 \ \mathrm{and}\ 3$ light
quarks as well as the ``(2+1)'' case. Right panel: $T$ axis scaled
by the value of $\Tc$ for each curve.} \label{fig:qcdPressure}
\end{figure}

Figure \ref{fig:qcdEnergyDensity} shows recent results for the
energy density $\epsilon$ as a function of temperature in lattice
QCD simulations with $N_F$ = 0, 2 and 3 light quarks as well as
two light and a heavier (strange) quark (2+1 flavour QCD). The
pressure $p$ is shown in Figure \ref{fig:qcdPressure}a for QCD
with different number of flavours as well as for the pure SU(3)
gauge theory. The curves clearly reflect the strong change in the
number of degrees of freedom when going through the transition. In
the high temperature limit ($T \gtrsim 1.5\ \Tc$), we expect both
$\epsilon$ and $p$ to asymptotically approach the Stefan-Boltzmann
free gas limit Eq.~\ref{eq:energyDensityQGP}, indicated by the
arrows in the figures.  From the figure, it is evident that even
at $4\Tc$ the Stefan-Boltzmann limit is not reached.  This has been taken as indication of
a significant amount of interactions among partons in the high
temperature phase, with only a logarithmic approach to the free-gas behaviour.

In addition,
the dependence of $\Tc$ on the number of partonic degrees of freedom is clearly
visible in Figure \ref{fig:qcdPressure}a.
It is therefore striking that $p/p_{SB}$ is almost flavour independent when
plotted in units of $T/\Tc$ as shown in Figure \ref{fig:qcdPressure}b.

The most recently reported results on the value of $\Tc$ from lattice calculations
are found to be, for one particular choice of actions (improved staggered fermion
action) \cite{karsch:01}

\begin{eqnarray}
 N_F  =  2 & : & \Tc = 173 \pm 8 \mev \ \nonumber  \\
 N_F  =  3 & : & \Tc = 154 \pm 8 \mev \ .
\label{eq:tc}
\end{eqnarray}
The results also suggest that the transition temperature in (2+1)-flavour
QCD is close to that of 2-flavour QCD.

The behaviour of the different actions (improved staggered,
Wilson, Clover, etc.) currently studied in lattice QCD should show
agreement in the vicinity of the phase transition.  In this
regime, the correlation lengths become large and cut-off effects
in the calculations become less important.  One can therefore
compare calculations made with different actions. In particular,
the recent results for the energy density at \Tc\ yield

\begin{equation}
\epsilon_c \simeq (6 \pm 2) \Tc^4  = 0.70 \pm 0.23 \ \gev/\fm^3.
\label{eq:critEnergyDensity}
\end{equation}
as shown in Figure \ref{fig:qcdEnergyDensity} by the vertical line at $T = \Tc$.
It is important to stress that these values refer to \textit{initial}
energy densities, when the medium exists at the early stages of QGP evolution,
and must be translated into a \textit{final} energy density that
can be measured in an experiment using detected particles.  Bjorken~\cite{bjorken:83}
introduced a relation to address this question.  One can experimentally
estimate the energy density
achieved in a nucleus-nucleus collision via the energy
measured in the central rapidity region, $d\Et/dy$, divided by the effective
interaction volume, ($A_{T} \cdot \tau$).  In this case, we estimate the
energy by a product of the particle multiplicity $(\dndy |_{y=0})$, times the mean
transverse mass of the particles, (\meanmt).

\begin{equation}
\epsilon = \frac{\frac{dN}{dy}\arrowvert_{y=0} \cdot \meanmt}{A_{T} \cdot \tau} \ .
\label{eq:bjEnergyDensity}
\end{equation}
In this expression,  the factor $A_{T}$ is the transverse area of
overlap in the collision.  The quantity $\tau$ is less accurately
defined.  It is normally interpreted as the parton
\textit{formation time}, \ie\ the time needed to pass from the
initial hadronic environment to the partonic degrees of freedom.
Usually, this time is taken as 1 $\fm/c$ at SPS in order to
compare different experiments. However, there is currently no real
consensus as to what is the appropriate formation time $\tau$ to
use at RHIC, although if anything there are arguments that it
should be smaller than 1 $\fm/c$ at high energies
~\cite{rischke:00} because it should take less time to equilibrate the
system. An estimate of the energy density using this
relation should then be at best considered a lower limit (with respect to
this formation time).

For reference, in the case of STAR, this relation can be calculated
for a given centrality selection, we choose the 5\% most central collisions,
using the total charged multiplicity and the transverse momentum distribution for
charged particles (Sec.~\ref{sec:eta}, and ~\ref{sec:pt} respectively).
The transverse area $A_{T}$ can be computed in
a geometrical model that reproduces the measured multiplicity
distribution (Sec. ~\ref{sec:glauber}).  Equation~\ref{eq:bjEnergyDensity}
is a very accessible experimental observable, but there are additional caveats
associated with its interpretation of which the formation time is but one example.
We discuss the applicability and additional uncertainties associated
with equation~\ref{eq:bjEnergyDensity} in Section~\ref{sec:eta}.

\subsection{Heavy Quark Effective Potential}\label{sec:latticeVeff}

In the lattice approach, an interesting quantity one can calculate is the
effective potential between two heavy quarks, for it is the example
which most cleanly illustrates
the modification of the behaviour of the coloured fields at high temperature.
From the previous
section, we delineated the behaviour of the strong coupling
constant: at large momenta or small distances it should be small
and at small momenta or large distances the coupling rises as
$1/q^2$ (asymptotic freedom).  This leads to an effective
potential that rises linearly with $r$, the distance between the
coloured constituents, and can be calculated in the lattice.

\begin{figure}[htb]
  \centering
  \mbox{\subfigure[]{\includegraphics[width=0.53\textwidth]{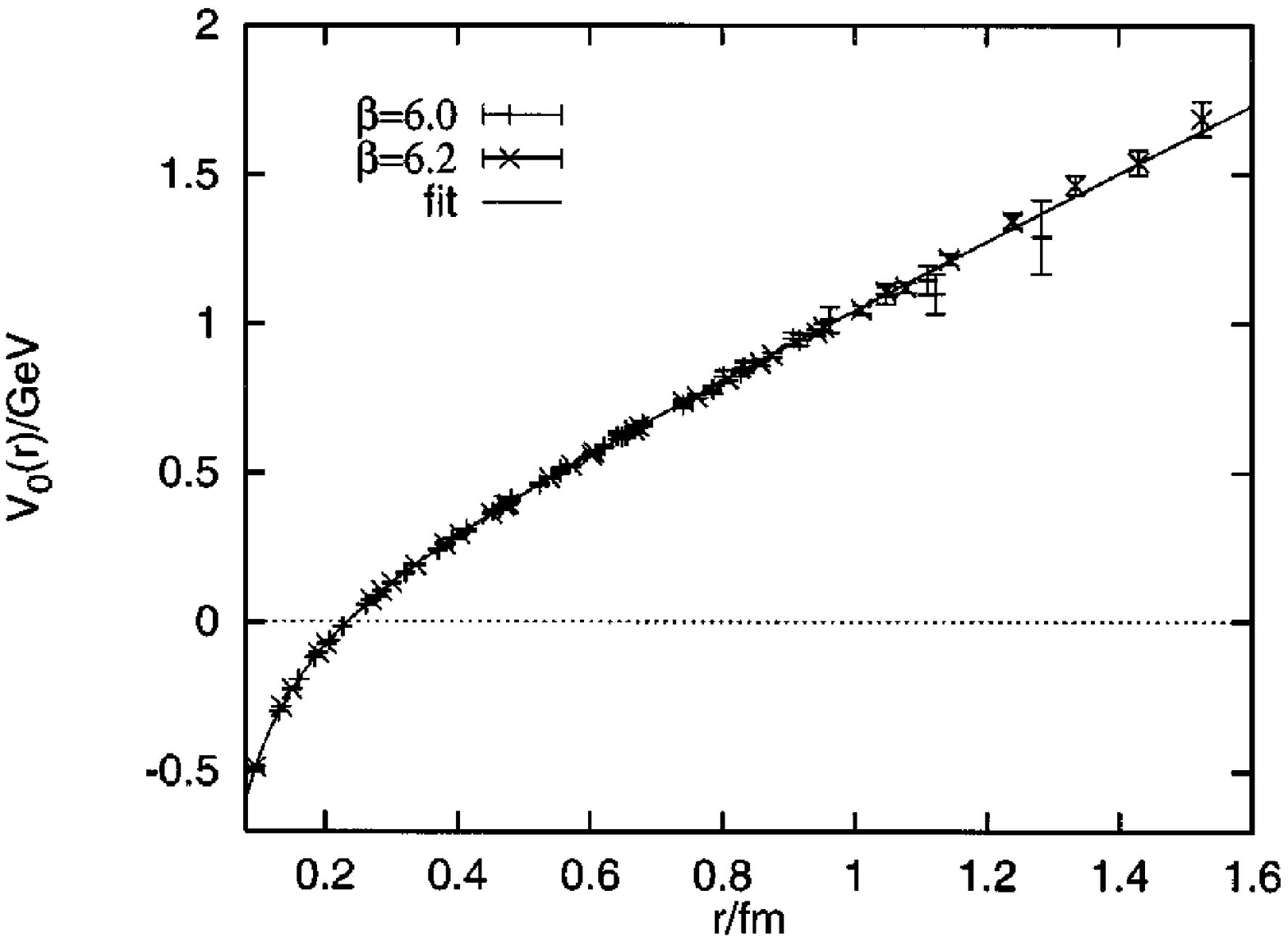}}\quad
    \subfigure[]{\includegraphics[width=0.42\textwidth]{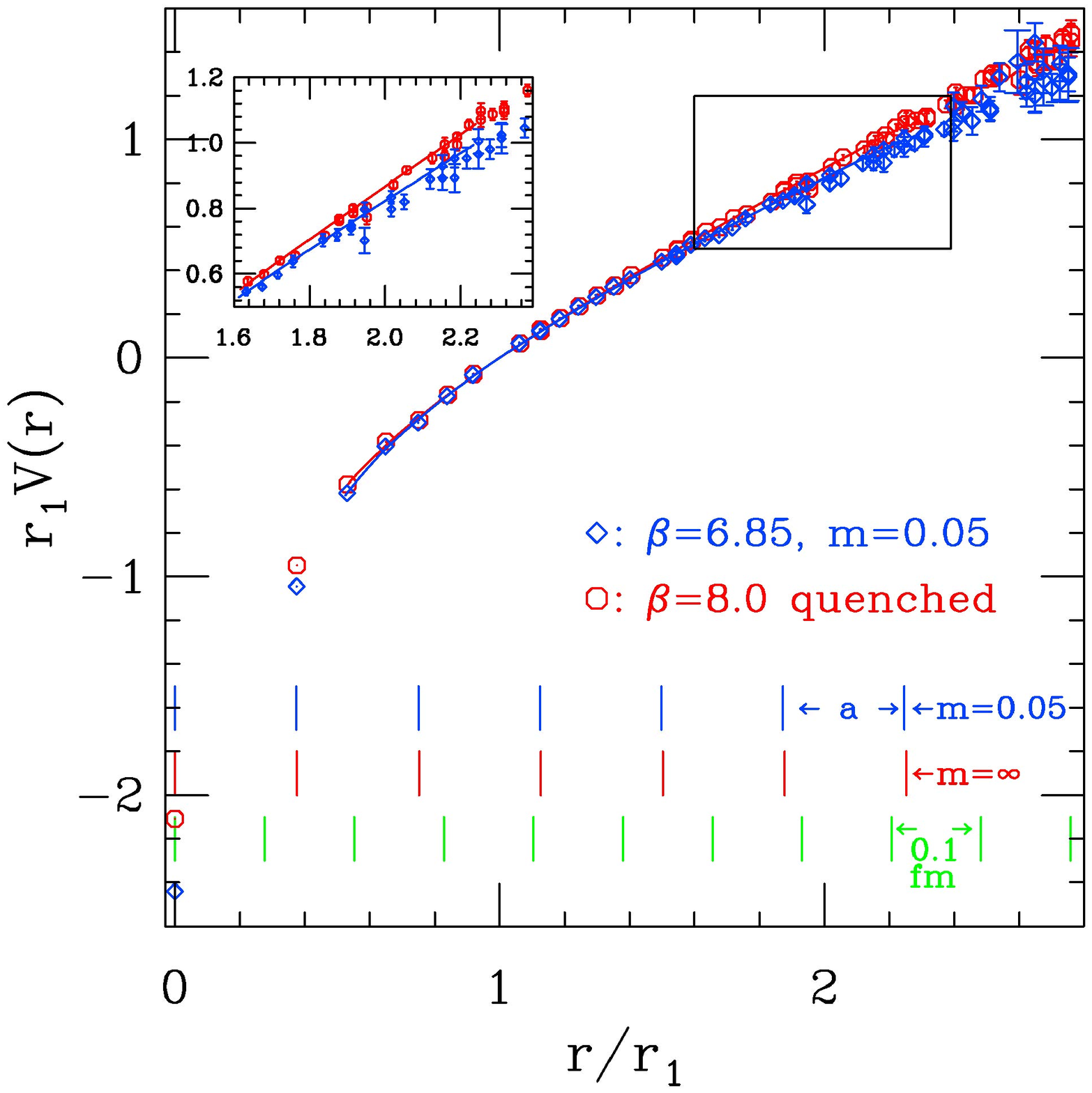}}}
  \caption[The heavy quark effective
potential as a function of distance $r$ in lattice QCD.]{The heavy quark effective
potential as a function of distance $r$ in lattice QCD.  The left panel shows
a quenched ($N_f=0$) calculation using Wilson loops, in physical units.  The right panel shows a
comparison between a quenched calculation (red) and one including quark loop effects (blue)
showing nice agreement.  Vertical bars near the horizontal axis show
the scale of the lattice spacing $a$, and intervals in physical
units 0.1 \fm\ wide for reference. Figures are from Refs.~\cite{bali:97} (a) and \cite{bernard:00}(b).
}\label{fig:qcdVeff}
\end{figure}

Figure ~\ref{fig:qcdVeff} shows the results of a lattice
calculation for the heavy quark potential as a function of
distance.
We see the potential is very weak at small distances and the
expected linear rise at large distances.  The slope of the linear
rise is usually called the \textit{string constant}, since it can
be thought of as the tension of a spring.

One can test the idea of deconfinement at high temperature in
lattice calculations in a straightforward way.  As we raise the
temperature, one can study the behaviour of the effective
potential.  We expect that near the critical temperature, the
linear piece of the potential should be modified and weakened.
At high temperature, the energy cost to create light $q\bar{q}$ pairs
from the vacuum is reduced (this is related to the vanishing
of the quark condensate and the restoration of chiral symmetry in
the $T>T_c$ phase of QCD).  These light $q\bar{q}$ pairs then act as screening colour
charges around the heavy quark pair, thus weakening the potential.

\begin{figure}[htb]
  \centering
  \includegraphics[width=0.8\textwidth]{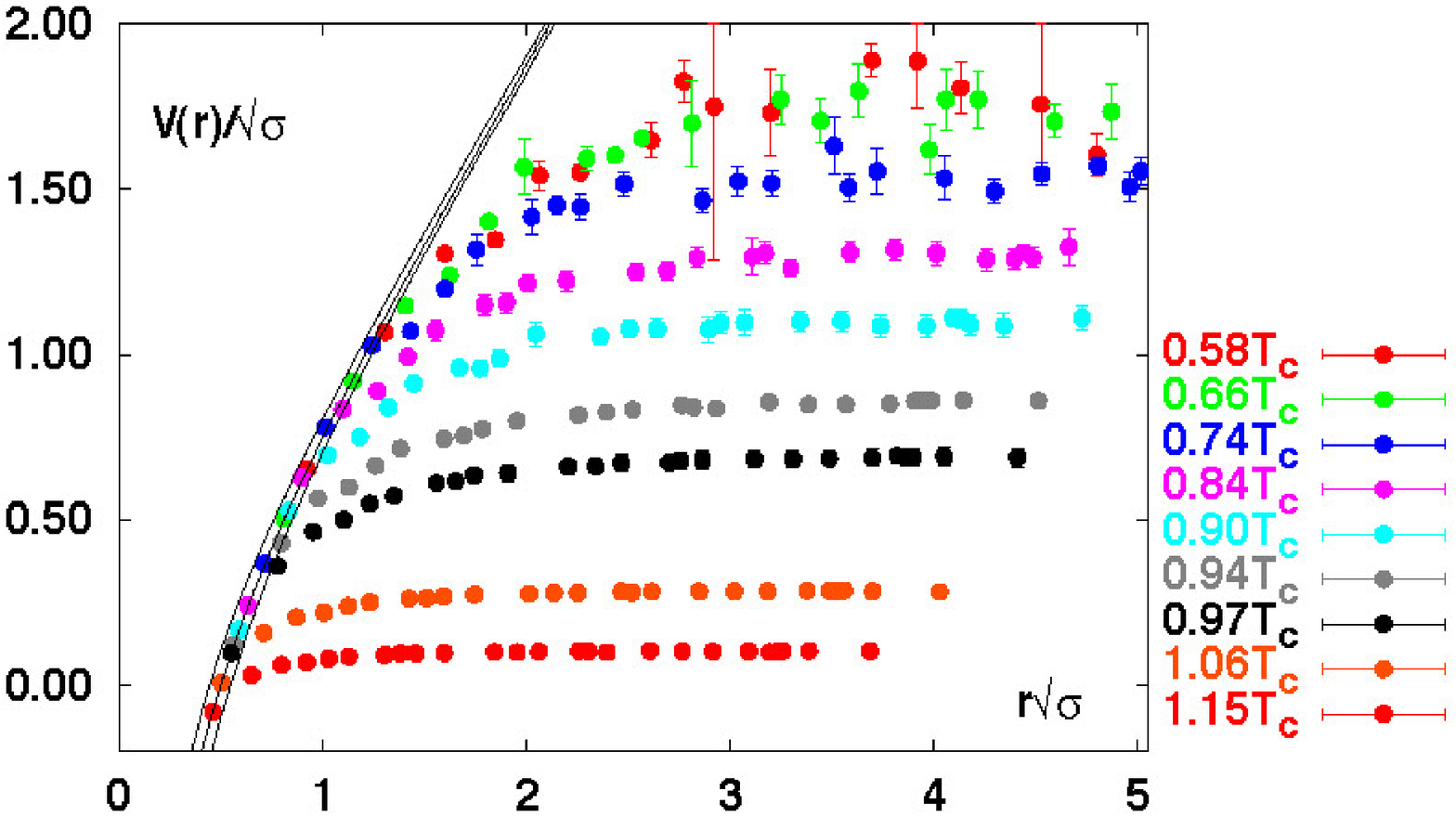}
  \caption[Heavy quark potential at different temperatures.]{The heavy
quark effective potential for different temperatures, taken from Ref.~\cite{Karsch:2000kv}.
The linear
rise of the potential is weakened as one approaches the critical
temperature.  The solid curves show the Cornell potential $V(r) = -\alpha/r + \sigma r$
with $\alpha = 0.25 \pm 0.05$, which is used to normalize the finite temperature
free energies at the shortest distance available, $r=1/(4T)$.}
\label{fig:qcdVeffvsT}
\end{figure}

Figure ~\ref{fig:qcdVeffvsT} shows the heavy quark potential for
different temperatures in the region near $T_c$.
It is clear that there are important modifications to the strength
of the potential at high temperatures.  We see that as the
temperature increases, the strength systematically decreases.  The
potentials in Figure ~\ref{fig:qcdVeffvsT} also do not show the
steep rise as in the quenched case.  This is an expected
phenomenon that had proved elusive in lattice calculations.  As
the quark-antiquark pair separate, we expect the formation of a
tube of flux, or string, which should break in the presence of
light quark-antiquark pairs.  The results in the figure show that
the string does in fact break in the confining phase at non-zero
temperature. The weakening of the potential at high temperature
has important consequences for heavy quark bound states.  In
particular, it is this behaviour that is the basis for the concept of
J/$\psi$ suppression in a deconfined medium ~\cite{matsui:86}.

We next need to address how we can measure the properties of
excited hadronic and partonic matter in the laboratory and how can
we hope to see a signal of QGP formation. We therefore proceed to
discuss the experimental side of QGP physics.

%
%

\chapter{Experimental Search for the QGP}\label{ch:expSearchQgp}

The use of beams of heavy elements allows us to distribute the
incoming energy over an extended region of space, large compared
to the size of one nucleon, with the hope of creating conditions
that are suitable for the formation of a QGP. One can identify the
following phases, shown schematically in
Fig.~\ref{fig:spacetime1d}:
\begin{figure}[htb]
  \centering
  \mbox{
  \subfigure{\includegraphics[width=.17\textwidth]{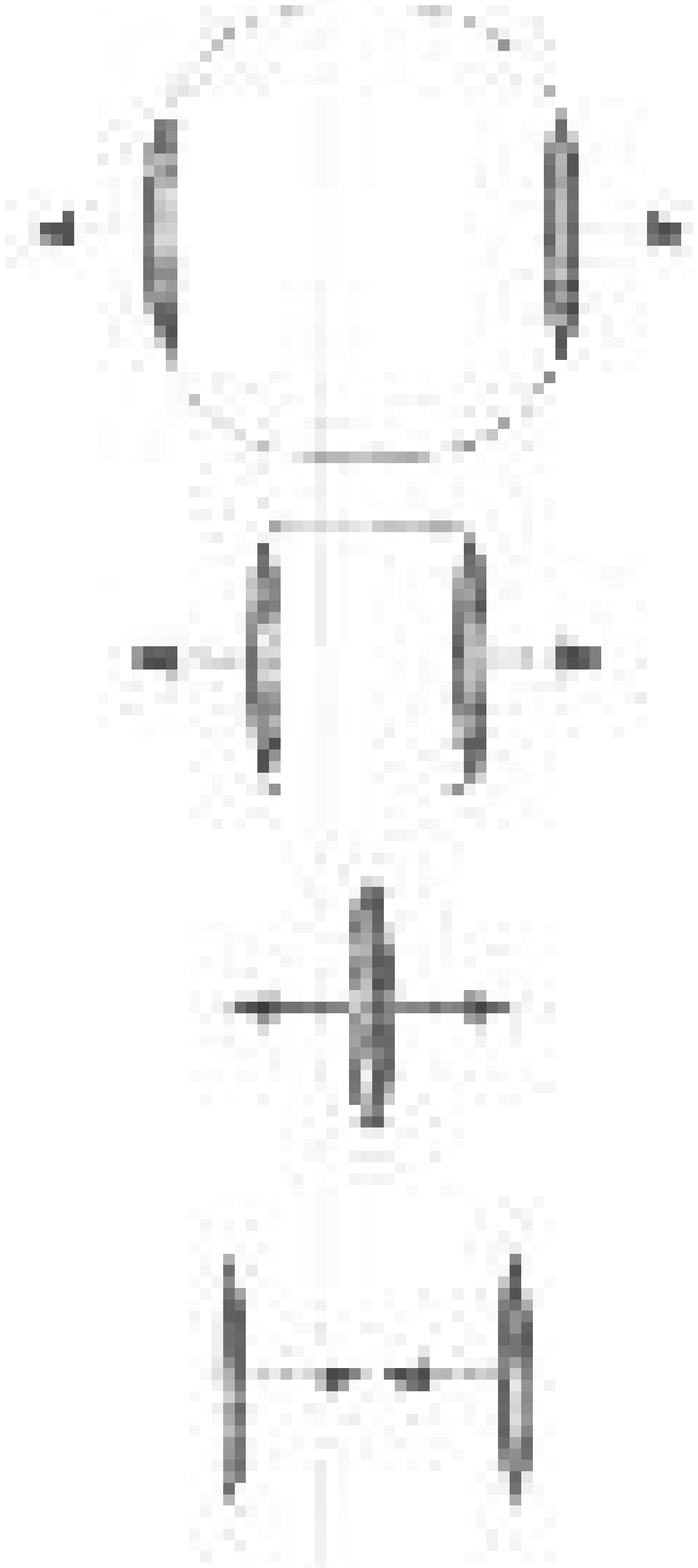}} \quad
  \subfigure{\includegraphics[width=.6\textwidth]{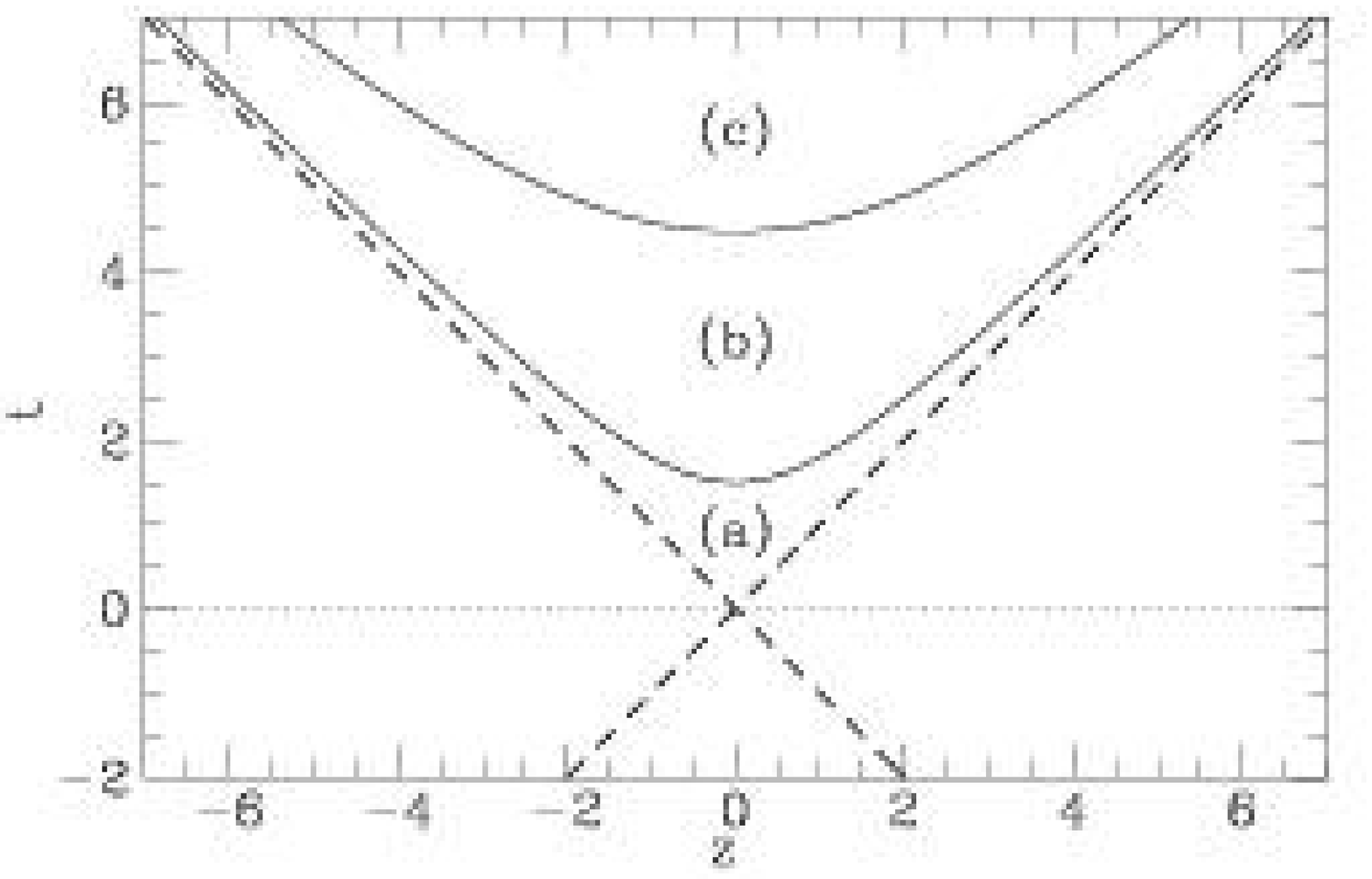}}}
  \caption[Schematic 1-D space-time evolution of a heavy ion collision]
  {1-D Space-time picture of the evolution of an ultra-relativistic nuclear
  collision, distinguishing 3 regions: (a) Pre-equilibrium state,
  (b) thermalized QGP, (c) hadronization stage. (see text).  The
  $z$ axis denotes the beam direction, and the dashed lines
  indicate the colliding nuclei near the light-cone region.
  }\label{fig:spacetime1d}
\end{figure}

\textit{i)} The interpenetration of the nuclei with partonic
interactions at high energy.  This stage features the creation of
\eg\ high-\pt\ jets, $c\bar{c}$ pairs or other products of high
momentum transfer scattering processes on the parton level.  In
addition, large cross-section soft nucleon nucleon scattering
between the two highly Lorentz-contracted nuclei help redistribute
a fraction of the incoming kinetic energy into other degrees of
freedom. The small cross-section hard processes are used as
experimental probes for the hot and dense zone.

\textit{ii)} The interaction of the particles of the system,
driving it towards chemical and thermal equilibrium. Partons
materialize out of the highly excited QCD field.  If QGP forms,
the quark mean free path at energy densities $\epsilon = 2\
\gev/\fm^3$ is $\lambda \approx 0.5\ \fm$, and individual
parton-parton scattering is expected to play a role in
thermalizing the system during this early stage as the nuclear
dimensions are larger than $\lambda$. The interactions also
originate the development of collective flow.  Rapid expansion,
mainly along the longitudinal direction, lower the temperature of
the system eventually reaching the cross over temperature $\Tc$.
Direct photon signals from the QGP are generated from collisions
of charged particles during the expansion stage, although there
are also photon signals from the hot hadron gas stage. Final
formation of charmonium states (J/$\psi$, $\psi\prime$) from the
initial $c\bar{c}$ pairs happens also during this stage.

\textit{iii)} The hadronization stage and the `freeze-out' of the
final state particles is then reached at $T<\Tc$, as the system
cools down so that there is not enough energy in each collision to
further change the different species' populations or
\textit{ratios} (\textit{chemical} freeze-out, $T \sim 160-170\
\mev$). Eventually, the energy is small enough and the system
dilute enough such that the interactions cease and the
\textit{momentum spectra} do not change further (\textit{kinetic}
freeze-out, $T \sim 120\ \mev$). To obtain information of the
different stages, experimentally one must start from the
measurement of final state particles.  Global observables are
useful to determine the initial conditions such as centrality,
initial volume, and possibly energy density.

There are some observables that also provide information from the
early stages. Signals such as direct photons and dilepton pairs
that originate during early times are interesting since they are
little disturbed by the hadronic final state.  The caveat in
studying such signals is that they typically have a much smaller
cross section compared to hadronic observables.

There are also a class of hadronic observables that are thought to
be sensitive to early times.  In particular, the measurement of an
azimuthal anisotropy in the emission of particles (with respect to
the \textit{reaction plane}, \ie\ the plane formed by the beam
direction and the direction of a vector connecting the center of
the two colliding nuclei) is one example. In non-central
collisions, an initial spatial anisotropy would result in pressure
gradients which drive the emission of particles, producing a
modulation in the azimuthal distribution of particles with respect
to the reaction plane. This effect is typically measured by a
Fourier analysis of the azimuthal particle distribution, where the
2nd Fourier component is called \textit{elliptic flow} (in
reference to the picture of particles in a fluid moving under the
influence of the initial pressure gradient) and is denoted as
$v_2$.  The STAR experiment \cite{Ackermann:2000tr} already
measured the charged particle elliptic flow signal at low \pt, and
from these results we already see evidence of significant
collective behaviour during the early stages of collisions at RHIC
energies.

\section{QGP Signatures}

Many different effects have been proposed as possible ways to
detect experimentally the formation of the QGP state of matter.
These range from the earliest and most na\"{\i}ve studies which
involved plotting mean \pt\ (\meanpt) as a function of particle
multiplicity as a quick way to look into the structure of the $T -
\epsilon$ phase diagram \cite{vanHove:82}, to the more current
searches for modification of particle properties (enhancement,
suppression or medium-induced changes in mass or width), and the
statistical studies of charge (or other observable) fluctuations
event-by-event. While some probes give information primarily about
`surface' effects, like the hadrons' \pt\ distributions discussed
here, strangeness production and particle interferometry which
reveal final state information; some others deal with deeper
`volume' effects which are sensitive to early times after the
collision. These include most `hard' processes, \ie\ those which
involve large momentum transfers, for example charm and beauty
production ($J/\psi$, D mesons, $\Upsilon$), jets and high-\pt\
particle production. Usually, the attention is on `volume' probes
that address the features of the QGP itself, \ie\ those probes
which provide more direct information from the hot and dense phase
of the reaction and are not influenced much by the high number of
hadrons produced in the collision (\eg\ by hadronic rescattering).
This is the ideal case, but the experience has been that one must
carefully study the modification of the proposed signals by
conventional nuclear means (\ie\ non-QGP).  It is also important
to emphasize that all of the different clues must be investigated
as a function of the associated particle multiplicity or
equivalent probe giving information on global characteristics. One
would like to understand the onset of all of the observed signals
in terms of the different handles at our disposal. Experimentally
we can vary the collision centrality (selecting on multiplicity or
transverse energy), we can collide different nuclear species (a
more controlled variation of system size) and we can vary the
centre-of-mass energy.  The experimental data will in turn help to
constrain the theoretical model parameters and input (\eg\
equation of state, expansion dynamics and collective flow, size
and lifetime of the system for hydrodynamical models; and initial
parton densities, parton mean-free path and cross section, nuclear
shadowing of initial parton distributions, and amount of parton
energy loss in the plasma for perturbative approaches). For
example, using hadron and electromagnetic spectra from several SPS
experiments, attempts have been made to constrain the equation of
state and initial conditions in a hydrodynamical approach
\cite{sollfrank:97}.

In the following we will briefly mention some of the experimental
signals that have been proposed to probe the system created in
heavy ion collisions. For a recent review on QGP signatures, see
Ref.~\cite{bass:99}.

\subsection{Charmonium Suppression}

The arguments for charmonium suppression were laid out in Section
\ref{sec:latticeVeff}.  Essentially, the weakening of the heavy
quark effective potential with increasing temperature, or
alternately viewed as the Debye screening of free colour charges
in a QGP, is responsible for the breakup of charmonium states
\cite{matsui:86}. Excited states of the $c\bar{c}$ system, such as
$\psi\prime$ and $\chi_c$, are easier to dissociate due to their
larger radii, and are expected to dissolve just above $\Tc$.  The
smaller J/$\psi$ becomes unbound at a higher temperature, $T
\gtrsim 1.2 \Tc$.  Similar arguments apply to the dissociation of
the heavier $b\bar{b}$ bound states, but they require much shorter
screening lengths to dissolve, \ie\ greater temperature and energy
densities \cite{Karsch:1991wi}.  The $\Upsilon$ state may dissolve
above a temperature of $\sim 2.5 \Tc$, while the $\Upsilon'$ could
also dissolve near \Tc.

``Anomalous'' J/$\psi$ suppression has been reported by the NA50
collaboration for central \PbPb\ collisions at SPS which has been
heralded as evidence for QGP formation \cite{Gonin:1996wn,
Abreu:1999hx, Abreu:2000ni}.  There are nuclear effects, such as
the breakup of the J/$\psi$ by hadronic comovers, which also
suppress the measured cross section in nucleus-nucleus
collisions~\cite{Capella:2000zp}.  This issue is still one of
intense debate, with many journal publications devoted to the
topic.  There are already some model predictions on J/$\psi$
suppression at RHIC and LHC energies based on normal nuclear
effects, such as \pt\ broadening and nuclear shadowing of the
parton distribution functions \cite{Gerland:2000vj}.  Such an
analysis lead for example to \textit{anti-shadowing} at RHIC
energies, \ie\ the modifications introduced by cold nuclear matter
enhance quarkonium production at RHIC (at LHC the opposite
occurs).  It has become clear that several mechanisms need to be
considered in order to fully comprehend quarkonium production and
suppression: the underlying physics description at different
energies (hadronic at AGS and SPS, partonic at collider energies),
nuclear effects such as shadowing, interaction with comovers,
J/$\psi$ production due to the decay of higher mass resonances,
and the signals from \pA\ data.  For the most recent developments,
one can see for example the talks in the dedicated J/$\psi$
session at Quark Matter '01~\cite{qm:01}, where suppression from
the comover mechanism \cite{Capella:2000zp}, a model based on \Et\
fluctuations for the most central events to explain the NA50
\PbPb\ data \cite{Blaizot:2000ev}, and even enhancement of
J/$\psi$ production in deconfined matter at RHIC energies
\cite{Thews:2000rj} were discussed.

\subsection{Jet Quenching}

The colour structure of QCD matter can be probed by its effect on
the propagation of a fast parton.  The mechanisms are similar to
those responsible for the electromagnetic energy loss of a fast
charged particle in matter: energy may be lost either by the
excitation of the penetrated medium or by radiation . The QCD
analog of this effect indicates that the stopping power of the
Quark-Gluon Plasma should be higher than that of ordinary matter
\cite{Gyulassy:1994hr}. This effect is called \textit{jet
quenching} \cite{jetq:92,jetq:98} and has several consequences
that could be observable in experiments.  Of more direct relevance
to particle spectra, a comparison of the transverse momentum
spectrum of hadrons compared to appropriately scaled distributions
from \pp\ or \ppbar\ collisions should show a suppression at
high-\pt\ ($\gtrsim 4$ \gevc).  In addition, a quark or gluon jet
propagating through a dense medium will not only lose energy, it
will also be deflected. This will destroy the coplanarity of the
two jets with respect to the beam axis. The angular deflection in
addition leads to an azimuthal asymmetry.  One can then perform
angular correlations among high \pt\ particles to study the energy
loss effects of the partons in the medium.

\subsection{Medium Effects on Hadron Properties}
The widths and masses of the $\rho$, $\omega$ and $\phi$
resonances in the dilepton pair invariant mass spectrum are
sensitive to medium-induced changes, especially to possible drop
of vector meson masses preceding the chiral symmetry restoration
transition.  The CERES data from S+Au and Pb+Au collisions at SPS
showed an excess of dileptons in the low-mass region 0.2 < $M$ <
1.5 \gevcc, relative to \pp\ and \pA\
collisions~\cite{Ullrich:1996wt,Drees:1996wf}, which has been the
subject of active discussion.  Although the CERES data can be
explained by a hydrodynamic approach assuming the creation of a
QGP~\cite{Srivastava:1996wr}, alternative scenarios have also
provided explanations. These have included for example microscopic
hadronic transport models incorporating mass shifts of vector
mesons, and calculations involving in-medium spectral functions
(coupling the $\rho$ with nucleon resonances) without requiring a
shift in the $\rho$ mass \cite{Rapp:1997fs}.  With the addition of
a TPC to the CERES experiment~\cite{Agakishiev:1999ka}, the
resulting increase in resolution (and statistics) should help
verify or falsify some of the conflicting hypotheses on the origin
of the low-mass enhancement in the dilepton spectrum.

\subsection{Direct Photons and Thermal Dileptons}

The detection of radiation from a high temperature QGP would be an
ideal signal to detect, as black body radiation is one of the most
directly accessible probes of the temperature of a given system.
In the quark-gluon phase, the gluon-photon Compton process $gq
\rightarrow \gamma q$ is the most prominent process for the
creation of direct (thermal) photons (with additional
contributions from the $q\bar{q} \rightarrow \gamma g$
annihilation process).  Unfortunately, a thermal hadron gas with
the Compton scattering reaction $\pi \rho \rightarrow \gamma \rho$
(and pion annihilation $\pi \pi \rightarrow \gamma \rho$) has been
shown to `shine' as brightly as a QGP (or even brighter still)
\cite{Kapusta:1991qp}.  However, a clear signal of photons from a
very hot QGP possibly formed at RHIC could be visible at
transverse momenta in the range 2--5 \gevc
\cite{Strickland:1994rf,Srivastava:1992dm,Chakrabarty:1992pn}.
However, it is also possible that flow effects can prevent a
direct identification of the temperature and the slope of the \pt\
distribution.  WA98 has observed a direct photon signal \PbPb\
collisions at SPS \cite{Aggarwal:2000th}.  Comparing the results
to \pA\ data, they observe an enhancement for central collisions,
suggesting a modification of the photon production mechanism.

In addition, dileptons can also carry similar information as
photons on the thermodynamic state of the medium (\textit{thermal
dileptons}). Since dileptons interact only electromagnetically,
they can also leave the hot and dense reaction zone basically
unperturbed.  The difficulty of this type of signal is that one
does not have a significant feature, such as a mass peak.  One has
to analyze a spectrum which is a convolution of several
complicated backgrounds on top of the (small cross-section)
signal.  At CERN-SPS, the expectation is that the contribution of
hadronic backgrounds to the dilepton spectrum will dominate over
the QGP radiation. The main backgrounds are, at low masses: pion
annihilation, resonance decays, $\pi-\rho$ interactions. At high
masses, the Drell-Yan process dominates at SPS. At RHIC energies
there is an additional charm contribution above 2 \gevcc. There is
only a small window, 1 < $M$ < 1.5 where the rates for a plasma
(at very high temperatures, $T \approx 500$ \mev) may be dominant.
This signature has proved to be a difficult experimental
observable, but there is a continued effort to improve the
sensitivity of the measurements: a study of the \pt\ dependence of
various mass windows might perhaps help to disentangle the
different contributions to the spectrum.

\subsection{Strangeness Enhancement}

In hadronic reactions, the production of particles containing
strange quarks is normally suppressed due to the high mass of the
$s$-quark ($m_s \simeq 60-170$ \mevcc) compared to $u$ and $d$
masses. In the presence of a QGP, the temperature of the order of
the $s$-quark mass and the rapid filling of the phase space
available for $u$ and $d$ quarks should favor the production of
$s\bar{s}$ pairs in interactions of two gluons
\cite{Rafelski:1982ii,Koch:1986ud}. This should be reflected in an
enhancement of the production of multi-strange baryons and strange
antibaryons if a QGP is formed compared as compared to a purely
hadronic scenario at the same temperature.  Important observables
in this respect are the yields and ratios of strange hadrons
(mesons, strange and multi-strange baryons and their
antiparticles) which allow the determination of the relative
strangeness equilibrium.  To account for incomplete chemical
equilibration, a strangeness fugacity $\gamma_s$ is introduced in
a thermochemical approach.  The particle ratios can be calculated
assuming either a hadron gas scenario or a QGP and a comparison
can be made of the values thus extracted in conjunction with other
model parameters such as $T$ and $\mu_B$.

Because strange hadrons interact strongly, their final-state
interactions must be modelled in detail before predictions and
comparisons of strange particle yields can be done.  It is also
important to stress that an understanding of the enhancement
mechanism present in \pA\ collisions is crucial in order to
interpret the signals in \NN\ collisions.

STAR is currently addressing several of these topics.  The large
acceptance of the detector coupled with precise tracking allows
for the reconstruction of the decays of strange
particles~\cite{caines:01}. Studies of high-\pt\ hadron
spectra~\cite{dunlop:01} and angular
correlations~\cite{Ackermann:2000tr} are well suited for STAR as
the detector has full azimuthal coverage. Additional detector
components for future runs, specifically the completion of a
barrel electro-magnetic calorimeter, will permit studies of
dilepton production and J/$\psi$ suppression.

In all cases, the specific observables that are expected to be
sensitive to deconfinement have to be correlated with the global
characteristics of the collision in order to better understand
their systematics.  For example, critical in the debate of the
$J/\psi$ results has been the dependence of the effect on the
measured transverse energy (\Et) of the collision, an observable
similar to multiplicity in that it is correlated with the
collision geometry, and the determination of the \textit{number of
participants} (see Sec. \ref{sec:glauber}) from the measured \Et.
It is therefore essential to understand the global observables
involved in the systematics of any QGP signature. In the following
chapter, we therefore turn our attention to these global hadronic
observables, and their relationship to the collision geometry.

%
%

\chapter{Global Observables and Charged Hadron Spectra}
\label{ch:spectraReview}

The search for the new state of matter has not been an easy one.
There appears to be no simple and unambiguous experimental
signature of plasma formation, and one of the main lessons we have
learned in this field is that an understanding of QGP formation
and an elucidation of its properties will only come about through
systematic studies. It is necessary to measure nucleus-nucleus
(\NN) collisions at various centre-of-mass energies, to use
different beam species, and to make comparisons with reference
data. These comparisons are especially important since from
proton-proton (\pp) collisions one can measure basic processes in
a cleaner environment, and from proton-nucleus (\pA) collisions
one gains insight into the modification of the basic processes by
the presence of normal nuclear matter.  These are required to
understand any signal in \NN\ collisions. For a recent review on
hadronic particle production in nucleus-nucleus collisions from
SIS to SPS energies, see \cite{stroebele:99}.

Through global distributions one gains insight into the `kinetic
freeze-out' stage of the system produced in the collision when
hadrons no longer interact and their momenta no longer change.
These final-state measures supply information that constrain the
possible evolutionary paths of the system and can help establish
conditions in the early, hot and dense phase of the collision.

Typically, the first information studied in heavy ion collisions
comes from the observed particle distributions, both the
transverse momentum (\pt) and rapidity ($y$) distributions. The
spectral shapes are intimately related to the underlying collision
dynamics.  The expected behaviour of these distributions from
scenarios consistent with a phase transition can be tested. While
these may not be sufficient to completely answer the question as
to whether a quark-gluon plasma is found, they are most certainly
a necessary first step to provide consistency with any given
scenario, be it QGP or other.  The studies of the proposed QGP
signatures so far, however, have not provided unambiguous evidence
for quark-gluon plasma formation. It is the general belief that
the proposed experimental programme of high energy heavy-ion
collisions at the Relativistic Heavy Ion Collider (RHIC) at
Brookhaven National Lab will yield key pieces of the puzzle. In
the analysis presented here, we focus on charged particle
distributions from the first collisions measured by the Solenoidal
Tracker at RHIC (STAR) experiment.  These measurements will serve
as a baseline for studies of QGP signatures and as a guidance for
theoretical models. We now give some background for the
distributions that we will present here.

\section{Particle Multiplicity}
The negatively charged hadron (\hminus) particle multiplicity
distribution ($\Nhminus$) yields information on both the impact
parameter and energy density of the collision. It is not possible
to directly measure the impact parameter of the collision, so one
must use an indirect measure. The event multiplicity is one of the
observables that is correlated to the impact parameter.  The idea
is simple.  Each of the nucleons in the nuclei that participate in
the collision produces (on average) a certain number of particles.
We can calculate in a geometrical model the average number of
nucleons that participate  in the collision (\npart) at a given
impact parameter $b$.  We can thus obtain a statistical mapping of
$\langle\Nhminus\rangle \rightarrow \langle\npart\rangle
\rightarrow \langle b \rangle$.  The \textit{number of
participants} (\npart) is also called \textit{number of wounded
nucleons} \cite{Bialas:1976ed}, and we will use them
interchangeably here. The scaling of the multiplicity with the
number of participants is typically thought of as a reflection of
the particle production due to low momentum transfer
(\textit{soft}) processes. There are refinements to this model. At
high energy, it is expected that there will be an increased
particle production from large momentum transfer (\textit{hard})
processes. Hard process cross sections in \pA\ collisions, \eg\
the \pt\ distributions at very high \pt, are found to be
proportional to the number of elementary nucleon-nucleon
collisions, which we call the \textit{number of binary collisions}
(\ncoll). Some recent models~\cite{dima:00} include for example
the assumption that the particle production is derived from a
linear combination of the soft and the hard processes, \ie\ a
linear combination of \npart\ and \ncoll.

To make the distinction between the two quantities \npart\ and
\ncoll\ clear, since the two are related and sometimes lead to
confusion, we describe the concepts in more detail. \npart\ refers
to the number of nucleons that were hit, or that interacted in
some sense, which is why they are also sometimes called
\textit{wounded nucleons}. For a head-on ($b=0$) \AuAu\ collision
assuming the nucleus to be a hard sphere, or rather, to be a bag
filled with hard spheres, we then find simply $\npart = 197+197 =
394$.  There will be deviations from this, as one introduces a
more realistic density profile for the Au nucleus.  In addition,
the distribution of nucleons in the nucleus is not always the
same, there are volume fluctuations and the nuclei have Fermi
motion.  This is typically introduced in the models by adding a
parameter to represent the size of the fluctuations. It is also
common to find in the literature references to the \textit{number
of participant pairs} and comparisons made this way. If one
normalizes the particle production ``per participant pair'', it is
straightforward to compare to $pp$ or \ppbar\ data which can be
thought of as the limit of 1 participant pair. Again, for a
head-on \AuAu\ collision assuming hard spheres, the number of
participant pairs is just $\npart/2 = 197$.

\ncoll\ refers to the number of elementary nucleon-nucleon
collisions.  It includes all participating nucleons (\ie\ $\ncoll
\geq \npart$).  The difference can be thought of in the following
simple picture.  Let us follow one particular nucleon through the
collision as if it were a billiard ball and do its accounting. If
it does not interact at all, it does not count for either \npart\
or \ncoll\ purposes, and we call it a \textit{spectator} nucleon.
If it interacts, then we count it once for \npart\ purposes and
that is the end of our \npart\ accounting using this nucleon. We
of course also count it for the purposes of \ncoll\ at this point.
The difference is that there are can still be other nucleons in
its path, and if it interacts \emph{again}, we increase our
\ncoll\ counter. We do this for every time our original nucleon
collides, whereas our \npart\ counter remains at 1.  Since each
nucleon in a nucleus can interact many times as it ``punches
through'' the other nucleus, it is evident that $\ncoll \geq
\npart$.  The corresponding simple limit for \ncoll\ for the hard
sphere case is $\ncoll \propto \npart^{4/3}$.  This can be seen
from the following argument.  It is easy to see that \npart\ is
proportional to A (\eg\ $\npart\ = 2A$ for central collisions).
Every nucleon that counts for \npart\ must also count for \ncoll.
The number of additional collisions that count for \ncoll\ have to
do with the number of additional nucleons (from the target
nucleus) that lie in the path of our original nucleon (from the
projectile nucleus). As seen from the target nucleus, the path of
the projectile nucleon is a straight line parallel to the beam
axis. A nucleon that punches through the center of the target
nucleus will cross a length $2R$ of the target nucleus, where $R$
is the nuclear radius.  If the nucleus is not in the center, the
length will still be proportional to $R$.  For the hard sphere,
the nucleon density is constant and therefore the additional
collisions that sum up to \ncoll\ are then proportional to $R$.
Since the nuclear radius is proportional to $A^{1/3}$, we arrive
at $\ncoll \propto A \cdot A^{1/3} = A^{4/3} \propto
\npart^{4/3}$.

It is important to stress the limitations of our mapping from
\Nhminus\ to impact parameter.  First of all, it relies on the
accuracy of the simplified model relationship between the number
of participants and particle production.  In addition, the
relationship is statistical only, \ie\ we cannot experimentally
measure the impact parameter of a given event.  We can only
measure the multiplicity of each event, and for a given ensemble
of events compute the mean multiplicity. Then we can relate this
to an \emph{average} number of participants, since for every
event, even if we keep the number of participants fixed, there
will still be multiplicity fluctuations. Furthermore, even keeping
the impact parameter fixed does not fix the number of participants
either, since we expect fluctuations in the initial configuration
of the nucleons in the nucleus for every event due to Fermi motion
as well as small variations in the size of the nucleus.  We
therefore expect that such a statistical map will probably work
best for the central collisions (\ie\ we know that the
highest-multiplicity events \emph{must} come from central
collisions) and be less reliable for peripheral collisions where
fluctuations will dominate.

The multiplicity distribution \Nhminus\ belongs to the most global
class of observables (or rather, of the most \textit{integral}
sort of observables, as we measure the cross section integrated
over azimuth, over \pt\ and for a wide pseudorapidity slice). The
information that can be derived from the multiplicity distribution
is basically related to whether there are significant deviations
from the simple geometrical picture of the collision, commonly
referred to as the Glauber model (see \eg\ \cite{wong:94}). We now
discuss the relationship between the nuclear geometry and the
final multiplicity in this model in more detail.

\subsection{Glauber Model}\label{sec:glauber}

\begin{figure}
  \centering
  \includegraphics[width=.8\textwidth]{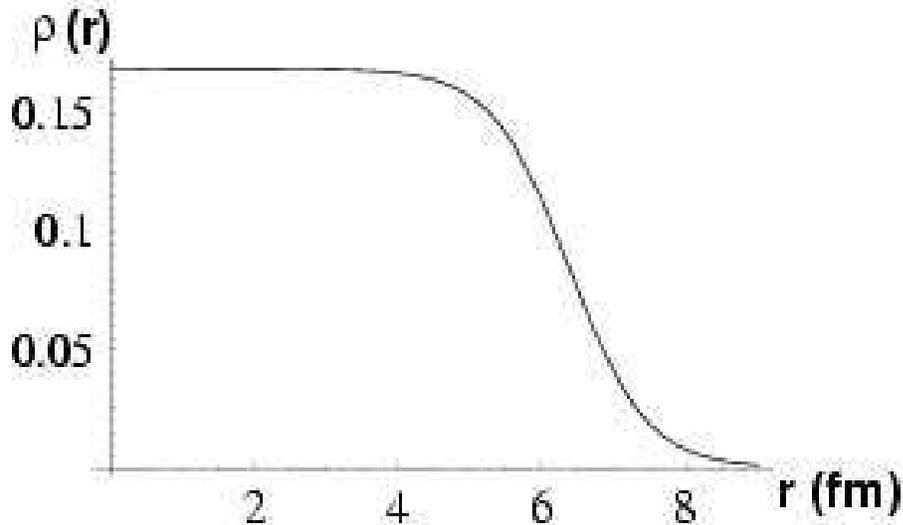}
  \caption{The Woods-Saxon Density profile for the Au nucleus.}
  \label{fig:WoodsSaxonDensity}
\end{figure}

To clarify the meaning of the geometrical model, which can have
different \textit{a priori} assumptions in different
implementations in the literature, we discuss here the
characteristics we consider. The starting point is to assume that
the nuclei are composed of discreet and point-like nucleons. We
distribute the nucleons according to the Woods-Saxon spherically
symmetric density profile
\begin{equation}\label{eq:WoodsSaxonDensity}
  \rho(r) = \frac{\rho_0}{1+e^{\frac{r-r_0}{c}}} \ .
\end{equation}
For the case of Au, the parameters are  $r_0 = 6.38\ \fm$, $\rho_0
= 0.169 \ 1/\fm^3$ and $c = 0.535\ \fm$, obtained from $eA$
scattering \cite{dejager:74}. The density is shown in
Fig.~\ref{fig:WoodsSaxonDensity}.
With these parameters, we get a total number of nucleons in the Au
nucleus of
\begin{equation}\label{eq:WoodsSaxonIntegral}
  \int_0^\infty \rho(r)\ 4\pi r^2 \ dr = 196.6 \ \ .
\end{equation}
It is also common in the literature to take a different value of
$\rho_0$ such that the integral over the density is normalized to
unity.  We can then interpret $\rho(r)$ as the probability of
finding a baryon in the volume element $d^3x = d^2bdz$ at the
position $(\vec{b},z)$.  With this convention, when calculating
quantities such as the number of binary collisions, we see that
there will be factors of $A$ every time this integral appears.

The next assumption is that the interaction probability is just
given by the $pp$ cross section, neglecting effects like
excitations and energy loss. At these energies, the $pp$ and the
$\ppbar$ cross sections are very similar in value, and can
therefore be used interchangeably.
Unfortunately, there are no measurements of either cross section,
\spp\ or \sppbar, at $\sqrts = 130\ \gev$. The experiments UA5
\cite{ua5x:86} and UA1\cite{ua1:90} at CERN-SPS have reported the
\ppbar\ cross section at $\sqrts = 900\ \gev$. Although UA5 took
data at $\sqrts = 200\ \gev$, in Ref.~\cite{ua5x:86} they only
measure the ratio of the cross sections at the two different
energies and use a parameterization given in ~\cite{amos:85}
(Eq.~\ref{eq:sigmaScalingS}) to get a value for the total cross
section.  Then, based on the ratio $\sigma_{el}/\sigma_{tot}$ of
elastic to total cross section, they obtain a value for
$\siginel$.
\begin{eqnarray}
 \sppbar(\sqrts) & = & C_1 E^{-\nu_1}+C_2 E^{-\nu_2}+ C_3 + C_4 \ln^2(s)\ \label{eq:sigmaScalingS}
\\
 \sppbar(p) & = & A + Bp^{-n} + C\ln^2(p) + D\ln(p) \ .
\label{eq:sigmaScalingP}
\end{eqnarray}
For Eq.~\ref{eq:sigmaScalingS}, the important part at energies
above $\sim 100\ \gev$ is the $\ln^2(s)$ term.  The first two
terms are used to describe the data at lower energies and the
difference between $pp$ and $p\bar{p}$ collisions, $E$ is the beam
energy. At $\sqrt{s} = 130$ \gev\ and above, the \pp\ and \ppbar\
cross sections are very similar.  The parameterization found in
the Particle Data Book ~\cite{pdg:00} is given in
Eq.~\ref{eq:sigmaScalingP} in terms of the laboratory momentum
$p$. They quote parameters for the total and for the inelastic
cross section.  The relevant numbers are found in Table
~\ref{tab:sigmappbar}. At the energy of the RHIC 2000 run, we
obtain a value of $\sppbar(130\ \gev) \cong 40.35\ \mathrm{mb}$.
\begin{table}
\centering
\begin{tabular}{|c|r|r|} \hline
                & \sqrts\ = 200 \gev\ & \sqrts\ = 130 \gev\ \\
                \hline
  $\sigma_{\mathrm{tot}}$& 52.40 mb& 49.26 mb\\
  $\sigma_{\mathrm{el}}$ & 10.66 mb&  8.91 mb\\
  \siginel\              & 41.74 mb& 40.35 mb\\ \hline
\end{tabular}
  \caption[The total, elastic and inelastic cross sections in proton-antiproton collisions.]{The \ppbar\ total, elastic and inelastic
cross sections
  obtained from the parameterization from the Particle Data Book ~\cite{pdg:00}.
  The errors on the values are on the order of 1\%.
  }
  \label{tab:sigmappbar}
\end{table}
For head on collisions $b=0$, as we said before, in the hard
sphere limit the number of participants will just be $2A$ where
$A$ is the mass number of the nucleus.  We need a prescription to
calculate the overlap at any given impact parameter.  This is
given by the \textit{nuclear overlap integral}, $\TAA$, which is a
calculation of the overlap of the density profiles (in cylindrical
coordinates, where the $z$ direction is the beam direction) of two
specific nuclei at a given impact parameter $b$:
\begin{equation}\label{eq:Taa}
  \TAA(b) = \int d^2 s\ dz_1\ dz_2\
  \rho_1(\vec{s},z_1)\cdot\rho_2(\vec{s}-\vec{b},z_2)
\end{equation}
\begin{figure}[htb]
  \centering
  \includegraphics[width=.7\textwidth]{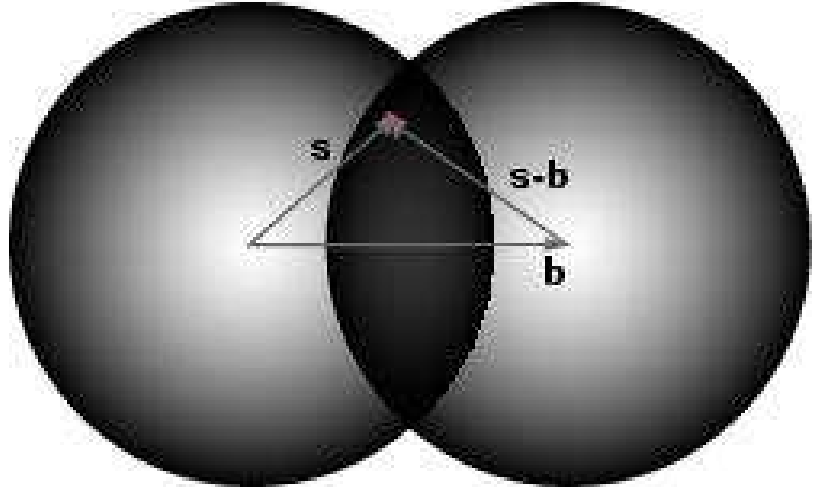}
  \caption[Schematic of collision geometry]{Schematic of the collision geometry for the calculation
   of the overlap integral \TAA, Eq.~\ref{eq:Taa}. The area of overlap
   is the ``football'' shaped region
   in the middle of the two spheres.  The $z_1$ and $z_2$ coordinates
   are perpendicular to the plane of the paper.}\label{fig:Taa}
\end{figure}
We limit ourselves to the case of symmetric \NN\ collisions.  The
coordinate system is shown schematically in Fig.~\ref{fig:Taa}.
For a given impact parameter, we calculate the product of the
densities of each nucleus at a given point $\vec{s}$ and integrate
over all space. We normalize the integral such that
\begin{equation}\label{eq:TaaNorm}
  \int \TAA(\vec{b}) d^2b = 1 \ .
\end{equation}
Since $d^2b$ is an element of area, \TAA\ then has units of
inverse area. With this definition, we can obtain the probability
of having $n$ interactions at a given impact parameter
\begin{equation}\label{eq:TaaProbability}
  P(n,b) = \left( \begin{array}{c}
    A^2 \\
    n \
  \end{array}
\right)\left(\TAA(b)\spp\right)^n\left(1-\TAA(b)\spp\right)^{A^2-n}
\ .
\end{equation}
The first terms takes care of the combinations of choosing $n$
nucleons out of $A^2$, the second term is the probability of
having exactly $n$ collisions and the third term is the
probability of  having exactly $A^2-n$ misses. The total hadronic
cross section for \AuAu\ collisions is then found to be:
\begin{equation}\label{eq:TaaSigmaAuAu}
  \sigma_{\mathrm{AuAu}} = \int d^2b\ \left[
  1-(1-\TAA(b)\spp)^{A^2} \right]= 7.2 \ \mathrm{barn}
\end{equation}
which can be read as 1 minus the probability of not having any
collision ($n=0$) at a given impact parameter (\ie\ the
probability of having at least one interaction at each $b$)
integrated over all impact parameters.

The mean number of binary collisions $\langle \ncoll \rangle$ and
the mean number of participants $\langle \npart \rangle$ are
obtained from \TAA\ at a given impact parameter as
\begin{eqnarray}
 \langle \ncoll \rangle(b) & = & \spp \cdot A \cdot A \cdot \TAA(b) \ \nonumber  \\
 \langle \npart \rangle(b) & = & 2A \int d^2s\ \TAA(\vec{s})\ \left\{1-\left(1-\TAA(\vec{s}-\vec{b})\spp\right)^{A^2}\right\} \ .
\label{eq:TaaNpartNcoll}
\end{eqnarray}
Since the definition of the overlap integral takes care of
counting interactions, it is not surprising that the number of
binary collisions is simply proportional to $\TAA$.  In the above
definition of \ncoll, the factor $A^2$ comes from our choice of
normalization.

\begin{figure}
  \centering
  \mbox{
  \subfigure[]{\includegraphics[width=.45\textwidth]{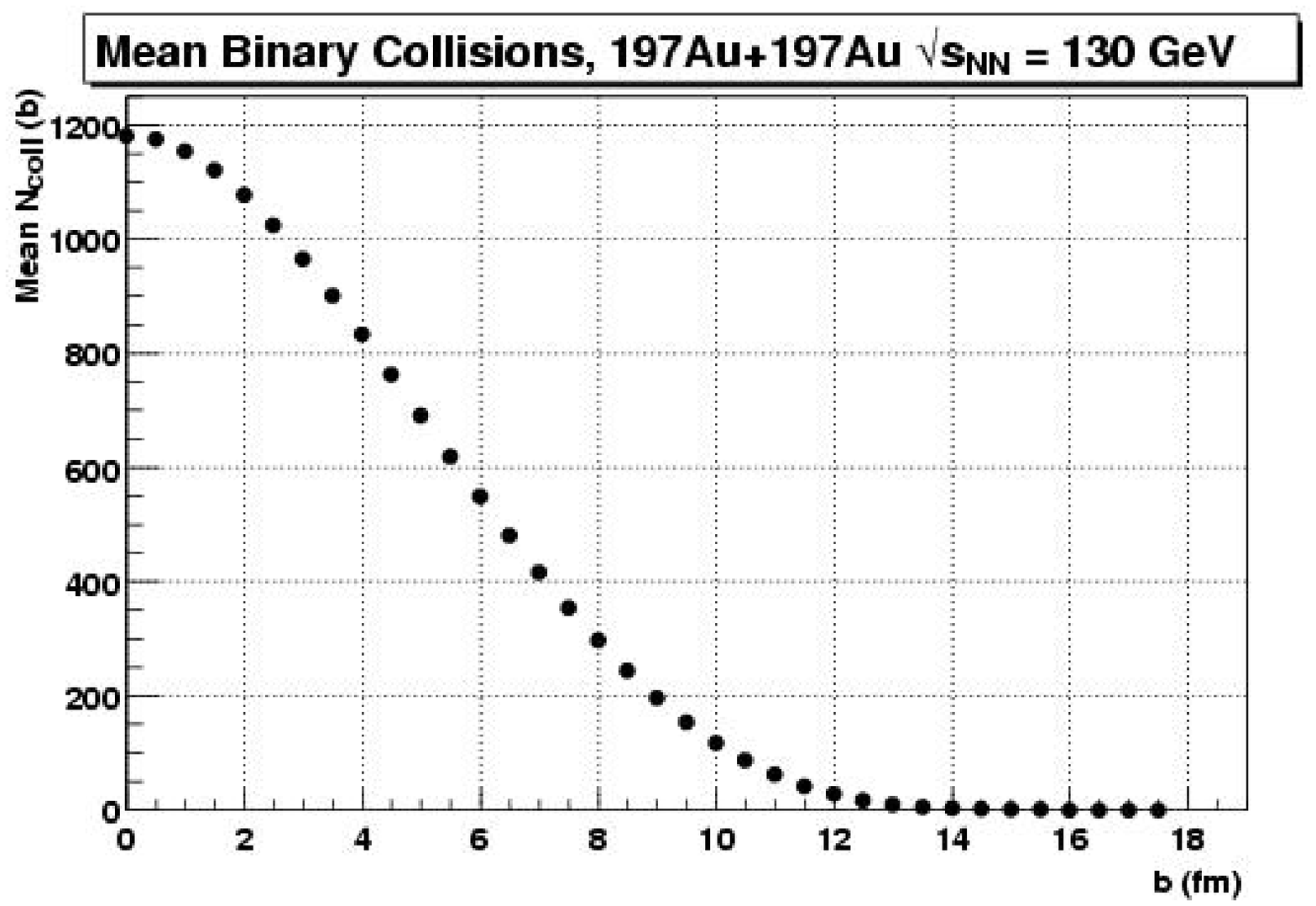}}
  \quad
  \subfigure[]{\includegraphics[width=.45\textwidth]{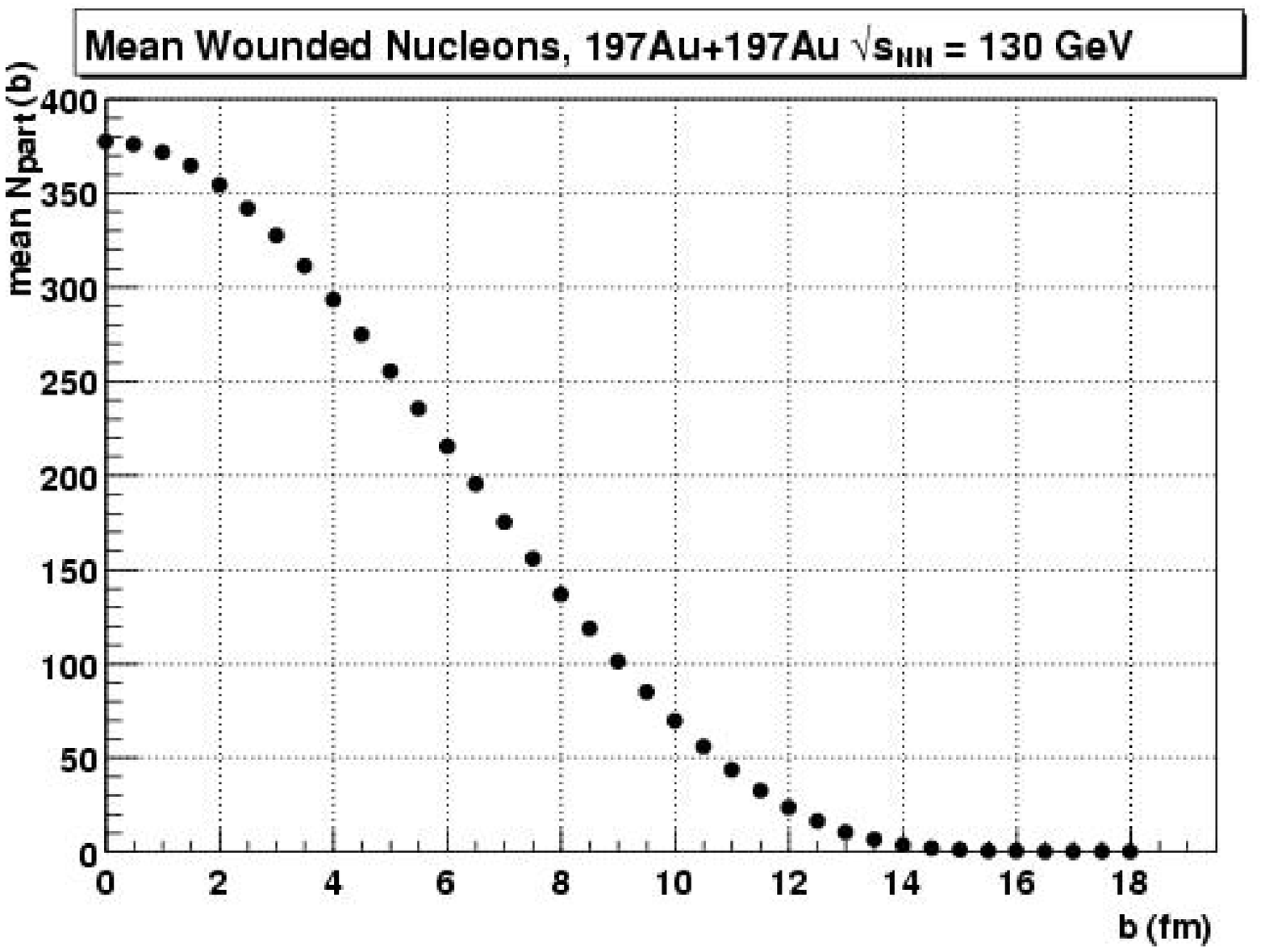}}}
  \caption[Mean \npart\ and mean \ncoll\ \vs\ impact parameter]{$\langle \npart\ \rangle$ and $\langle \ncoll\ \rangle$ as a function
of impact parameter.}
  \label{fig:npartvsb}
\end{figure}
Fig.~\ref{fig:npartvsb} shows the resulting statistical relation
between the impact parameter and \ncoll\ (\npart) in the left
(right) panel.  For large impact parameter, both \npart\ and
\ncoll\ are close to zero.  For the most central collisions,
\npart\ is found to be $\sim 394$ as expected.  We see that the
shapes are similar, although the overall scale is the different.

\begin{figure}[htb]
  \centering
  \includegraphics[width=.7\textwidth]{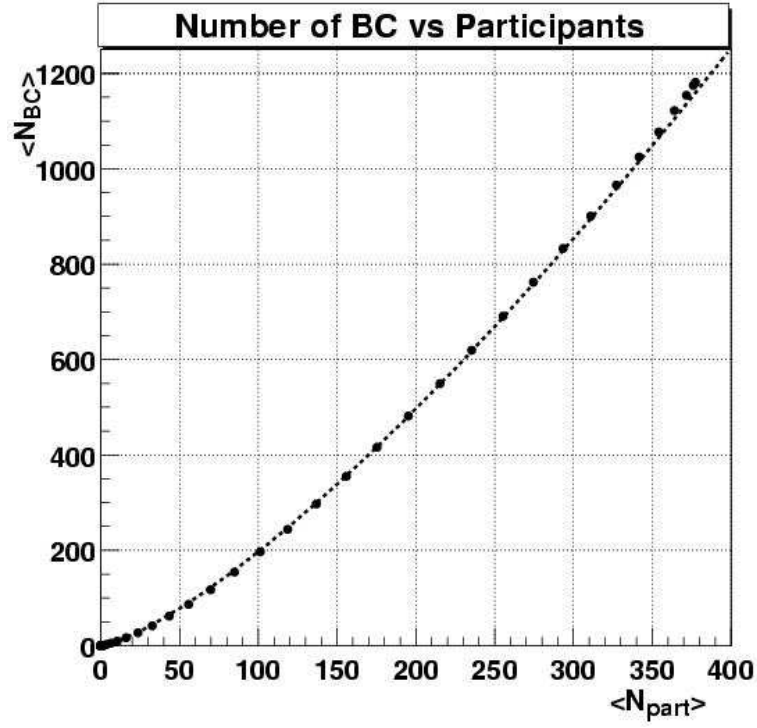}
  \caption[Mean \ncoll\ correlated to mean \npart.]
  {$\langle \ncoll\ \rangle$ correlated to $\langle \npart\ \rangle$.
  The curve is the limit $\ncoll = (\npart/2)^{4/3}$ .}
  \label{fig:ncollnpart}
\end{figure}

Indeed, Fig.~\ref{fig:ncollnpart} shows the statistical relation
between \npart\ and \ncoll\ as a function of impact parameter. The
dotted curve corresponds to $\ncoll = (\npart/2)^{4/3}$.
The above definitions and relations are the basis of the
geometrical model.  We obtain a relationship between the \npart\
and \ncoll\ as a function of impact parameter.  These are still
not directly measurable quantities in experiments.  In a fixed
target environment, one can try to estimate \npart\ by placing a
hadron calorimeter to measure the forward-going energy, and thus
approximately determine the spectator nucleons.  In the RHIC
environment, the closest we can get to such a scheme is to place
calorimeters in the forward-backward regions, called \textit{zero
degree calorimeters} (ZDC's) at RHIC, which detect the spectator
neutrons (see Section ~\ref{sec:trigger}).  So there has to be
another indirect step to model the particle production based on
\npart, \ncoll, and \TAA. There is not a unique way to do this. We
will describe the ``eikonal'' approach to particle production
given in Ref.~\cite{Kharzeev:1997yx} which was used already in the
analysis of PHOBOS data \cite{dima:00}.

We assume that each participant contributes to particle
production, \ie\ $\langle \Nhminus(b) \rangle = q \langle
\npart(b) \rangle$ where $q$ is a scale factor.  This approach has
worked well at low energies, but at higher energies the hard
processes contribute to particle production as well. Therefore, in
\cite{dima:00}, the assumption is that particle production scales
as a linear combination of \npart\ and \ncoll. This model also
attempts to incorporate the multiplicity measured in $pp$
collisions. The particle production is then given by:
\begin{equation}\label{eq:KNhminusMean}
\langle \Nhminus(b) \rangle = q\cdot
n_{\mathrm{pp}}\left(\sqrt{s}\right) \left((1-x)\cdot
\frac{\left\langle \npart(b) \right\rangle}{2} + x \cdot
\left\langle \ncoll(b) \right \rangle \right)
\end{equation}
The idea is that $x$ is a number between $0-1$ which gives the
fraction of particle production that scales as $\sim \ncoll$. We
also need the expected particle production per participant from
$pp$ collisions.  It is measured at 200 \gev\
\cite{ua5:86,ua1:90}. The number of charged particles in the
collision per unit of pseudorapidity at mid-rapidity is reported
in Ref.~\cite{ua1:90} to scale as
\begin{equation}\label{eq:multScalingRootS}
   d\mathrm{N_{ch}}/d\eta(\sqrts) = -0.32 + 0.55 \ln(\sqrts/\gev)\ \ ,
\end{equation}
which at 130 \gev\ yields $n_{\mathrm{pp}} =
d\Nhminus/d\eta|_{\eta=0} = 2.36/2 = 1.18$ .

The next assumption is to pick the statistical distribution to
model the fluctuations about the mean value.  One can choose \eg\
a Poisson distribution \cite{baym:87,jackson:87} or a Gaussian
distribution \cite{dima:00,Bialas:1976ed}.  We follow the Gaussian
prescription here, we then require an additional parameter to get
the variance for the Gaussian, which is taken as simply
\begin{equation} \label{eq:KNhminusSigma}
\sigma_g^2 = a\cdot q^2 \cdot \langle \Nhminus(b) \rangle \ .
\end{equation}
The parameters of the model are thus $a$, $x$ and $q$.

We are finally in the position to calculate the multiplicity
distribution, which is the actual experimental observable.  This
is done by convoluting the various Gaussian distributions obtained
for each impact parameter, weighted by the appropriate interaction
probability and integrating over all impact parameters:
\begin{equation}\label{eq:KNdsdhminus}
  \frac{d\sigma_{\mathrm{AuAu}}}{d\Nhminus}(\Nhminus) =
  \int d^2b \ (1-P(0,b)) \  \frac{1}{\sqrt{2\pi}\sigma_g}\ e^{\frac{N_{h^-}-\langle N_{h^-}(b) \rangle}{2\sigma_g^2}}
\end{equation}

\begin{figure}[hbt]
  \centering
  \includegraphics[width=0.47\textwidth]{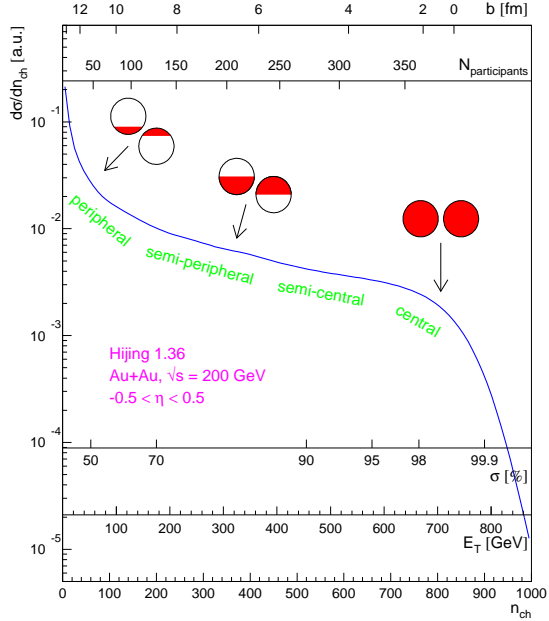}
  \caption[Schematic multiplicity distribution and related variables.]{Schematic of a multiplicity distribution from
  the \textsc{hijing} model, showing the relationship between the shape of the distribution
  and the underlying collision geometry (see text).
  Figure prepared by T. Ullrich.}\label{fig:hijingdsigmadnch}
\end{figure}

From such a picture, we expect roughly the following behaviour as
a function of impact parameter.  The cross section is largest for
very peripheral collisions ($b \sim 12\ \fm$), dropping rapidly at
first ($b \sim 12 - 10\ \fm$) and then falling more slowly in the
region of mid-central collisions ($b \sim 8-4\ \fm$), eventually
reaching a limit for the most central collisions ($b \lesssim 3\
\fm$). A schematic curve of the cross section as a function of
multiplicity obtained from the \Hijing\
model~\cite{hijing:91,hijing:94} is given in Figure
\ref{fig:hijingdsigmadnch}. The multiplicity is the lowest
ordinate axis.  A related experimental observable is the
transverse energy $\Et$ in the next axis.  The percentage of the
hadronic cross section is then given.  The ordinate axes at the
top of the figure are the non-measurable quantities which one can
relate in the geometrical model to the multiplicity: $\npart$ and
the impact parameter $b$.
In this model, we obtain for the 5\% most central \AuAu\ events
$\langle \npart \rangle/2 = $ 172 participant pairs \cite{dima:00}
and $\langle \ncoll \rangle = $ 1050 using $\siginel\ = 40.35$ mb.

\section{Kinematic Variables: \pt, \mt, $y$ and $\eta$}

The next observable after a simple measurement of the charged
multiplicity is to look at the momentum distribution of particles.
Particle spectra are often treated separately in the longitudinal
and transverse directions.

For the transverse direction one normally employs the transverse
momentum (\pt) or for identified particles the transverse mass
\begin{equation}\label{eq:mt}
  \mt = \sqrt{\pt^2 + m^2}
\end{equation}
where $m$ is the mass of the particle.

It is convenient to treat longitudinal momenta using the rapidity
\begin{equation}\label{eq:rapidity}
y = \frac{1}{2} \ln\left(\frac{E+p_{z}}{E-p_{z}}\right)
\end{equation}
where $E$ and $p_z$ are the energy and longitudinal momentum of
the particle. The use of rapidity guarantees that the shape of the
corresponding distribution is independent of the Lorentz frame.
Under a Lorentz transformation from a reference system $R$ to a
system $R'$ moving with velocity $\beta_z$ with respect to $R$ in
the longitudinal direction, the rapidity $y'$ in the $R'$ frame is
related to $y$ in the $R$ frame only by an additive constant: $y'
= y - y_\beta$, where $y_\beta$ is the rapidity  of the moving
frame
\begin{equation}\label{eq:rapidityFrame}
y_\beta = \frac{1}{2}
\ln\left(\frac{1+\beta_z}{1-\beta_{z}}\right)
\end{equation}

For the incident energy of the Au beams at RHIC ($\gamma=70$), the
initial rapidity of each Au beam is $y_{\mathrm{beam}}=\pm 4.94$,
since the beams are symmetric. The rapidity of the centre-of-mass
system, called \textit{mid-rapidity}, is $y=0$; \ie\ the
centre-of-mass reference frame is the same as the laboratory frame
for the collider geometry.

In the limit of a particle whose mass is much smaller than its
momentum, the \textit{pseudorapidity} variable $\eta$ is often
used.  It is defined as
\begin{equation}\label{eq:pseudorapidity}
y = \frac{1}{2} \ln\left(\frac{E+p_z}{E-p_z}\right)
  \approx \frac{1}{2} \ln\left(\frac{|\vec{p}|+p_z}{|\vec{p}|-p_z}\right)
  = \ln \left(\sqrt{\frac{1+\cos \theta}{1-\cos \theta}}\right)
  = -\ln\left(\tan \theta/2\right)
  \equiv \eta
\end{equation}
For the case of negative hadron distributions, since no particle
identification is performed, this is the variable of choice since
one only needs to measure the angle $\theta$ of the detected
particle relative to the beam axis (polar angle in spherical
coordinates, also called \textit{dip angle} in the helix
parameterization commonly used in tracking). Sometimes in the
literature \hminus\ distributions are also presented under the
assumption that all particles are pions, \eg\
\cite{ceretto:98,na49:99}.  For the present work, the \hminus\
distributions are given in pseudorapidity since this is the actual
measured observable, the identified $\pi$ distributions are given
in rapidity.

The momentum distributions are usually presented in terms of the
invariant cross-section
\begin{equation}
E\frac{d^3\sigma}{d^3p} = \frac{d^3\sigma}{d\phi dy \pt d\pt} =
\frac{d^2\sigma}{2\pi dy \pt d\pt} = \frac{d^2\sigma}{2\pi dy \mt
d\mt}
 \label{eq:invariantCrossSection}
\end{equation}
The double differential in Eq.~\ref{eq:invariantCrossSection} is
obtained from integration over $\phi$, and the last equality
follows from the definition of transverse mass, Eq.~\ref{eq:mt}.

It is important to understand the difference in shape of the $y$
and $\eta$ distributions that arises simply from the change of
variables. In particular, the Jacobian $\partial y/\partial \eta$
characterizes the difference between a distribution given as
$d^2N/d\pt d\eta$ and one given as $d^2N/d\pt dy$.  From the
relation $p_z = \mt \sinh y = \pt \sinh \eta$, the Jacobian at
fixed \pt\ is
\begin{equation}\label{eq:jacobianBeta}
  \frac{\partial y}{\partial\eta} = \frac{\pt \cosh \eta}{\mt \cosh y} =
  \frac{|\vec{p}|}{E} = \beta  \ .
\end{equation}
From this relation we again see that $\eta$ approaches $y$ for
highly relativistic particles.  The Jacobian in terms of the
variables (\pt, $\eta$) and (\mt, y) is
\begin{eqnarray}
\label{eq:jacobianMtY}
  \frac{\partial y}{\partial\eta}(\mt,y) & = &  \sqrt{1-\frac{m^2}{\mt^2\cosh^2y}} \\
\label{eq:jacobianPtEta}
  \frac{\partial y}{\partial\eta}(\pt,\eta) & = & \frac{(\pt/m)^2\cosh \eta}{\sqrt{(\pt/m)^2\cosh^2\eta +1}}  \ .
\end{eqnarray}
We can then infer from Eq.~\ref{eq:jacobianMtY} that in the region
$y \approx 0$, there is a small depression in the pseudorapidity
distribution $dN/d\eta$ relative to $dN/dy$.  At high energy,
where $dN/dy$ has a plateau shape, this leads to a small dip at
mid-rapidity for $dN/d\eta$, with the $dN/d\eta$ yield being
smaller by a factor of approximately
$\sqrt{1-m^2/\langle\mt\rangle^2}$ relative to $dN/dy$.

\section{Dynamics from the Kinematics}

To understand the dynamics of relativistic heavy-ion collisions,
it is essential to have information on certain basic aspects of
the collision dynamics. The basic idea is that particle spectra
act as kinematic probes. By analyzing the rapidity and transverse
momentum distributions of the particles produced in the collision,
one can study the energy density $\epsilon $, pressure $P$, and
entropy density $s$ of the hadronic matter formed in the
collision, as a function of temperature $T$ and baryochemical
potential $\mu_B$ \cite{Rischke:2001bt}. From Lattice QCD, the
expectation is to observe a rapid rise in the effective number of
degrees of freedom, expressed by the ratios $\epsilon /T^4$ or
$s/T^3$ over a small range of temperatures \cite{blum:95}.
Experimental observables which are thought to be related (\eg\ in
hydrodynamics) to $T$, $s$, and $\epsilon$ are the average
transverse momentum \meanpt, the hadron rapidity distribution
\dndy\ \cite{Landau:1953gs,Shuryak:1980tp}, and the transverse
energy \detdy , respectively \cite{vanHove:82}.  (The average
transverse momentum, or the related slope of the transverse
momentum distributions, actually reflect not just the temperature
in such models but also the transverse expansion of the system.)
If there is a rapid change in the effective number of degrees of
freedom due to deconfinement in the medium, the first hope was
that one might see vestiges of this effect in a plot of \meanpt\
as a function of \dndy, which would be expected to show a rise,
then a characteristic saturation of \meanpt\ while the mixed phase
persists and then a second rise when the underlying matter
undergoes a structural change to its coloured
constituents\cite{vanHove:82}.  This simple picture however has
several caveats. We discuss here a few of them. We already stated
that it is much too na\"{\i}ve to identify \meanpt\ or the inverse
slope parameter of a transverse momentum spectrum directly with
the temperature of the system. It is also simplistic to assume
that the exponential shape arises in an identical manner as that
of a Boltzmann gas that reaches thermal equilibrium through a
series of internal collisions of the particles in the system.
Statistical thermal models~\cite{Hagedorn:1965st} have even been
applied to \ep\ collisions in fits to hadronic particle
spectra~\cite{Becattini:1996if}. The observed spectra for \ep\
collisions certainly do not come about because of a hadronic
rescattering of the final state particles, but are a the result of
a sampling of the available phase space, \ie\ the particles are
already \textit{born} into equilibrium~\cite{Stock:1999hm}.
Furthermore, if the system formed in \NN\ reaction does in fact
thermalize through collisions, the final state particles observed
in the experiments, and in particular the momentum spectra, will
only reflect the coolest phase of the evolution of the system.
Stated another way, if thermalization occurs the effective
temperature extracted from the spectrum (even if we assume we can
take care additional collective effects such as flow which modify
the \pt\ distributions) will probably not carry information of the
hot initial phase, smearing any structure in a \meanpt\ \vs\
\dndy\ plot. In addition, a visible flat structure in the `$T$
\vs\ $\epsilon$' diagram necessitates a significant duration of a
mixed phase, an effect that probably requires the presence of a
strong first-order phase transition. However, lattice simulations
currently favour a more smooth cross over, perhaps a second order
transition. In this respect, critical phenomena in the form of
increased event-by-event fluctuations are possibly a more robust
observable with respect to the existence of a phase transition.
Even in this case, fluctuation studies would necessitate probing
the region near a possible critical point in the $T - \mu_B$ phase
diagram, and current efforts to shed light as to the presence and
location of a critical point indicate it is in the large $\mu_B$
region~\cite{Fodor:2001pe}.

Rapidity and transverse momentum distribution also allow us to
address properties of the particle emitting source. The degree to
which the incoming nuclei are stopped by the collision is
reflected in the rapidity distributions of produced particles as a
shift with respect to beam rapidity.  At sufficiently high
energies for example, a picture due to Bjorken \cite{bjorken:83},
assumes that the mid-rapidity region should undergo an idealized
hydrodynamic longitudinal expansion.  The charged particles found
at mid-rapidity would be mainly \textit{produced} particles, the
energy of the incoming nuclei would be so great that the collision
among the nucleons would be insufficient to stop the nuclei, so
the incoming nuclei would essentially go right through each other.
The incoming baryons are then found very close to their initial
rapidity.  More importantly, the rapid longitudinal expansion
would have as a result that the rapidity distribution of produced
particles should be flat around mid-rapidity.

This contrasts the picture found at AGS energies, where a
significant amount of stopping is observed and the mid-rapidity
region is net-baryon rich. This is reflected in the rapidity
distribution of charged particles which is peaked at mid-rapidity,
as are the net-proton distributions ($p-\pbar$).   Full stopping
as in the model of Landau \cite{Landau:1953gs} is expected then to
work at low energies, and the rapidity distribution of produced
particles at mid-rapidity is found to more closely follow a
Gaussian shape \cite{lacasse:96,Liu:1998rt,akiba:96}.  Since the
observed width of the distribution is significantly narrower than
that observed for lighter systems, this has been understood as
evidence for strong baryon stopping.

From the rapidity distributions, we can therefore obtain
information for example, as to whether the source is spherically
symmetric or elongated, whether it is static or expanding, and the
degree of longitudinal and transverse expansion.  The momentum
distributions, when paired with elliptic and transverse flow
studies, can also test if there is significant collective
behaviour in the system and approach the question of whether the
system reaches thermal equilibrium. Experimentally, one can test
the hypothesis of boost invariance by measuring $\eta$ and $y$
distributions. If the system is boost invariant to a certain
extent and $dN/dy$ is flat as a function of rapidity in some phase
space range, the \hminus\ pseudorapidity distribution should be
well approximated by Eq.~\ref{eq:jacobianPtEta}.  The rapidity
distribution of identified particles should simply be constant as
a function of $y$.    The slope parameter of $\pi$ distributions
have also been found to show a significant rapidity dependence at
lower energies \cite{lacasse:96,Liu:1998rt,akiba:96,Jones:1996xc}.
We can measure the rapidity dependence of the $\pi$ spectra at
RHIC and further test the hypothesis of boost invariance. These
are of course not sufficient conditions to establish boost
invariance, but they are necessary in case it does hold.

\section{Overview of Transverse Momentum Spectra}
The transverse momentum spectra possess many interesting features
in \NN\ collisions. The use of the transverse mass $\mt$ is
sometimes preferred because experimentally the cross-section
$(1/\pt)(d\sigma/d\pt)$ of a given particle species is better
described by an exponential in \mt\ rather than in \pt\
\cite{guettler:76}. There are contributions to the spectrum that
come from the various physics processes of interest. In order to
extract correct information about the collision, it is necessary
to take these into account. Even now, our understanding of all the
features of the spectra is incomplete.  A brief overview of the
main features of the \pt\ distributions will be discussed.

A general feature emerging from measurements of transverse
momentum distributions in central collisions of heavy nuclei is
that the invariant distributions $d^2 N/\mt d\mt dy$ are
approximately exponential, i.e. $d^2 N/\mt d\mt dy \propto
\exp(-\mt/T)$. One approach to analyze the \pt\ distributions has
been to treating the system as a hadronic gas, and in particular
to use the measured \meanpt\ to estimate the freeze out
temperature $T$ of the gas and its transverse flow velocity
\betat. These also have inverse slope constants $T$ increasing
linearly with the mass of the particle under study. This has been
interpreted as evidence for collective transverse flow: since $\pt
= m\gamma\beta_{\perp}$, if there is a common flow velocity
$\betat$ superimposed on the random thermal motion of particles
$\beta_\perp^{\mathrm{thermal}}$, the slope constant $T$, which is
proportional to \meanpt\ increases linearly with $m$. It is
important to note, however, that in these models, $T$ and \betat\
are coupled: a higher $T$ can compensate for a lower \betat\ and
vice versa.  This approach relies only on the analysis of \pt\
spectra.  It must also be noted that in a plot of $T$ \vs\ mass,
it is difficult to make meaningful comparisons of slope parameters
when they are obtained from different phase space regions, and
there is the danger of obtaining a very different slope simply by
fitting in a different part of the \pt\ spectrum.  Therefore,
rather than concentrating on the slope parameters, more realistic
models incorporate additional observables such as two-particle
correlation data \cite{appelshauser:98} and attempt to describe
both the observed \pt\ distributions and the observed correlations
to obtain $T$ and $\betat$.

The slope of the \pt\ distribution is also seen to increase in
going from \pp\ to \NN.  The spectrum of the pion, due to its low
mass, is not as affected as the spectra of heavier particles by a
given collective flow velocity.  Thus, the pions are good probes
for studying thermal properties at freeze-out.  These features of
spectra can be measured at RHIC energies, and the linear scaling
of spectra with the mass of the particle checked. The scaling
could very well be reduced given that RHIC energies are a
significant leap with respect to SPS energies, and the collision
process can occur in a short enough time scale such that particles
involved might not have time to develop a significant flow
velocity.  On the other hand, an increase in the radial flow (and
also of the elliptic flow $v_2$) would signal an significant
amount of collective behaviour. Uncertainties remain, and they
will only be clarified once we measure spectra and compare them
with the various scenarios: from superposition of \pp\ collisions
governed by measured hadronic cross sections; to relativistic
microscopic models based on \textit{string} formation and string
fragmentation (where a \textit{string} is represented by the color
flux tube from a quark and a diquark. In \textit{longitudinal}
exchange string formation, the excitation of the string originates
from a stretching of the original partons in the hadron caused by
a large longitudinal momentum exchange from the hadronic
collision.  \Hijing, \textsc{fritiof} and \textsc{rqmd} use this
scheme.); to hydrodynamics, which can be thought of the limit
where the mean free path of the constituents is zero and the
system is a fluid.

It is also known that there are significant deviations from a
purely exponential spectrum, both at low and high-\pt\
\cite{schukraft:91}. The high-\pt\ region (at SPS, $\pt \gtrsim
1.5\ \gevc$) shows an enhancement present already in \pA\
collisions known as the Cronin effect~\cite{cronin:79}.  This
enhancement shows a strong dependence on both target and
projectile mass, and a weak rapidity and \sqrts\ dependence. It is
thought to be partly due to multiple low-momentum scattering at
the partonic level inside nuclear matter.  Perturbative QCD
calculations for \pA\ have been done~\cite{lev:83} and agree
reasonably with data. However, a complete quantitative
understanding of this effect has not yet been achieved. The CERES
collaboration measured spectra in \pA\ collisions, found a similar
effect when comparing to \pp, and reached similar conclusions.
Their S+Au data indicated that the onset of this enhancement
occurred at lower \pt\ and that the overall enhancement was
larger.  This was taken as evidence that in \NN\ collisions, there
is an increased number of scattering processes per parton compared
to \pA\ collisions, leading to increased thermalization of the
collision system.  In addition, the shape of the high \pt\
spectrum was very close to being exponential.  These two
observations suggested that a large degree of thermalization is
reached very early for central S+Au collisions at $\sqrtsNN \sim
20\ \gev$.

The low-\pt\ region shows an enhancement as well.  This
enhancement, however, shows different systematic behaviour than
the high-\pt\ excess. The underlying physics that give rise to
this enhancement are different than in the high-\pt\ region. In
fixed target experiments, the low-\pt\ enhancement was observed to
have a strong target dependence but weak projectile dependence, as
well as a strong rapidity dependence.  These effects are not yet
completely understood.  There have been several attempts at
explaining the enhancement (at least qualitatively). Resonance
decays are thought to contribute significantly in this region
\cite{sollfrank:90,brown:91}, and their contribution to the
low-\pt\ enhancement would also be consitent with the observed
strong dependence on target and rapidity. Produced particles can
also rescatter in the target nucleus, and the ensuing cascade
would yield additional soft pions and resonances. In addition, at
the lowest \pt\ one should expect that the treatment of a
radiating gas based on Boltzmann statistics must break down, and
one should really use either Bose-Einstein or Fermi-Dirac
distributions to fit the data. For pions, a Bose-Einstein fit will
give a natural low-\pt\ enhancement compared to a Boltzmann fit to
the spectrum. Other processes can also contribute to the low-\pt\
region, but our understanding of the enhancement is by no means
complete at this point. Studying this region at RHIC energies will
certainly help to shed light on the subject.

To fully understand the non-hadronic effects that may arise in an
\NN\ collision, studying the particle distributions, in order to
obtain basic information about the collision dynamics, is a
necessary step. Several approaches have been proposed to describe
the observed transverse momentum spectra, and to obtain different
types of information from them. In the absence of any collective
behaviour, the spectra should be a simple geometrical
superposition of nucleon-nucleon scatterings. A deviation of the
particle distributions observed in \NN\ collisions from those
observed in \pp\ or \ppbar\ collisions  is already strong evidence
for some form of collective behaviour.  We discuss two models that
have almost diametrically opposite assumptions.

In hydrodynamic models, the colliding system is treated as a
fluid, and can be thought of as the limit of zero mean free path
among the constituent partons. The transverse momentum
distributions are obtained from a transverse flow velocity profile
of the component particles of the fluid. The profile will depend
in general on the initial conditions (centrality, initial energy
density) as well as the underlying equation of state (including or
not a phase transition) as a function of time.  This models
typically yield the strongest signals for collective effects such
as elliptic flow an longitudinal expansion.  They should work best
at low transverse momentum, and their signals can be tested for
example by measuring the centrality dependence of the elliptic
flow signal.

In partonic cascade models, the energy deposition in the collision
is broken up into scattering of the constituent partons of the
colliding nuclei. They rely on the framework of perturbative QCD,
following the perturbative interactions among partons until they
reach thermal equilibrium. These models predict a very rapid
thermalization, mainly due to radiative energy degradation, which
yields a transverse momentum distribution that is exponential to a
high degree.  In all these theoretical models, the shape of the
particle distribution yields information to the underlying
collision dynamics.  Since they rely on perturbative calculations,
they should work best for hard processes such as jets and
high-\pt\ momentum spectra.

There are, of course, other observables that need to be studied to
provide further insight into the full picture of the reaction
process.  These include studies of strangeness production, HBT
correlation analyses, and fluctuations of observables on an
event-by-event basis.  These and other questions are also being
addressed in STAR. The particle distributions are, however, a very
basic building block of the collision picture, and one of the
first that can and has to be addressed in the initial stage of the
experiment.

In previous heavy-ion experiments at lower energies, the study of
the particle spectra has helped to provide guidance into
interesting aspects of the collision dynamics, and has also
provided a means to compare results with model calculations
\cite{braun-m:98}.  The usefulness of studies with particle
spectra is best illustrated with a few examples.

\section{Previous Studies Using \mt\ Spectra}
At the BNL-AGS, E877 \cite{lacasse:96} measured hadron spectra
from Au+Au collisions at 10.8 \gevc\ per nucleon.  They measured
proton and pion \mt\ distributions as a function of $y$ and
performed exponential fits to extract inverse slope constants from
the spectra. For both pions and protons, the spectra showed
deviations from the simple exponential, \eg\ an excess for pions
was observed at $\mt-\mpi < 0.2\ \gevcc$, with the excess
increasing systematically as one approached mid-rapidity. They
compared the \AuAu\ inverse slope constants extracted from the
measured \mt\ spectra with a lighter system, Si+Al. The slope
parameter in the \AuAu\ case systematically increased when going
from the fragmentation region to central rapidities.  The data
indicated that the maximum the observed slope parameter would be
reached around mid-rapidity (\eg\ the measured inverse slopes for
protons at $y/y_{\mathrm{beam}} = 0.7$ were $\Teff \sim 150\ \mev$
for Si+Al and $\Teff \sim 250\ \mev$ for \AuAu). This could be
interpreted as being due to a larger collective transverse flow
component in Au+Au compared to that of Si+Al.  The proton rapidity
distribution was also measured for both systems, being much
narrower in Au+Au. This provided information as to the degree of
stopping in the collision.  The greater stopping observed in the
Au+Au system was interpreted as being a consequence of the smaller
surface to volume ratio as well as the increased average number of
rescatterings. E877 also measured pion spectra in Au+Au
collisions. The rapidity distributions were well matched by
calculations from RQMD, an event generator based on hadronic
rescattering using a mean field approximation~\cite{Sorge:1995dp}.
The \mt\ spectra, and specifically the low-\mt\ region ($\mt -
\mpi < 0.2\ \gevcc $) showed an enhancement above the pure
exponential. In addition, the deviations were seen to be
systematically larger for $\pi^-$ than for $\pi^+$.  This
asymmetry was also observed to systematically decrease as a
function of rapidity, an effect seen also by the E866
collaboration\cite{videbaek:95}.  This was interpreted as due to
the different Coulomb potentials seen by the different charge
types at freeze-out.  (Coulomb interactions are also relevant to
the interpretation of particle correlation studies.) The study of
the influence of the Coulomb effect on the shape of the particle
spectra provided E877 with a different approach to determine the
spacetime particle distribution at freeze-out.

At the CERN-SPS, several experiments have measured charged hadron
spectra. NA49 \cite{appelshauser:98} in central Pb+Pb collisions
measured \mt\ spectra of charged pions. They also measured
two-particle correlation functions to deduce parameters from the
collision dynamics. As mentioned earlier, in a typical fit to an
\mt\ distribution, one can trade lower values of \betat\ for
higher values of $T$.  NA49 applied a model including longitudinal
and transverse expansion \cite{Heinz:1996qu,Chapman:1995nz} using
both the \hminus\ and deuteron transverse momentum spectra as well
as results from two-particle correlations.  They were thus able to
constrain the region for the freeze-out temperature $T$ and the
transverse flow velocity \betat\ to $T = 120 \pm 12\ \mev $ and
$\betat = 0.55 \pm 0.12$ \cite{appelshauser:98}. Comparison of the
charged hadron spectra to predictions of a simple hydrodynamical
model\cite{braun-m:98}, yields similar values for $T$ and \betat\
assuming a quark-gluon plasma first order phase transition at a
temperature
 $\Tc \sim 160\ \mev\ $ (where $\Tc$ was fixed by the lattice results
 then available).
Under these assumptions, the data favor a relatively low kinetic
freeze-out temperature compared to the critical temperature,
consistent with the picture from Fig.~\ref{fig:spacetime1d}.

CERES \cite{ceretto:98} in Pb+Au collisions measured \mt\ spectra
of identified pions and negative hadrons as well as `net proton'
($h^{+} - h^{-}$) spectra. Fourier analysis studies of the
azimuthal charged particle distribution with respect to the
reaction plane were performed.  The amplitude of the first
harmonic, $v_1$, is associated with \textit{directed flow}.  The
second harmonic coefficient, $v_2$, is called \textit{elliptic
flow}. The pseudorapidity dependence of these two coefficients was
studied as a function of centrality, giving strong evidence for
collective behaviour in these collisions.  In addition, a
comparison of the `net proton' spectrum from CERES with
random-walk models allowed one to conclude that these models were
not suitable to describe the flow-like features observed in the
transverse momentum spectra.

For many of these measurements, a large acceptance detector such
as STAR is ideal.  We now proceed to discuss the STAR experimental
setup for the RHIC summer 2000 run.

%
%
\chapter{The STAR Experiment}
\label{ch:StarExperiment}

The main thrust of the STAR detector is dedicated to the heavy ion
collision programme.  It must cope with the large event
multiplicities associated with central heavy ion collisions, and
with the large interaction rates expected at RHIC.  The collider
was designed to deliver a luminosity of $\mathcal{L} = 2 \times
10^{26}$ cm$^{-2}$ s$^{-1}$ for \AuAu\ collisions at \sqrtsNN\ =
200 \gev. STAR is also designed with several other physics
programmes in mind.  It has the capability to measure the products
from ultra-peripheral heavy ion collisions in order to study
coherent photon and pomeron interactions from the intense electric
fields originated by the Au nuclei \cite{seger:91,Roldao:2000ze}.
There will also be \pp\ and \pA\ collisions at RHIC to obtain
information on nuclear parton distribution functions and for heavy
ion reference data. In addition, there is a strong programme of
polarized \pp\ collisions starting in 2001 which aims to measure
the contributions from the gluon (and from sea-quarks) to the
total spin of the proton \cite{Bunce:2000uv}.

This presents a challenge, since STAR has to deal with a wide
dynamic range in several areas. For heavy ion collisions, the
number of particles to be reconstructed in the detector vary from
less than 10 for peripheral \AuAu\ events to as many as $\sim
1000$ for central collisions. The interaction rates are large:
RHIC reached its goal for the first year of delivering 10\% of the
design luminosity.  The collider is designed to operate with 60
bunches in each ring. The time between beam crossings assuming all
60 bunches are filled and taking into account the perimeter of the
RHIC rings (3.83 Km) is $\simeq 0.21\ \mu$s.

The rates will be even larger for polarized \pp\ collisions ($2
\times 10^{32}$ cm$^{-2}$ s$^{-2}$). Coupled to the large amount
of information recorded in the TPC, STAR must handle a large rate
to tape (20 Mb/s for the 2000 run, double the number for 2001). As
this would be the first year of RHIC operations, this also
presented an opportunity: a wide range of available energies would
now fall under experimental scrutiny. The excitement of getting
the first glimpse at the data taken by the experiments was felt
throughout the community, and was well matched with the struggle
to understand the systematics of the different detector systems.

\begin{figure}[htb]
\begin{center}
\includegraphics[width=.7\textwidth]{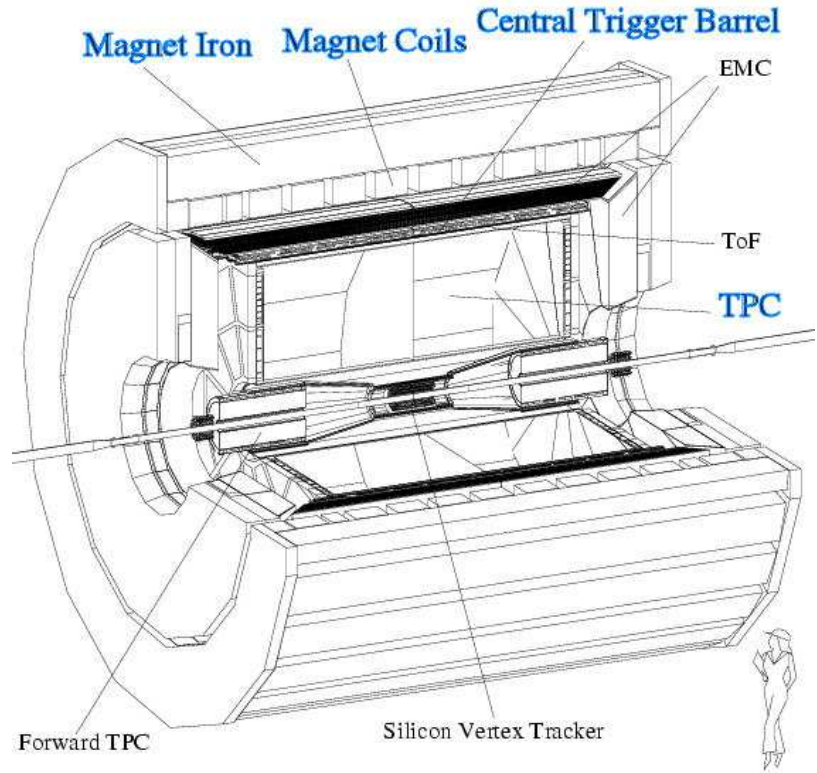}
\caption{The STAR Detector}
\label{fig:stardetector}
\end{center}
\end{figure}

The STAR experimental setup for the year 2000 run centered around
charged particle tracking using a large acceptance Time Projection
Chamber (TPC). The trigger detectors were a pair of hadron
calorimeters placed in the very forward/backward direction and a
barrel of scintillator slats surrounding the TPC.  In addition, a
Ring Imaging \v{C}erenkov detector (RICH) placed at mid-rapidity
provides particle identification for high momentum particles.  The
STAR experiment with the main detector subsystems which are in
place for year 2001 data is illustrated in
Fig.~\ref{fig:stardetector}. In place for the year 2000 were the
Central Trigger Barrel, the Zero Degree Calorimeters, the magnet,
the TPC and the RICH (the ZDC and the RICH are not shown, as the
ZDC is far from the detector center and the RICH is located under
the TPC and obstructed from view in the figure).
\section{Magnet}
Of critical importance to the experiment, the STAR magnet is
designed to provide a very uniform field parallel to the beam
direction.  It is constructed as a large solenoid which must house
the main subsystems of STAR: the TPC, central trigger barrel, and
RICH.  The strength of the field can be tuned from 0 up to 0.5 T,
the strength used for the 2000 run was set to $B_z = 0.25$ T.
Having a uniform field is desirable for it greatly simplifies the
track model used in the pattern recognition in the offline
analysis: all charged tracks follow a helical trajectory to first
order.  The field uniformity at full strength was better than $\pm
40$ Gauss in the radial direction and $\pm 1$ Gauss in the
azimuthal direction. The magnetic field was mapped before TPC
installation to a precision of 1-2 Gauss for all components of the
field.  This allows a calculation of the distortion effects on
tracks due to field inhomogeneities to a precision of $\sim
200-300\ \micron$.

\section{STAR Time Projection Chamber}
The main tracking detector for STAR is a large Time Projection
Chamber (TPC) with complete azimuthal acceptance. With the
magnetic field of 0.25 T, the \pt\ acceptance for charged
particles starts is $\pt \geq 50\ \mevc$. Particle identification
is achieved by measuring the ionization energy loss \dedx,
applying topological cuts, or reconstructing invariant masses of
pairs of particles.

\begin{figure}[htb]
\begin{center}
\includegraphics[width=.8\textwidth]{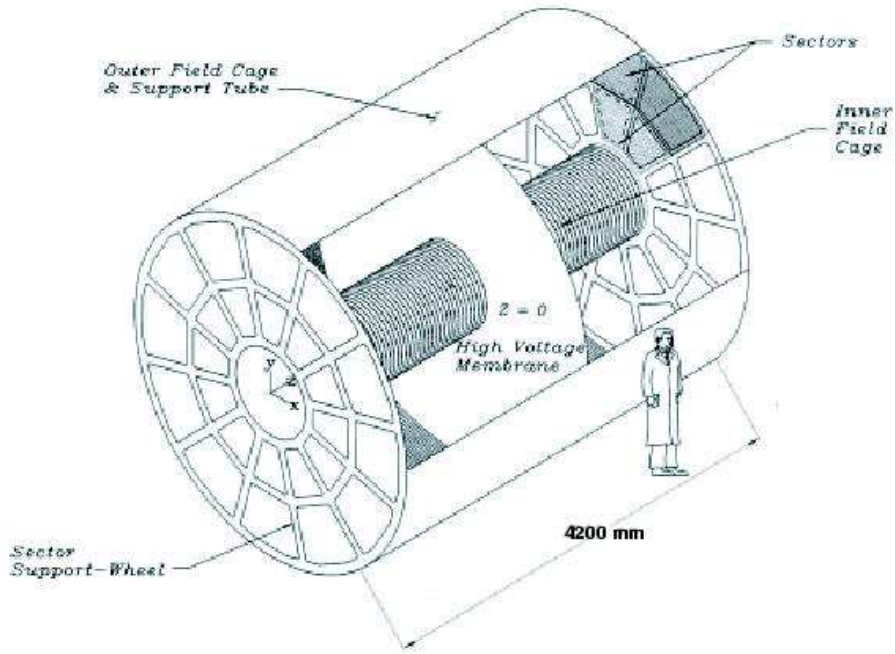}
\caption[The STAR Time Projection Chamber]{Schematic view of the
STAR Time Projection Chamber.} \label{fig:tpc}
\end{center}
\end{figure}

The TPC tracking volume is 4 meters in diameter and 4.2 meters
long. It is shown schematically in Fig.~\ref{fig:tpc}.
The inner radius of the tracking volume starts at 50 cm. The outer
radius subtends a pseudorapidity interval of $\pm 1$ unit and the
corresponding range for the inner radius is $\pm 2$ units. Thus,
the useful tracking volume extends out to $\sim \pm 1.5$ units due
to a requirement of having at least 10 padrow hits in order to
reasonably reconstruct a track.  For the analysis discussed in
this work, we focused on the tracks crossing the entire tracking
volume to reduce systematic effects from variations in acceptance.
The tracking volume is split in two along the beam direction as
shown in Fig.~\ref{fig:tpc}.  The two halves are divided by a high
voltage cathode located at the center of the TPC (z=0), labelled
``high voltage membrane'' in the figure, but called simply the
\textit{central membrane} among the collaboration. The TPC gas is
a mixture of 90\% Ar and 10\% methane (P10).  Part of the
considerations to choose P10 were that the gas should not
attenuate the drifting electrons and it must be pure enough to
prevent other modes of electron loss due to attachment on oxygen
and water molecules.  This means that the oxygen concentrations
should be kept below a hundred parts per million.  These stringent
standards have already been met by P10 in the past~\cite{blum:93}.

The path of a track crossing 150 cm of this mixture is equivalent
to 1.17\% of a radiation length. The tracking volume is surrounded
by an electrostatic field cage which is built with 11.5 mm wide
rings to divide the voltage evenly, achieving a uniform decrease
from -31 kV at the central membrane to 0 V at the ground wires.
Secondary electrons produced by charged tracks ionizing the gas
will drift in the $E$ field away from the central membrane to the
nearest endcap. The field cages are very thin: 0.62\% radiation
lengths for the inner field cage and 1.26\% radiation lengths for
the outer field cage. The drift velocity of the electrons was
measured to be 5.44 $\pm$ 0.01 cm/$\mu$s, where the 0.01 here
denotes the size of variations over a period of several days.  For
a given run, the drift velocity was monitored to a precision of
$\sim 0.001$ cm/$\mu$s.  For such a large TPC, the gas and
magnetic field strength are also important to maintain the
diffusion of the drifting electrons to acceptable levels. The
transverse diffusion in P10 at full field (0.5 T) is $\sigma_T
\approx$230 \micron / $\sqrt{\mathrm{cm}}$ or equal to a width of
3.4 mm after the drift; and this sets the scale for the dimensions
of the pads under the anode wires, described below. Similarly, the
longitudinal diffusion is $\sigma_L \approx 360\
\micron/\sqrt{\mathrm{cm}}$. For a cluster that drifts the full
length of the TPC, this translates into a longitudinal width
$\sigma$ = 0.52 cm , corresponding to a drift time of $\sim$95 ns.
This determines the electronic sampling rate of 10 MHz, or every
100 ns, in order for a simple three point tracking and clustering
algorithms to work well.

\begin{figure}[htb]
\begin{center}
\includegraphics[width=.8\textwidth]{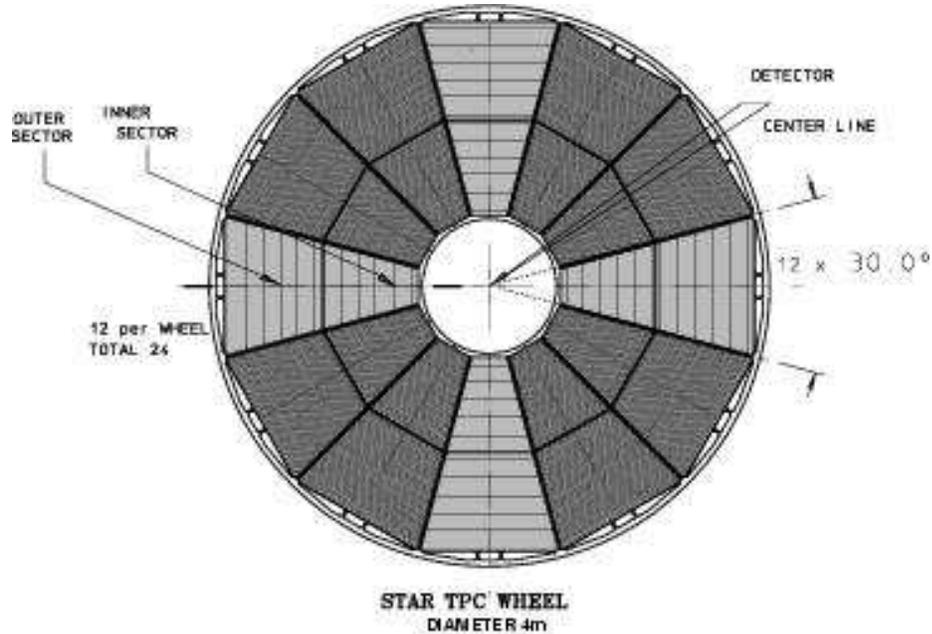}
\caption[End view of the STAR TPC]{End view of the STAR TPC, where
we see the 12 sectors covering the full azimuthal range.  }
\label{fig:tpcSide}
\end{center}
\end{figure}

The drifting electrons are amplified by a grid of wires on each
end of the TPC and the pulses are read out on small pads placed
behind the anode wires.  Each endcap of the TPC is segmented in
the azimuthal direction like the face of a clock into 12 pad
arrays, called \textit{sectors}. This is illustrated
Fig.~\ref{fig:tpcSide} which shows the end view of the TPC.
The beam ($z$ axis) goes through the center of the TPC,
perpendicular to the page in this view.

\begin{figure}[htb]
\begin{center}
\includegraphics[width=.8\textwidth]{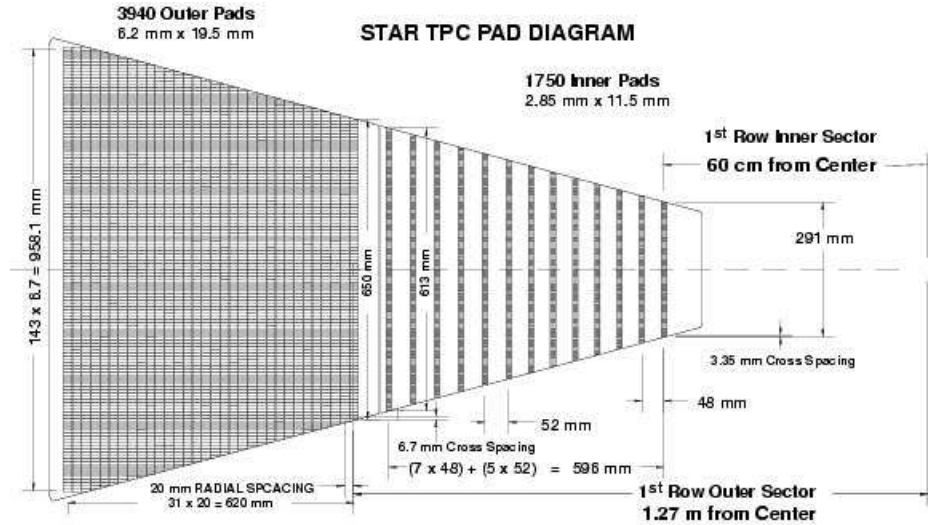}
\caption[Pad diagram of a TPC Sector]{Schematic of a full TPC
sector with pads. The first 13 rows comprise the inner sector, and
rows 14-45 are the outer sector pads.} \label{fig:tpcSector}
\end{center}
\end{figure}

The geometry of one of the sectors is shown in
Fig.~\ref{fig:tpcSector}.
Each sector is divided into two parts, an \textit{inner} and an
\textit{outer} sector. The inner sector has 1,750 pads, each pad
measures 11.5 mm in the radial direction and 2.85 mm in the
tangential direction.  The inner sector pads were made as small as
practical in order to provide closer space points for an improved
two-track resolution. The size is limited by the diffusion limit
of the TPC. They are grouped into 13 pad rows located between 60
cm and 116 cm from the beam line, measured from the center of the
padrow. The outer sector has 3,940 pads, and the corresponding
dimensions are 19.5 mm (radial) and 6.2 mm (tangential). The outer
sectors spans 32 pad rows between 127.2 cm and 189.2 cm radius.
The large outer sector pads completely cover the area under the
anode wires. The gap between adjacent pads in the same row is 0.5
mm for both the inner and the outer sector.

\begin{figure}[htb]
\begin{center}
\includegraphics[width=1\textwidth]{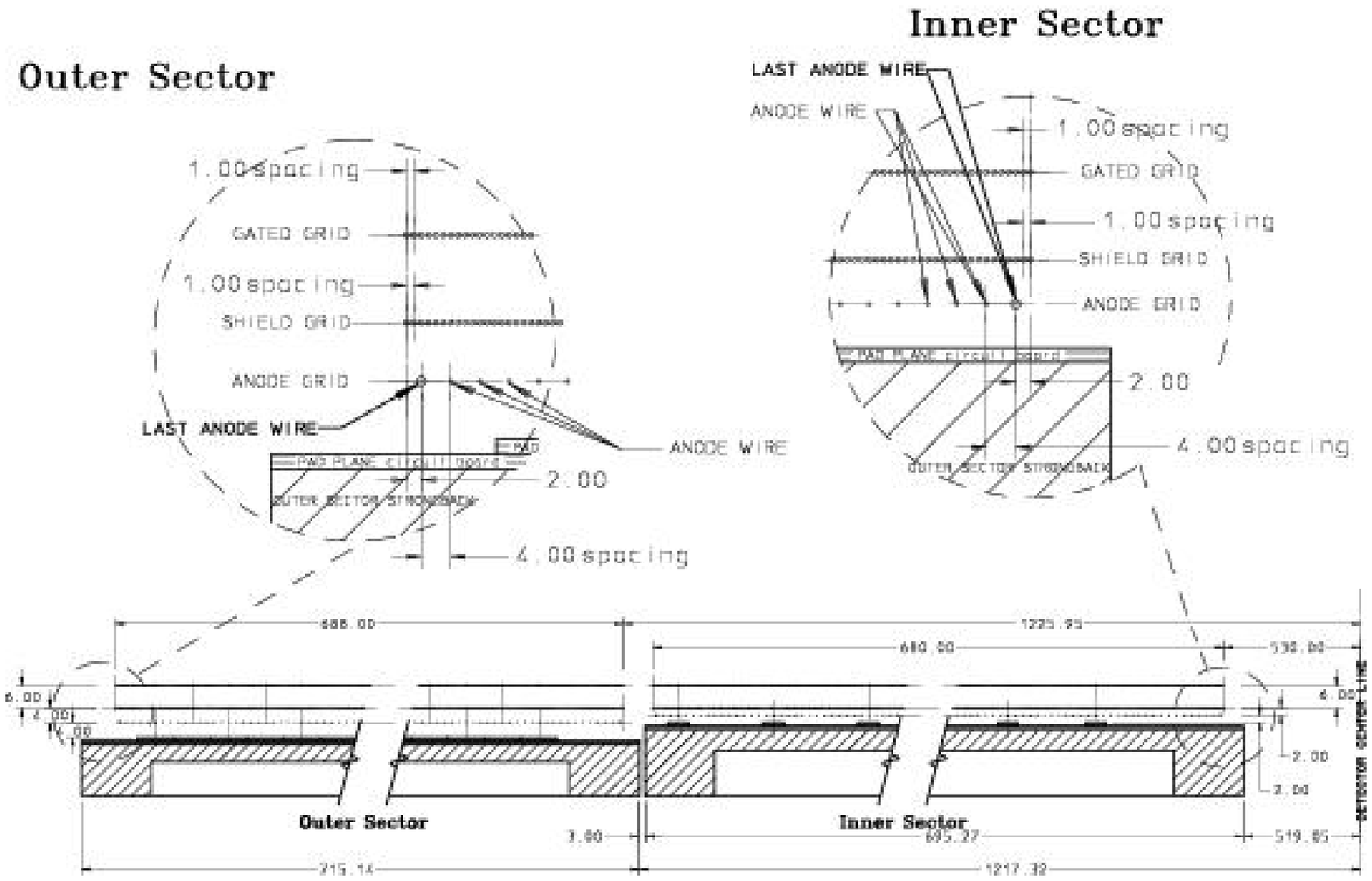}
\caption[Wire geometry of a TPC Sector]{Side view of the inner and
outer sector of the STAR TPC showing the wire geometry. All
distances are shown in mm.} \label{fig:tpcWires}
\end{center}
\end{figure}

The wire geometry is depicted in Fig.~\ref{fig:tpcWires} where we
see a cross-sectional view of the wire layers and of the pad
plane. There are 3 layers of wires terminating each side of the
chamber. The first layer is the \textit{gating grid}. The wires of
the gating grid are arranged to terminate the field cage voltage
and it is therefore used to prevent ionization from reaching the
anode wires of the chamber when no trigger is detected.  The
gating grid also helps minimize the space charge effects due to
the presence of positive ions with very low mobility.  The second
layer is the ground plane, 6 mm below the gating grid, which
captures the ions from the amplification region. The third layer
is the anode plane, 2 mm (4 mm) below the ground plane in the
inner (outer) sector, where gas amplification is achieved. The
induced signal is picked up on the pad plane located 2 mm (4 mm)
under the anode wires in the inner (outer) sector.  In order to
keep the signal to noise ratio constant at 20:1 for both pad
sizes, the anode voltages were set to achieve a gain of 1100 on
the outer sector and 3000 on the inner sector.

Combining the inner and outer sector, there are a total of 5690
pads per sector which corresponds to a total of 136,560 channels
for all 24 TPC sectors.   The signal measured on the pads is
amplified using custom CMOS integrated circuits ~\cite{klein:96}.
These consist of a pre-amplifier and shaper circuit followed by a
Switched Capacitor Array (SCA) and a 10 bit Analog-to-Digital
Converter (ADC). These are responsible for sampling the arrival of
the electrons into at most 512 time bins and to digitize the
signal on the pads.  Including the segmentation in the time
(drift) direction, the TPC can be thought of as a large 70
megapixel 3D digital camera.  The position of the particle along
the drift direction is then reconstructed by converting from time
bin to position by knowing the drift velocity.  The position
resolution along the drift direction for the summer 2000 run was
on the order of 500 \micron.  The front-end electronics are
bundled into groups of 16, with the pre-amp, shaper and buffer in
one chip, and the SCA and ADC in another chip.  Two groups of 16
channels make up each front-end board, making a total of 32
channels and 4 chips per board.  The data are sent to the Data
Acquisition system (DAQ) over optical fibers. The full TPC DAQ
rate is designed to be read out once per second for central \AuAu\
collisions, which produces a rate to tape of $\sim$20 Mbytes/s.

For a more complete description of the STAR TPC see
~\cite{star:99,starcdr:92}. More detail on the TPC operation will
be presented in Chapter \ref{ch:DetectorSimulation} where we
describe the physical processes of the TPC that were simulated for
Monte Carlo studies.  For more information on drift chambers, the
standard reference is \cite{blum:93}.

\section{Trigger Detectors} \label{sec:trigger}
The trigger detectors are an array of scintillator slats (CTB)
arranged in a barrel at the outer diameter of the TPC; and two
hadronic calorimeters (ZDCs) at $\pm 18$ m from the detector
center and at close to zero degrees relative to the beam axis. The
ZDC units subtend an angle of $\sim 2.5$ mrad from the interaction
point.  The calorimeters are designed to measure neutrons emitted
from nuclear fragments of the spectator matter, \ie\ matter that
did not interact in the collision. In contrast to fixed target
experiments, where one can measure all spectator nucleons in the
forward direction, at RHIC the proton trajectories will be
deviated by the dipole magnets in the beam line.  The same dipole
magnets that are used to steer the beam head on at each
interaction region will bend the charged fragments away from the
ZDCs.  Protons of the same momentum as the Au ions are actually
deviated by a \textit{larger} angle, since they have a larger
charge-to-mass ratio. Ergo the ZDC signal is produced by neutrons.
Nevertheless, they still provide possibly the best determination
of the collision centrality at RHIC, and are used for triggering
(along with some other observable based on multiplicity) in all
four experiments. This can serve as a standard for comparisons
among the different results.

During the Summer 2000 run, RHIC delivered collisions between Au
nuclei at $\sqrtsNN \ = 130 \ \gev$. The data presented here are
from a minimum-bias sample, triggered by a coincidence of signals
above threshold in both ZDCs with the RHIC beam crossing. The ZDC
threshold was set to ensure efficient detection of single
spectator neutrons.  The efficiency of the ZDC coincidence trigger
for central events was measured using a high-threshold CTB
trigger.  The trigger efficiency was found to be above 99\% for
the entire range of multiplicities reported in this study. The CTB
was used to trigger on central events, as the signal is correlated
to the multiplicity at mid-rapidity.  The threshold for the
central trigger using the CTB was set to obtain the events with
the 15\% highest CTB signals. The maximum luminosity achieved in
2000 was $\sim 10\%$ of the design luminosity.  This translates
into a hadronic interaction rate of $R = \sigmahad \times
\mathcal{L} \simeq (7.2 \times 10^3\ \mathrm{mb}) \times (0.02\
\mathrm{mb}^{-1} \mathrm{s}^{-1}) = 144\ \mathrm{Hz}$.  For higher
luminosities, the trigger will certainly play an important role.

\section{RICH}
A Ring Imaging \v{C}erenkov detector (RICH) was placed outside the
CTB.  The RICH covers a smaller area, 1 m $\times$ 1 m and is
designed to provide high precision velocity measurements for
enhanced particle identification at high momentum.  The momentum
information is provided by the TPC.  One then uses this
information to search for \v{C}erenkov photons in 3 different
annulus regions. For tracks with a perpendicular trajectory to the
RICH, the regions will be circles of different radii depending on
the particle's velocity, which for a known momentum uniquely
determines the mass of the particle.  In the presence of the
magnetic field, the trajectory of the tracks will not be normal to
the RICH detector, so the \v{C}erenkov photons will not be found
in a circular region, but one can still determine the bounds of
the search region analytically.  The RICH detector is not used in
this analysis, but it is the subject of another STAR thesis
\cite{horsley:01}. For more information on the RICH detector see
\cite{lasiuk:01}.

\begin{figure}[htb]
\begin{center}
\includegraphics[width=.75\textwidth]{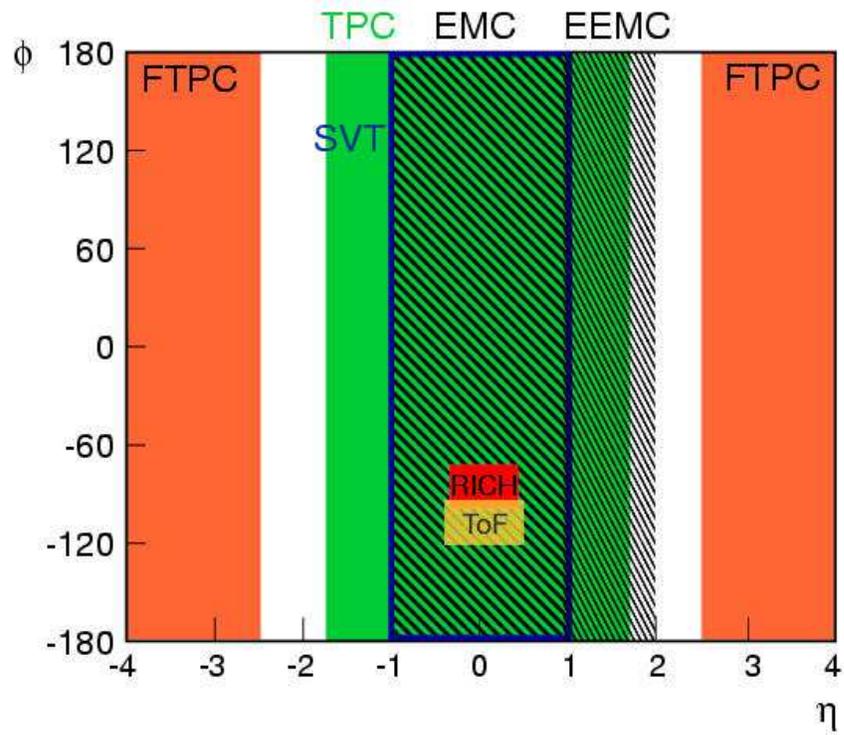}
\caption[Hermeticity of STAR including all detectors]{Hermeticity
of STAR, including all detectors.  Note that the RICH occupies
real estate that will used for the full barrel EMC when completed,
so this plot only illustrates their coverage.}
\label{fig:hermeticity}
\end{center}
\end{figure}

The detector upgrades planned for the upcoming STAR runs include a
Silicon Vertex Tracker (SVT), an Electromagnetic Calorimeter to
cover the barrel at mid-rapidity around the TPC and one of the
endcaps, a Time-of-Flight patch replacing one of the CTB slats,
two Forward TPCs ($2.5 \leq |\eta| \leq 4$) and additional
Beam-Beam Counters for the trigger system.  The hermeticity of the
STAR detector when all the components are installed is shown in
Figure~\ref{fig:hermeticity}.


%
%
\chapter{Reconstruction and Calibration}
\label{ch:ReconstructionCalibration}
 We discuss here the STAR
offline reconstruction and analysis software relevant for the
results presented in this work. The offline simulation and
reconstruction software is a major component of the experiments,
as we rely on it
 to reconstruct the collision event to a sufficient extent that the physics goals
and the physics capabilities of the sub-detectors can be realized,
 to sufficiently evaluate and
visualize the results to determine their adequacy and correctness,
and to generate acceptance and reconstruction efficiency tables,
among other things. The offline software includes all major
sub-detectors.  We will focus on the reconstruction done in the
TPC, as this was the main detector used in STAR during the first
run and is the relevant component of the results discussed here.
Then we discuss the STAR global chain, where vertex finding and
propagation of tracks to find the primary track candidates is
done. The other major detector sub-system in the year 2000 run was
the RICH, and it is discussed in another dissertation
\cite{horsley:01}.  The original design for the STAR offline
simulation and analysis software is given in
Ref.~\cite{bossingham:97}.

\section{Event Reconstruction in the TPC}
The task of the TPC reconstruction software is to reduce the raw
data taken in an event (nearly 7 million ADC values) to lists of
meaningful quantities such as space points, particle tracks,
vertices, etc.  It must employ the best knowledge of the relevant
calibration parameters for each run. These include for example the
drift velocity of the electrons in the gas, trigger time offsets,
temperature, pressure, magnetic and electric fields.
\begin{figure}[htb]
  \centering
  \includegraphics[width=.7\textwidth]{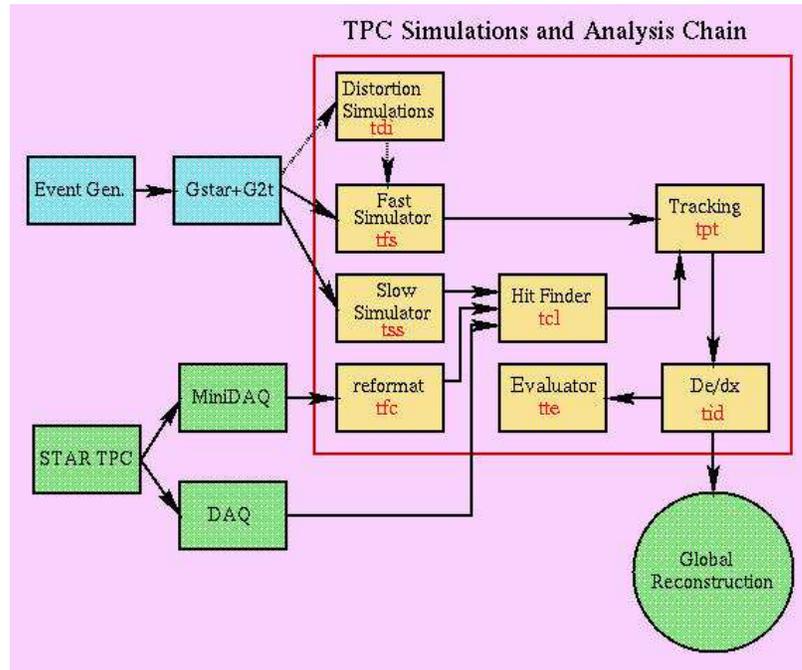}
  \caption{Schematic of the STAR TPC Software chain.}\label{fig:tpcRecoFlowChart}
\end{figure}

The procedure employed here is typical of that employed in many
experiments.  A flow diagram of the different software modules is
depicted in Fig.~\ref{fig:tpcRecoFlowChart}.  The chain is
designed to cope with simulated and real data using the same code.
We focus on the real data reconstruction.  The details of the
detector simulation (labelled ``slow simulator'' in
Fig.~\ref{fig:tpcRecoFlowChart}) are discussed in Chapter
\ref{ch:DetectorSimulation}.

One starts by examining the pixel information for every padrow
independently in order to find the location where a charged
particle has crossed the row, which we call a \textit{hit}.  A
tracking algorithm then links together series of hits to find
particle trajectories through the TPC volume. Information
extracted from the ionization or from the topology of the tracks
is then used for identification.  We now discuss the
reconstruction, emphasizing the two parts that are important for
the charged hadron distributions discussed in this work: the
determination of the momenta and the identification via \dedx.

\subsection{Cluster Finding}

The hit reconstruction software converts raw pixel data into
reconstructed space points.  The raw pixel data is arranged into
time ordered sequences for each pad in a row, where only pixels
above a threshold are taken into account.  At this stage, one must
be careful to apply gain corrections and remove channels tagged by
DAQ or previous offline runs as bad, noisy or dead.  Any relative
timing corrections between pads are also applied.  The cluster
finding algorithm then looks for contiguous pixels in a 2D space
given by the time direction and the pad-row direction. After the
2D cluster finder stage, a pass is made to find single hits and
multiple hits in a cluster.  The single-hit finder assumes that
the cluster is made by the crossing of a single track and
estimates the centroid of the cluster.  The multiple-hit finder
tries to find the local maxima in a cluster and may deconvolute
the cluster into individual hits. Deconvolution of close hits is
of critical importance for two-track resolution and \dedx\
measurement. The fitted hits are then transformed into TPC space
points, taking into account the drift velocity, trigger time
offsets, sector geometry, and electronics shaper response. The
space points then contain information on the position of the hit
in the STAR global coordinate system and on the energy deposited
by the track.

\subsection{Track Finding}
The detection of charged particle tracks is, of course, the
\emph{raison d'\^{e}tre} of a Time Projection Chamber. The track
finding algorithm performs the critical task of converting the
reconstructed space points into particle tracks, and of
determining their 3-momentum.  Efficient tracking and good
momentum resolution are an essential part of the analysis
presented here, and especially for analyses such as high-\pt
spectra and resonance decay reconstruction.

The tracking algorithm starts with overlapping hits in the outer
pad-rows as the track seeds.  The first stage of the algorithm
consists of a follow-your-nose approach which produces a
collection of track segments, \ie\ collections of space points
associated with a track, either complete or partial. Once we have
the segments, we perform an initial segment fitting, which
incorporates multiple Coulomb scattering and energy loss in the
gas. Outlier space points, \ie\ hits that were assigned to the
track in the segment finding stage but whose position lies far
from the allowed track extrapolation error after the fitting
stage, may be removed at this point. A further extrapolation of
the tracks for purposes of segment joining is then executed,
including also multiple Coulomb scattering and the effects of
energy loss. The tracking parameters are updated for the joined
segments and a clean-up of fragments is done in the final
filtering step.

\subsection{Particle Identification: \dedx}

\begin{figure}[htb]
  \centering
  \includegraphics[width=.7\textwidth]{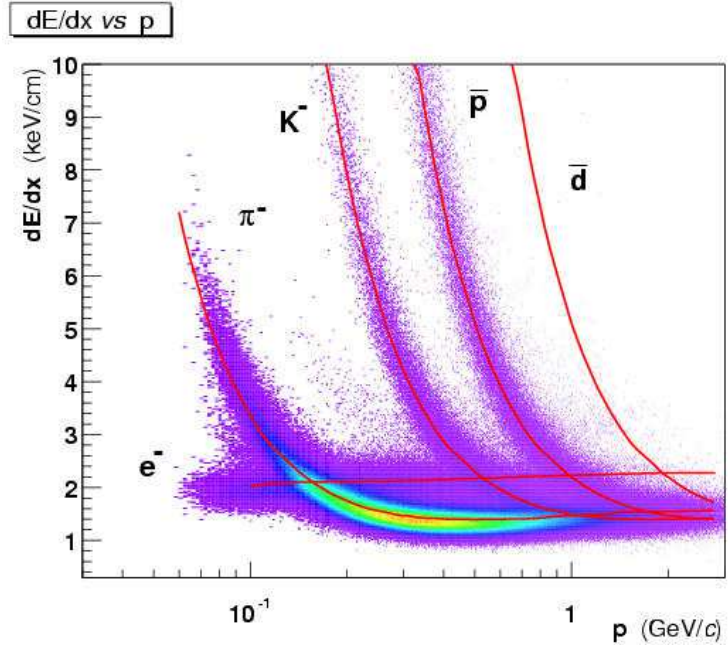}
  \caption[\dedx\ \vs\ $p$]{Particle identification via \dedx\ in the TPC.  The plot shows the
  truncated mean \dedx\ value
  for negative tracks as a function of the reconstructed momentum.  The expected
  ionization energy loss curves
  for $e^-$, \piminus, $K^-$, anti-protons and anti-deuterons are also shown.}
  \label{fig:dedxvsp}
\end{figure}

The TPC allows us to separate pions using their ionization energy
loss. The charge collected for each hit on a track is proportional
to the energy loss of the particle. We parameterize the particle's
trajectory in order to obtain a path length in the gas which
corresponds to each segment, \ie\ the deposition that corresponds
to a specific TPC hit. For a track crossing the entire TPC we thus
obtain 45 \dedx\ samples, which are distributed according to the
Landau probability distribution.  One of the properties of this
distribution is that its tail dies off very slowly, and the
dispersion of values around the mean is very large (infinite in
theory). A typical procedure in order to reduce fluctuations from
the long Landau tails is to truncate the distribution.  In the
case of STAR, we used 70\% truncation, \ie\ the highest 30\%
ionization values were discarded.  Using the remaining values, a
\textit{truncated mean} is computed, and this becomes the basis
for any analysis using identified particles in the TPC. The
measured truncated mean for negatively charged primary particles
is illustrated in Fig.~\ref{fig:dedxvsp}, shown as a function of
the momentum.

The curves are the Bethe-Bloch parameterization used in the
analysis for the different particle hypotheses.  We see that at
the lowest momentum, the pions have a greater ionization energy
loss than the electrons which are already in the saturation region
of the curve. The pions cross over the electron band at $\sim 150\
\mevc$ reaching a minimum at about $300-400\ \mevc$.  The pions in
the relativistic rise merge with the Kaons, which are still in the
$1/\beta^2$ region, at about 1 $\gevc$. However, due to the width
of the bands, a single \piminus\ peak is only discernible up to
$\sim 750$ \mev.

\subsubsection{\dedx\ Resolution}

\begin{figure}[htb]
  \centering
  \includegraphics[width=.5\textwidth]{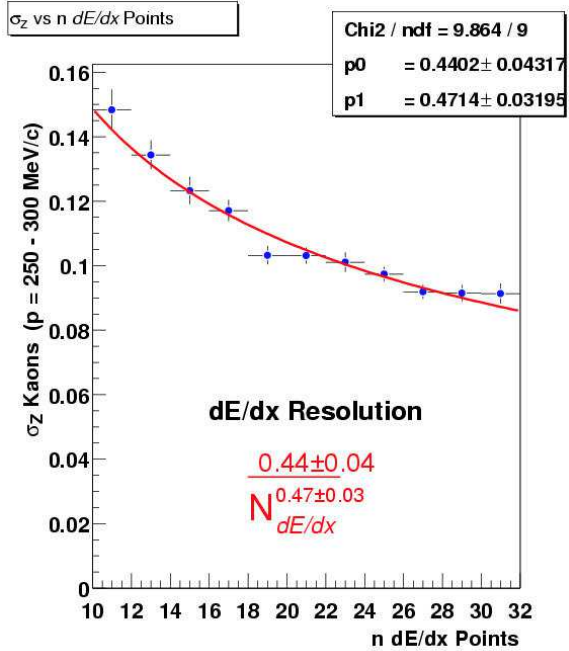}
  \caption[\dedx\ resolution]{The \dedx\ resolution as a function of
  the number of \dedx\ samples.}\label{fig:dedxResolution}
\end{figure}

The \dedx\ resolution for this analysis was found to be $8.8\%$
for tracks with 45 fit points (31 \dedx\ samples after
truncation). This is illustrated in Fig.\ref{fig:dedxResolution}.
A widely used empirical expression \cite{allison:80} describes the
resolution at minimum ionization, for $N$ \dedx\ samples of length
$h$ cm in argon gas with up to 20\% CH$_4$ at a pressure of $P$
atmospheres:
\begin{equation}\label{eq:dedxResolution}
\frac{\sigma_{dE/dx}}{\langle dE/dx \rangle_{\mathrm{trunc}}} =
\frac{0.47}{N^{0.46}} (Ph)^{-0.32} = \frac{0.38}{N^{0.46}}
\end{equation}
For the last equality, typical operation of the STAR TPC is at 1
atm and we assume the same radial pad length of $h = 1.95$ cm
(outer sector pads) for all ionization samples.
Equation~\ref{eq:dedxResolution} then yields an estimate of the
best achievable resolution of $\sim7.8\%$ for tracks with 31
\dedx\ points. Improvements such as the removal of noisy pads,
sector-by-sector calibration, and drift-distance dependence are
expected to bring the resolution closer to this expected design
value.  These refinements are however not critical for the
analysis presented here.

This is the first stage of the track finding and fitting
pertaining to the TPC only.  The information is then passed to the
\textit{global reconstruction} software, discussed next.

\section{Global Event Reconstruction}
There are two additional steps to arrive at the final tracks used
in the analysis.  The connection of the space point information
among different detectors is the next step in tracking, \ie\ a
\textit{global} track fit.  This step becomes indispensable when
including SVT or any additional information from a tracking
detector. Finally, after the primary vertex is found, this acts as
an additional measured point.  If a track's trajectory goes near
the primary vertex, a subsequent \textit{primary} track fit is
then performed. For the analysis presented here, as we focus on
production of primary charged hadrons, primary tracks are used.

\begin{figure}[htb]
  \centering
  \includegraphics[width=.7\textwidth]{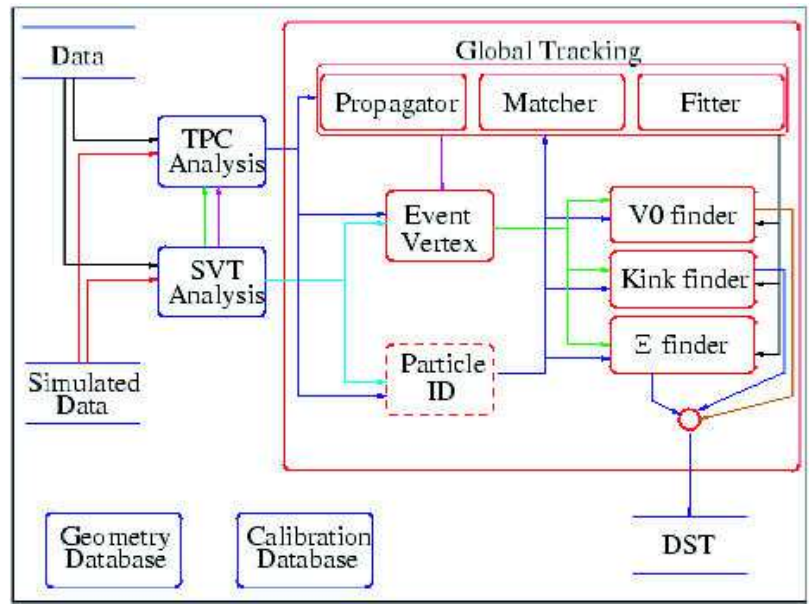}
  \caption{Schematic of the STAR Offline Software chain.}\label{fig:globalRecoFlowChart}
\end{figure}

The global event reconstruction offline software correlates all
the tracking, timing and energy deposition information from each
detector in STAR and produces the overall event characteristics of
the triggered events into a Data Summary Tape format for use in
all the physics analyses.  We will discuss the major components
here in roughly sequential form, although one sometimes performs
several iterations among different modules.  This is the case for
example if one uses PID information in global track propagation,
but would also want tracking information when calculating track
segment lengths for the calculation of the specific ionization in
a PID module.  The flow of the global software chain is sketched
in Fig.~\ref{fig:globalRecoFlowChart}.  The SVT is shown in the
figure, although it was not yet installed for the 2000 run; we
will therefore not discuss it here.  The main objective of the
global part of the chain however, is to combine the information
from all STAR sub-detectors and was therefore exercised with
simulated data to ensure readiness for real data.

\subsection{Global Tracking}\label{sec:globalTracking}
The task of a global tracking module is two-fold.  At its core, it
is in charge of matching track and hit information from the
different detector sub-systems.  Once this is done, it must refit
the newly matched tracks taking into account all available
information.  The role of the global tracking software was
somewhat special for the 2000 run, since only 1 detector subsystem
provided track information.  In principle, one could then simply
use the same TPC tracking parameters.  However, since this would
only be the case once, it was decided to actually perform a
separate fitting routine at this level in order to work in a mode
similar to that of having several sources of space point
information.

The global tracking then really performed an independent track fit
from the TPC.  It used a 3D helix model and incorporated energy
loss and multiple scattering in the TPC gas when doing the track
propagation.  The errors assigned to the space points were given
by a parameterization, based on real data, of the hit residuals as
a function of the track crossing angle with respect to the pad-row
direction.  This is also a case where one needs an iteration in
order to have information flowing both ways: the tracking needs
the errors in order to weigh each space point for a minimization
routine to obtain the best track parameters, yet the hit errors
depend on the crossing angle which is known once the track
parameters are set.

\subsection{Primary Vertex Finding}\label{sec:vertexFinding}

The offline reconstruction finds a primary vertex for each event
by propagating the global tracks through the field towards some
reference point which is close to the estimated vertex.  One can
choose a start point towards which to propagate the tracks either
in the transverse plane or along the beam line.  Since the
interaction point is much better determined in the transverse
plane  (RMS $\sim 0.5$ mm) than along the beam axis (RMS $\sim 90$
cm). The method at the heart of the routine is a $\chi^2$
minimization of the perpendicular distances from the track vectors
to a point. We can see the basis for this method in a simplified
scenario where all the tracks are straight. In the transverse
plane, the real vertex can only be a few millimeters away from the
central axis. After propagating the helices to their point of
closest approach to the reference point, we can then approximate
them as straight lines around this point as in a Taylor series.
This has the advantage of leading to an analytical solution. We
consider then an arbitrary vector $V_i$, with a unit vector
$\hat{e}_{V_i}$. The distance from a point $P = (x_0, y_0, z_0)$
(in cartesian coordinates) to the vector $V_i$ is the norm of the
vector obtained through the cross product of the unit vector
$\hat{e}_{V_i}$ with any line connecting the point $P$ to any
point $Q$ along the line of the vector $V_i$, $Q = (x_i, y_i,
z_i)$.  The line segment $\overrightarrow{PQ}$ is
$(x_0-x_i)\hat{x}+(y_0-y_i)\hat{y}+(z_0-z_i)\hat{z}$, and the
distance from $P$ to $V_i$ is
\begin{equation}\label{eq:vertexDistancePVi}
  |d_i| = |\overrightarrow{PQ} \times V_i| \ .
\end{equation}
A least squares minimization must employ a merit function
($\chi^2$) to assess the quality of the fit.  We take the partial
derivatives of the merit function with respect to the fit
variables, and set them equal to zero in order to solve for the
optimum values of the fit variables. For the vertex finder, the
merit function is the summation of the squares of the
perpendicular distances from the vectors $V_i$ to a point
$(x_0,y_0,z_0)$, with the appropriate weight given by the
uncertainty in the propagation of the track,
\begin{equation}\label{eq:vertexMeritFunction}
  \chi^2 = \sum_{i=0}^{N} \frac{d_i^2}{w_i^2} \ .
\end{equation}
By taking the partial derivatives with respect to $x_0$, $y_0$ and
$z_0$ and equating them to zero, we are left with a system of
three equations for three unknowns, the coordinates of the vertex
$x_v$, $y_v$ and $z_v$.

The vertex finder then has two main components.  One algorithm is
used to extrapolate a helix and calculates the coordinates of the
point of closest approach from a given point in the \textit{bend}
plane, defined relative to the $\vec{B}$ field which is along the
beam axis. The \textit{transverse} plane is defined perpendicular
to the beam axis. The other task is to perform the least square
fit to find the common vertex. Additional iterations can be
performed such that the linearization is done as close to the
actual vertex as possible, and rejecting outlier tracks in the
fit.  The details of the vertex finding algorithm are found in
Ref.~\cite{cebra:92}

The vertex resolution for high multiplicity events is
approximately 150 \micron, both perpendicular and parallel to the
beam axis. The vertex finding efficiency is 100\% for events with
more than 50 primary tracks in the TPC acceptance, decreasing to
60\% for those with fewer than 5 primary tracks.  We discuss the
vertex finding efficiency in more detail in
Section~\ref{sec:vertexEfficiency} as it is an important part of
the analysis of the multiplicity distribution.

\subsection{Primary Track Fit}\label{sec:primaryFitting}

\begin{figure}
  \centering
  \mbox{
  \subfigure[]{\includegraphics[width=.4\textwidth]{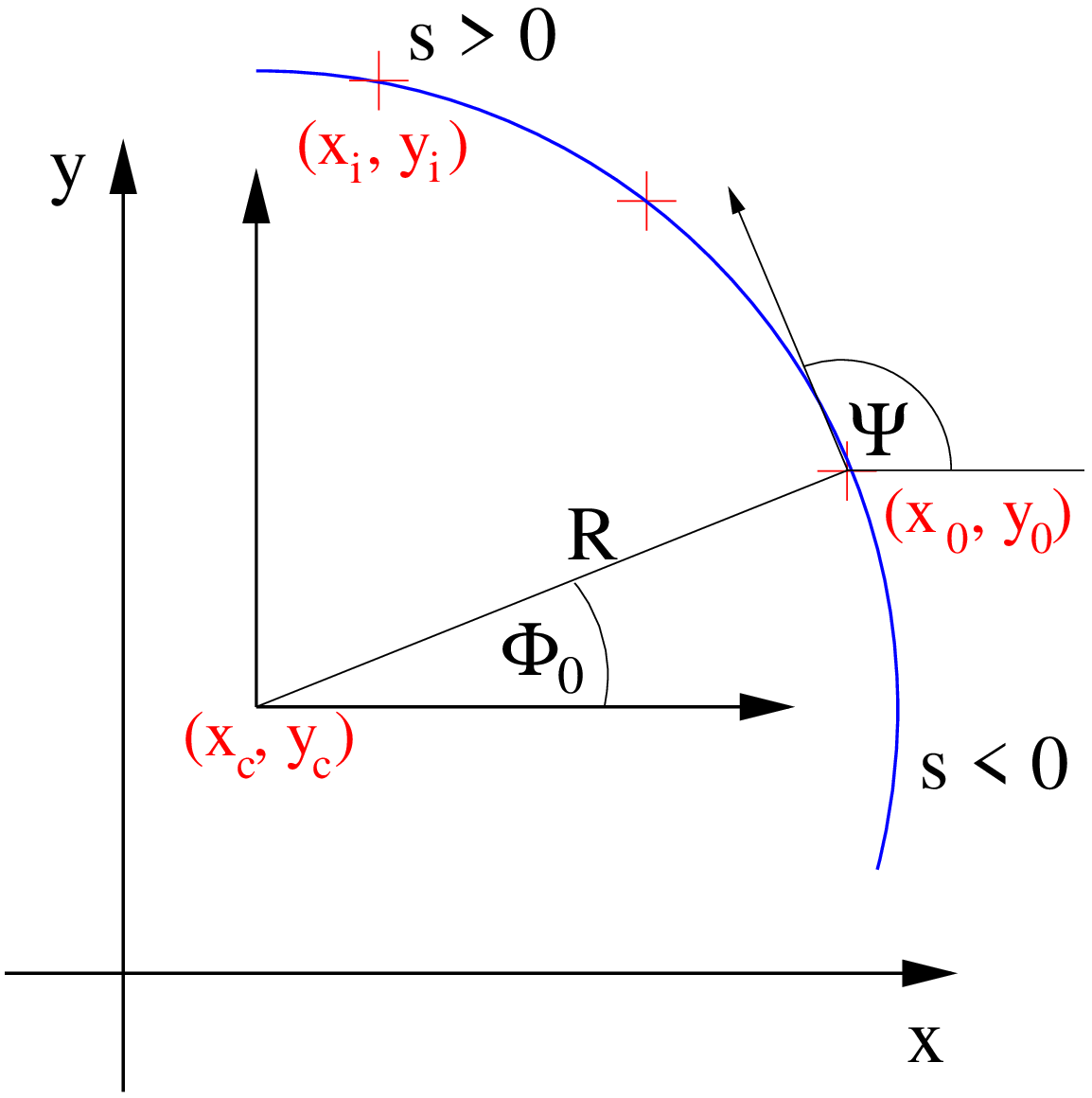}} \quad
  \subfigure[]{\includegraphics[width=.4\textwidth]{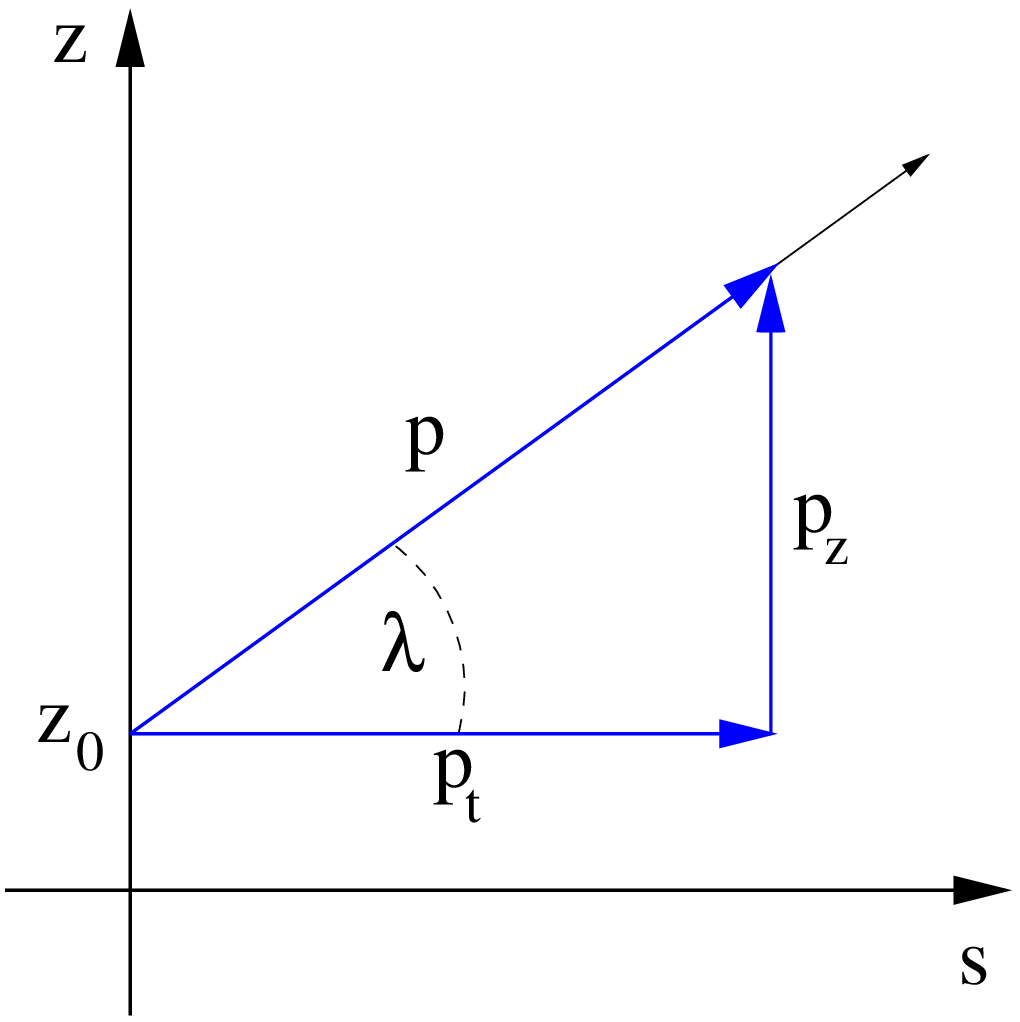}}}
  \caption[Planes for primary track helix fit]
  {Projection of a helix onto the transverse plane (a) and
  onto the bend plane (b)}
  \label{fig:helixPlanes}
\end{figure}

Once the vertex has been found, global tracks whose distance of
closest approach to the vertex is less than 3 cm  are chosen for a
refit using the vertex as an additional space point, yielding as a
result different parameters associated with \textit{primary}
tracks. For high multiplicity events, the error associated to the
primary vertex is much smaller than the error associated to TPC
space points, so even though we are only adding one additional
point to the track it can significantly improve the momentum
resolution for tracks that actually come from the vertex. This is
the reason for choosing the primary track parameters for the
analysis presented here. The track model in the STAR geometry and
magnetic field can be to first order represented by a helix, which
can be parameterized as
\begin{eqnarray}
    x &=& x_0 + R  \left( \cos(\Phi(s)) - \cos(\Phi) \right) \nonumber \\
    y &=& y_0 + R  \left( \sin(\Phi(s)) - \sin(\Phi) \right)
\label{eq:helixParam}
\end{eqnarray}
where $R$ is the radius of curvature, $\Phi = \psi + \pi/2$ and
$\Phi(s) = \Phi - (s/R)*\cos(\lambda)$, and $s$ is the path length
of the helix (see Fig.~\ref{fig:helixPlanes}). The radius of
curvature is related to the transverse momentum through
\begin{eqnarray}
    R &=& q / c \label{eq:radiusCurv}\\
    c &=& K\frac{qB}{\pt} \label{eq:curvature}
\end{eqnarray}
where $q$ is the charge of the particle, $c$ is the curvature, and
$B$ is the magnetic field. With the following units: $B$ in
KGauss, \pt\ in \gevc, and $q$ in units of the proton charge; the
value of $K$ to yield a curvature $c$ in cm is $K=0.000299$. The
parameters that define the helix ($x_0$, $y_0$, $\psi$, $\lambda$,
$q$, $c$) are obtained by two simultaneous 2D fits. We separate
the fits in the bend plane and in the transverse plane, shown in
Fig.\ref{fig:helixPlanes}.
In the transverse plane, the space-points of a helix will project
onto a circle, so the fit in the transverse plane ($x$ and $y$
coordinates in the STAR global coordinate system) is to a circle.
In particular, the curvature of the track is obtained from such a
fit, which relates to the transverse momentum as in
Eq.~\ref{eq:curvature}. With the knowledge of the curvature, we
can now focus on the fit in the bend plane.  We can think of the
helix as a string winding around a cylinder.  If we cut the
cylinder and unfold it into a plane, the helix segments will be
mapped into straight lines in this plane. We then perform a
straight line fit in the bend plane $z = (\tan \lambda)s + z_0$
where the coordinates for the fit are the path length ($s$) of the
helix and the beam axis coordinates ($z$). The parameter $\lambda$
is called the \textit{dip} angle in the helix parameterization,
and it is identical to the polar angle $\theta$ in the definition
of pseudorapidity, Eq. \ref{eq:pseudorapidity}.

It is possible that secondary and decay tracks that are fit with
the constraint of going through the primary vertex will have worse
track parameters than those obtained from the global track fit.
This is especially important for high-\pt\ tracks, since biasing
the tracks towards the primary vertex can artificially increase or
decrease the curvature. For the majority of the tracks however,
and in particular for true primary tracks, the resolution is
improved by the primary track fit.  Tracks for which the primary
fit fails are flagged and not used in the analysis.

\subsubsection{Momentum Resolution}

\begin{figure}[htb]
  \begin{center}
  \includegraphics[width=.7\textwidth]{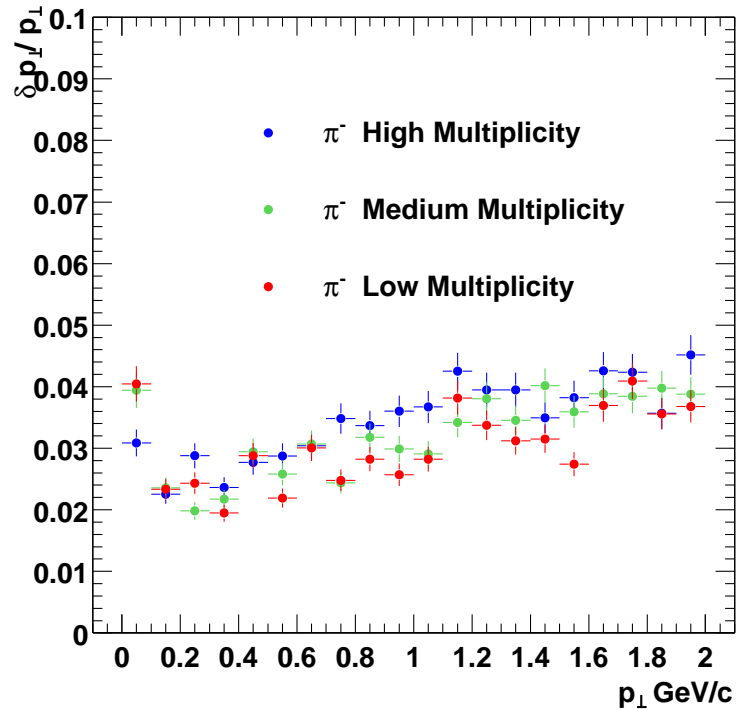}
  \caption{$\pi^-$ momentum resolution}
  \label{fig:ptresol}
  \end{center}
\end{figure}

The \pt\ momentum resolution for \piminus\ for various
multiplicities is shown in Fig.~\ref{fig:ptresol}.
This is obtained from embedding pions into real events using the
TPC simulation discussed in the next section.  The effects of the
momentum resolution in the measurement of the \pt\ spectra is
discussed in Section~\ref{sec:momentumResolution}.

\subsection{$V_0$'s, $\Xi$'s, and Kinks}
A specific part of the main STAR reconstruction code is dedicated
to the topological identification of neutral and charged particle
decays. Neutral particles such as the \kzeros\ and $\Lambda$ can
be reconstructed by identifying the secondary vertex, commonly
called $V_0$ vertex, of their charged daughter decay modes,
$\kzeros \rightarrow \pi^+\pi^-$ and $\Lambda \rightarrow
p+\pi^-$.  The more complicated ``cascade'' decay schema of the
$\Xi$ and $\Omega$ baryons can also be reconstructed by a series
of topological pattern recognitions using the $\Lambda$'s found in
the first iteration through the decays: $\Xi^- \rightarrow \Lambda
+ \pi^-$ and $\Omega^- \rightarrow \Lambda + K^-$.

The charged kaon and pion one-prong decays found in the TPC volume
are also reconstructed at this level ($K \rightarrow \mu + \nu$).
The topology signature of this decay is that of a track which
decays in flight while propagating in the gas of the TPC, with the
daughter charged track following a slightly different direction,
and is called a \textit{kink} decay.

These steps are important for finding strange hadrons, \ie\
$\lambda$ and $\Xi$ candidates using the $V_0$-finder algorithm
and charged kaons using the kink-finder algorithm.  In particular,
the identification of charged kaons through the kink decay
technique extends the available phase space to high momenta, where
the \dedx\ measurement no longer has any resolving power.  For
recent results from the analysis of strange hadrons in STAR, see
\eg\ \cite{caines:01}.

The secondary vertices are then reconstructed using the global
tracks, propagating them towards the primary vertex and requiring
that the $V_0$ vertex of the track pair should not be close to the
primary vertex.  In addition, when combining the pair momenta to
obtain the momentum of the parent, one requires that the parent
should point back to the primary vertex. This is the last
important part of the global reconstruction. Additional pieces of
the code mainly organize the reconstructed information into the
STAR Data Summary Tape (DST) to be written to disk for further
analysis. Although the main tuning of cuts to find secondary
vertices is done at the analysis level, the initial stages of
$V_0$ finding are done at the reconstruction level since part of
the input are the space points, which are not saved to disk to
minimize data volume. We turn our attention now to the TPC
simulation.

%
%
\chapter{Detector Simulation}
\label{ch:DetectorSimulation}
 This analysis relies heavily on a realistic simulation of
the response of the TPC.  An estimation of the backgrounds and
tracking efficiency is best done through such a simulation, and it
is therefore necessary to include a significant amount of physics
details. The primary function of the TPC Response Simulator (TRS)
\cite{lasiukTRS:99,gongTSS:95} is to reproduce the electronic
signals produced by tracks ionizing the gas of the TPC. The
physical processes that are implemented are the drift of the
ionized electrons in the gas, the amplification of the signal in
the sense wires, the induction on the readout pads, and the
response of the readout electronics which produce digitized data.
While the software was designed to cope with varying levels of
detail in each of the different simulation stages, we describe
here the implementation that was used for the analysis stage in
the past year.

\section{Detector Response Overview}
The algorithmic basis for the response simulations is that
developed in detail by the ALEPH TPC group, described in Ref.
\cite{blum:93}, the standard reference for drift chambers.  The
TRS package contains parameterizations of the TPC response at
various levels of detail.

As part of the requirements, TRS was designed to deal with two
sources of ionization of the TPC gas: charged tracks and laser
events. The charged tracks create clusters whose distribution of
amplitudes (\ie\ number of electrons in a cluster) is Landau-like,
whereas the laser creates uniformly distributed ionization,
resulting in Poisson distributed cluster amplitudes.

For the simulations used in the analysis, the magnetic field was
taken to be uniform with a value of 0.25 T in the $z$ direction,
and parallel to the electric field.  During the analysis of the
year 2000 data, several corrections needed to be applied to the
raw data to take into account various distortion effects and
modifications to the idealized geometry.  These modifications were
necessary for analyses such as the reconstruction of resonance
decays and tracking of high \pt\ particles.  However, for the
analysis of low momentum charged tracks up to 2 \gevc, these
effects were negligible.  The analysis was repeated every time new
distortion and geometry corrections were added to the
reconstruction chain and the results did not vary significantly.
For example, a comparison of the measured raw multiplicity in the
midrapidity region before and after distortion corrections using
different event samples showed that they were consistent to better
than 1 \%. Therefore, it was sufficient to do the simulations in
this idealized scenario.

There are four main processes that are essential for the
simulation to reproduce the behaviour of a TPC.

\begin{itemize}
  \item \textit{Ionization Transport} -- charge transport of the
  ionization and its deposition in the active region of the
  readout chambers.
  \item \textit{Charge Collection} -- electron/ionization
  collection on the sense wires of the multi-wire proportional
  chamber (MWPC).
  \item \textit{Analog Signal Generation} -- charge induction on
  the pad plane and generation of the time evolution of the analog
  signals on the pads.
  \item \textit{Digital Signal Generation} -- conversion of the
  analog signals into ADC counts.
\end{itemize}

We now discuss briefly each of the processes.

\section{Ionization Decomposition and Transport}

The input into the simulation of the TPC response is the charge
ionization left by tracks in the sensitive volumes of the detector
generated by GSTAR \cite{gstar:00}, the GEANT \cite{Brun:1978fy}
implementation of the STAR detector. The ionization transport
takes the charge deposited in these active areas of the detector
and transports it through the field cage structure to the read-out
plane.  The GEANT simulation provides the amount of energy
deposited ($dE$) over a given path-length ($ds$) by a particle
with momentum $\vec{p}$. Given the average ionization potential of
the gas, one can calculate the total number of electrons such that
the transport can be done at the segment level ($dE$) or to break
each segment into parts with the limiting case of transporting
each ionization electron individually.  This provides a mechanism
which allows the possibility to distinguish between a detailed
microscopic simulation and a macroscopic parameterization. In
addition, one can also vary the length of the segment that is
transported (and subsequently processed). Thus, the granularity of
the simulation can be adjusted to provide different levels of
detail.

The ionization is distributed on the pad plane according to
distributions which characterize the effects of diffusion.  The
role of the charge transporter is to alter the $x$ and $y$
positions according to transverse diffusions, to alter the $z$
position to reflect the drift time with longitudinal diffusion
folded in, and to alter the amount of charge that reaches the
read-out plane.

\section{Charge Collection and Amplification}

Once the electrons arrive at the read-out plane, they must be
collected by the individual anode/sense wires of the multi-wire
proportional chamber (MWPC).  It is here that the avalanche
process multiplies the signal of several tens of electrons to
several $10^3-10^5$ electrons, depending on the potential of the
wires.

The basic principle of avalanche creation in a proportional wire
is as follows.  As an electron drifts toward the wire it travels
in an increasing  electric field.  In the vicinity of the wire,
the field $\vec{E}$ at radius $r$ is given by the linear charge
density $\lambda$ on the wire
\begin{equation}\label{eq:EfieldWire}
  \vec{E} = \frac{\lambda}{2\pi\epsilon_0}\frac{\hat{r}}{r} .
\end{equation}
The electron's path is directed towards the wire.  The trajectory
also terminates in the wire in the presence of a magnetic field if
the electric field is sufficiently strong.  Once the electric
field surrounding the electron is strong enough, the electron will
have enough energy to ionize the gas and still pick up sufficient
energy between collisions to ionize the gas further, starting an
avalanche. The number of electrons continues to grow until all
electrons reach the wire.  The process develops over as many
mean-free paths as there are generations of electrons, typically
50-100 \micron, and takes less than 1 ns for most gases.  Since
the signal is proportional to the number of electrons collected,
the name ``proportional wire''  was adopted.  For a review see
\cite{blum:93} and \cite{knoll:79} which also contains a list of
texts.

The distribution of avalanches started by single electrons can be
modelled as a stochastic process (see \cite{blum:93}, p.144).  A
simple description is given by the Yule-Furry process, after the
authors who first used it to describe biological population
growth. The idea was applied to proportional avalanches by Snyder
and Frisch (see \cite{byrne:62} for references), and is briefly
described in what follows.

We start from the idea that as the avalanche (or other stochastic
process) develops, the variation of the number $n$ of ions (or
electrons) is followed as a function of some parameter $t$.  This
parameter can be thought of as the time, but it is a bit more
general in the sense that we need only require that its increase
describes the progress of the process in the order given by the
causal sequence of events.  Therefore, $t$ need only be a
monotonic function of time.  We are interested in the probability
$P(n,t)$ that at some `time' $t$ there are $n$ electrons in the
avalanche.  The initial condition is that the avalanche starts
with a single electron, \ie\
\begin{equation}\label{eq:y-fInit}
  P(1,0) = 1  \ .
\end{equation}

We suppose that the probability for the birth of one electron in
any interval $\Delta t$ is proportional to the number of electrons
$n$ with some proportionality constant $\lambda$.  The solution to
the corresponding differential equation giving the probability
distribution at a given $t$ is
\begin{equation}\label{eq:y-fSolution}
  P(n,t) = e^{-\lambda t} (1 - e^{-\lambda t})^{n-1} \ .
\end{equation}
The mean $\bar{n}$ and the variance $\sigma^2$ are
\begin{equation}\label{eq:y-fMean}
  \bar{n} = \sum_{n=1}^{\infty} nP(n,t) = e^{\lambda t}
\end{equation}
\begin{equation}\label{eq:y-fVariance}
  \sigma^2 = \sum_{n=1}^{\infty} n^2P(n,t) - \bar{n}^2 = e^{\lambda
  t}(e^{\lambda t} - 1) \ .
\end{equation}
Rewriting \ref{eq:y-fSolution} with \ref{eq:y-fMean} the
distribution becomes
\begin{equation}\label{eq:y-fSolution2}
  P(n)=\frac{1}{\bar{n}}\left(1-\frac{1}{\bar{n}}\right)^{n-1} \ ,
\end{equation}
where the `time' variable no longer appears.  In the limit
$\bar{n} \rightarrow \infty$, which is appropriate for avalanches,
we obtain
\begin{equation}\label{eq:y-fSolution3}
  P(n) = \frac{1}{\bar{n}} e^{-n/\bar{n}} \ ,
\end{equation}
\begin{equation}\label{eq:y-fSigma2}
  \sigma^2 = \bar{n}^2 \ .
\end{equation}

We see that the Yule-Furry process has an exponential signal
distribution.  Subsequent theoretical refinements to this simple
expression which were made in order to take into account effects
like the asymmetric growth of the avalanche profile as well as
saturation effects.  One such refinement is the Byrne process,
which introduces the idea that a fluctuation to larger $n$ in the
first part of the avalanche reduces the rate of development in the
second part (so an additional parameter has to be introduced,
whose effect is to suppress the small amplification factors). The
Byrne process includes the Yule-Furry process as a special case
when the additional parameter approaches zero.  The solution to
the differential equations for this process is given by the Polya
(or negative binomial) distribution function.

Exponential behaviour  at low to moderate gas gains (\ie\ $<10^4$)
is observed experimentally.  In parallel plate geometry however,
slight deformation from exponential shape is observed at gas gains
above $10^5$.  This is probably due to self-saturation effects
which become important in the presence of space charge. Since the
STAR TPC is generally operated at low gas gains, the simple
exponential is found to be acceptable for the simulations.

It is important to note that in any drift chamber operation, the
effect of the fluctuations in gas gain is to simply degrade the
attainable space-point resolution, and for this purpose it is not
critical to use the exact functional form of the avalanche yields.
The effect of degradation is the process that the simulation of
the gas-gain fluctuation is attempting to reproduce, and the
simple exponential should be sufficient in this regard.

Once the charge has been amplified, we can proceed to calculate
the amount of charge induced on the cathode pad plane.

\section{Analog Signal Generation}

This simulation step has three main parts:

\begin{itemize}
  \item Determination of the charge induced on single pads from
  the charge collected on the anode wires
  \item Sampling of the induced charge signals in time according to the
  electronics response (\ie\ pre-amplifier and shaper)
  \item Distribution of the \textbf{analog} charge into time bins
\end{itemize}

\subsection{Charge Induction}

The charge induced on a grounded pad plane by a point charge $q$
located a distance $d$ above the plane can be calculated by the
method of images.  The charge density $\sigma$ on the plane is
given by:
\begin{equation}\label{eq:chargeDensityPlane}
  \sigma(x,y) = \frac{1}{2\pi}\frac{qd}{((x-x_o)^2+(y-y_o)^2+d^2)^{3/2}}
\end{equation}
where the charge is located at the position $(x_o,y_o)$.  However,
the typical geometry of a MWPC is a bit more elaborate. The
proportional wires normally form a plane between \textit{two}
grounded cathode planes.  In addition, the charge is not
point-like, but in the form of a line of charge with density
$\lambda$.  We can simplify and locate it in the middle of the
parallel plates.  In order to calculate the charge density induced
on the pad plane we thus have to include all higher-order
multi-pole terms.  The images are alternately negative and
positive, situated at $z_n = \pm (2n+1)d$, $(k=1,2,\cdots)$ and
$z_o = d$.  The total charge density is obtained by integrating
over $y$ and doing the sum
\begin{equation}\label{eq:chargeDensityPlates}
  \sigma(x) =
  -\frac{\lambda}{\pi}\sum_{n=0}^{\infty}(-1)^n \frac{(2n+1)d}{(x-x_o)^2+(2n+1)^2d^2}
  = -\frac{\lambda}{4d \cosh\left(\frac{\pi (x-x_o)}{2d}\right)}
\end{equation}
which is called the \textit{Endo} function. The derivation was
done without taking into account the limited extent of the pads in
the y direction, \ie\ for the limiting case of zero width to
length ratio $w/l$.

The effect of finite geometry of segmented cathodes can be taken
into account with the addition of another parameter. We can
rewrite Eq.\ref{eq:chargeDensityPlates} as
\begin{equation}\label{eq:chargeDensityPlates2}
  \sigma(x) = C_1 \frac{1-\tanh^2\left(\frac{\pi (x-x_o)}{4d}\right)}{1+\tanh^2\left(\frac{\pi (x-x_o)}{4d}\right)}
\end{equation}
where $C_1$ is a normalization constant.  Eq.
\ref{eq:chargeDensityPlates2} can be used to generalize the Endo
function to take into account finite geometry effects due to
segmented cathode pads, \ie\ $w/l>0$.  By introducing the constant
$C_2$
\begin{equation}\label{eq:chargeDensityGatti}
  F(x) = C_1 \frac{1-\tanh^2\left(\frac{\pi (x-x_o)}{4d}\right)}{1+C_2 \tanh^2\left(\frac{\pi (x-x_o)}{4d}\right)}
\end{equation}
we arrive at the generalized solution to the distribution of
charge induced on a grounded pad plane, and is usually dubbed the
\textit{Gatti} function.

A comparison of the Gatti and Endo functions, along with a
Gaussian (for comparison purposes) is given in
\begin{figure}
  \centering
  \includegraphics[width=.6\textwidth]{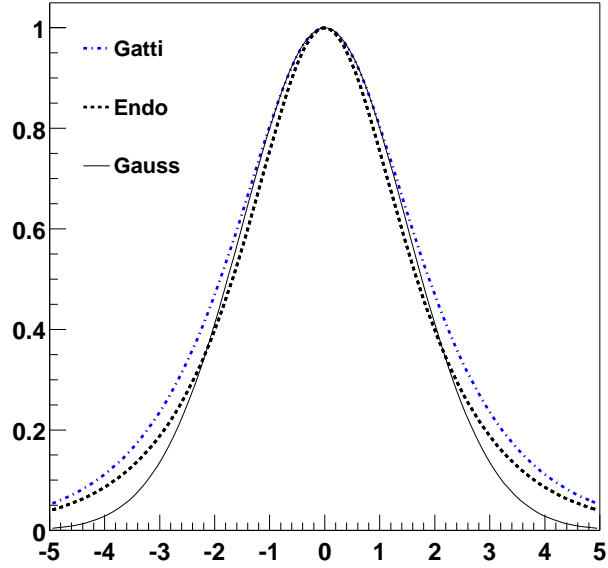}
  \caption[Pad response functions]{Comparison of Gatti, Endo and Gaussian functions as profiles of the pad-response function.}
  \label{fig:padResponseFunctions}
\end{figure}
Figure \ref{fig:padResponseFunctions}.  For reference, the
functions were normalized to 1 at $x=0$ for the plot and the
parameters were $\sigma=1.5$ for the Gaussian, $d=2$ for the Endo
function and $d=2$, $C_2 = 0.5$ for the Gatti function.

For the case of the TPC, the quantity of interest is the total
amount of charge ($Q$) induced per pad, so we have to integrate a
given pad-response function, which gives the charge
\textit{density}, over the area of a pad:
\begin{equation}\label{eq:totalChargeFromPRF}
  Q = \frac{1}{2\pi}\int_{y_l}^{y_u}\int_{x_l}^{x_u}
  \sigma(x,y)dx dy
\end{equation}

The advantage of using such functions is that they allow the
production of longer tails which have non-Gaussian
characteristics.  The tails are an important feature to
understand, as they determine the efficiency of the ionization
collection which is of relevance in the study of \dedx\
resolution.  As an additional case, the pad response function was
measured for the STAR TPC pads in the directions transverse to and
parallel to the anode wires.  This allows us to calculate the pad
response as a convolution of the transverse diffusion function
with a parameterization of the measured pad response function.

\subsection{Sampling of Signal in Time and Shaper Response}

Once the amount and centroid of the charge distribution on each
pad is determined, this charge can be sampled in time, thereby
emulating the response of the analog electronics. The signals are
generated by superimposing each analog signal from each of the
avalanches that induce a signal on the pad plane. This permits one
to vary the shaping time of the electronics independently of the
width of the pad response function.  In reality, the time
evolution of the signal that develops on the wires is almost
entirely due to the motion of positive ions away from the wire.
This produces a signal with a long tail. For the STAR geometry,
the duration of this signal is on the order of $\sim 62\ \mu$s. In
order to make the detector faster, the signal is differentiated
after a characteristic time -- the shaping time of the
pre-amplifier.  The trade-off is that only a fraction of the total
charge is seen by the downstream electronics, \ie\ the ADC. The
fraction $F$ of the charge is given by
\begin{equation}
F = \frac{\ln \left(1 + \frac{t_m}{t_o}\right)}
         {\ln \left(1 + \frac{t_s}{t_o}\right)}
         \label{eq:FracShapingTime}
\end{equation}
where $t_m$ is the length of time the long, undifferentiated
signal would persist ($\sim 62\ \mu$s), $t_o$ is the
characteristic time of the development of the signal ($\sim 1\
\mu$s) and $t_s$ is the shaping time of the pre-amplifier ($\sim
180$ ns).

For STAR this means that the signal shape is dominated by the
shaping properties of the electronics. Therefore, the long time
response of the chamber is parameterized in the electronics
processing section of the simulator, instead of modelling the
motion of the positive ions. The shaper response function for the
case of one stage differentiation and two stage integration used
in STAR is
\begin{equation}\label{eq:shaper}
  g(t,\tau) = \Phi(t)\left(\frac{t}{\tau}\right)^2e^{-t/\tau}
\end{equation}
where $\Phi(t)$ is the step function, and the time constant $\tau$
is 55 ns for the STAR electronics.  This function is then
convoluted with a Gaussian to parameterize the effect of a
longitudinally diffused cluster to yield the final response of the
shaper.

Once we obtain the functional form of the signals induced on a pad
over the read-out period of the TPC electronics, we can distribute
the analog signal into discrete time bins.  This sampling
simulates the behaviour of the Switched Capacitor Array (SCA) in
the front end electronics.  Essentially, this entails integrating
the amount of charge in a time interval $\Delta t$, which is
determined by the SCA sampling frequency.  Simulated chamber noise
and electronic noise are also added at this point.

\subsection{Digital Signal Generation}

Once the analog charge is distributed into time bins on the pads,
we can digitize the signal.  This step is done via a simple
conversion from voltage to ADC counts. This is the final stage in
the TPC simulation, and the pixel values are then used as input to
the reconstruction chain in exactly the same way as the raw data.

\section{Embedding}\label{sec:embedding}
One of the main purposes of the detector simulation response is to
obtain estimates of tracking corrections to the raw particle
yields.  It is of great importance to have a reliable detector
simulation as detailed in the previous section.  To make the
environment as realistic as possible, we use a procedure commonly
known as \textit{embedding}. The idea is to take a real event and
embed into the raw data file the signal from a few simulated
tracks at the level of ADC counts.  The simulated signal is
obtained via a GEANT Monte Carlo simulation of the energy
deposited by the tracks in the detector volume followed by a
simulation of the detector response as detailed in the previous
section.  The final simulated ADC signal is then convoluted with
the raw data from the real event and then fed into the STAR
reconstruction software chain. Since we have the information from
the GEANT simulation, we can associate the reconstructed tracks to
the corresponding Monte Carlo track.  The association is done at
the hit level, since from the reconstruction side the hits are the
``seed'' from which the track parameters are built.

The embedding procedure combines the advantages of having a very
realistic event environment along with having a controlled track
population. From the comparison of reconstruction output to
simulated input we can obtain information about many interesting
performance diagnostics of the detector.

\begin{figure}
  \centering
  \includegraphics[width=.6\textwidth]{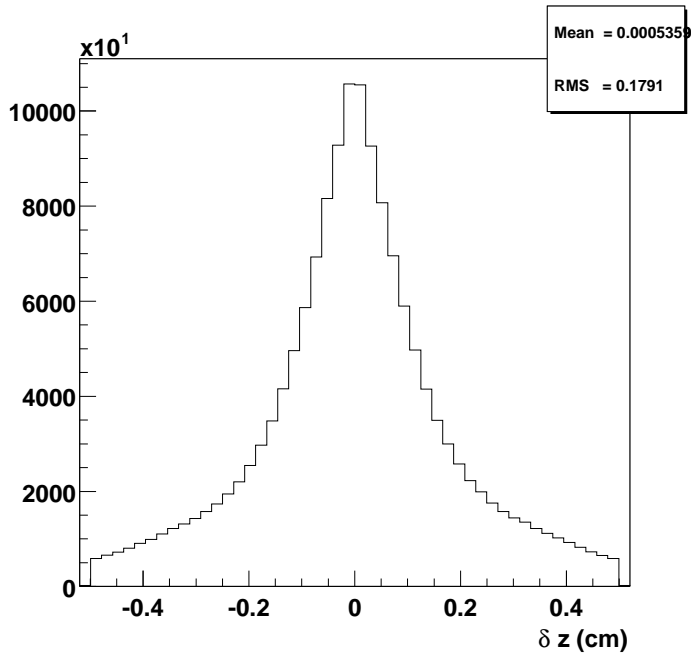}
  \caption{TPC hit position resolution in the drift direction.}
  \label{fig:hitResolution}
\end{figure}

Figure \ref{fig:hitResolution} shows the position resolution for
hits in the TPC in the drift direction.  The RMS width is found to
be 180 \micron, a similar analysis in the padrow direction yields
an RMS width on the order of 120 \micron.
Once we have a match between the TPC hits, we can use this
information to make a matching between tracks.  For every
reconstructed track, we scan through its reconstructed space
points and find their corresponding Monte Carlo hits.  These
simulated hits have the necessary information of the track that
generated them, and so form the basis for the track match.  It is
this track match information that we use for the determination of
the reconstruction efficiency.  We discuss the corrections to the
raw spectra obtained through the embedding procedure and through
background simulations in the next chapter.

%
%
\chapter{Analysis of Charged Hadron Spectra}
\label{ch:nchAnalysis}
\section{Event-wise Studies}
In this section we delineate the relevant parts of the analysis
that was performed to obtain the multiplicity distribution.  This
is also the baseline for any other analysis.

\subsection{Trigger and event selection}
We use the ZDC and the CTB for triggering. All the runs that we
used for this analysis were recorded with a minimum bias trigger,
with no pre-scale. The minimum bias trigger consisted of a
coincidence requirement between the East and West ZDC's. This
trigger proved to be better than 99\% efficient.  At the beginning
of the run, however, this had yet to be established. The main
concern was that there might be an inefficiency in this trigger
for the most central collisions, where very few spectator nucleons
remain to produce a coincidence in the calorimeters.  Therefore, a
high CTB signal, indicative of a high multiplicity in the
mid-rapidity region, was used as an additional trigger condition.
The ZDC trigger thresholds were set such that a single nucleon
hitting the calorimeter would generate an acceptable signal.  The
CTB threshold was set to accept the highest $\sim 30\%$
multiplicities.

\begin{figure}[htb]
  \centering
  \includegraphics[width=0.7\textwidth]{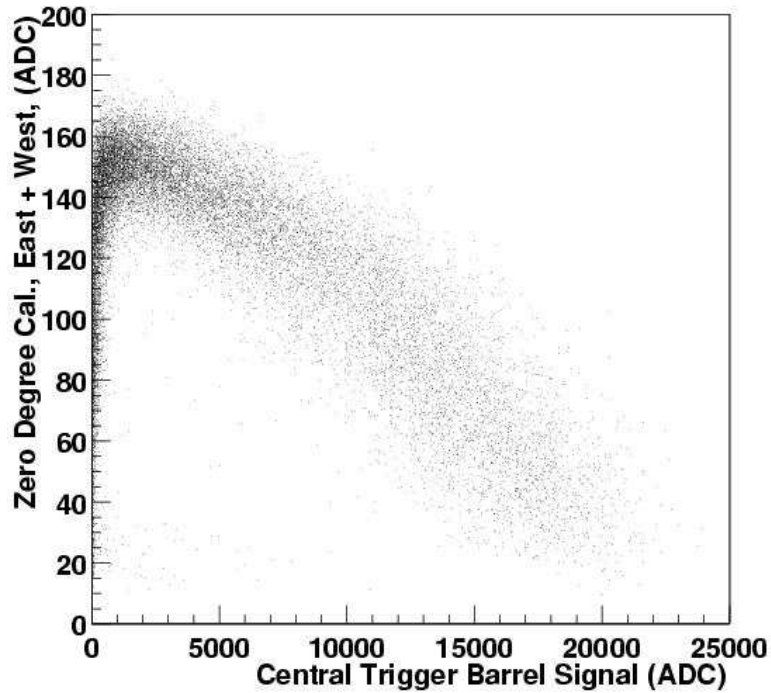}
  \caption[The ZDC \vs\ CTB Trigger Signal]
  {The ZDC (sum of East and West calorimeters)
  signal is plotted \vs\ the corresponding
  CTB signal.  The most central events have the highest CTB signal (high multiplicity)
  and a low ZDC signal (very few spectator neutrons).}
  \label{fig:zdcvsctb}
\end{figure}

A plot of the ZDC \vs\ CTB trigger signals for a subset of the
events used in this analysis is shown in Fig.~\ref{fig:zdcvsctb}.
The peripheral collisions are in the lower left corner, where both
the CTB signal and the sum of the East and West ZDC signals are
small. This is indicative of low multiplicity at mid-rapidity and
a scarce number of dissociation neutrons.  In the collider
geometry, neither the excited nucleus nor the dissociation protons
reach the calorimeters because their trajectories are bent by the
beam optics.  As the overlap of the colliding nuclei (\ie\ the
\textit{centrality}) increases, more neutrons reach the
calorimeters and the ZDC signal increases. Likewise, the
multiplicity at mid-rapidity increases. After a certain point, the
collision is sufficiently central that few neutrons reach the ZDC
while the multiplicity continues to increase.  The ``boomerang''
shape observed in Fig.~\ref{fig:zdcvsctb} is therefore the result
of the correlation between impact parameter $b$ and multiplicity
on the CTB side, and of a dual behaviour on the ZDC side:
correlation between $b$ and number of neutrons at high impact
parameter, and anti-correlation for central collisions.

An important input into this analysis is the \textit{hadronic
cross section} in \AuAu, \sigmahad.  This value is necessary to
normalize the multiplicity distribution. The calculated value is
$\sigmahad$ = 7.2 barn from Eq.~\ref{eq:TaaSigmaAuAu}, but it is
desirable to have a measurement confirming this expectation. The
trigger requirement of a ZDC coincidence  is not only sensitive to
the hadronic cross section, but also to the mutual Coulomb
dissociation of the two Au nuclei. In this process, there will be
a pair of correlated forward- and backward-going neutrons, but no
measurable tracks in the mid-rapidity region. The ZDC's therefore
are sensitive to the sum of the hadronic + mutual Coulomb
dissociation cross sections, which we denote by $\sigma_{xn,xn}$,
\ie\ the cross section in which at least one neutron is detected
in each ZDC detector. Since the same ZDC's are used in all
interaction zones at RHIC, this combined cross section can be
measured by all 4 RHIC experiments independently. The STAR
preliminary result, measured by the van der Meer scan technique
\cite{zhangbu:00} is $\sigma_{xn,xn} = 8.9 \pm 0.3_{\mathrm{stat}}
\pm 0.7_{\mathrm{sys}}$ barn. This value has also been measured in
an independent analysis by the RHIC accelerator crew
\cite{drees:01}.  The value reported in a calculation
\cite{baltz:98} is $\sigma_{xn,xn} = 10.9 \pm 0.6$ barn, and seems
to be outside of the allowed range of the systematic uncertainty
in the measurements.  Attempts to resolve this discrepancy are
underway.   In order to arrive at a measured hadronic cross
section, \sigmahad, it is also necessary to disentangle the
contributions from the hadronic and the Coulomb processes.  Since
a $\ppbar$ inelastic collision at \sqrts\ = 130 \gev\ produces on
average $\sim 2.4$ (see Eq.~\ref{eq:multScalingRootS}) charged
particles per unit pseudorapidity at mid-rapidity and we expect to
identify the vertex for hadronic events, to first order this
fraction can be measured as
\begin{equation}
\label{eq:fractionCrossSectionMeasurement}
  \frac{\sigmahad}{\sigma_{xn,xn}} =
  \frac{\mathrm{Events\ with\ Vertex}}{\mathrm{Total\ Triggered\ Events}}
\end{equation}

This simple ratio must be corrected for the vertex-finding
efficiency and acceptance that reduce the counts in the numerator
and for backgrounds such as beam+gas that generate a trigger and
are therefore counted in the denominator. The vertex finding
efficiency is important in this analysis and for the determination
of the shape of the \hminus\ multiplicity distribution at low
multiplicity.

\subsection{Vertex Acceptance}

The vertex reconstruction for the summer 2000 run was undertaken
essentially up to the limits of the TPC acceptance, \ie\
$|\zvertex| < 2$ m where \zvertex\ denotes the position of the
primary vertex along the $z$ direction parallel to the beam.
Although it is possible to try to obtain a vertex as long as there
are tracks in the TPC, this was not pursued for the following
reasons. If an event occurred outside of this region, the TPC sees
only the forward- or the backward-going particles.  This also has
the disadvantage of reducing the accuracy of the vertex
determination, since the $z$ position of tracks going
perpendicular to the beam axis contribute significantly to the
determination of the vertex. In addition, the CTB in this case is
not triggered on multiplicity at mid-rapidity, but rather on
multiplicity forward or backward of mid-rapidity with all tracks
having a large dip angle. To avoid these systematic effects, the
vertex finder was set to abort when it determined that the vertex
lay outside the TPC limits.

\begin{figure}[htb]
  \centering
  \includegraphics[width=0.9\textwidth]{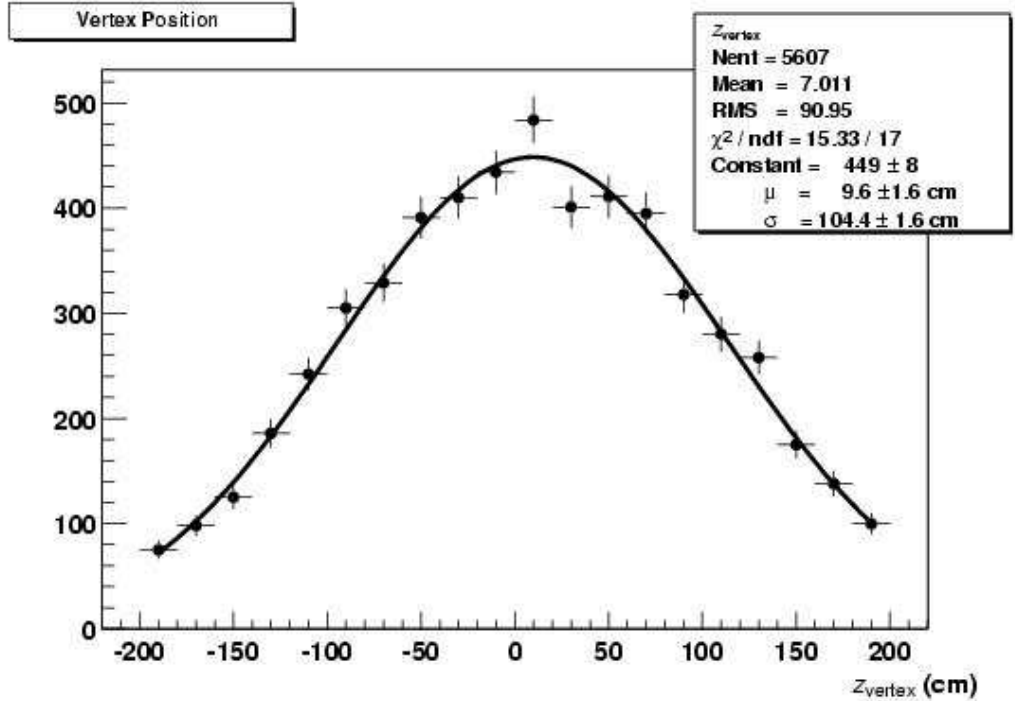}
  \caption[The \zvertex\ distribution for 10K triggered events.]
  {Distribution of reconstructed vertex $z$ positions, \zvertex, for 10,000 triggered
  events.  The limits of the TPC are at $\pm$2 m from the center.
  The acceptance for events used in the multiplicity analysis was placed
  at $\pm 95$ cm.}
  \label{fig:zvtx}
\end{figure}

Due to the large size of the beam diamond, the standard deviation
of the vertex Z position was $\sim$ 100 cm. This results in valid
collision events, even at high multiplicity, without a vertex
determination. In addition, in order to keep a flat acceptance for
tracks with $|\eta| < 0.5$, we restricted the multiplicity
analysis to events with a vertex between $\pm 95$ cm, about a $\pm
1 \sigma$ cut. To correct for the vertex acceptance, we assume
that the distribution is Gaussian and from the fit parameters
estimate the fraction of events that lie outside our acceptance
cut. As an example, the $\zvertex$ distribution for 10K triggered
events from one of the runs taken in September 2000 is shown in
Fig.~\ref{fig:zvtx}.

For the distribution in Fig.~\ref{fig:zvtx}, the acceptance
correction factor is 1.57. Analyzing more runs for a total of 166K
triggered events, we find an acceptance correction factor of 1.68,
(a $\pm 1\sigma$ cut for a Gaussian distribution would yield a
correction of 1.46, so our cut is tighter). With the vertex
efficiency and acceptance correction, we can then determine the
numerator in Eq.~\ref{eq:fractionCrossSectionMeasurement}.  For
the denominator, we must estimate the backgrounds to our
minimum-bias trigger.

\subsection{Vertex Efficiency}\label{sec:vertexEfficiency}

We know that our offline vertex reconstruction is not 100\%
efficient, therefore we need to correct for this effect when
measuring the multiplicity distribution. The vertex reconstruction
is based on tracing the path of reconstructed tracks back to a
common space point. Since tracking is done first, before finding
the vertex there is no \textit{a priori} knowledge of which tracks
are primary and which ones are not.  The vertex-finding therefore
depends on all \textit{global tracks} (Sec.
~\ref{sec:globalTracking}) found in the event. We will then
characterize the efficiency as a function of the number of global
tracks (\Nglobal) in the event.  The vertex-finding efficiency
(\effvtx) affects mainly the low multiplicity events.   The reason
is that these events provide the vertex-finding algorithm with
very little information to work with, \ie\ very few global tracks.
We find that for events with more than $\sim 100$ global tracks,
the efficiency is  $\sim 100\%$. To confirm the findings based on
software, a visual analysis of $\sim 100$ events was performed,
with a resulting lower bound on the efficiency for events with
\Nglobal\ = 100 of \effvtx\  > 98\%.

\begin{figure}[htb]
  \centering
  \includegraphics[width=0.8\textwidth]{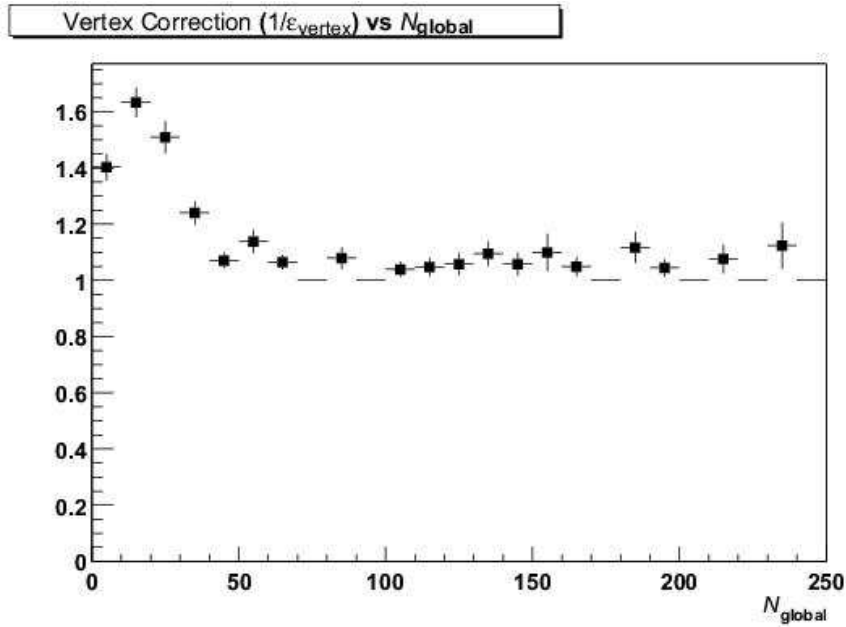}
  \caption[Vertex efficiency correction vs. \Nglobal]
  {Vertex finding efficiency correction factor as a function of the
  number of global tracks, \Nglobal.}
  \label{fig:vtxeffcorr}
\end{figure}

The more important part is the efficiency at low \Nglobal. To get
a handle on this number, we used \Hijing\ events generated with
large impact parameter $b$ = 12 - 20 fm. At these impact
parameters, \Hijing\ should be a reasonable model, since basically
only geometry and the $pp$ cross section play a role and nuclear
specific effects are not expected to influence the results. These
events were then processed through the STAR reconstruction chain
and a record was kept of the number of events generated and
whether the vertex was or was not found for each event. In this
way, we obtain a plot of the vertex efficiency correction factor
(1/\effvtx). The correction depends on \Nglobal\ and is
illustrated in Fig.~\ref{fig:vtxeffcorr}.

To take this correction into account in the \hminus\ multiplicity
distribution, the correction in Fig.~\ref{fig:vtxeffcorr} was used
as a weight for each event depending on the number of global
tracks. Since the \hminus\ multiplicity distribution refers to
negative primary tracks instead of global tracks, this correction
affects primarily the region \Nhminus\ < 5 and is negligible
beyond \Nhminus\ = 10. These are the first two bins of the final
distribution. However,  most of the cross section is found in
these bins, and the shape of the distribution is sensitive to
these values at low multiplicity. To figure out what is the
fraction of the hadronic cross section that we actually see, we
proceed as follows.  We choose events within a $\pm 95$ cm range
along the $z$ axis of the beam direction.  We can then count the
the raw number of events with vertex in our multiplicity
distribution.  We compare this to the number of vertex events
corrected for efficiency, \ie\ to the number of events in the
multiplicity distribution appropriately weighted with the
correction from Figure~\ref{fig:vtxeffcorr}.  From these two
numbers, we see that the fraction of the hadronic cross section
available to the offline analysis is $94.9 \pm 0.5\ \pm 4\ \%$ of
\sigmahad, where the systematic uncertainty comes from the
uncertainty in the estimation of the vertex efficiency at the
lowest multiplicities based on the simulations.

\subsection{Trigger backgrounds}
We now focus on estimating the contribution to the events in the
denominator in Eq.~\ref{eq:fractionCrossSectionMeasurement} that
are not part of the total $\sigma_{xn,xn}$ cross section, e.g.
beam-gas events. However, knowing which events are background so
that we can take them out of our total trigger sample, is not
trivial.

To investigate this issue, we focus on events for which the vertex
was not found. To study systematic effects and variations, from
the September data, we used specific minimum bias runs with good
statistics, and stable detector operation. The run with the most
statistics in this period has $\sim$ 77K events (the events in
Fig.~\ref{fig:zvtx} is from this sample). This was used along with
a few other runs to compare backgrounds and other systematic
effects.

For these background studies, we compared 4 runs with different
luminosity conditions spread over several weeks.  Two of these
(August 2000 data) had intensities in RHIC which were about a
factor of 3 lower than in the other two (September 2000) runs. We
are interested in the different intensities because the hadronic
interaction rate grows as the product of the intensity in each
beam, while the beam+gas background grows only as the sum. This
type of background should then have a different contribution in
runs with differing beam intensities.

\begin{figure}[htb]
  \centering
  \mbox{\subfigure[]{\includegraphics[width=0.45\textwidth]{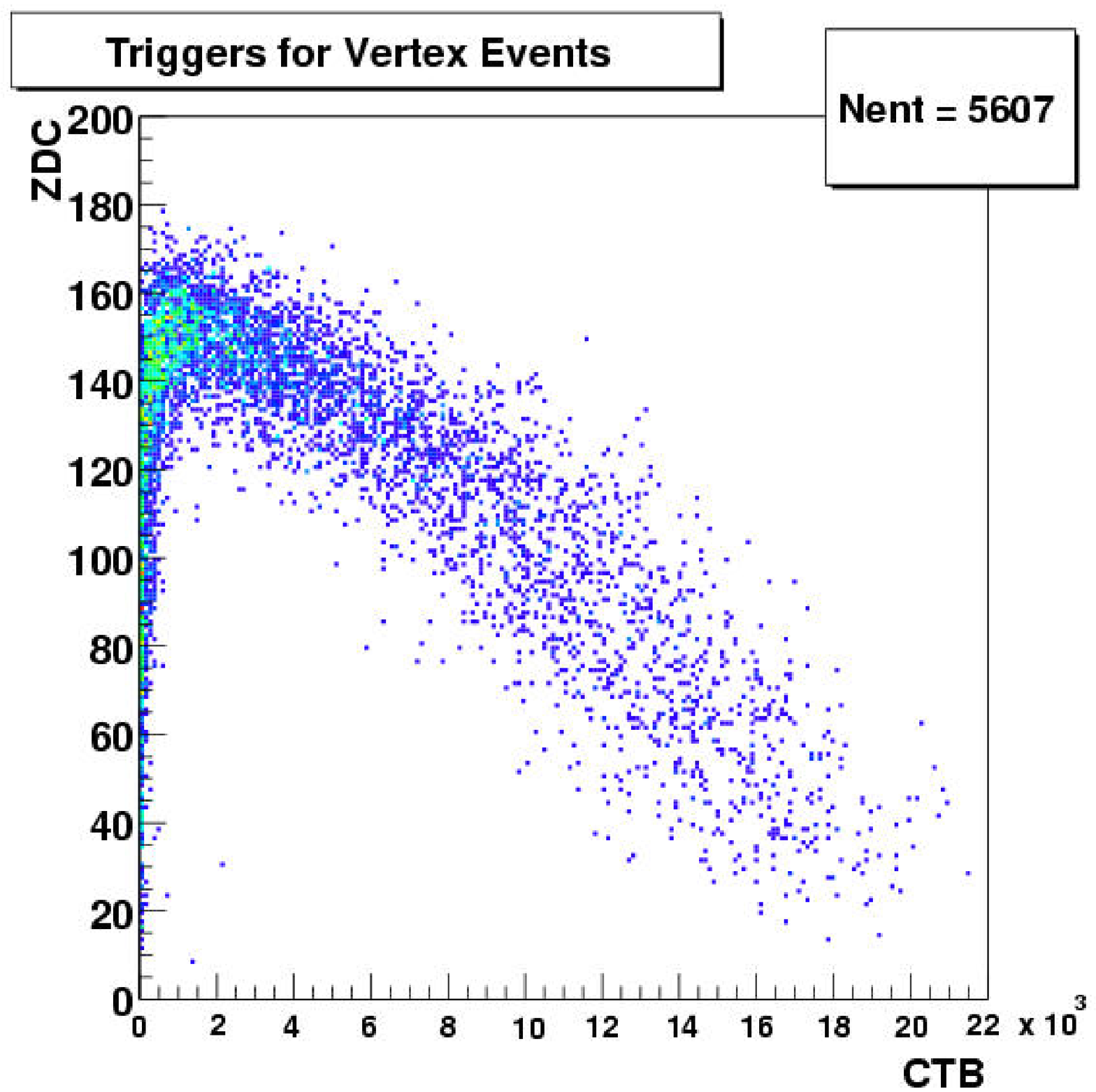}}\quad
    \subfigure[]{\includegraphics[width=0.45\textwidth]{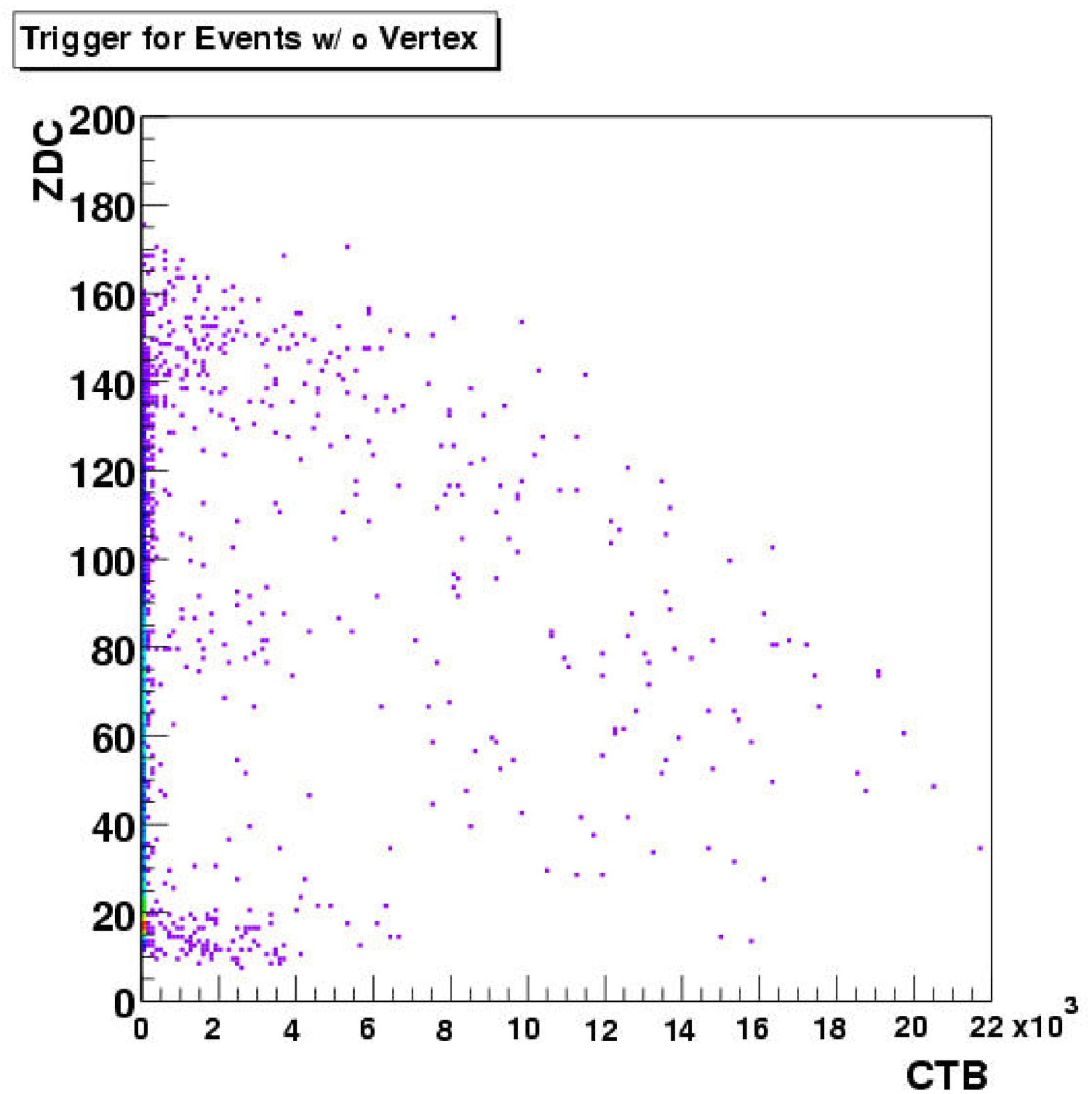}}}
  \caption[Trigger signal for events with and without a reconstructed vertex.]
  {The trigger signal for events with a valid reconstructed vertex
  (a) shows the expected shape.  There are additional event
  classes that show up when plotting the trigger signal for events
  without a reconstructed vertex (b).}
  \label{fig:triggerVtxNoVtx}
\end{figure}

To illustrate the difference between the events with and without
vertex, a plot of the trigger information from 10K events from one
of the runs is shown in Fig.~\ref{fig:triggerVtxNoVtx} Events with
a reconstructed vertex are shown in the left panel; the right
panel illustrates the trigger signals for events without a
reconstructed vertex.

For events in Fig.\ref{fig:triggerVtxNoVtx}(a), we see again the
very clean ``boomerang'' band. These represent valid hadronic
events for the determination of \sigmahad. For the events without
a reconstructed vertex, there are several regions to understand.
The events which lie in the ``boomerang'' are most likely good
events that did not have a vertex because the collision occurred
outside the bounds of the TPC, where the acceptance of the offline
vertex finder terminates. This does not present a problem.
However, in the 2001 run a cross check will be done using timing
in the ZDC's. There are two other regions that are most probably
background and have been the subject of further scrutiny:
\begin{itemize}
\item ZDC < 30 (low ZDC region)
\item ZDC $\sim$ 80 (mid ZDC region)
\end{itemize}

\begin{figure}[htb]
  \centering
  \mbox{\subfigure[]{\includegraphics[width=0.45\textwidth]{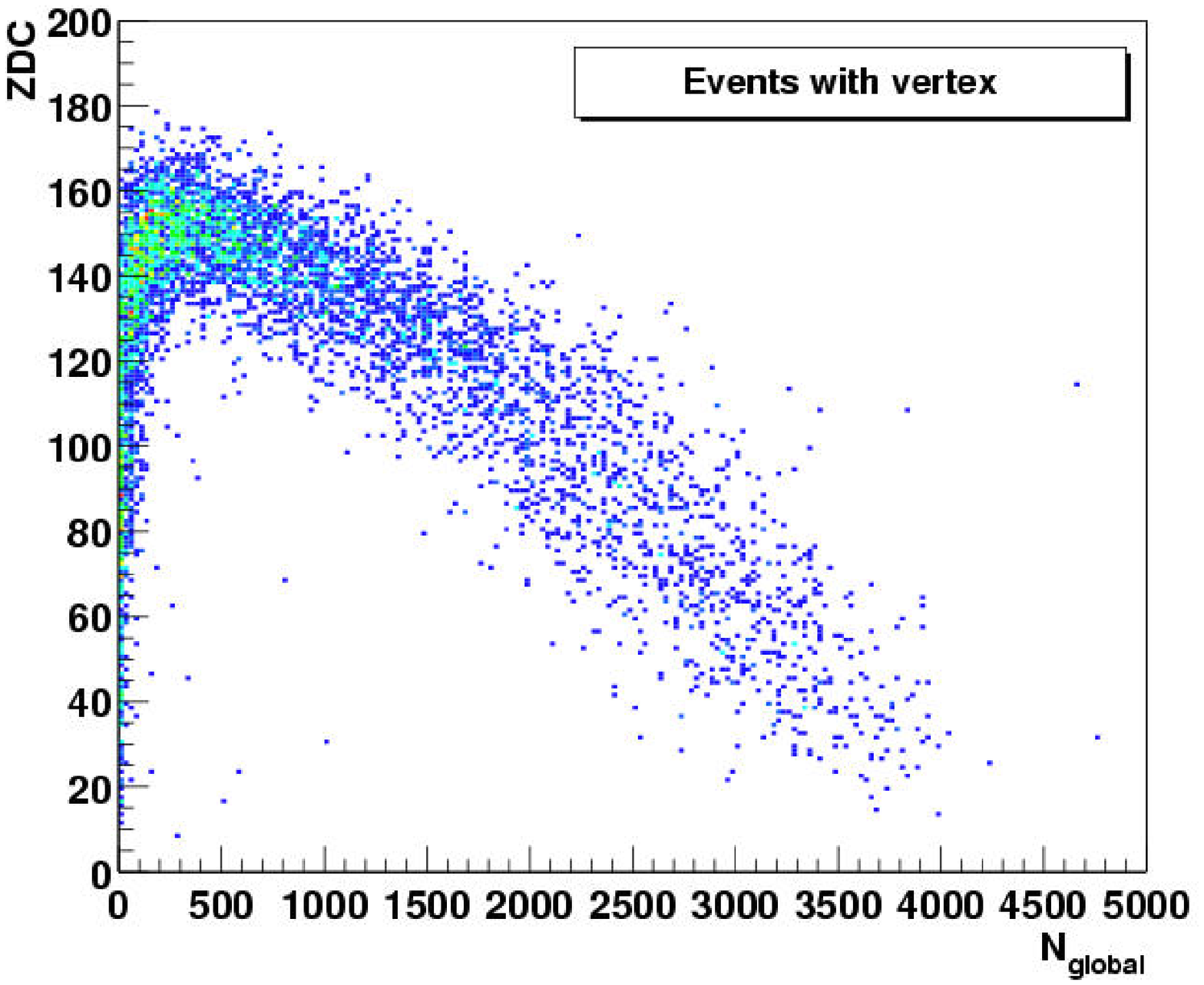}}\quad
    \subfigure[]{\includegraphics[width=0.45\textwidth]{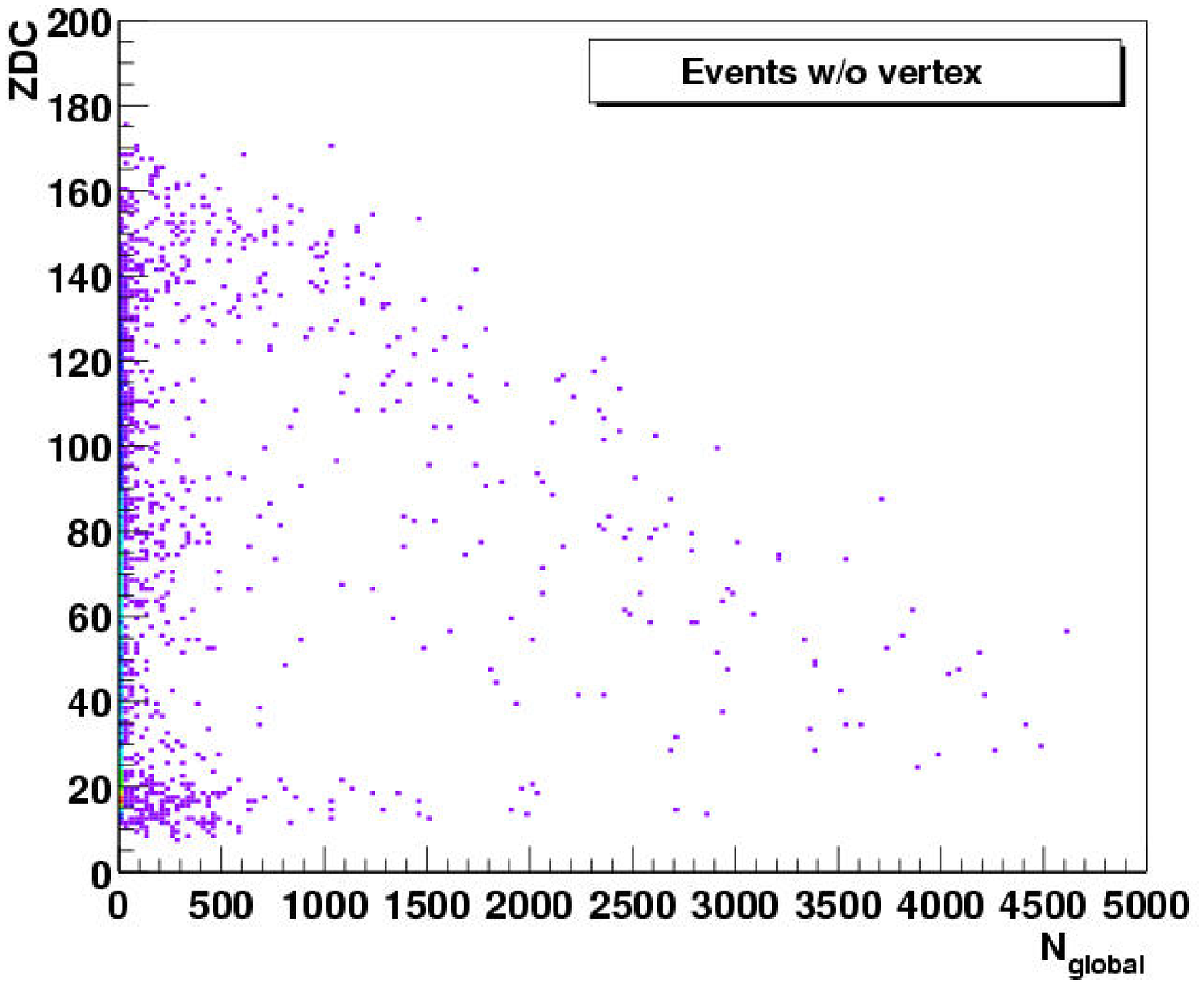}}}
  \caption[ZDC \vs\ \Nglobal\ for events with and without a reconstructed vertex.]
  {The ZDC signal \vs\ the number of global tracks for events with a valid reconstructed vertex
  (a) and without one (b).  The same structure as in Fig.~\ref{fig:triggerVtxNoVtx} is
  seen, indicating that there is definitely an increased number of
  charged tracks in the detector for the events in the two regions
  outside the ``boomerang''.}
  \label{fig:zdcNglobal}
\end{figure}

Both of these regions have events whose trigger signals extend
along the CTB axis. If the multiplicity is increasing for these
events, we should see also an increase in the number of global
tracks. The same thing is found if we plot the ZDC signal \vs\
\Nglobal for the events with and without vertex, shown in
Fig.~\ref{fig:zdcNglobal}.

We see that there are events that have $\sim$ 1000 global tracks,
do not have a reconstructed vertex and lie in a different region
than the normal hadronic events.

We studied the characteristics of these two bands to try to
understand their origin. We investigated other differences between
these bands and the rest of the ``good'' triggers to shed light on
their nature.  We expect that in the region \Nglobal $\leq 1$ most
of the events are of electromagnetic origin. In any case, there
can be no vertex reconstructed for events with less than 2 tracks.
By selecting non vertex events with \Nglobal>1 we focus on the
possible backgrounds the we're aiming to understand.

We find that the band at low ZDC has both East and West ZDC
triggers just above threshold, and the band in the mid ZDC region
has one of the calorimeters just above threshold and the other one
high. For the ZDC $\sim 80$ region, this is just the opposite
behaviour than is seen in the events with a found vertex, where
the ZDC signal is relatively symmetric between East and West
ZDC's. It is possible that a beam+gas event could generate such an
asymmetric ZDC signal with some charged tracks seen in the TPC.
Asymmetries in tracking were also found, but this can also
indicate simply an interaction outside the TPC volume.  Several
hypotheses were proposed, but there was insufficient information
to unambiguously discern the nature of the background.  In
particular, we would like to know its contribution to the region
where the valid events lay. Without this knowledge, we can remove
only the background contribution outside the ``boomerang'' region.
We can therefore only produce a lower limit on the fraction in
Eq.~\ref{eq:fractionCrossSectionMeasurement}.  The numerator
depends on the efficiency and acceptance corrections already
discussed.  For the purpose of the calculation of the fraction, we
used events within $\pm 2$ m to reduce the acceptance correction
($6\%$).  For the denominator, we start with the total triggers.
The background is obtained via the number of events without vertex
that have 1 or more global tracks. Since some of these events are
valid events that were not found because of efficiency or
acceptance, we need to subtract the number of events we added to
our numerator due to these two corrections from the background
estimate.  This is a lower bound since we don't take into account
the background for events without global tracks. From 10K events
from the low luminosity runs and from 15K events from the high
luminosity runs we obtain

\begin{eqnarray}
\mathrm{Low}\ \mathcal{L} & : & \frac{\sigmahad}{\sigma_{xn,xn}}
\geq 70.0 \pm 0.7\ \%
\nonumber \\
\mathrm{High}\ \mathcal{L} & : & \frac{\sigmahad}{\sigma_{xn,xn}}
\geq 73.5 \pm 0.6\ \% \ .
 \label{eq:fractionCrossSectionLowBound}
\end{eqnarray}
The errors are statistical, which is enough to compare the two
numbers since they were obtained in the same way.   We can see
that there is a difference beyond statistical.  However, this was
expected as the low-luminosity runs systematically have a greater
background contribution that our procedure does not account for.
For the high-luminosity sample, this background is reduced and the
lower limit on the fraction is higher.  Using the fraction from
the high-luminosity events and the current measured value for
$\sigma_{xn,xn} = 8.9 \pm 0.5$ barn, we obtain a lower limit on
the hadronic cross section $\sigma_{\mathrm{AuAu}} \gtrsim 6.5$
barn, consistent with the calculated value of 7.2 barn. There
should be an error of $\pm 0.4$ barn in this estimate which comes
basically from the cross section measurement (the statistical
error on the fraction is much smaller). This error in turn is
dominated by the uncertainty in the measurement of the beam
currents in the RHIC ring, which is on the order of 3\%. The
systematic uncertainty on the fraction is the only missing piece,
but as discussed, at this point we can only give a lower bound.

It is clear that with only the ZDC and CTB information at the
lowest multiplicities, it becomes increasingly difficult to
separate the background events such as beam+gas collisions.  It
would therefore be of great advantage to obtain an estimate of the
interaction point without having to rely on tracking, but rather
on trigger signals, \eg\ on timing between the arrival of the East
and West ZDC signal.  In addition, improved phase space coverage
for the trigger would drastically reduce uncertainties in the
background estimates and help to provide a more complete topology
of the valid low multiplicity hadronic events. For the 2001 run
ZDC timing will be implemented; and starting with the 2001 $pp$
run, additional detectors in the form of scintillator slats
covering from the pseudorapidity region $2 < |\eta| < 4$ will be
part of the STAR trigger as well.   For the analysis of the
multiplicity distribution presented in Sec.~\ref{sec:Multiplicity}
we will use the calculated value of the \AuAu\ hadronic cross
section of $\sigmahad = 7.2$ barn for the normalization.

To obtain the multiplicity distribution, one has to obtain the
corrected number of negative hadrons for every event.  The
analysis done here relies on tracking.  We first obtain a raw
$dN/d\pt d\eta$ distribution as a function of $\eta$ and \pt\ for
every event. Then several corrections are applied to this raw
distribution as discussed in the next section.  Finally, the
corrected $dN/d\pt d\eta$ distribution is integrated in the range
$|\eta|<0.5$, $0.1 < \pt < 2\ \gevc$ to obtain a corrected
multiplicity.  One can therefore also obtain \pt\ and $\eta$
distributions with this procedure.  Since the algorithm relies on
tracking corrections, we now turn our attention to this subject.

\section{Tracking Studies}\label{sec:trackingCorrections}

Particle production was studied through the yield of primary
negative hadrons, which are mostly \piminus\ with an admixture of
\kminus\ and \pbar.  The \hminus\ distribution includes the
products of strong and electromagnetic decays. Negatively charged
hadrons were the main focus of the work in order to exclude
effects due to participant nucleons which would show up in the
positively charged hadron sample. Charged particle tracks
reconstructed in the TPC were accepted for this analysis if they
fulfilled requirements on number of points on the track and on
pointing accuracy to the event vertex.

The main goal of this analysis is to determine the corrected yield
of primary particles coming from the collision. What we have in
the final state are measured tracks, both primary and secondary,
in our detectors. There are several losses and backgrounds that
need to be corrected: acceptance, decay losses, track
reconstruction efficiency, contamination due to interactions in
material, misidentified non-hadrons, and the products of weak
decays.  The appropriate corrections were obtained mainly by use
of the embedding technique. Backgrounds were determined by either
fully simulated events or by direct measurement when possible. All
corrections were calculated as a function of the uncorrected event
multiplicity. The corrections used in the determination of the
final spectra are obtained in the following order:
\begin{dingautolist}{192}
\item Geometrical acceptance and decay losses\spacing{.5}
\item Reconstruction efficiency
\item Track merging
\item Momentum resolution
\item Track splitting
\item Electron background
\item Weak decay and secondary interaction background
\item Ghost tracks
\end{dingautolist}\spacing{1.2}
We now discuss the most important corrections in more detail.

\subsection{Acceptance}
\begin{figure}[htb]
  \centering
  \mbox{\subfigure[]{\includegraphics[width=0.45\textwidth]{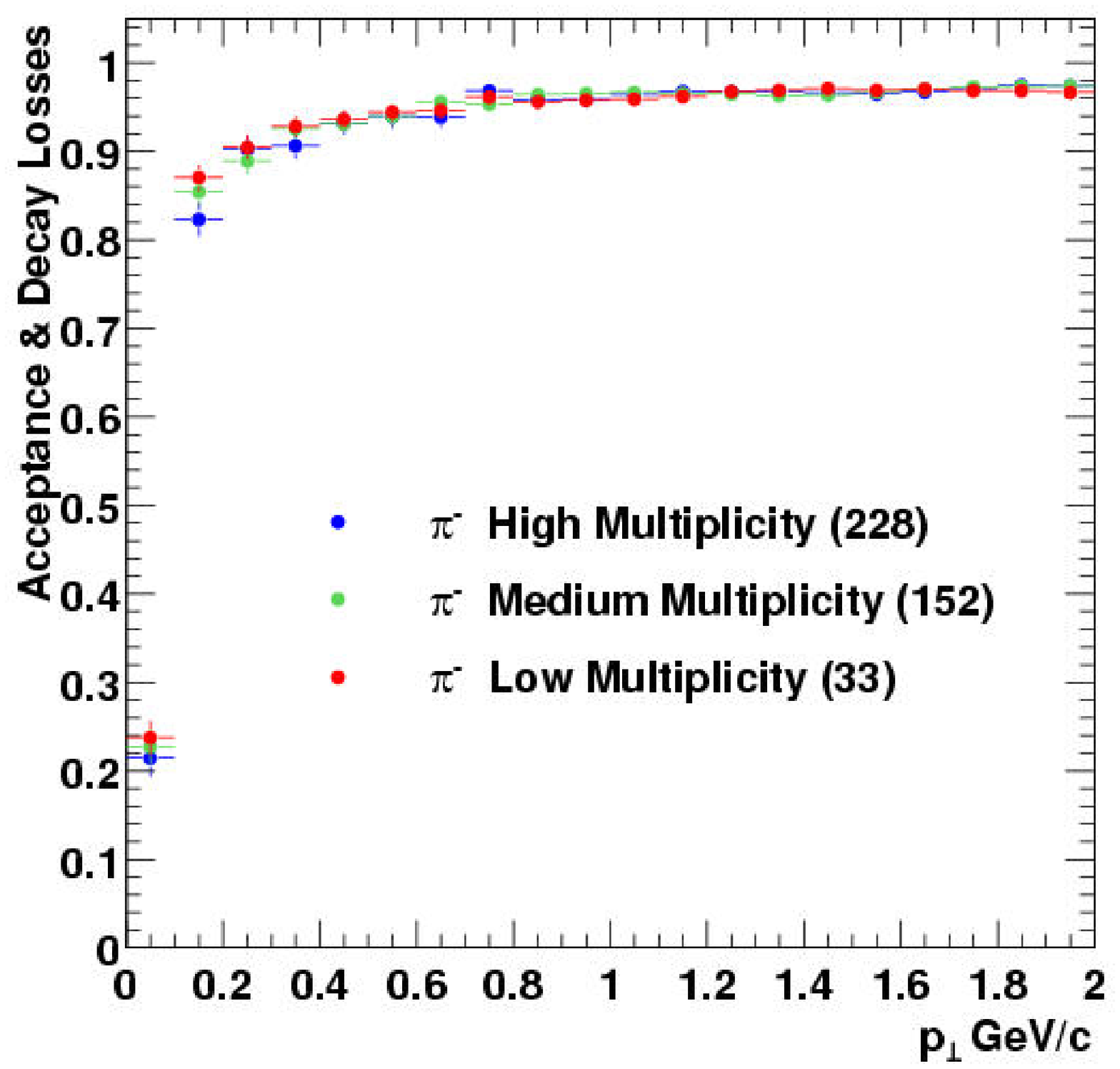}}\quad
    \subfigure[]{\includegraphics[width=0.45\textwidth]{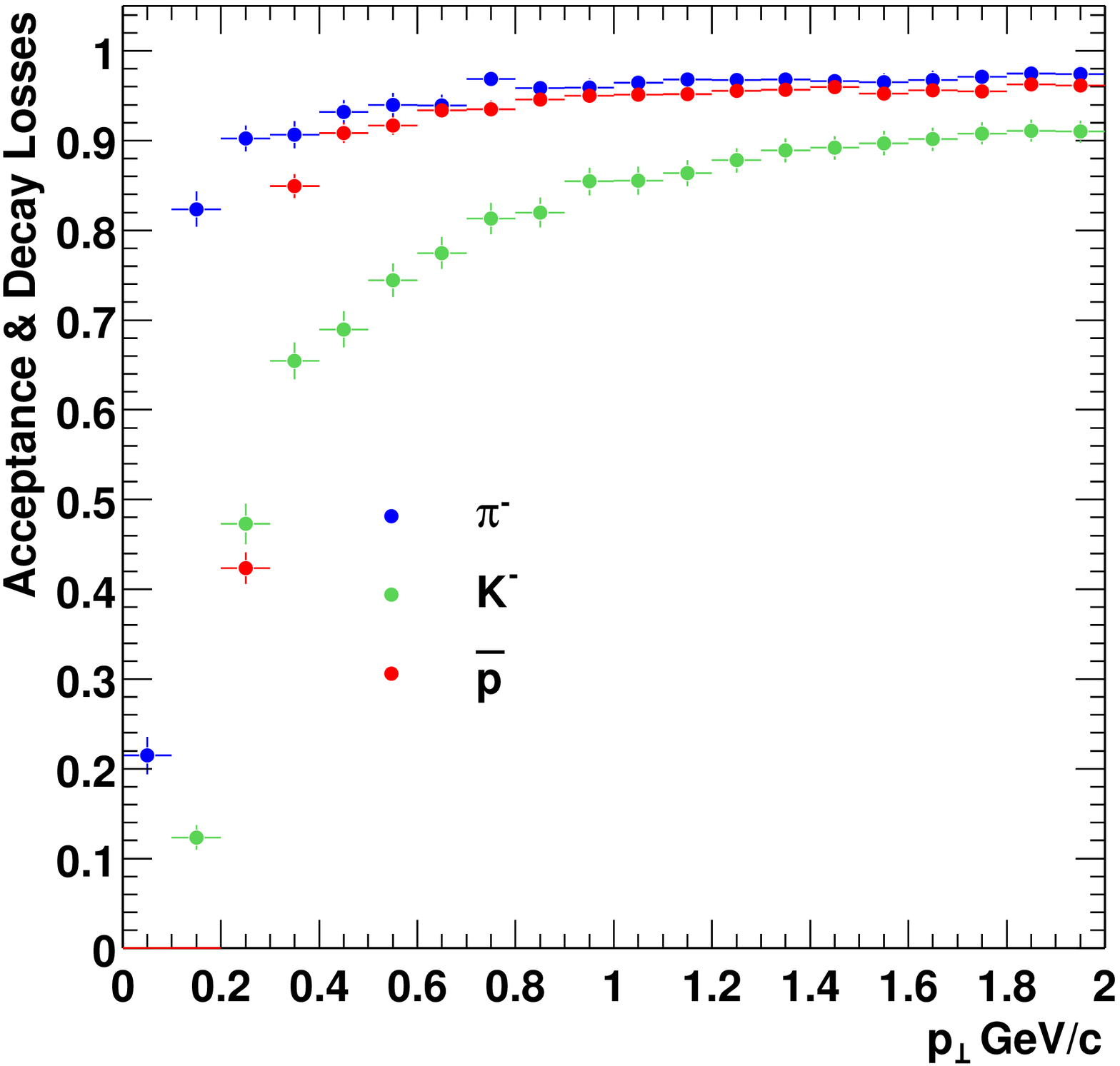}}}
  \caption[Acceptance Correction]{The acceptance correction for the \hminus\ analysis.  Left panel:
  \piminus\ acceptance for 3 multiplicity selections.  Right panel: Acceptance for \piminus\, $K^-$ and
  $\bar{p}.$}\label{fig:acceptance}
\end{figure}

The acceptance correction takes care of two different things. What
we really focus on here is whether or not a charged particle makes
a measurable signal in the detector such that it is possible to
reconstruct it afterwards.  If the charged particle does not
deposit energy in any sensitive volume of the detector, it will be
lost.  This is the more common definition of the geometrical
acceptance.  In addition, a charged particle may also leave no
signal if it decays in flight before it reaches the detector.  In
the STAR Monte Carlo simulation implemented for this work, these
two losses were taken into account simultaneously by adopting the
following definition of acceptance: a track is accepted if it
leaves at least 10 Monte Carlo hits in the TPC.  The
reconstruction code then at least has the possibility to find the
track.  This correction can be calculated in a full Monte Carlo
simulation and also using embedding, and should be completely
independent of the event multiplicity.  There should be a
dependence on particle type, as the decay characteristics are
included in our definition.  This is illustrated in
Fig.~\ref{fig:acceptance}.
The left panel is the acceptance for negative pions in three
multiplicity bins. The numbers correspond to the mean raw
negatively charged multiplicity ($\langle \hminus
\rangle_{\textrm{raw}}$) in each of the 3 bins: $\langle \hminus
\rangle_{\textrm{raw}} = 223$ for the high multiplicity, $\langle
\hminus \rangle_{\textrm{raw}} = 152$ for medium, and $\langle
\hminus \rangle_{\textrm{raw}} = 33$ for the low multiplicity bin.
The plot shows that the correction is essentially independent of
the multiplicity, except perhaps at the lowest \pt. The right
panel is the acceptance for the three particle species which make
up the \hminus\ distribution. We see that the $K^-$ acceptance is
lower than that for the pions, which is expected since $c\tau =
3.7$ m for kaons, and it is 7.8 m for \piminus.

The acceptance is on average 90\% for tracks within the fiducial
volume having $\pt>200$ \mevc.  We see that the acceptance rapidly
drops at low \pt.  For the \hminus\ analysis, we accept tracks
which have $0.1 < \pt\,< 2$ \gevc\ and $|\eta|<1.0$.  This
accounts for the majority of the produced particles at
mid-rapidity, as we find the yield beyond \pt\ = 2 \gevc\ to be
only 1\% of the total yield.  A motivation to go as low in \pt\ as
possible was driven by the fact that for 2001 the magnetic field
would be set to the design operating value of 0.5 T for the bulk
of the data taking, instead of the 0.25 T used in 2000. This
raises the low-\pt\ acceptance of the TPC, so essentially the 2000
data would give us access to the lowest \pt.  Nevertheless, to
study systematic effects the multiplicity analysis was carried out
three times with various low-\pt\ cutoffs: \pt>0 (no cutoff, large
systematics expected), \pt>0.1 \gevc\ and \pt>0.2 \gevc\ yielding
consistent results.

\subsection{Tracking Efficiency}
\label{sec:efficiency} 
\begin{figure}[htb]
  \centering
  \mbox{\subfigure[]{\includegraphics[width=0.45\textwidth]{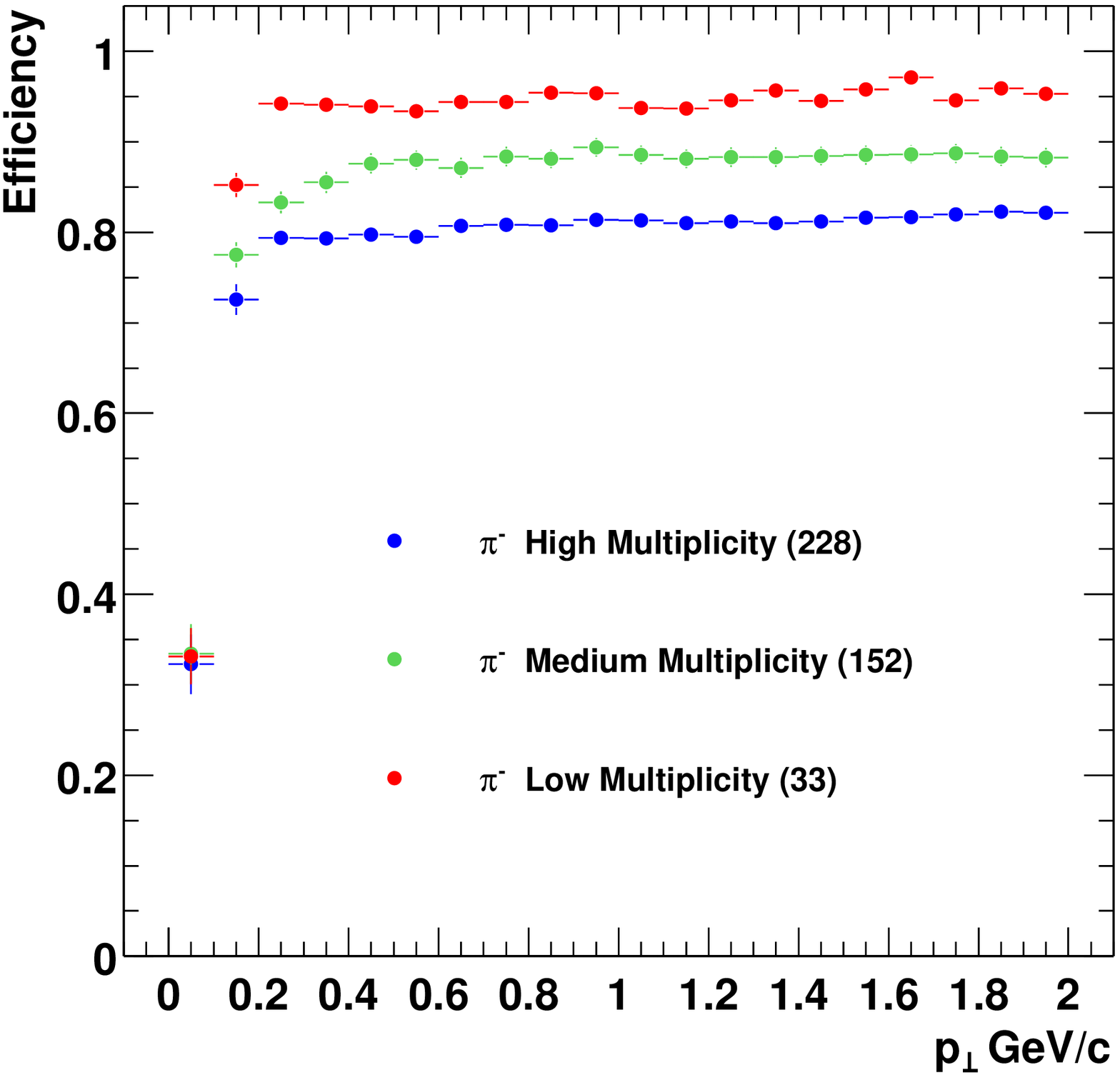}}\quad
    \subfigure[]{\includegraphics[width=0.45\textwidth]{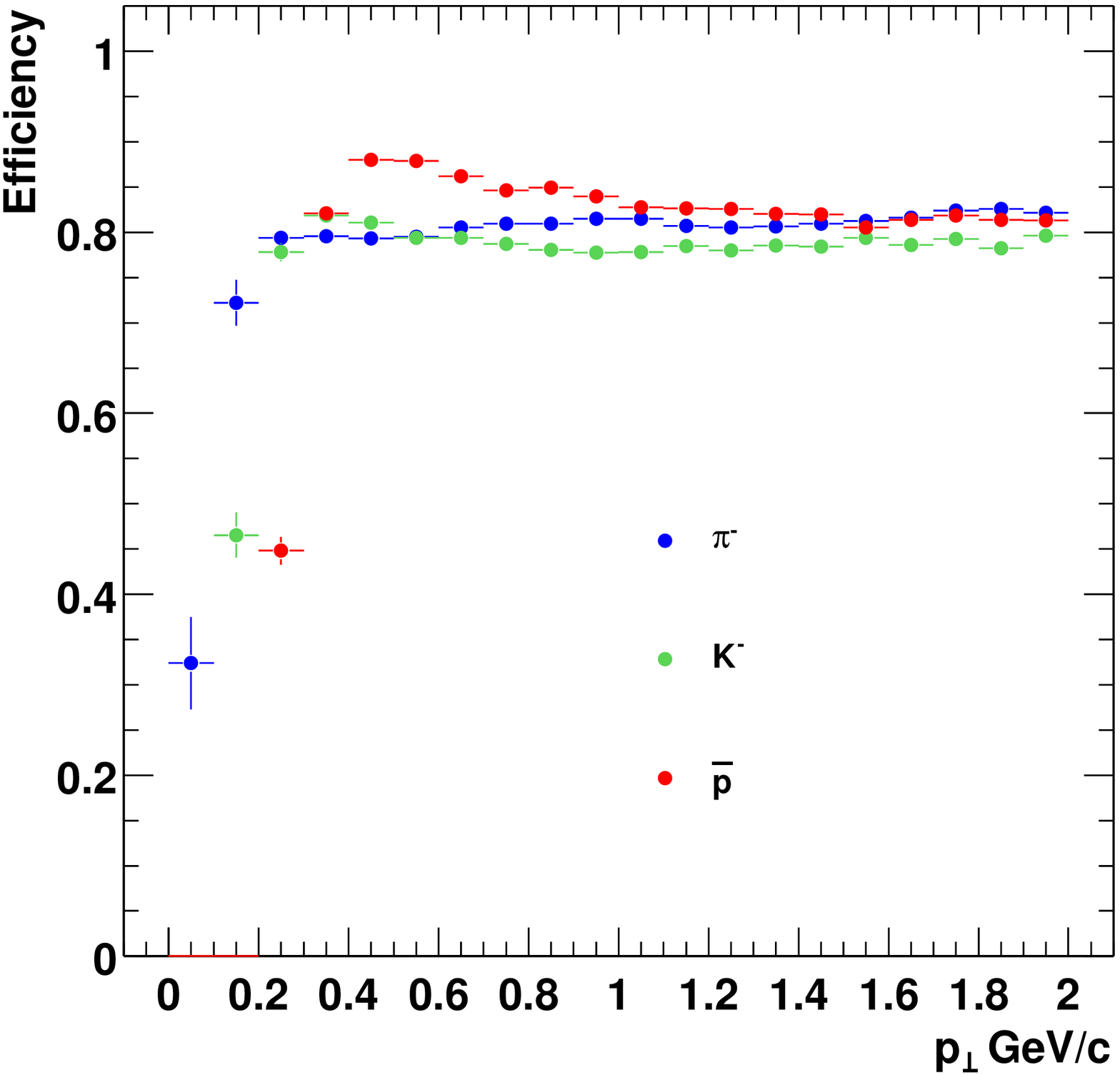}}}
  \caption[Reconstruction Efficiency \vs\ \pt.]{The reconstruction efficiency for the \hminus\ analysis.  Left panel:
  \piminus\ efficiency for 3 multiplicity selections.  Right panel: Efficiency for \piminus, $K^-$ and
  $\bar{p}$ at high multiplicity.}\label{fig:efficiency}
\end{figure}

Once a track has made it into the detector,
we focus on the question of how likely it is to be found by the
offline software chain. The reconstruction efficiency was
determined by embedding simulated tracks into real events at the
raw data level, reconstructing the full events, and comparing the
simulated input to the reconstructed output. This technique
requires a precise simulation of isolated single tracks, achieved
by a detailed simulation of the STAR apparatus based on GEANT and
a microscopic simulation of the TPC response discussed in
Chapter~\ref{ch:DetectorSimulation}. The multiplicity of the
embedded tracks was limited to 5\% of the multiplicity of the real
event in the same phase space as the simulated data, thereby
perturbing the real event at a level below the statistical
fluctuations within the event sample. Fig.~\ref{fig:efficiency}
shows the reconstruction efficiency obtained from embedding as a
function of \pt\ for a slice at mid-rapidity.
The tracking efficiency varies depending on \pt\ and the
multiplicity of the event. For the lowest multiplicity events, the
\piminus\ efficiency is $\gtrsim 95\%$ in the region \pt\ > 400
\mevc. Going to the high multiplicity events degrades the
efficiency to $\sim 80\%$.  However, since the difference between
the multiplicities in the high and low bins is about a factor of
7, a reduction of only $\sim 15\%$ in the efficiency is a
significant achievement for the offline reconstruction. In
general, for all multiplicities and all 3 particle species, in the
region $\pt > 200$ \mevc\ we always have an efficiency greater
than 80\%.

\begin{figure}[htb]
  \centering
  \includegraphics[width=.6\textwidth]{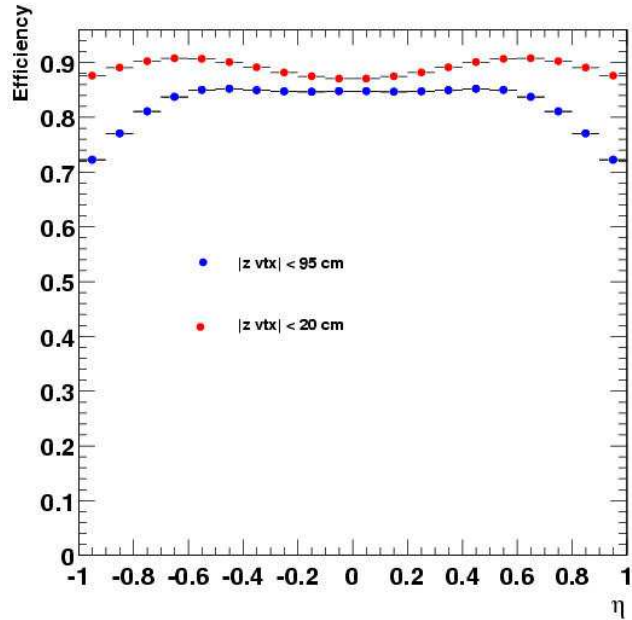}
  \caption[Reconstruction Efficiency \vs\ $\eta$.]{The \piminus\ reconstruction efficiency
  as a function of pseudorapidity.  Two different vertex selections are shown, illustrating
  the influence of the choice of vertex on the efficiency due to the detector geometry.}
  \label{fig:efficiencyEta}
\end{figure}

The pseudorapidity dependence of the tracking efficiency for
\piminus\ is shown in Fig.~\ref{fig:efficiencyEta}.
The two data sets in the figure are for selections of events with
vertex $z$ position, \zvertex, within $\pm 95$ cm (blue points)
and $\pm 20$ cm (red points) of the center of the TPC. The wide
vertex cut allows us to use most of the available data. The
interest in studying such variations in efficiency lies in the
observation that a particular phase space region, for a given
vertex position, will be measured in a different region of the
detector. For example, tracks with $\eta$
> 0.5 will not cross all the available padrows in the TPC when the
vertex is in a position \zvertex\ > 95 cm. This is reflected as a
drop in the efficiency.  By making a tighter \zvertex\ cut, one
can probe a wider phase space region with the guarantee that
tracks have the chance to cross all TPC padrows.  For a vertex cut
of |\zvertex| <  20 cm, tracks within $|\eta|<0.8$ will cross the
entire TPC, and one can see a higher efficiency for these tracks
in the figure (red points).

\begin{figure}[htb]
  \centering
  \mbox{\subfigure[]{\includegraphics[width=0.45\textwidth]{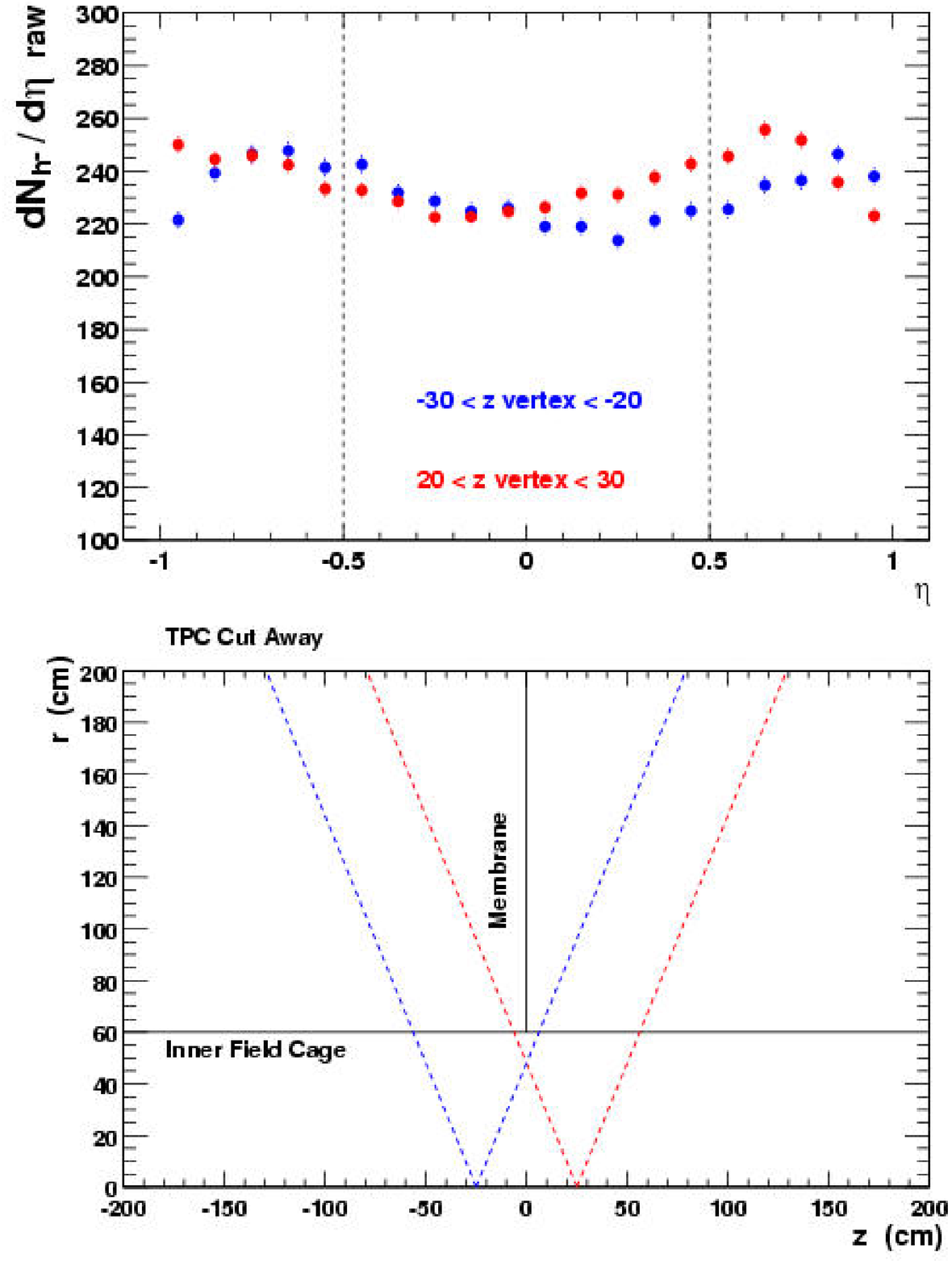}}\quad
  \subfigure[]{\includegraphics[width=0.45\textwidth]{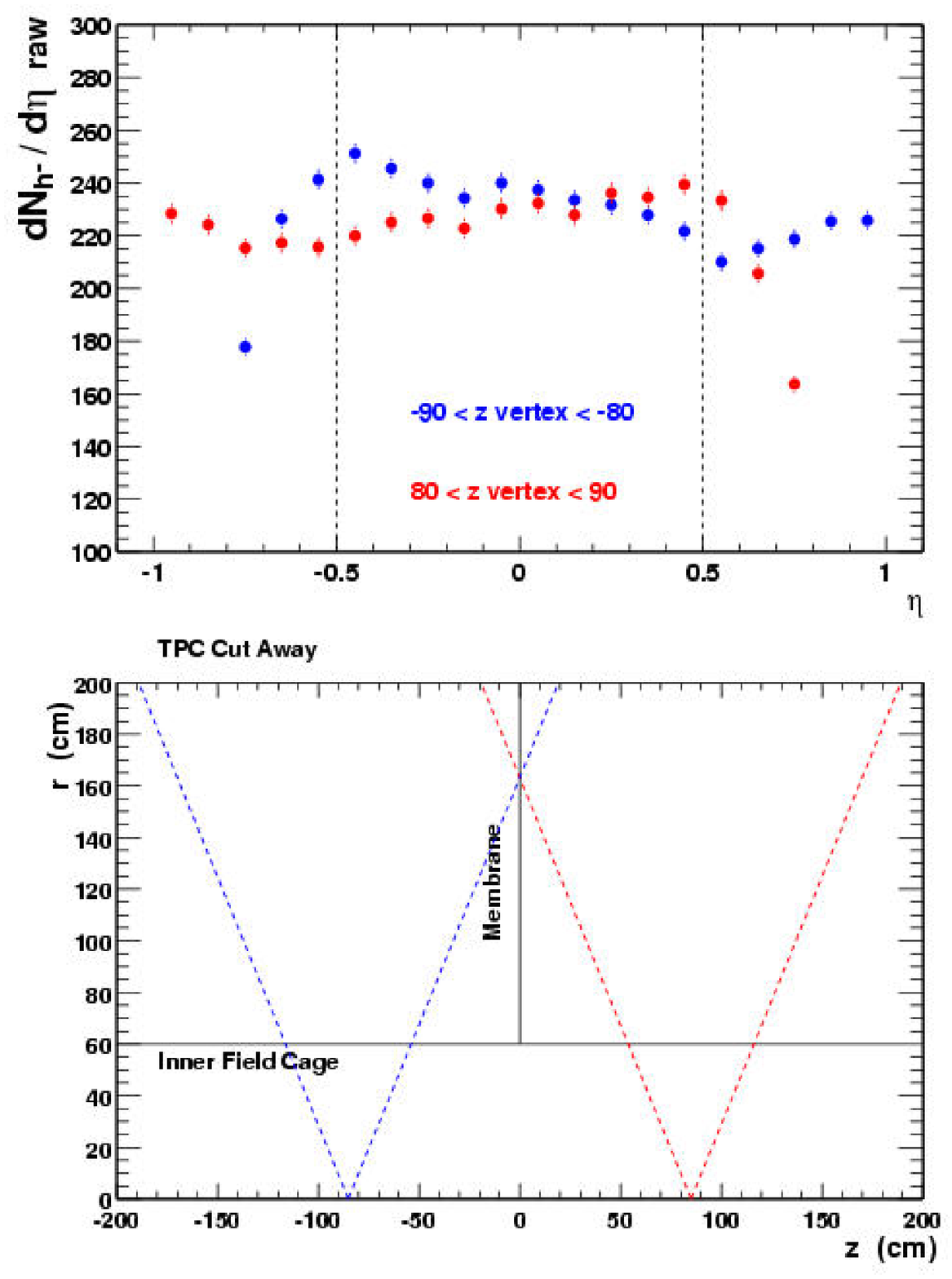}}}
  \caption[Raw yield dependence on \zvertex.]{Raw yield dependence on
  \zvertex. The left panel is for the vertex choice [20, 30] cm
  (red) and [$-30$, $-20$] cm (blue), right panel is for the
  interval [80, 90] cm (red) and [$-90$, $-80$] cm (blue).
  Dashed lines in the bottom panels represent tracks with $\eta = \pm 0.5$,
  corresponding to the limits given by the vertical dashed lines in the top panels.}
  \label{fig:etavtxRaw}
\end{figure}

Furthermore, even when tracks cross the entire TPC, we also expect
a variation in efficiency for a given $\eta$ slice as a function
of the vertex position.  This arises from diffusion effects, as a
track crossing the detector in the region near the central
membrane will produce space points that must drift across the
entire chamber, and their clusters will be wider than those for
tracks near the pad plane.  We can test the adequacy of our
detector simulation by studying the raw and corrected yield in a
given $\eta$ bin as a function of the vertex position, and hence
as a function of the position of the track in the detector. In
addition, since the TPC volume is separated into two identical
halves separated at $z=0$ and set up such that the ionization
drifts toward the wires (located at $\pm 2.1$ m), the behaviour
for a given choice $\eta$ and \zvertex\ should be the same as for
$-\eta$ and $-\zvertex$.  This is illustrated in
Fig.~\ref{fig:etavtxRaw} which shows the raw $\eta$ distribution
for 10 cm slices in the choice of \zvertex\ interval: [20, 30] cm,
[80, 90] cm; and the corresponding symmetric \zvertex\ selections
[$-30$, $-20$] cm and [$-90$, $-80$] cm.
The top panels show the raw distributions as a function of $\eta$,
left for the [20, 30] cm interval and right for the [80, 90] cm
region. In the bottom panel a sketch of the detector geometry is
given, as a slice in the $r$ and $z$ coordinates of the TPC
cylinder. The central membrane is at $z=0$, the drift distance is
2.1 m from the membrane to the wires. The first pad row at 60 cm
and the last pad row at 200 cm.  The lines in this panel represent
the trajectories of tracks originating from the given vertex
position at $\eta=\pm 0.5$.  We can see that there are significant
systematic variations in the raw yield depending upon where the
event happened.  As expected, the raw yield begins to decrease
significantly once tracks do not cross the entire TPC.  Such
systematics must be taken into account in the efficiency
corrections.

\begin{figure}[htb]
  \centering
  \mbox{\subfigure[]{\includegraphics[width=0.45\textwidth]{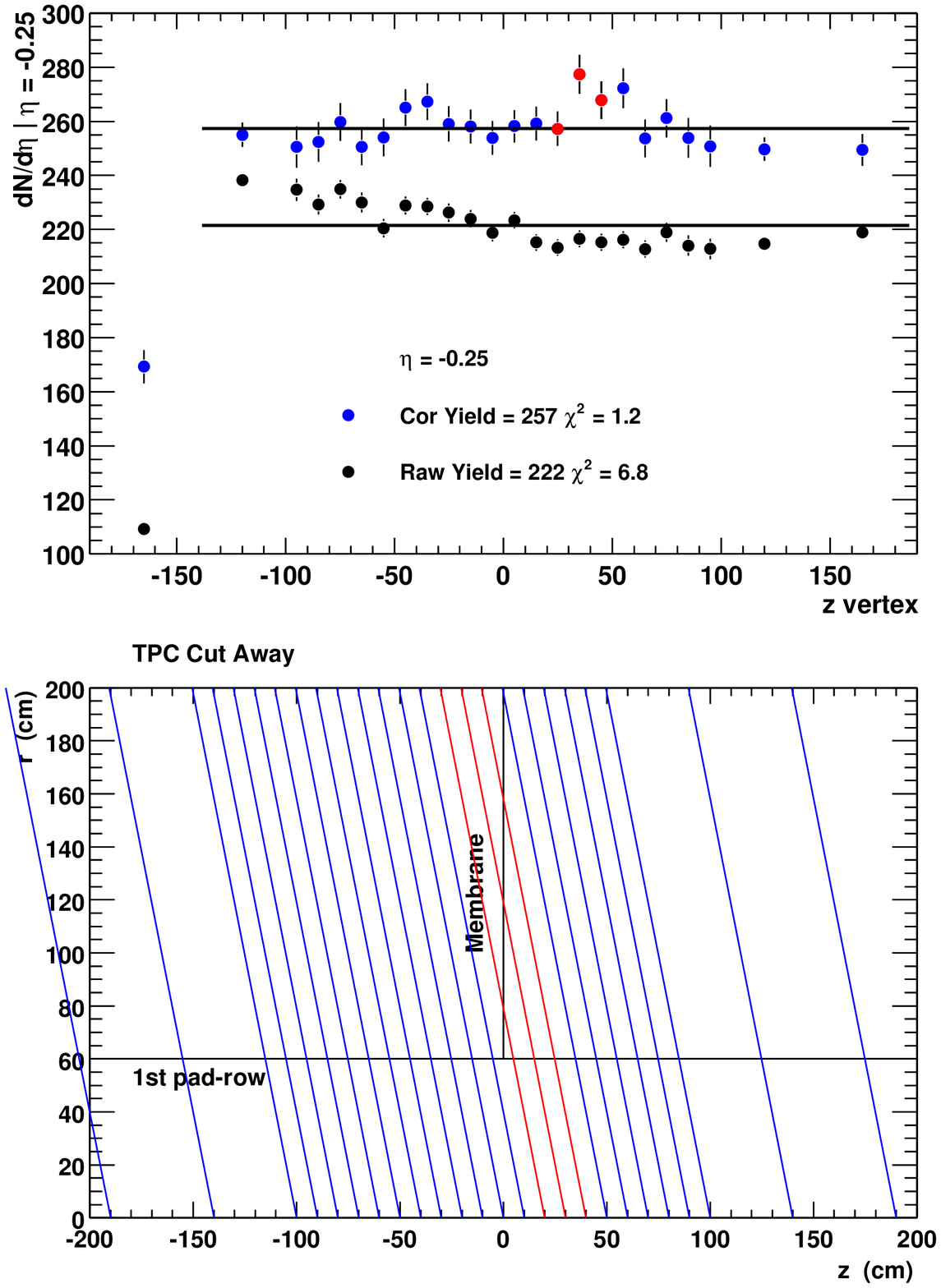}}\quad
  \subfigure[]{\includegraphics[width=0.45\textwidth]{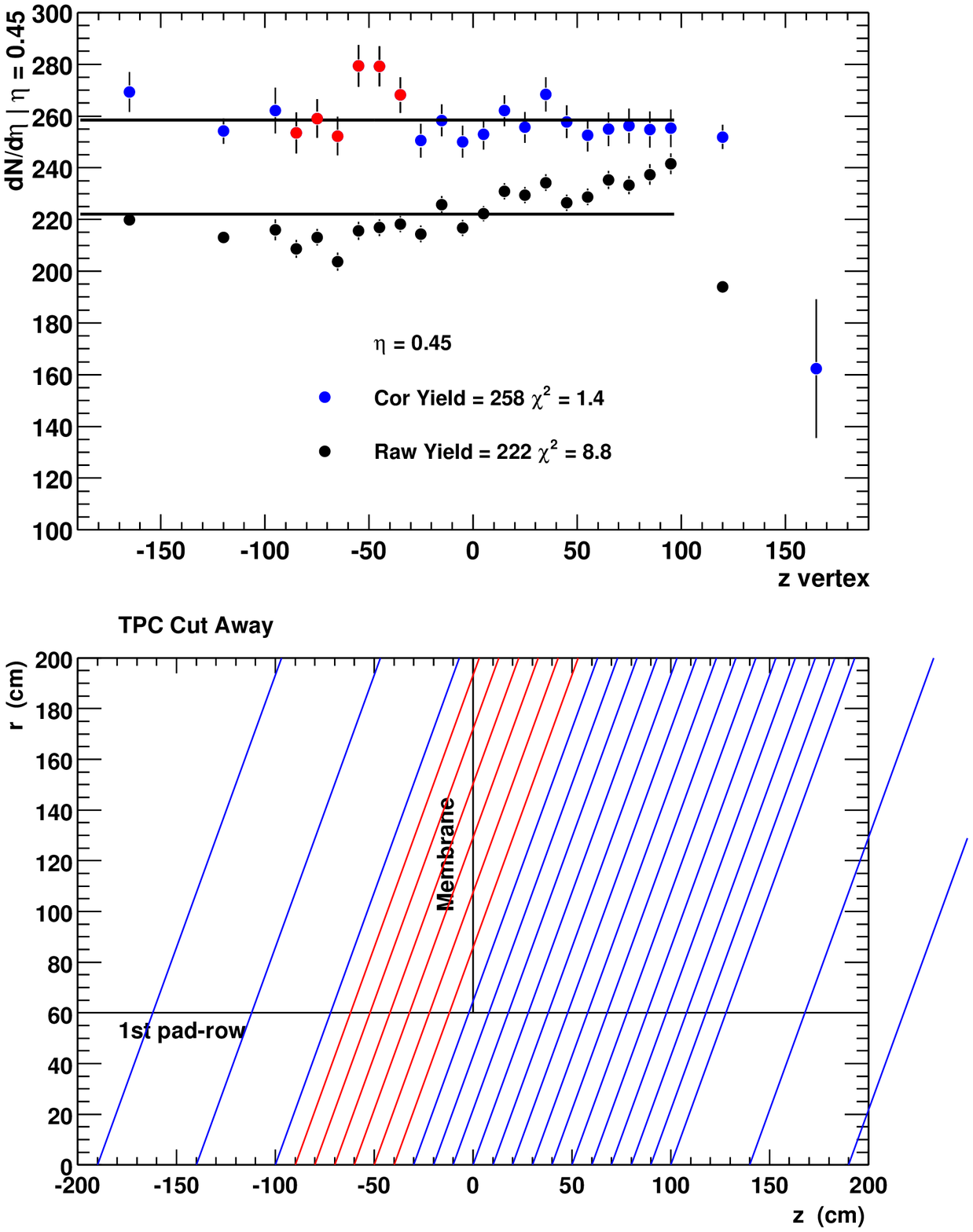}}}
  \caption[Corrected yield dependence on \zvertex.]
  {Raw and corrected yield for $-0.3<\eta<-0.2$ (left) and $0.4<\eta<0.5$(right) as a function
  of \zvertex.  Data points in each of the upper plots correspond to tracks crossing the detector
  region illustrated schematically in the lower plots.}
  \label{fig:etavtxCorr}
\end{figure}

Since the corrected yield we report must be independent of any
detector effects, it is easier to study the adequacy of the
corrections by focusing on a single $\eta$ bin and comparing the
raw and corrected yield for the bin as a function of the vertex
position. This is illustrated in Fig.~\ref{fig:etavtxCorr}
which shows the raw and corrected yields for two choices of
$\eta$, $-0.3<\eta<-0.2$ (left) and $0.4<\eta<0.5$ (right), as a
function of \zvertex. The lower panel shows the geometry of the
relevant tracks as they cross the detector for the different
vertex positions.  The steps in \zvertex\ were done in 10 cm bin
sizes between $\pm 100$ cm (20 bins), and 4 wider bins with limits
$[-190, -140],\ [-140, -100],\ [100, 140],\ [140, 190]$ in order
to get enough statistics (see the \zvertex\ distribution,
Fig.~\ref{fig:zvtx}). We again see that there are systematic
effects in the raw yield. We can identify two general trends.  For
tracks that do not cross the entire TPC, the yield drops rapidly.
For the remaining tracks, the raw yield is the lowest for tracks
closest to the central membrane and increases as the tracks get
closer to the pad planes on either side of the TPC. For reference,
a fit to the raw data assuming a constant yield is done (black
points) with a resulting poor $\chi^2$; 6.8 (left) and 8.8
(right). The corrected yield is essentially free of \zvertex\
systematics, except for two cases. In the events closest to the
edges of the TPC, where the acceptance is varying rapidly and most
of the tracks in the event lay outside the TPC, the tracks that do
not cross the entire TPC show a corrected yield that is lower than
the rest.  This is expected, as tracks with very little
information left in the detector will be difficult to reconstruct.
We also found an additional systematic effect in the efficiency
correction traced back to the simulation of tracks crossing the
central membrane. This effect causes a slight over correction of
the yield for such tracks, as can be seen by the corrected yield
for these tracks which is coloured differently in the figure.
Therefore, to prevent these systematic effects from appearing in
the final pseudorapidity distribution, we included only events
within $\pm 100$ cm of the center of the TPC, excluded tracks
going through the membrane, and excluded tracks which did not
cross the entire TPC.  Each data point in the final \hminus\
pseudorapidity and \pt\ distributions is then obtained as the
average of each of the $\sim 20$ independent measurements of the
corrected yield obtained for each of the \zvertex\ bins.

\begin{figure}[htb]
  \centering
  \mbox{\subfigure[]{\includegraphics[width=0.45\textwidth]{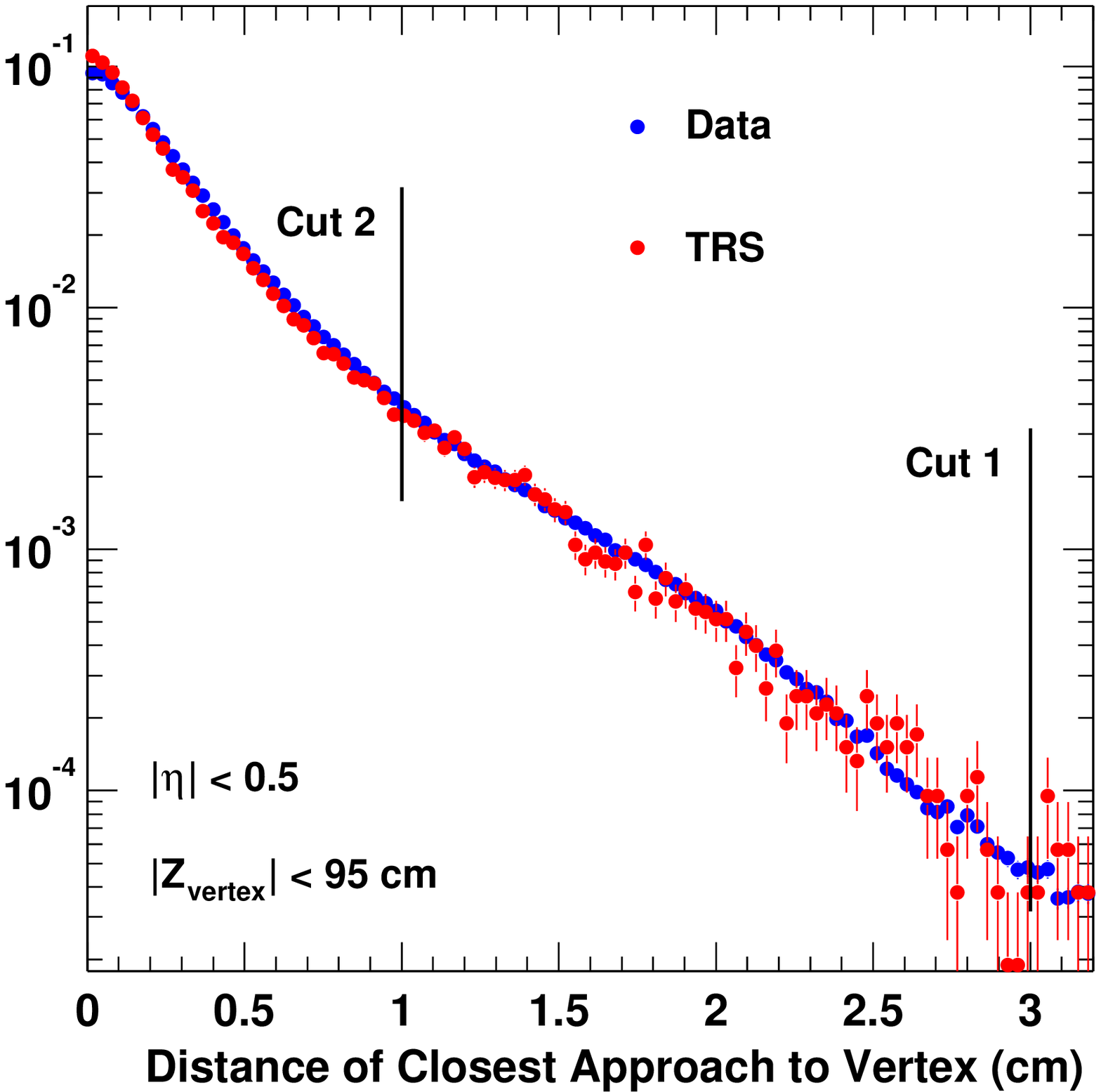}}\quad
  \subfigure[]{\includegraphics[width=0.45\textwidth]{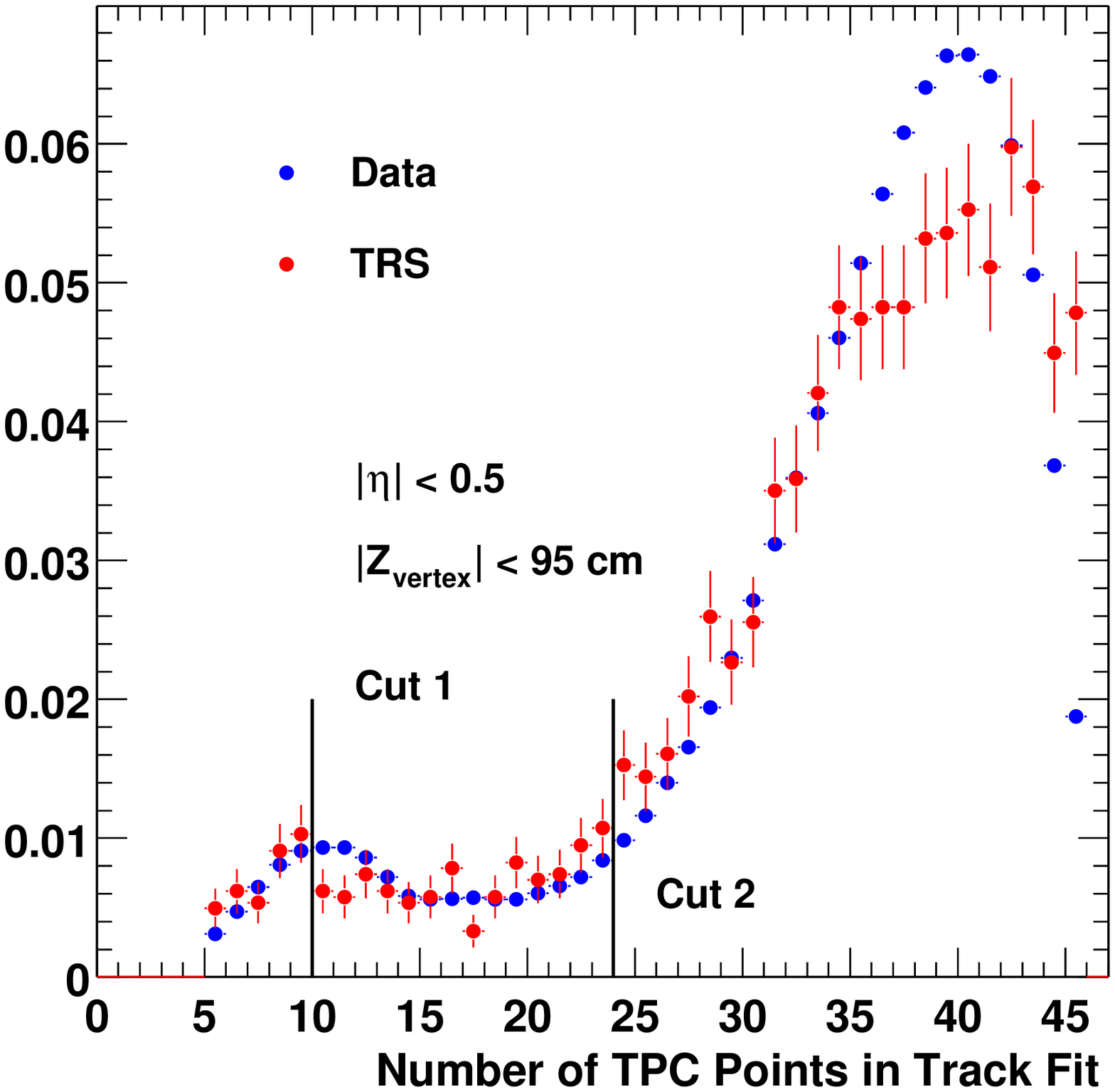}}}
  \caption[Distance of closest approach and Fit point distributions.]
  {Comparisons of the distance of closest approach (a) and number of fit points (b) distributions
  for simulated and real data.}
  \label{fig:dcaFitPts}
\end{figure}

Since the most important correction for \pt\ spectra and yields is
the efficiency, this is where we concentrated a large fraction of
the studies of systematic effects. The systematic uncertainty due
to the corrections was estimated in two ways.  In order to ensure
consistency of the results, we studied the variation in the final
spectra due to a large variation in the track quality cuts. The
distributions of the cut variables are given in
Fig.~\ref{fig:dcaFitPts} from both data and simulations; the left
panel shows the distance of closest approach distribution and the
right panel shows the number of fit points distribution. We varied
the selection based on number of fit points from 10 to 24 and the
selection based on distance of closest approach from 3 cm to 1 cm.
These choices are labelled ``cut 1'' and ``cut 2'' in the figure.
The different corrections for the more stringent set of cuts (cut
2) were then recalculated. The final multiplicity distribution
obtained in both cases was compared and they were in agreement to
$\sim 1\%$. To study the sensitivity to the cuts a small variation
in the track quality cuts was then made, accepting tracks with 23
fit points. A corrected spectrum using the correction factors
calculated for tracks with 24 fit points was then applied.  This
yields then a corrected distribution that is systematically higher
than the measured value, as one is then over-correcting the raw
data.  The yield at high multiplicity obtained via this systematic
effect was found to be 6.4\% above the measured yield.  This is
the main contribution to the systematic uncertainty in the total
particle yield for the analysis presented here.

\subsection{Backgrounds}
\begin{figure}[htb]
  \centering
  \mbox{
  \subfigure[]{\includegraphics[width=0.45\textwidth]{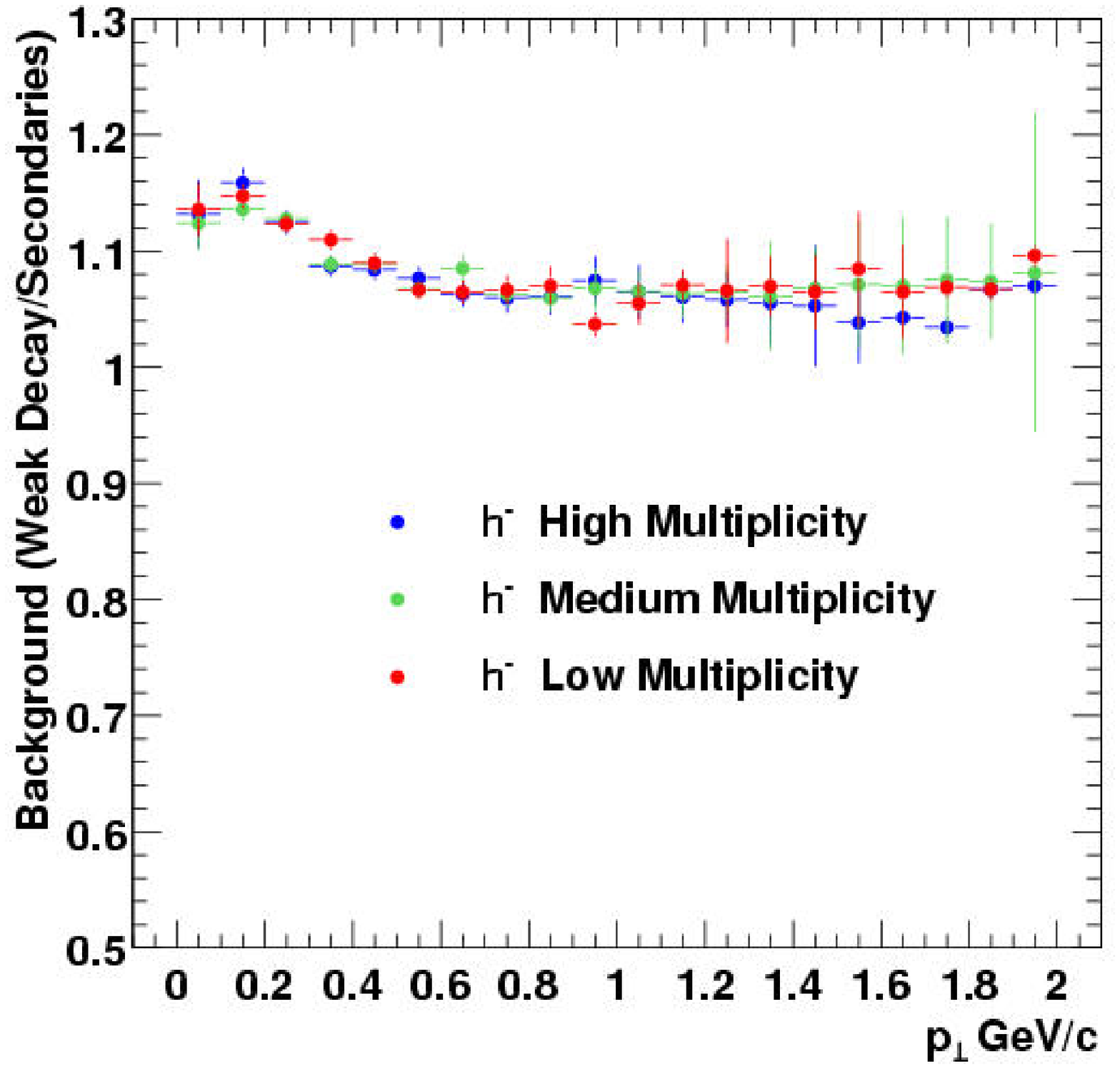}}\quad
    \subfigure[]{\includegraphics[width=0.45\textwidth]{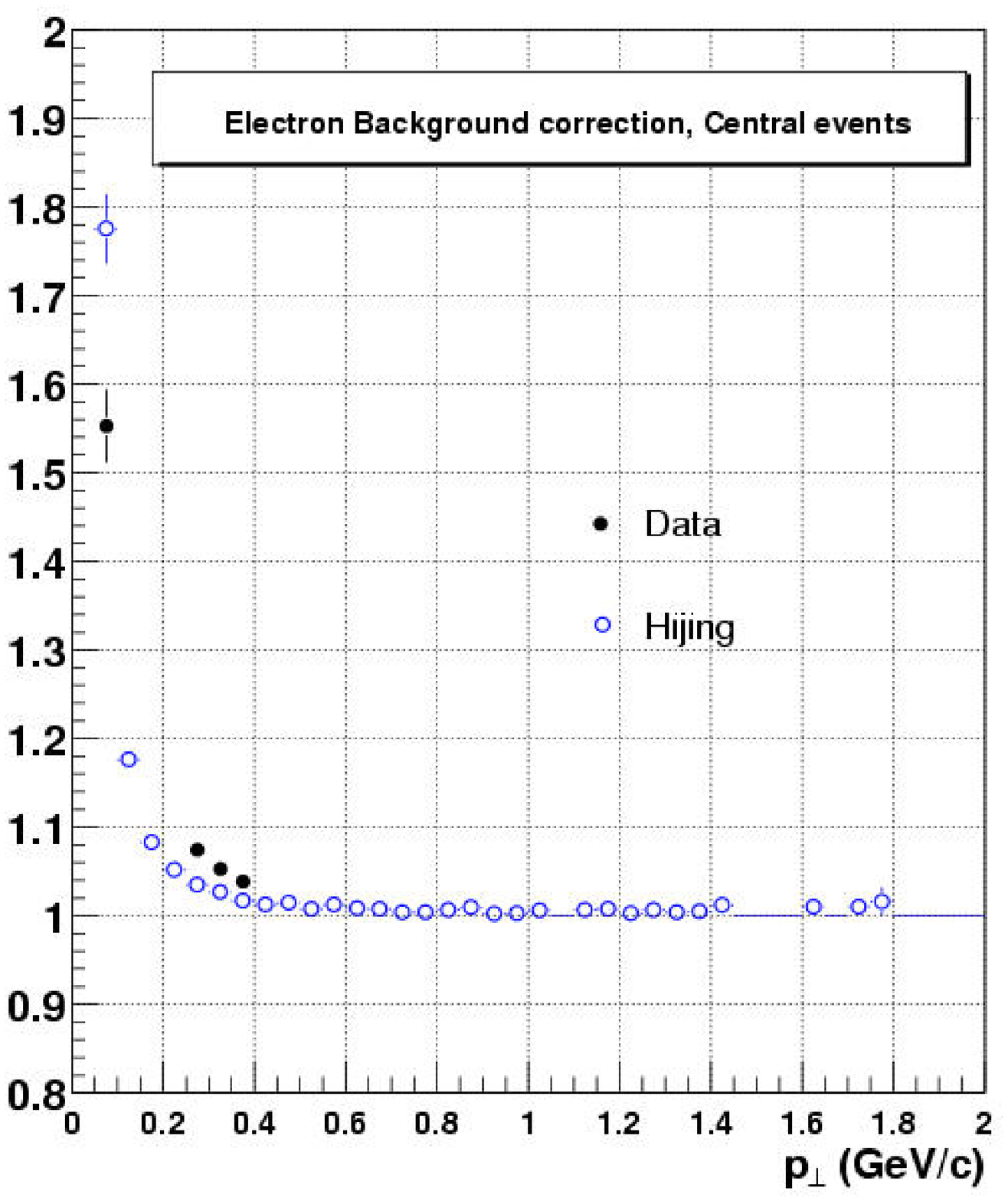}}}
  \caption[Background corrections \vs\ \pt.]
  {Background correction to the \hminus\ spectrum coming from weak decays and secondary
  interactions in the detector material (a) and electrons (b) as a function of \pt.
  The corrections were obtained
  via \Hijing\ events processed through the STAR offline simulation and reconstruction chain.
  }
  \label{fig:background}
\end{figure}

Instrumental backgrounds due to photon conversions and secondary
interactions with detector material were estimated using the
detector response simulations mentioned above, together with
events generated by the \Hijing\ model \cite{hijing:91,hijing:94}.
The simulations were calibrated using data in regions where
background processes could be directly identified. The measured
yield also contains contributions from the products of weak
decays, primarily \kzeros, that were incorrectly reconstructed as
primary tracks that must also be accounted for in the background
correction. Figure~\ref{fig:background} (left panel) illustrates
the background fraction of the raw signal coming from secondary
interactions and decays.  The shape of the background correction
is independent of multiplicity. We see that the hadronic
background is approximately 15\% at \pt\,= 100 \mevc, decreasing
with increasing \pt\ to a constant value of $\sim 7\%$. The
average fraction of hadronic background tracks in the uncorrected
sample is 7\%.  The error on the background correction increases
with \pt\ because statistics are limited, but there is also a
systematic uncertainty associated with this correction.  Since the
background depends on the yields of various particles, in
particular \kzeros\ and $\bar{\Lambda}$, differences between the
model and the data will systematically affect the correction. An
estimate of the background correction at \pt\ = 2 \gev\ assuming
the ratios $\bar{\Lambda}/\pbar \sim 0.8$ and $\pbar/\piminus \sim
1$ would yield a value of $\sim 1.15$ instead of $1.07$. This is
the main source of systematic uncertainty in the hadronic
background correction.

The right panel of Figure~\ref{fig:background} shows the electron
background obtained from \dedx\ data (filled points) and from
\Hijing\ simulations (hollow points).  The shape of the electron
background is exponential, so we can use the shape and normalize
to the data in order to obtain the final electron correction.
Since the slope of the exponential can affect the final slope of
the \pt\ distribution, we can also do a correction using the
\Hijing\ points to study the variation in the \pt\ spectrum.  We
find that the choice of the slope in the electron correction
changes the corrected value of \meanpt\ by 10 \mevc\ and is the
main contribution to the systematic uncertainty of the
determination of \meanpt.

\subsection{Comparison of TPC Halves and Sector ($\phi$) Dependence}
From the design of the TPC, the tracking volume is essentially
divided into two independent halves and it is not \textit{a
priori} obvious that one can combine them without introducing
systematic effects. Therefore, one of the necessary studies in the
evaluation of the reliability of the results is to compare the raw
yields in the different halves.  However, one has to be careful in
this comparison to isolate the possible differences.  Because of
the design of the TPC, we expect to see differences in the raw
yields even within the same half of the TPC simply due to an
increased drift length.  As an example, we can take tracks close
to midrapidity, which are emitted at an angle close to $90^\circ$.
As the event vertex is not a fixed quantity in the z axis for the
different events, tracks at midrapidity coming from events with
different vertex z positions will have different drift distances,
and thus we expect systematic differences in the raw yield at
fixed $\eta$ for varying vertex z positions, as was discussed in
Section ~\ref{sec:efficiency}.  To isolate differences between the
East ($z<0$) and West ($z>0$) halves of the TPC, we therefore have
to make sure that we take event and track samples that are related
by the transformation $\eta^{\mathrm{East}} = -
\eta^{\mathrm{West}}$ and $z_{\mathrm{vertex}}^{\mathrm{East}} = -
z_{\mathrm{vertex}}^{\mathrm{West}}$. In addition, as each TPC
half is made up of 24 independent sectors to cover the full
azimuthal range, it is also interesting to do a comparison of the
raw azimuthal yields to make sure there are also no systematic
effects.  In doing this comparison, one can also separate the raw
yields from positively and negatively charged tracks, for there
might be distortions that affect these tracks differently. The
tracks selected must also satisfy requirement of having at least
10 fit points (``cut 1'' in Fig.\ref{fig:dcaFitPts}(b)).  This cut
is also used for the \hminus\ analysis.
\begin{figure}[htb]
\subfigure[]{\includegraphics[width=.48\textwidth]{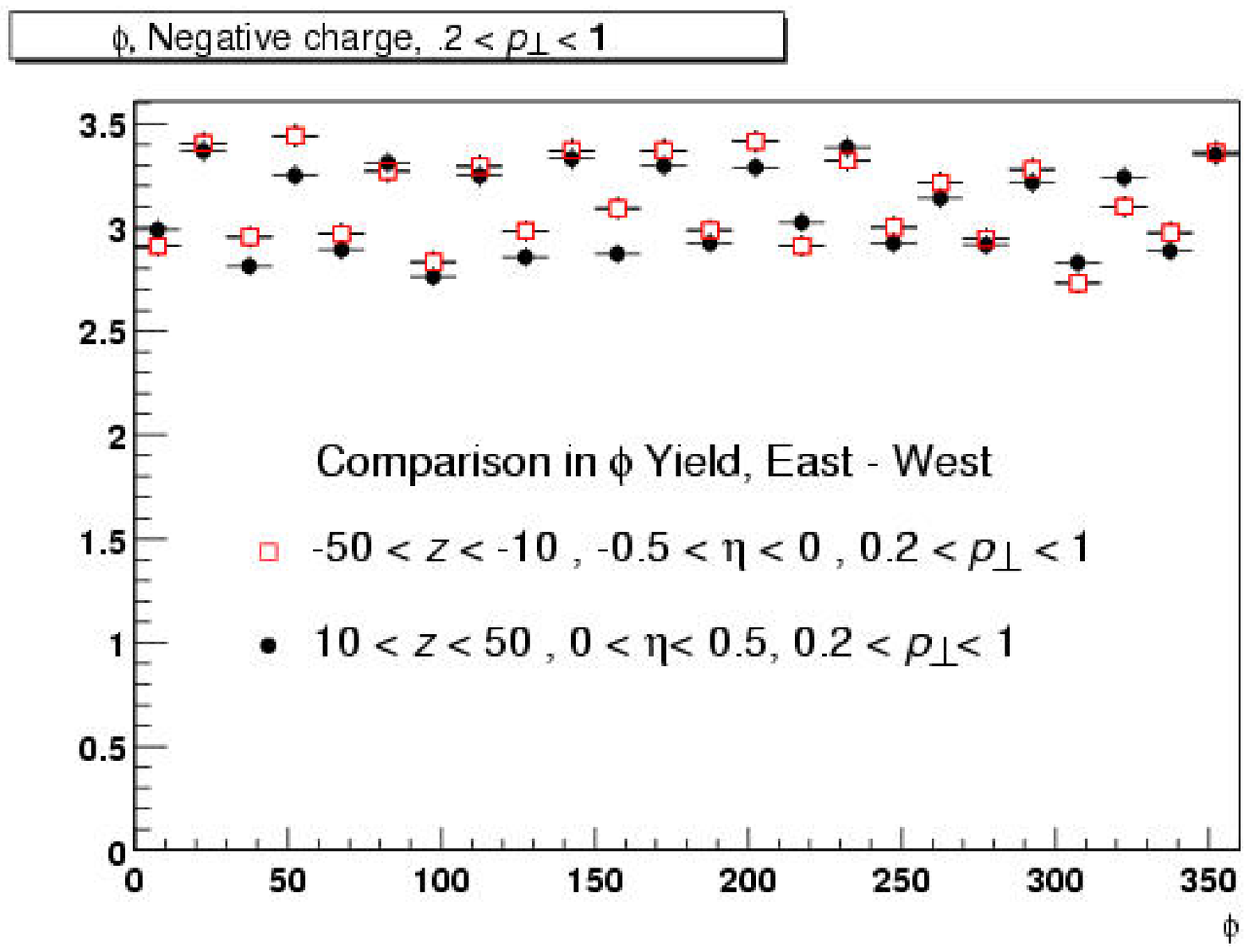}}\quad
\subfigure[]{\includegraphics[width=.48\textwidth]{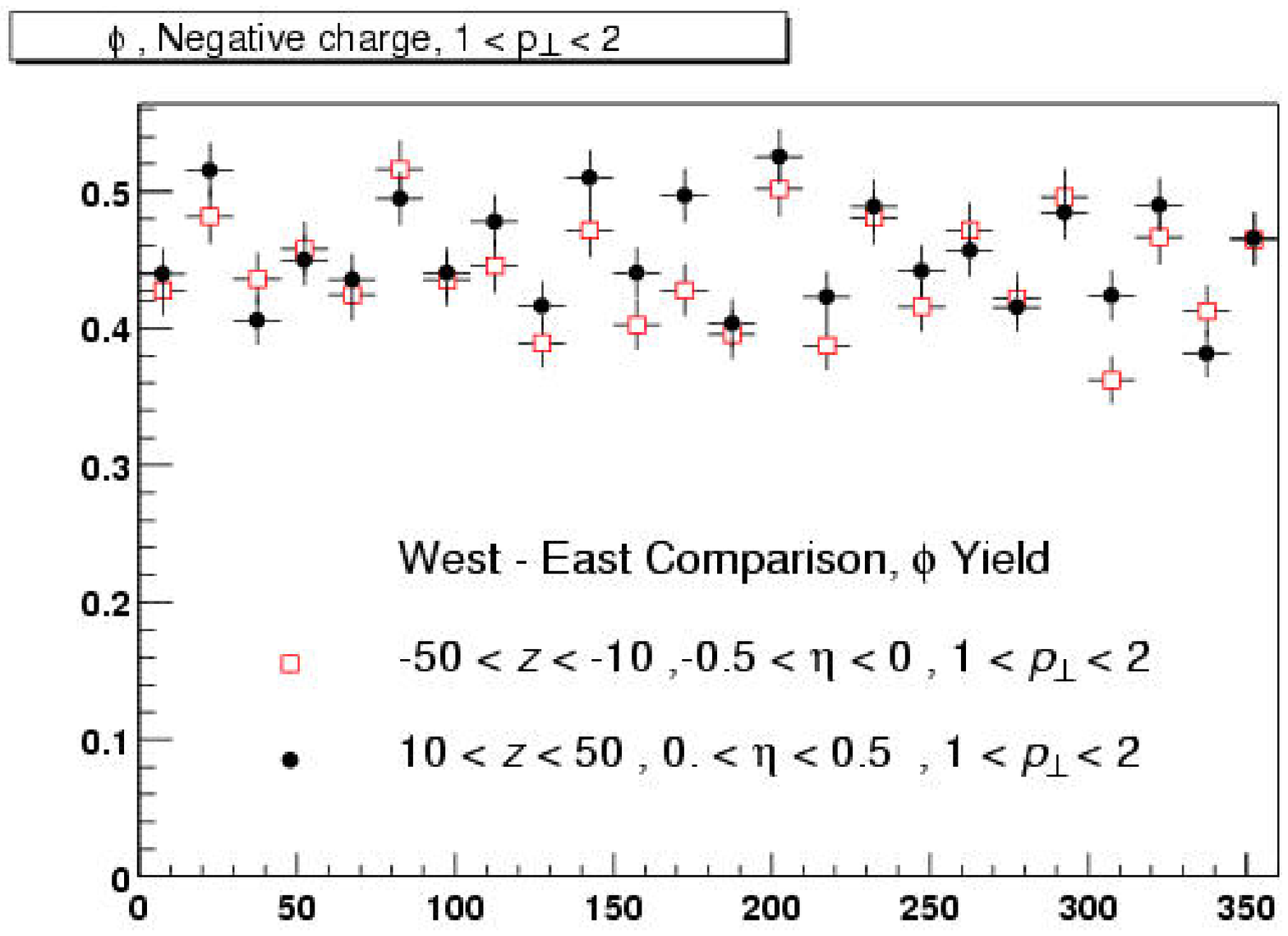}}
\caption[Raw $\phi$ distribution for the two TPC halves for
\hminus.]{Raw $\phi$ distribution of East and West half of the TPC
    for negatively charged
    particles. Left panel: low \pt\ (0.2 - 1 \gevc). Right panel: moderate \pt\ (1 - 2 \gevc)}
\label{fig:eastwestnegch}
\end{figure}

Figure \ref{fig:eastwestnegch} shows the raw azimuthal
distribution for negatively charged tracks in the two halves of
the TPC.  The East half track sample, shown as the hollow square
data points, was obtained from events with $-50 <
z_{\mathrm{vertex}} < -10$ cm and from tracks with $-0.5 < \eta <
0 $.  Similarly, for the West half the events had a vertex
selection of $10 < z_{\mathrm{vertex}} < 50$ cm and tracks were
selected according to $0 < \eta < 0.5 $.  The left panel shows the
yields for the low \pt\ region, 0.2 < \pt\ < 1 \gevc.  A similar
plot showing the raw azimuthal distribution in the range 1 < \pt\
< 2 \gevc\ is shown in the right panel.  The samples are
normalized per event to isolate the differences in the raw yields.

\begin{figure}[htb]
\begin{center}
\includegraphics[width=.5\textwidth]{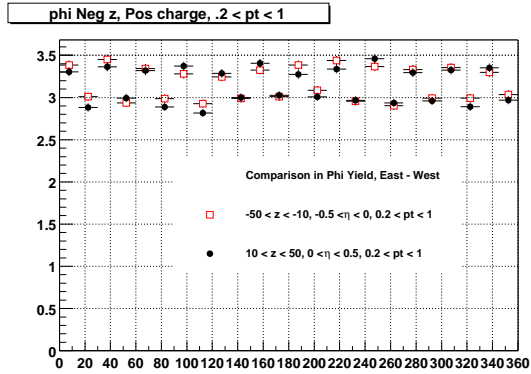}
\caption[Raw $\phi$ distributions at low \pt\ for \hplus.]{Raw
$\phi$ distribution of East and West half of the TPC for low
transverse momentum (0.2 - 1 \gevc) positively charged particles.}
\label{fig:eastwestposchlowpt}
\end{center}
\end{figure}

The size of the bins in the azimuthal direction is $15^\circ$, or
half a sector. We see that there is a periodic structure to the
distribution in both cases. This comes about because charged
tracks will curve in the magnetic field. Therefore, depending upon
their charge sign and entrance point to a sector (\ie\ their
azimuthal angle), will have a trajectory that is either fully
contained in a sector boundary or that crosses a sector boundary.
For a specific charge sign, tracks in one side of the sector will
be more easily reconstructed than on the other side, giving the
structure seen in Figure \ref{fig:eastwestnegch}. This effect is
more pronounced for low momentum (\ie\ large curvature) tracks.
The high yield bins for the negative charge tracks should be the
low yield bins for the positive ones.  The low momentum positively
charged tracks are shown in Figure \ref{fig:eastwestposchlowpt}.
Indeed we see the expected change, for this case the first bin is
a high yield bin (compare to Fig.~\ref{fig:eastwestnegch}).

\begin{figure}[htb]
\begin{center}
\subfigure[]{\includegraphics[width=.4\textwidth]{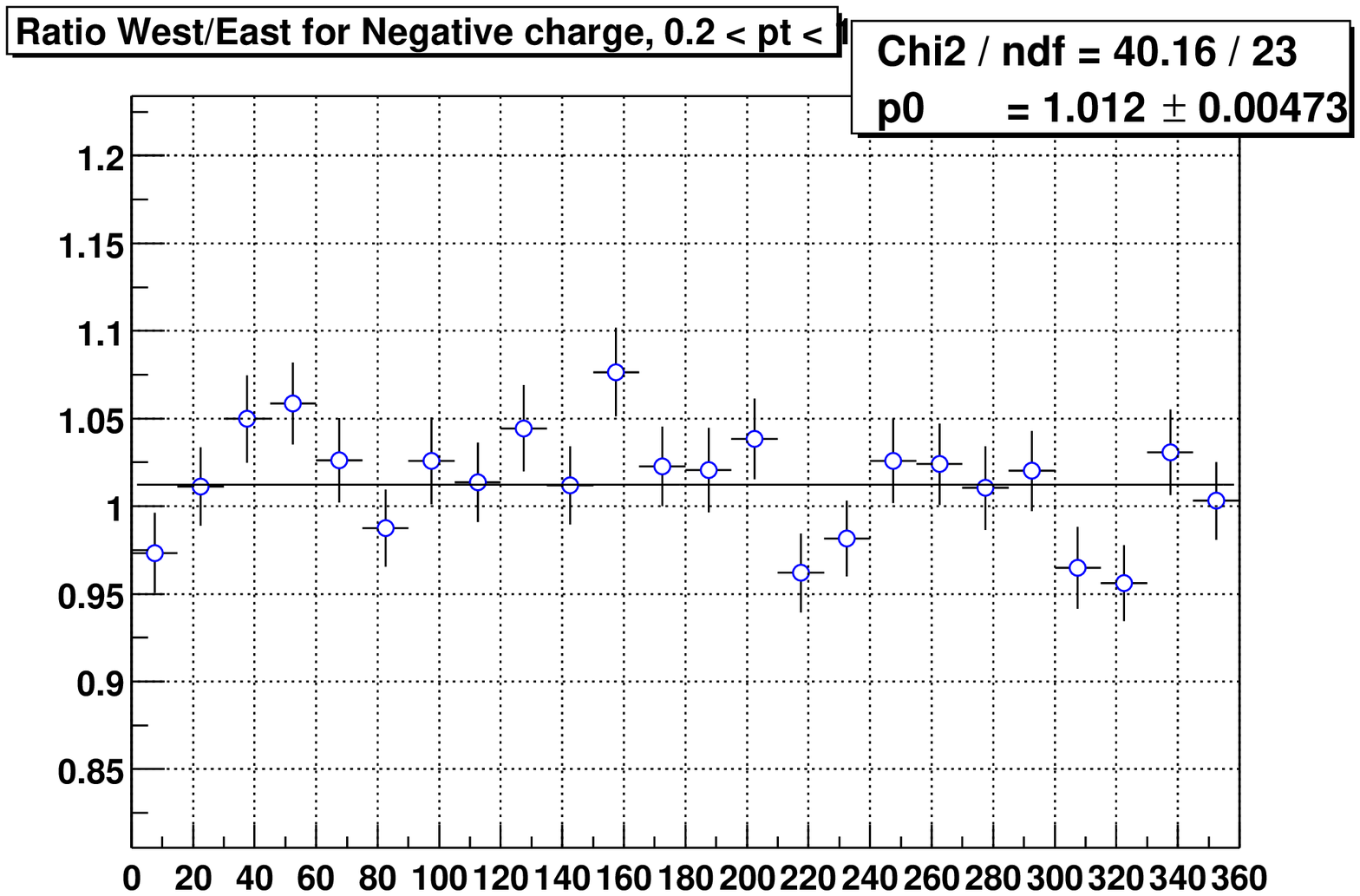}}
\subfigure[]{\includegraphics[width=.4\textwidth]{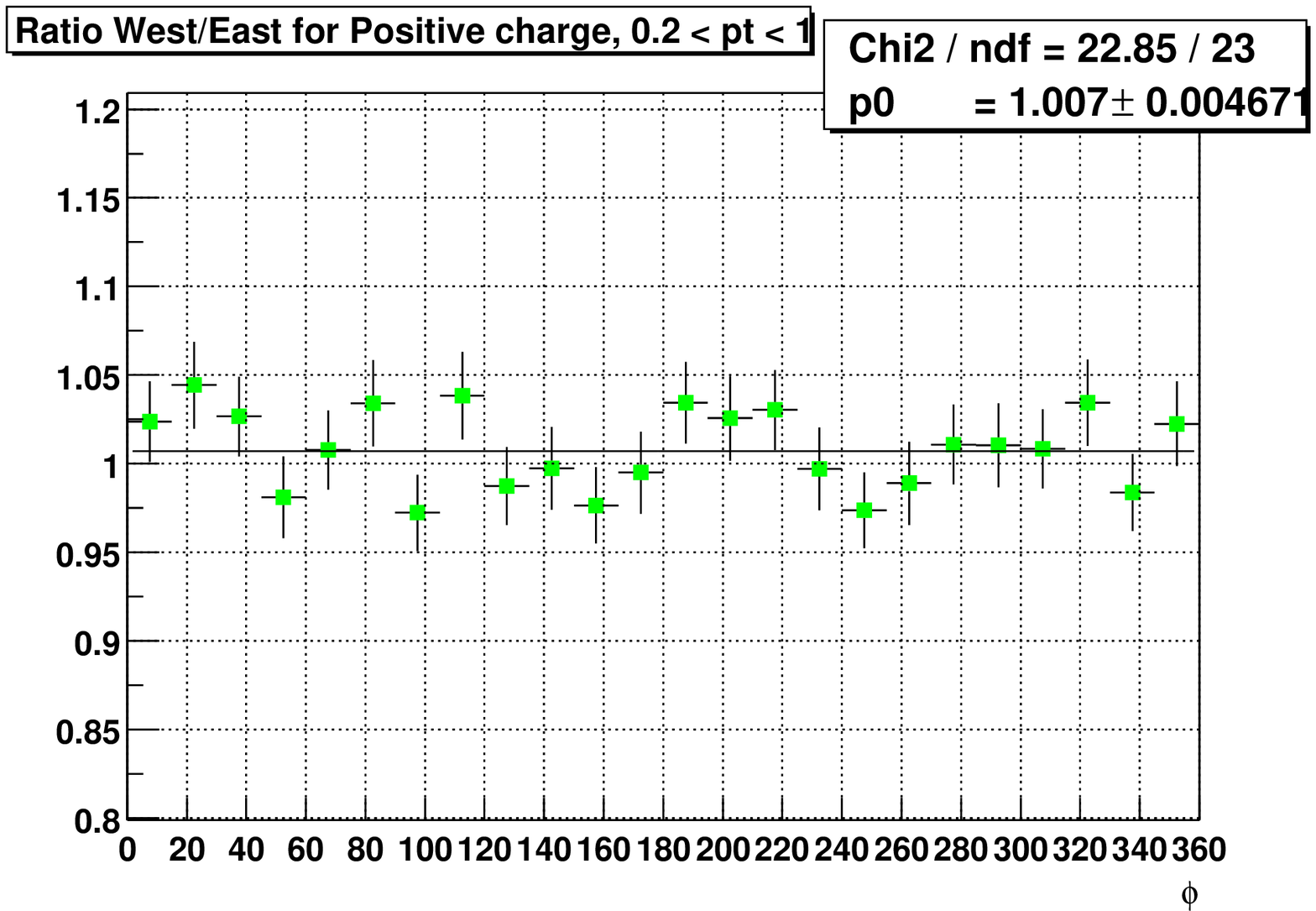}}
\caption[Ratio of $\phi$ distributions for \hminus\ and
\hplus.]{Ratio of raw $\phi$ distribution of East and West half of
the TPC for low transverse momentum (0.2 - 1 \gevc) negatively (a)
and positively (b) charged particles.}
\label{fig:eastwestlowptratio}
\end{center}
\end{figure}

To focus on changes between East and West halves of the TPC, we
make a ratio of the previous histograms.  Figure
\ref{fig:eastwestlowptratio} shows the ratio for low \pt\ negative
(a) and positive (b) tracks.  The average difference in both cases
is on the order of 1\%.

We can also make a comparison of the $\hminus/\hplus$ ratio in
both halves of the TPC.  Since we know that there will be
systematic differences as a function of $\phi$ due to the
curvature effect mentioned before, we can simply make a ratio of
the raw yields integrated over $\phi$.  For the East half of the
TPC we find $\hminus/\hplus = 0.983 \pm 0.005$ and for the West
half a similar analysis yields $\hminus/\hplus = 0.988 \pm 0.005$
where the errors are statistical only.  We conclude that for the
purposes of this analysis the two halves of the TPC yield
sufficiently similar results.

However, we will still separate the corrections to the raw yields
in other variables as mentioned in Sec.~\ref{sec:efficiency}.  We
expect variations in the raw yields due to different track
geometries and event topology. In summary, we expect that tracks
that do not cross the entire tracking volume of the TPC to be more
difficult to reconstruct than those tracks that do. In addition,
it is much easier to reconstruct the tracks when there are not
many tracks in the detector, so we must see a decrease in the
tracking efficiency in a high multiplicity environment.
Furthermore, low momentum tracks coming from the interaction
vertex will have a difficult time reaching the tracking volume.
The radial distance at the center of the first TPC sensitive pad
row is 60 cm, which along with the magnetic field of 0.25 T places
an effective low \pt\ acceptance cut-off for primary tracks of
$\sim 50\ \mevc$.  Finally, as mentioned before even tracks in
identical regions of phase-space will have different raw yields
depending upon where the event vertex was placed because they will
sample a different detector geometry. Therefore, we have divided
all the tracking corrections according to
\begin{itemize}\spacing{0.6}
  \item the phase-space cell occupied by the tracks,
  \item the multiplicity of the event, and
  \item the position of the primary vertex.
\end{itemize}\spacing{1.2}
The previous corrections are the most important ones in the
present analysis.  The following corrections to the spectra and
yields were also studied. Each of them was found to produce
changes of less than 1\%.

\subsection{Momentum resolution}\label{sec:momentumResolution}

The momentum resolution is momentum dependent.  It is well known
that for a \pt\ distribution, the resolution of the detector will
introduce a change in the slope of the observed spectrum. This can
be seen from the following simple argument.  The effect of a
finite momentum resolution is that a certain number of tracks will
be reconstructed with the wrong momentum, and will therefore be
counted in an incorrect bin. It is possible for a track to be
reconstructed with a lower or a higher \pt.  So for a given \pt\
bin, there will be a loss of particles to adjacent bins and a gain
of particles from adjacent bins.  The magnitudes of these fluxes
compared to the yield in the given bin are the important
quantities.  From the rapid decrease in cross section with
increasing \pt, we expect that the feeding of particles from the
lower to the higher \pt\ bins will be higher than the flux in the
opposite direction, and hence the net effect is to flatten the
spectrum to some extent.  This effect becomes important when the
momentum resolution is of the same order as the size of the \pt\
bin. It is therefore necessary to quantify this effect in any
momentum analysis.

For the results presented here we concentrate on  $\pt \leq 2$
\gevc. We find the momentum resolution to be better than 4\% as
illustrated in Fig.~\ref{fig:ptresol}. For a \pt\ bin width of 100
\mevc\ we expect the correction due to momentum resolution to be
less than 1\% in the full range of \pt. This effect was quantified
in two different analyses. The first approach relies on the
embedding procedure, and consists of dividing the phase space into
bins and using the track matching between simulated and
reconstructed tracks from. We can keep track of the \pt\ bin in
which a particle was created, and then see if it was reconstructed
in the same bin. We can thus know, for each phase space bin, what
percent of tracks are reconstructed correctly, what percent are
reconstructed in a different bins and correct for the resolution.

The only significant correction found this way occurs for the very
first \pt\ bin.  The reason is the following.  The raw yield in
each bin is composed of two parts: the tracks which were correctly
reconstructed in the same \pt\ bin as the one they were generated
in, let's call these the \textit{healthy} tracks; and the tracks
that are found by the reconstruction but placed in a different bin
than the one they were generated in, let's call these the
\textit{crippled} tracks.  Now let's focus on the region $\pt <
100\ \mevc$ where the efficiency drops very rapidly with
decreasing \pt. There will be very few healthy tracks in the first
\pt\ bin because the efficiency is low. There will be, however, a
considerable amount of crippled tracks in the first bin coming
from the feed-down from the next (high efficiency) \pt\ bins. The
crippled tracks are almost as numerous as the healthy population
for this bin.  We therefore must apply a correction to obtain a
realistic estimate of the initial healthy population in that bin.
For the region $\pt > 200\ \mevc$ the efficiency is independent of
\pt\ so we need only concentrate on the \pt\ dependence of the
parent distribution which we want to measure.

We can also calculate the expected correction based on a knowledge
of the momentum resolution, the bin size, and a given \pt\
distribution. One way to treat this problem is through an
iterative procedure, starting with a given input \pt\
distribution, doing a Monte Carlo study by smearing the tracks
with the measured \pt\ resolution and looking at the shape of the
resulting distribution.  We repeat the process until the output
distribution matches the one measured in the experiment.  The
approach we followed relies on a related method. Starting from a
given input \pt\ distribution we use the measured \pt\ resolution
as a function of \pt\ to construct a set of Gaussians (one for
each \pt\ bin).  We also used different functional forms to
parameterize the shape of the \pt\ resolution, \eg\ Lorenzian and
double-Gaussian, yielding similar results.  The area under each
Gaussian is the initial yield for each \pt\ bin, the mean is the
center of the \pt\ bin and the $\sigma$ is obtained from the
measured $\delta\pt$ \vs\ \pt\ curve. We can then figure out what
is the contribution of each \pt\ bin to any other bin in
principle.  In practice, each bin only contributes mainly to its
nearest and next-to-nearest neighbours. The observed yield in the
$i^{\mathrm{th}}$ \pt\ bin, $N_i^{\mathrm{obs}}$ is then:
\begin{equation}\label{eq:resYieldIthBin}
  N_i^{\mathrm{obs}} = \sum_{k=0}^\infty N_k
  \int_{p_\perp(i)}\frac{1}{\sqrt{2\pi}\sigma_k}e^{\frac{(p_\perp-p_\perp(k))^2}{2\sigma_k^2}}dp_\perp
\end{equation}
We can think of Eq.~\ref{eq:resYieldIthBin} as defining a matrix
equation
\begin{equation}\label{eq:resMatrix}
  N_i^{\mathrm{obs}} = \sum_{k=0}^\infty N_k C_{ki}
\end{equation}
where the elements of the matrix $C_{ki}$ are defined by the
integral of Eq.~\ref{eq:resYieldIthBin}, and are interpreted as
the percent contribution of bin $k$ to bin $i$, \ie\ the mean of
the Gaussian is the center of bin $k$ and the integration limits
are given by the upper and lower limits of the $i^{\mathrm{th}}$
\pt\ bin. The width of the Gaussian is given by the detector
resolution at \pt\ bin $k$. Therefore, the integral depends only
on quantities that are measurable (the resolution) or defined by
our analysis (the bin limits and bin center). This matrix then
embodies all our knowledge of the effects of resolution.

For a given \pt\ distribution, we can then calculate what the
observed distribution will be and obtain appropriate correction
factors for the effect of resolution.  We have performed this
procedure with different input distributions -- power-law
function, exponential in \pt\ and exponential in \mt\ with various
slope parameters -- and find that in all cases the correction for
any given \pt\ bin is less than 1\% in the range up to 2 \gevc\
with our given bin size. This is true for both the \hminus/\hplus\
analysis (bin size 100 \mevc, range 0.1 - 2 \gevc) and for the
\piminus\ analysis (bin size 50 \mevc, range 0.05 - 0.75 \gevc).
For higher \pt\ the resolution plays a more significant role,
where we expect a correction of 20\% or more at ~ 5 \gevc\ with a
1 \gevc\ bin size and the present magnetic field of $B = 0.25$ T.

An additional advantage of this resolution study is that, once we
construct the $C_{ki}$ resolution matrix, we can also use it to
recover the original yields $N_k$ by inverting the matrix, and no
iterative procedure is needed:
\begin{equation}\label{eq:resCkiInverse}
  N_k = \sum_{i=0}^\infty N_i^{\mathrm{obs}} (C_{ki})^{-1}
\end{equation}

\subsection{Track splitting}
For a loose track selection based on the number of points
reconstructed in the TPC, it is possible to overestimate the track
yield in the presence of split tracks. This effect can come about
for example when the track-finding algorithm fails to recognize
two track segments as belonging to a single particle.  This can
happen typically when there are gaps in the track pattern, such as
tracks crossing sector boundaries, tracks crossing the central
membrane of the TPC, or tracks crossing a region of the TPC where
the read out pad is noisy or dead.

To study the effect of splitting in this analysis, we used two
approaches that relied on the detector simulation.  An additional
study was made using only real reconstructed tracks. All three
results are in agreement both in the size of the overall effect
and on its $\pt$ dependence.  We will discuss the approach that
relies on full simulation, since this is the most
straight-forward.

In a full Monte Carlo generated event, we have all the information
about the input tracks.  Running the tracks through the GEANT
implementation of the STAR detector, we obtain among other things
the information on the TPC energy deposition left by the tracks.
This is the input to the microscopic simulation of the TPC
response yielding as output a simulated raw-data file that is then
passed through the STAR reconstruction chain.  A comparison of the
TPC space points that are found by the reconstruction algorithms
to the GEANT input is then performed.  As discussed in Section
\ref{sec:embedding}, the matching as implemented in this analysis
is based on spatial proximity. A feature of this matching
procedure is that it allows a many-to-many matching: if 2 hits are
very close together for example there will be a 2-to-2 hit match.
Each of the 2 Monte Carlo hits will be matched to the 2
reconstructed hits.

The match of simulated point to reconstructed point serves as the
footing for a simulated track to reconstructed track association.
Typically, the association is 1-to-1.  A \textit{split} track will
have a different topology. There will be one original Monte Carlo
track, but it will be associated to 2 (or more) reconstructed
tracks. The association algorithm allows a many-to-many type of
matching, so one must be careful to really single out a split
track from other matching topologies. We look for a single Monte
Carlo track matched to more than one reconstructed track, and
those reconstructed tracks are singly matched to the Monte Carlo
track.  Through this procedure, an estimate of the split track
population and was found to be also below the 1\% level.  In
addition, the analysis done with the more stringent requirement on
the number of fit points (``cut 2'' in
Fig.~\ref{fig:dcaFitPts}(b)) guaranteed that there are is no
double counting from split tracks.  The results obtained from both
analyses were found to be consistent.

\subsection{Track merging}

When 2 tracks lie very close together, the cluster finder might
not be able to resolve the 2 ionization peaks and produce a single
space point when there were originally 2 for each hit in the track
trajectory.  The losses due to this effect were estimated to be
less than 1\% for the most central collisions and negligible for
peripheral collisions.

\subsection{Ghost Tracks}

This correction takes into account possible cases where the track
finding algorithm might associate space-points from different
tracks and give us track parameters from a non-existent particle.
It was not known if this could be of importance in the high
multiplicity environment of a heavy ion collision at high energy.
It was found that the TPC occupancy was low enough that this did
not present a problem.  Essentially no such tracks were found in
all the simulation and embedding analyses even at higher simulated
multiplicities than those observed in the data.

Finally, all analyses were carried out with 3 different software
production versions.  As we understood the systematics of the
detector better, we incorporated our knowledge of the
calibrations, distortions and corrections to software bugs. The
first production was done in August while data was still being
taken.  This was followed almost immediately by a new production
one month later with improved calibrations. The data presented
here is produced with the calibrations and distortion corrections
processed in early 2001.  The biggest systematic effect observed
throughout this process was a 4.5\% change in the efficiency
obtained from the first simulations compared to the following two
software versions.  This was understood as coming from a more
realistic parameterization of the allowed space point errors used
during the track finding algorithm based on the measured residuals
obtained in the first analysis of the data.

%
%
\chapter{Results and Discussion: Charged Hadrons}
\label{ch:NchResults}
\section{Multiplicity Distribution}\label{sec:Multiplicity}
\subsection{Results}
\begin{figure}[htb]
\begin{center}
    \includegraphics[width=0.75\textwidth]{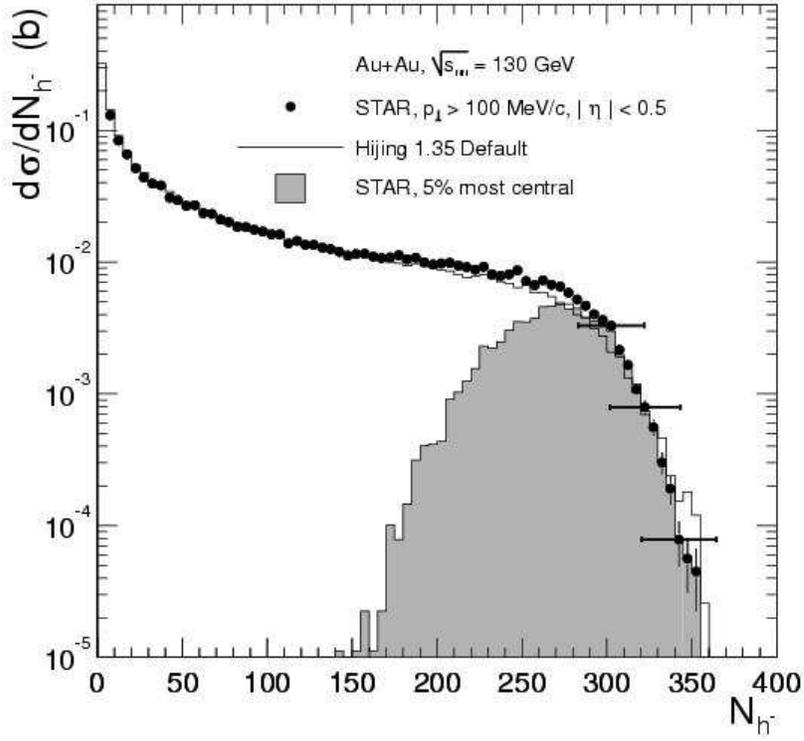}
       \caption[The \hminus\ multiplicity distribution]
       {Normalized multiplicity distribution of \hminus\ with $\pt > 100\
        \mevc$, $|\eta|<0.5 $ in \AuAu\ collisions at $\sqrtsNN\, = 130\ \gev$.
        The shaded
        area is 5\% most central collisions, selected by ZDC
        signal (see text).  The solid curve is the prediction from the
        {\sc hijing} model.}
        \label{fig:hminus}
    \end{center}
\end{figure}

Figure \ref{fig:hminus} shows the corrected, normalized
multiplicity distribution within $|\eta| < 0.5$ and $\pt > 100\
\mevc$ for minimum bias \AuAu\ collisions.  For this plot, a total
of 350K minimum bias triggers were analyzed, yielding 119,205
events with a reconstructed vertex in the region |\zvertex| < 95
cm.
The distribution for the 5\% most central collisions (360 mbarn),
defined using ZDC coincidence, is shown as the shaded area in
Fig.~\ref{fig:hminus}.  The definition adopted here was based on
the signals shown in Fig.~\ref{fig:zdcvsctb}, by selecting events
with ZDC sum $< 66$ ADC counts and a high CTB threshold placed at
8500 ADC counts to remove the ambiguity between central and
peripheral collisions.  The actual value of the ZDC sum threshold
is set by requiring the shaded area in Fig.~\ref{fig:hminus} to
equal 5\% of \sigmahad.  This is the event selection that was used
for the \hminus\ \pt\ and $\eta$ distributions for central events
presented in the following sections.

The data were normalized assuming a total hadronic inelastic cross
section of 7.2 barn for \AuAu\ collisions at $\sqrtsNN\,= 130\
\gev$, derived from Glauber model calculations. The first bin
(below \Nhminus\,= 5) is not shown, due to systematic
uncertainties in the vertex reconstruction efficiency and
background contamination. Its relative contribution to the total
cross section was estimated to be 21\% by normalizing the {\sc
hijing} multiplicity distribution to the measured data in the
region $5<\Nhminus<25$ (\ie\ the first three bins of
Fig.~\ref{fig:hminus}). This procedure relies on the assumption
that very peripheral interactions are well described by the
superposition of a few nucleon-nucleon collisions in the geometry
of a nuclear collision, and can therefore be accurately modelled
by {\sc hijing}.  A comparison of the actual yield measured in the
first bin (after the efficiency correction) to the shape predicted
by the \Hijing\ model indicates a relative difference of 10\%. The
systematic uncertainty on the vertical scale is therefore
estimated to be $\sim 10\%$ and is dominated by uncertainties in
the total hadronic cross section and the relative contribution of
the first bin.  The systematic uncertainty on the horizontal scale
is 6.4\% for the entire range of multiplicity and is depicted by
horizontal error bars on 3 data points only to maintain clarity in
the figure.

The shape of the \hminus\ multiplicity distribution is dominated
over much of the \Nhminus\ range by the nucleus-nucleus collision
geometry, consistent with findings from lighter systems and lower
energies. However, the shape of the tail region at large \Nhminus\
is determined by fluctuations and detector acceptance.  These
overall features are also observed in the \Hijing\ calculation,
shown as histogram in Fig.~\ref{fig:hminus} (although a more
detailed comparison indicates that, for the model, the shape of
the tail shows a different contribution due to fluctuations than
that found in the data).

\subsection{Discussion}

The negatively charged particle multiplicity distribution reveals
that the bulk features of the dynamics are dominated by the
geometry.  The shape shows the same features expected from a
Glauber picture of a superposition of nucleon-nucleon collisions.
The model proposed by Kharzeev and Nardi \cite{dima:00} attempts
to extract the contribution to the multiplicity distribution that
scales as either soft or hard processes, or more precisely, as
\npart\ and \ncoll.  Such scaling is expected to be \pt\
dependent, and therefore this study is better suited for a \pt\
distribution (see Sec.~\ref{sec:pt}).

\begin{figure}[htb]
  \centering
  \includegraphics[width=0.5\textwidth]{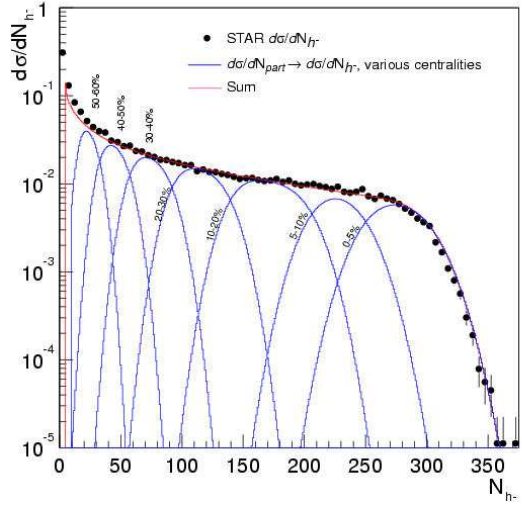}
  \caption[Multiplicity distribution with Glauber model fit.]
  {Fit to the multiplicity distribution $d\sigma/dN_{h^-}$ using
  a Glauber model (Sec.~\ref{sec:glauber})}.\label{fig:dsdhNpart}
\end{figure}

It is nevertheless useful to apply such models in order to obtain
a statistical determination of the number of participants for a
given event selection. One implementation of the Glauber particle
production model is illustrated in Fig.~\ref{fig:dsdhNpart}.
This was done by a Monte Carlo implementation, where one generates
events with random impact parameter, and then samples the
$d\sigma/db$ geometrical distribution to find the probability to
have an event at the given impact parameter.  When an interaction
takes place, we calculate \npart\ and \ncoll\ using
Eq.\ref{eq:TaaNpartNcoll}.  We can thus obtain a $d\sigma/d\npart$
distribution which we use as a weight in the final multiplicity
distribution (similar to the weight in Eq.~\ref{eq:KNdsdhminus}
given by the probability of no interaction). We can assume that
particle production is proportional to the number of participants
(Sec.~\ref{sec:glauber}) $\langle \Nhminus(b) \rangle = q \langle
\npart(b) \rangle$, or use a linear combination of \npart\ and
\ncoll\  \textit{\`{a} la} Kharzeev-Nardi \cite{dima:00}
(Eq.~\ref{eq:KNhminusMean}).  As mentioned before, we also choose
Gaussian fluctuations, Eq.~\ref{eq:KNhminusSigma}. The approach
works best from mid-central to central collisions, as the
fluctuations in peripheral collisions are large.
Figure~\ref{fig:dsdhNpart} shows the result of applying such a
model to the STAR multiplicity distribution.  The blue curves are
the distributions obtained by binning the $d\sigma/d\npart$
distribution to get successively $0-5\%$, $5-10\%$, $10-20\%$,
$20-30\%$, $30-40\%$, $40-50\%$, and $50-60\%$ of \sigmahad, which
result in the blue curves with near-Gaussian tails when
transformed to $d\sigma/dN_{h^-}$ because of the fluctuations. The
red curve is the convolution of all such distributions, giving a
fit to the multiplicity distribution. The size of the fluctuations
in Eq.~\ref{eq:KNhminusSigma} is found to be $a = 1.45 \pm 0.2$.
This parameter is constrained mostly by the shape of the terminus
of the distribution. For a Poisson distribution, this value would
be 1.  The presence of correlations makes this value exceed unity.
The origin of the correlations from the dynamics of the colliding
system is a combination of nucleus-nucleus geometry fluctuations,
nucleon-nucleon cross section fluctuations and hadron resonance
correlations~\cite{reid:01}.  In fact, this parameter for \Hijing\
turns out to be rather large, $a \gtrsim 4$, as is also apparent
in Fig.\ref{fig:hminus}. There are also fluctuations due to finite
detector acceptance. Since the TPC has a large acceptance, the
contribution from acceptance fluctuations is small. The shape of
the terminus of the distribution measured with a smaller
acceptance apparatus should be reveal a less steep shape of the
terminus of the distribution (\ie\ a larger $a$), \eg\ the PHENIX
Pad Chamber measurement \cite{Adcox:2000sp}.

\begin{figure}[htb]
  \centering
  \includegraphics[width=0.6\textwidth]{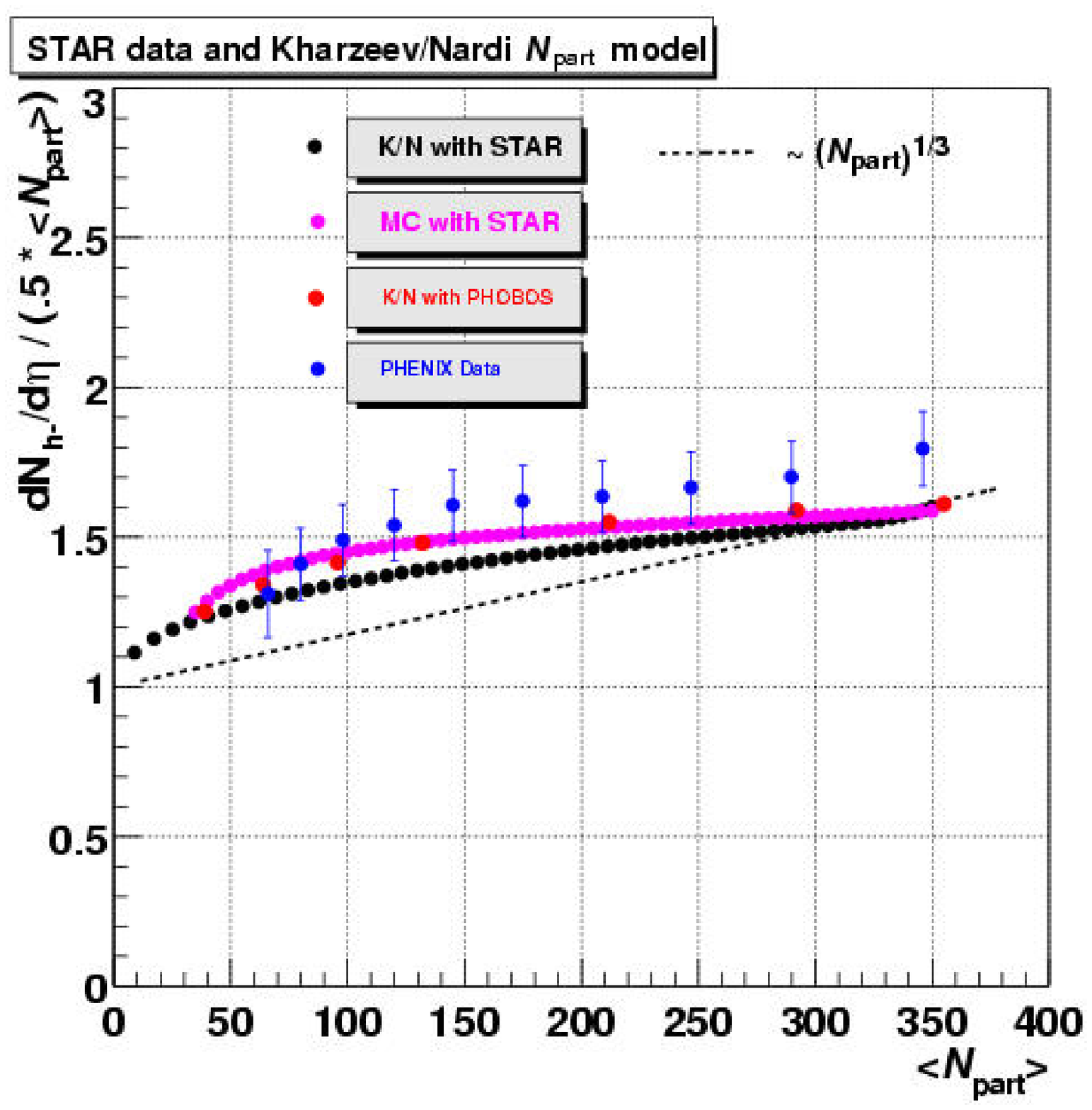}
  \caption{$dN/d\eta$ per participant pair \vs\ \npart.}
  \label{fig:dndetaPartKNMike}
\end{figure}

It is argued \cite{Wang:2000bf} that the centrality dependence of
the rapidity density per participant pair can distinguish between
models based on particle production from gluon saturation (EKRT
\cite{ekrt:00}) from those based on fixed scale pQCD such as
\Hijing. In the EKRT model, particle production is computed
assuming that the pQCD growth of low \pt\ gluons is only
integrated up to a certain saturation scale.  The saturation
requirement has as a consequence that the multiplicity per
participant decreases with centrality. For \Hijing, particle
production arises from two main contributions: (a) a component
arising from low-\pt\ hadron production, modelled as beam jet
string fragmentation, which is essentially proportional to \npart,
and (b) a contribution arising from mini-jet production, which is
directly proportional to \ncoll\ and to the averaged inclusive jet
cross section per nucleon-nucleon collision.  This is essentially
the same functional dependence as Eq.~\ref{eq:KNhminusMean}.  The
centrality dependence of the mid-rapidity density for \Hijing\ is
then
\begin{equation}\label{eq:dndetaParticipantsHijing}
  \frac{dN_{\mathrm{ch}}}{d\eta} = \langle \npart \rangle
  n_{\mathrm{soft}}+ f \langle \ncoll \rangle
  \frac{\sigma_\mathrm{jet}^AA(\sqrts)}{\sigma_{in}^{NN}}
\end{equation}
where $n_{\mathrm{soft}} \approx 1.3$ and $f \approx 1.2$. A plot
of $dN/d\eta/(\npart/2)$ \vs\ \npart\ should then na\"{\i}vely
grow as $\sim \npart^{1/3}$.  Using the Monte Carlo model just
discussed, and also the model from Kharzeev and Nardi
\cite{dima:00} applied to the STAR data, we obtain the centrality
dependence of $dN/d\eta$ shown in Fig.~\ref{fig:dndetaPartKNMike}.
The results from PHENIX~\cite{Adcox:2000sp} and from the Kharzeev
and Nardi analysis of the PHOBOS results ~\cite{dima:00} are also
shown.
 The growth with \npart\ clearly disfavours the saturation
picture from Ref.\cite{ekrt:00}. As pointed out in
Ref.\cite{dima:00}, the original saturation ideas pertain to the
behaviour of partons in the \emph{initial} wave function of the
nucleus; and when taking this approach the predictions
surprisingly turn out to be quite similar to the eikonal Glauber
approach discussed in Sec.~\ref{sec:glauber}. There are also
deviations from the simple $\npart^{1/3}$ dependence which can
arise from the dilute edges of the Wood-Saxon nuclear distribution
(Fig.~\ref{fig:WoodsSaxonDensity}) and from other medium effects
such as nuclear shadowing of the initial parton distributions
\cite{shadowing:00} or jet quenching\cite{jetq:92,jetq:98}.  The
similarity between the conventional approach and the high-density
QCD approach makes it difficult to distinguish which underlying
picture is correct. These effects are therefore better studied in
the differential \pt\ distributions.

\section{\pt\ Distribution}\label{sec:pt}
\subsection{Results}
\begin{figure}[htb]
  \centering
  \includegraphics[width=.6\textwidth]{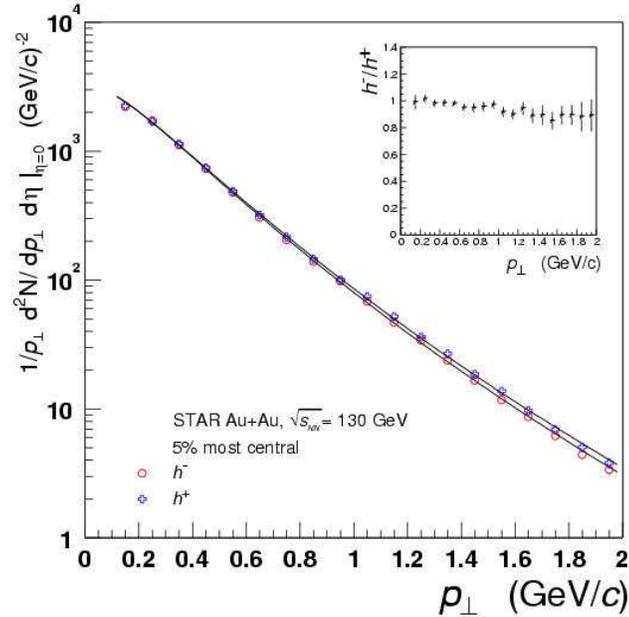}
  \caption[\hminus\ and \hplus\ \pt\ distributions.]
  {The \pt\ distributions for \hminus\ (open circles) and \hplus\ (open ``plus'' sign)
  in central collisions.
  The inset shows the ratio $\hminus/\hplus$ \vs\ \pt.  The data are fit
  to a power law in \mt\ to obtain the extrapolated yield $dN/d\eta$ (see text).}
  \label{fig:ptPlusMinus}
\end{figure}

Since the yield of charged hadrons at midrapidity depends on the
extrapolation of the \pt\ distribution outside the measured range,
we discuss the \pt\ spectrum first. The shape of the \pt\
distribution is well suited for more detailed scrutiny. Figure
\ref{fig:ptPlusMinus} shows the \pt\ distribution for positively
and negatively charged hadrons (\hplus\ and \hminus).
Statistical errors are smaller than the symbols. The correlated
systematic uncertainty is estimated to be below 6\%. The \pt\
distributions in this range of centre-of-mass energies are
customarily fit  by a QCD inspired power-law function of the form
$d^2\Nhminus/dp_\perp^2 d\eta = A\, (1+\pt/p_0)^{-n}$ where $A$,
$n$, and $p_0$ are free parameters. This power-law fit is useful
in comparison to other reference data such as \pt\ spectra from
\ppbar\ collisions \cite{ua1:90}.  For the extrapolation to \pt\ =
0, it is better to use a modified version of the power law, and
use \mt\ instead of \pt\ in order to better reproduce the data at
low \pt.  This is the fit shown in Fig.~\ref{fig:ptPlusMinus},
using a mass slightly larger than the \piminus\ mass, since the
\hminus\ spectrum is made not only of pions but includes \kminus\
and \pbar\ as well.  The ``mean mass'' of the \hminus\
distribution must therefore be $\langle m \rangle \gtrsim
m_{\pi}$. We studied the effect that such variations in the shape
of the extrapolating function have on the yield by comparing the
\mt\ and \pt\ power-law fits and also a simple exponential fit at
low \pt. The yield in the region $0-100\ \mevc$ was found to be
$6.8 \%$ using the \mt\ power law, the other fits varied from this
by about $\pm 1.5 \%$ which we take as an additional systematic
uncertainty for the extrapolated yield.  The choice of mean mass
does not influence the extrapolated yield obtained from the \mt\
power law, it was found to vary less than 0.5\% with mean masses
in the range $140 - 200\ \mevcc$.  The
$\chi^2/\mathrm{\textit{d.o.f.}}$ was minimized with a value of
$\langle m \rangle \simeq 187\ \mevcc$. In all cases, the yield in
the region $\pt > 2\ \gevc$ was $\lesssim 1\%$. We therefore use
an extrapolation factor  of $7 \pm 1.5 \%$ to obtain $dN/d\eta$.

\subsection{Discussion}
\begin{figure}[htb]
    \begin{center}
        \includegraphics*[width=0.6\textwidth]{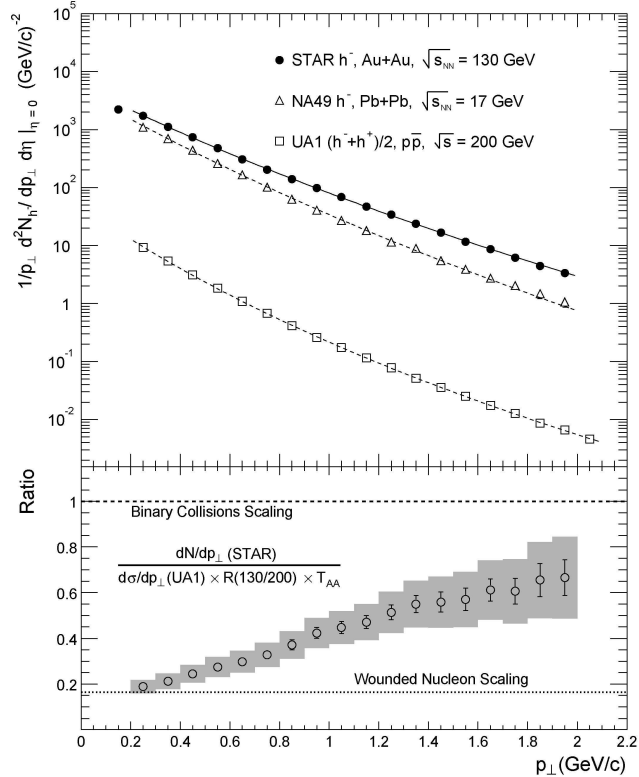}
        \caption[\hminus\ \pt\ distribution for central events.]{Upper panel: \hminus\ \pt-spectra for the 5\% most central \AuAu\ collisions
            at mid-rapidity ($|\eta| < 0.1$).  Central \PbPb\ data from
            NA49 at lower energy and \ppbar\ data from UA1 at a similar energy
            are also shown. The
            curves are power-law fits to the data. Lower panel: ratio
            of STAR and scaled UA1 \pt-distributions (see text).}
        \label{fig:pt}
    \end{center}
\end{figure}

We focus now on the comparison of the measured \pt\ spectrum in
\AuAu\ at RHIC to other energies and collision systems.  We will
focus more on \ppbar\ collisions. Figure \ref{fig:pt}, upper
panel, shows the transverse momentum distribution of negatively
charged hadrons for central \AuAu\ collisions at mid-rapidity
($|\eta| < 0.1$) within $0.1 < \pt < 2\ \gevc$.
The upper panel of Fig.~\ref{fig:pt} also shows the
\pt-distributions of negatively charged hadrons for central \PbPb\
collisions at \sqrtsNN\,= 17 \gev\ from NA49 \cite{na49:99} and
for minimum-bias \ppbar\ collisions at \sqrts\,= 200 \gev\ from
UA1 \cite{ua1:90}, fitted with the same function.  The NA49
distribution, which was reported in units of pion rapidity, was
transformed to units of pseudorapidity.  The UA1 invariant cross
section $E d^3 \sigma/d^3p$ reported in Ref.~\cite{ua1:90} was
scaled by $2 \pi / \siginel$, where $\siginel = 42$ mb for \sqrts\
= 200 \gev\ \cite{ua5x:86}.  The power law fits all three datasets
well. The mean \pt\ can be derived from the fit parameters as
$\meanpt = 2 p_0 / (n-3)$.  The fit to the STAR data gives $p_0 =
3.0 \pm 0.3\ \gevc$, $n = 14.8 \pm 1.2$, and $\meanpt = 0.508 \pm
0.012\ \gevc$.  The strong correlation of fit parameters $p_0$ and
$n$ must be taken into account when calculating the error on
\meanpt. The \meanpt\ from STAR is larger than that from both
central collisions of heavy nuclei at much lower energy
($\meanpt_{\mathrm{NA49}} = 0.429\ \gevc$) and nucleon-nucleon
collisions at a comparable energy ($\meanpt_{\mathrm{UA1}} =
0.392\ \gevc$).

Figure \ref{fig:pt}, lower panel, shows the ratio of the STAR and
UA1 \pt-distributions. Since the UA1 distribution is measured at
\sqrts\,= 200 \gev, $d\sigma/d\pt$ is scaled by two factors for
quantitative comparison to the STAR data at 130 \gev: \textit{(i)}
$R(130/200)$, the \pt-dependent ratio of invariant cross sections
for charged particle production in \ppbar\ collisions at \sqrts\,=
130 and 200 \gev, and \textit{(ii) } \TAA = $26\pm2$ mb$^{-1}$,
the nuclear overlap integral \cite{Eskola:1989yh} for the 5\% most
central \AuAu\ collisions.  $R$ varies from 0.92 at $\pt = 0.2\
\gevc$ to 0.70 at $\pt = 2.0\ \gevc$, and was derived using
scaling laws for \meanpt\ and $dN_{\mathrm{ch}}/d\eta$ as a
function of \sqrts\ \cite{ua1:90,pdg:00} together with the
extrapolation to 130 \gev of power-law parameterizations at
\sqrts\,= 200--900 \gev \cite{ua1:90}.  The shaded boxes show the
total error of the ratio, which is the linear sum of the errors of
the measured data, depicted by the error bars, and the systematic
uncertainty due to uncertainties in the scaling with \TAA\ and
$R$.

There are two simple predictions for the scaled ratio. In lower
energy hadronic and nuclear collisions, the total pion yield due
to soft (low \pt) processes scales as the number of participants
(or ``wounded'' nucleons) in the collision (see \eg\ Ref.
~\cite{Bialas:1976ed,na49:99}). The scaled ratio in this case is
0.164, assuming 172 participant pairs and a mean number of binary
collisions of 1050 ($=\siginel\, \TAA$, $\siginel\ = 40.35$ mb)
for the 5\% most central \AuAu\ events (Sec. \ref{sec:glauber}).
In contrast, if hadron production is due to hard (high \pt)
processes and there are no nuclear-specific effects, the hadron
yield will scale as the number of binary nucleon-nucleon
interactions in the nuclear collision and the value of the ratio
is unity.  There are important nuclear effects which will alter
the scaling as a function of \pt\ from these simple predictions,
including initial state multiple scattering \cite{cronin:79},
shadowing \cite{shadowing:00}, jet quenching
\cite{jetq:92,jetq:98}, and radial flow \cite{rflow:93}.  The
scaled ratio exhibits a strong \pt\ dependence, rising
monotonically with increasing \pt\ from Wounded Nucleon scaling at
low \pt\ but not reaching Binary Collision scaling at the highest
\pt\ reported. This is consistent with the presence of radial
flow, as well as the onset of hard scattering contributions and
initial state multiple scattering with rising \pt.  From $pA$
collisions at lower energy, the data tend to show an increase
beyond the scaling with binary collisions.  This ``Cronin'' effect
is thought to arise from initial state multiple scattering in the
nuclear medium, which broadens the \pt\ spectrum \cite{cronin:79}.
At SPS, the \pt\ spectra indicate a similar increase beyond binary
collision scaling for the region $\pt \sim 2\ \gevc$.  Although a
comparison of the \pt\ dependence of the data at different
\sqrtsNN\ is better done in $x_{\perp} = 2\pt/\sqrts$ rather than
in \pt, there are indications from preliminary
STAR\cite{dunlop:01} and PHENIX \cite{Zajc:2001va} hadron spectra
at high \pt that the scaling with \ncoll\ is not reached even for
\pt\ 2 - 6 \gevc. This effect has sparked active discussions in
the community \textit{vis-\`{a}-vis} the possibility of being
related to parton energy loss in a QGP, \ie\ jet quenching
\cite{Levai:2001dc}.

\section{$\eta$ Distribution} \label{sec:eta}
\subsection{Results}

\begin{figure}[htb]
    \begin{center}
        \includegraphics[width=0.75\textwidth]{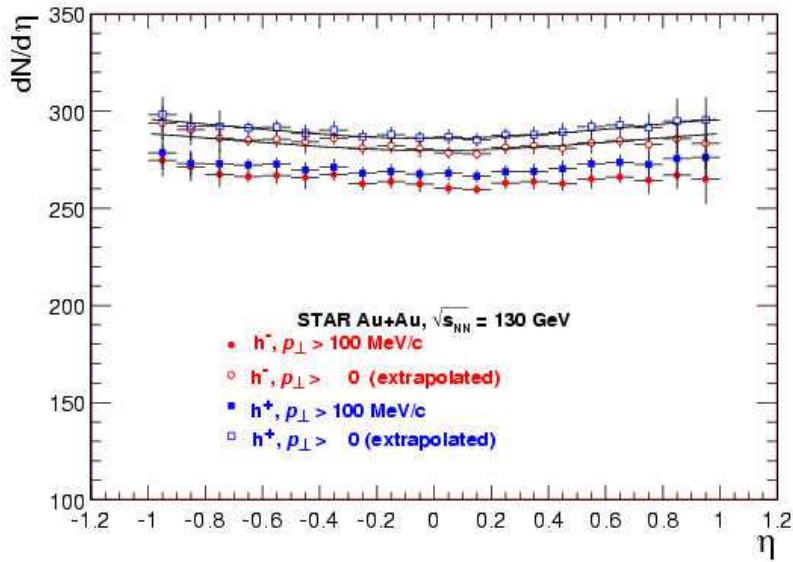}
        \caption[\hminus\ and \hplus\ $\eta$
        distribution]{\hminus\ (red circles)
        and \hplus\ (blue squared) pseudorapidity distribution
            from 5\% most central \AuAu\ collisions.  Data are integrated over
            $\pt\,> 100$ \mevc\ (filled symbols), and extrapolated
            to all \pt\ (open symbols).}
        \label{fig:dndeta}
    \end{center}
\end{figure}

By integrating the \pt\ spectra for different $\eta$ bins, we
obtain the $dN/d\eta$ \vs\ $\eta$ distribution. Figure
\ref{fig:dndeta} shows the normalized pseudorapidity distribution
of \hminus\ and \hplus\ for the 5\% most central collisions within
$|\eta| < 1.0$. The black points are the measured yield for $0.1 <
\pt\ < 2\ \gevc$ and the hollow points are the extrapolation for
all \pt.
The latter was obtained by fitting a power-law function in the
range $0.1 < \pt < 2\ \gevc$ and extrapolating to $\pt = 0$ in
order to estimate the content of the first \pt\ bin, as discussed
in Section~\ref{sec:pt}. The error bars indicate the uncorrelated
systematic uncertainties. The statistical errors are negligible.
The correlated systematic uncertainty applied to the overall
normalization is estimated to be below 6\% for $\pt > 100$ \mevc\
and 7\% for all \pt.

The \hminus\ density at midrapidity for $0.1 < \pt < 2\ \gevc$ is
$dN_{h^-}/d\eta|_{\eta = 0} = 261 \pm 1(\mbox{stat})\pm
16(\mbox{syst})$. Extrapolation to $\pt$ = 0 yields
$dN_{h^-}/d\eta|_{\eta = 0} = 280 \pm 1(\mbox{stat})\pm
20(\mbox{syst})$. Assuming an average of 172 participant pairs per
central \AuAu\ collision, this corresponds to $1.63 \pm
0.12$\,\hminus\ per participant nucleon pair per unit
pseudorapidity, a 38\% increase over the yield in \ppbar\
collisions extrapolated to the same energy \cite{ua5:86}
(neglecting isospin correction factors of order 1--3\%) and a 52\%
increase over \PbPb\ collisions at \sqrtsNN\,= 17\ \gev\
\cite{na49:99}. The corresponding \hplus\ density for $0.1 < \pt <
2\ \gevc$ is $dN_{h^+}/d\eta|_{\eta = 0} = 268 \pm
1(\mbox{stat})\pm 16(\mbox{syst})$, which after the extrapolation
to all \pt\ yields $dN_{h^+}/d\eta|_{\eta = 0} = 287 \pm
1(\mbox{stat})\pm 20(\mbox{syst})$.  The systematic uncertainty is
the same as for the \hminus\ since the analysis was done in the
same way.  It is a correlated systematic for both charge sign
yields, so it does not mean that the yields can be equal within
errors. In other words, the ratio \hplus/\hminus\ is $1.025 \pm
0.004$ (where most of the systematic uncertainties cancel and we
are left with a small statistical error).

The PHOBOS collaboration has reported a total charged multiplicity
density for the 6\% most central \AuAu\ collisions of
$dN_{ch}/d\eta|_{|\eta|<1} = 555 \pm 12(\mathrm{stat}) \pm
35(\mathrm{syst})$ \cite{phobos:00}. Analyzing positive charged
particles within the framework described above, STAR measures the
total charged particle density  $ dN_{ch}/d\eta|_{|\eta|<0.1} =
567\pm 1(\mathrm{stat}) \pm 34(\mathrm{syst})$ for the 5\% most
central \AuAu\ collisions. The PHOBOS centrality selection is
based on charged particles within $3 < |\eta| < 4.5$, while STAR
utilizes spectator neutrons in the ZDCs. A precise equality of the
two measurements is therefore not expected, due to the difference
in centrality selection.

\subsection{Discussion}

The $\eta$ distribution is almost constant within $|\eta|<1$,
exhibiting a small rise at larger $\eta$. This shape is expected
from a boost invariant source (i.e., constant in rapidity), taking
into account the transformation from $y$ to $\eta$.  As a
reference, the data are fit by the shape of the Jacobian
$dy/d\eta$ given in Eq.~\ref{eq:jacobianPtEta}, where the fit
parameters are (\textit{i}) the normalization constant (which
should be equal to $dN_{ch}/dy$ for a boost invariant source) and
(\textit{ii}) the dimensionless ratio $\pt/$m, or rather
\meanpt/$\langle$m$\rangle$, since the $\eta$ distribution is
already integrated over \pt\ and the mass should be a weighted
average over all 3 species. From the simple fit of the
distribution to Eq.~\ref{eq:jacobianPtEta}, we obtain the
parameters $dN_{h^-}/dy = 295 \pm 18$ and $dN_{h^+}/dy = 304 \pm
18$.  Using the value of \meanpt\ from the \pt\ distribution, we
can convert the other fit parameter into an average mass.  As
discussed previously, the value should be close to the \piminus\
mass since pions dominate, but we expect it to be somewhat larger
due to the contribution from kaons and anti-protons.  We obtain
$\langle m \rangle \cong 0.171\ \gevcc$, which is consistent with
this simple picture (and close to the value that minimizes the
$\chi^2$ for the power-law fit in \mt\ to the \hminus\ \pt\
distribution, $\langle m \rangle \cong 0.187\ \gevcc$). This
signifies that the system is very close to a boost-invariant
source. Measurement of the rapidity distribution of identified
particles is needed to test boost invariance at mid-rapidity, and
will be addressed in the \piminus\ analysis.  In particular, the
rapidity distribution and the rapidity dependence of the slopes
will be useful to address this question, as we find a slight but
systematic difference in the \piminus\ slopes as a function of
rapidity.

\begin{figure}[htb]
  \centering
  \includegraphics[width=.7\textwidth]{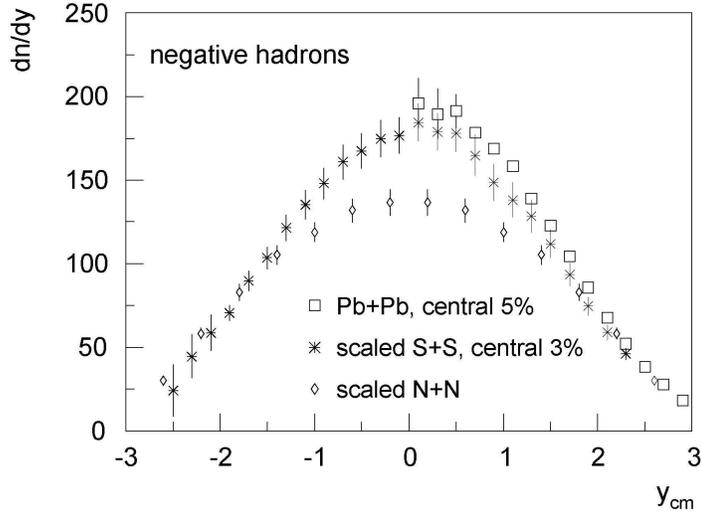}
  \caption[Lower energy $\eta$ distributions.]{$\eta$ distribution
  of negative hadrons at SPS (shown as $y_\pi$ in the
  center-of-mass).
  Data are from Ref.~\cite{na49:99}.}
  \label{fig:etaNA49}
\end{figure}

The approach to boost invariance is nevertheless quite striking as
compared to lower energy \PbPb\ collisions.
Figure~\ref{fig:etaNA49} shows the rapidity distribution of
charged hadrons at SPS energies.
The collision systems and beam energies are from central \PbPb\ @
$p_{\mathrm{beam}} = 158$ A\gevc\ (hollow squares), central S+S @
$p_{\mathrm{beam}} = 200$ A\gevc\ (asterisks),     and  minimum
bias nucleon-nucleon collisions are also shown (hollow rhombi).
The data are shown as $dN/dy$ \vs\ $y$ where $y$ is calculated
assuming the pion mass.  The shape of the distribution is peaked
around mid-rapidity in contrast to the relatively flat region we
find at \sqrtsNN\ = 130 \gev.  The forward and backward region
should show deviations from this behaviour, and it is interesting
to study the size of the flat mid-rapidity region.  The forward
TPC's in STAR will be in a position to answer this question.  From
other experiments, results from the PHOBOS collaboration
\cite{Back:2001bq} suggest that the plateau region extends up
$|\eta|\lesssim 1.5$. The ratio of positive to negative particles
$\hplus/\hminus\ = 1.025 \pm 0.004$ already indicate that the
mid-rapidity region is not net-baryon free, as has also been
reported by STAR~\cite{Adler:2001bp} and
BRAHMS~\cite{Bearden:2001kt} for the $\pbar/p$ ratio.

\subsubsection{Lower Limit on Energy Density}
With the \hminus\ multiplicity and \meanpt\ we can return to the
question of estimating the energy density.  As discussed in
Section~\ref{sec:CritTempEnergyDensity}, we can only obtain a
lower limit with these simple observables.  We already stated that
there is no reliable guidance as to the value of the
initial formation time of the system, $\tau$ in
Eq.~\ref{eq:bjEnergyDensity}.  We assume a value of 1$\fm/c$, similar
to the one adopted at SPS energies, although the system might
equilibrate faster at RHIC energies.
This is not the only assumption that goes into such an estimate.
The initial energy density is also modified, possibly by a large amount, by the longitudinal
expansion. Transversal expansion does not change $d\Et/dy$,  but is
not known how much longitudinal work is done by the system during the
expansion.  In this sense, it could be misleading to use the initial
formation time $\tau$ but a final $d\Et/dy$.  Since we measure
multiplicity instead of transverse energy, this might turn out to
be a better estimate since $dN/dy$ is approximately conserved during the
expansion.  However, it is more difficult to estimate the time dependence of
\meanpt.  The temperature of the system decreases with time, but the
collective flow increases.  In the best case, both effects would cancel, but it
is more likely that the build up of flow does not fully compensate the
decrease in temperature.  This also results in an estimate of the energy
density that is a lower limit with respect to \meanpt.
It is also a pre-requisite that the behaviour of the system in the
initial stages can be modelled via hydrodynamics.  Given a set of
energy-momentum conservation and particle number conservation equations
along with an equation of state, one can then calculate the energy
density, pressure and entropy density of the system.  If the entropy density
is approximately conserved during the expansion, and under the assumption that
the entropy density is approximately proportional to the particle multiplicity
in the final state \cite{Landau:1953gs,bjorken:83},
then one can estimate the energy density as in Eq.~\ref{eq:bjEnergyDensity}.

The estimate under these assumptions proceeds as follows.
For central collisions, we have $dN_{ch}/d\eta = 567$ and \meanpt = 0.511
\gevc.  We could also use $\langle \mt \rangle$ instead of \meanpt
because one should not neglect the masses.  In the best of all
worlds one actually would obtain the $\langle \mt \rangle$ for
each particle and do a sum weighing by the fraction each particle
contributes to the total \hminus\ distribution. Since we already
mentioned the larger uncertainties that go into the estimate, such
refinements can be ignored. In addition we do not know \meanpt\
for kaons and protons yet, nor do we know the ratios
\kminus/\hminus, and \pbar/\hminus to a degree that we can do
considerably better than the simple assumption. Since we measure
only the energy density carried by the charged hadrons, we also
use Bjorken's guess $\epsilon \simeq \epsilon_{ch} \times 1.5$.
The resulting equation is:
\begin{equation}
\epsilon \geq 1.5 \times\ \langle \mt \rangle \frac{dN_{ch}}{dy}
\frac{1}{\pi r_0^2 A^{2/3} \tau} \label{eq:bjEnergyDensityData}
\end{equation}
The parameters we enter into the formula are the following. We use
the same value of $\tau = 1$ fm to compare to SPS (although it
probably takes less time to equilibrate the system at RHIC
energies).  This also allows the estimate to be scaled to a
different formation time once it is better determined.

The radius ($r = r_0 \times A^{1/3}$) we use can be simply taken
as the one for $b=0$.  We must then extrapolate the multiplicity,
since it was measured for finite impact parameter.  In the
geometrical model, 5\% corresponds to roughly $b = 0 - 3$ fm.  For
head-on collisions $b=0$ the \hminus\ multiplicity increases with
respect to the 5\% most central by roughly 14\% ($\Nhminus \sim
320$ for $b=0$ from Fig.~\ref{fig:hminus}). In addition, we are
measuring $dN/d\eta$ not \dndy. We can use the estimate of \dndy\
from the fit to the $\eta$ distributions using the Jacobian
$dy/d\eta$.

We then have the parameters $\tau = 1$ fm, $A = 197$,  $dN_{ch}/dy
= (295+304)\times 1.14 = 682$,  $\langle \mt \rangle =
\sqrt{\meanpt^2+\langle m \rangle^2} = \sqrt{0.511^2+0.171^2} =
0.538$ which we can insert into Eq.~\ref{eq:bjEnergyDensityData}
obtaining
\begin{equation}
\epsilon \gtrsim \frac{1.5 \times 682 \times 0.538}{\pi \times
1.16^2 \times 197^{2/3}}
    = 3.9 \pm 0.4 \gev/\fm^3
    \label{eq:bjEnergyDensityNumbers}
\end{equation}
The $\sim 10\%$ systematic uncertainty is based on variations of
\meanpt\ and the \hminus, \hplus\ multiplicity estimates within
their systematic uncertainties.

This is actually somewhat small when one consider that NA49
reports 3.2 $\gev/\fm^3$ for Pb+Pb via calorimetry (d\Et/d$\eta$).
Using tracking they get however 2.6 $\gev/\fm^3$
\cite{guenther:97} for charged particles. The rather small
difference between 3.9 and the NA49 result 3.2 also points to the
fact that $\tau$ should indeed be smaller for RHIC energies. Other
scenarios, such as a denser core and a less denser shell, cannot
be probed with this simple method since this is just the average
density of the source volume. It is probably best to obtain
estimates from actual hydrodynamical simulations that reproduce
measured data (\eg\ particle yields and spectra, elliptic and
radial flow) and obtain from these calculation ranges of values
for the relevant parameters, \ie\ excluded regions of $\epsilon$.
However, for a mid-rapidity region with almost zero baryochemical
potential, Figure~\ref{fig:qcdEnergyDensity} already indicates
that at $T = 1.5\ \Tc$, the energy density in the deconfined phase
(for the case of 2+1 flavour QCD) is $\epsilon \approx 3.5\
\gev/\fm^3$, so our simple estimate would indicate a favorable
environment to reach the deconfined phase.

\section{Centrality dependence}\label{sec:hminusCentrality}

\subsection{Centrality selection}
\begin{figure}[htb]
\mbox{
\subfigure[]{\includegraphics[width=0.46\textwidth]{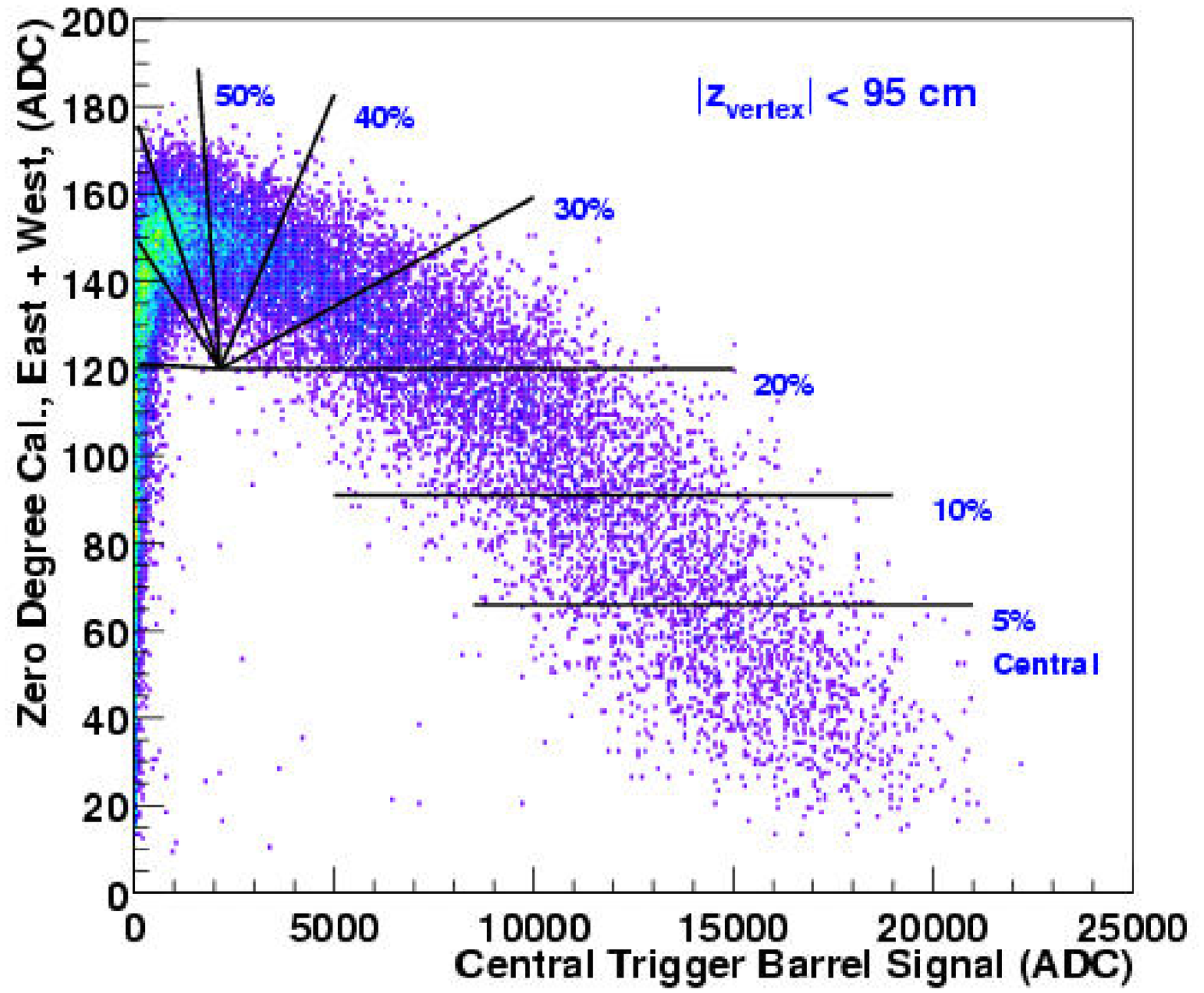}}\quad
     \subfigure[]{\includegraphics[width=0.44\textwidth]{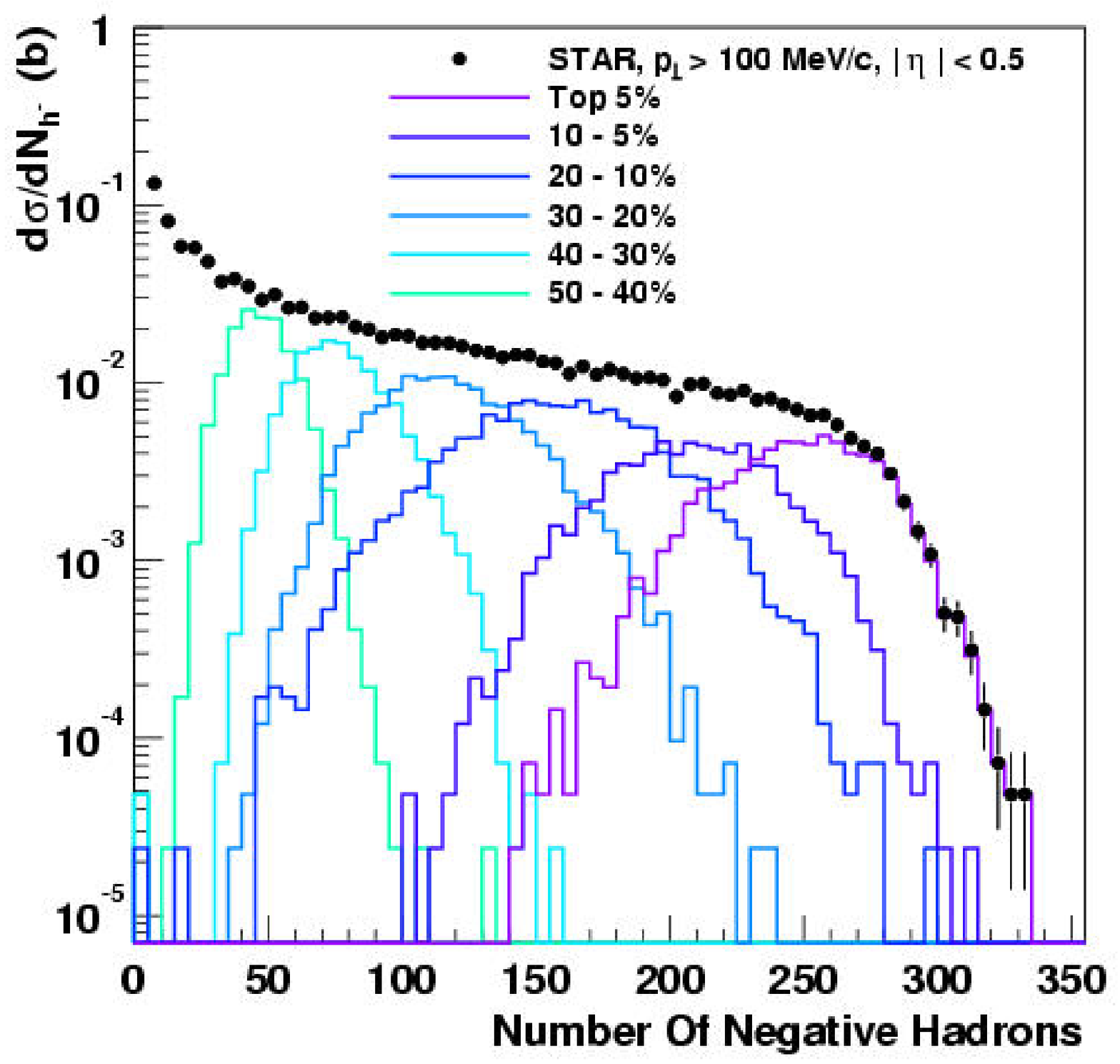}}}
  \caption[Centrality Classes for \hminus\ analysis.]{Centrality classes for \hminus\ analysis.
  Left panel: cuts in ZDC \vs\ CTB.  Right panel: Resulting classes in the \hminus\ multiplicity
  distribution.}
  \label{fig:centralityClasses}
\end{figure}

It is interesting to study how the distributions are affected by
the centrality of the collision.  As we mentioned in the
discussion of the Glauber model, it is not possible to
experimentally determine the impact parameter of the collision. In
practice, any study on centrality dependence relies on selecting
ensembles of events based on an observable correlated with the
impact parameter.  Typically these can be transverse energy
production, or the charged particle multiplicity of the event.
That is why the study of charged hadrons is of primary concern in
further studies of heavy ion collisions, such as strangeness
production.  The focus of the current work however, is the study
of charged hadrons, so to study the centrality dependence of
charged hadron production is somewhat ill-defined.  The approach
we followed for the centrality dependence relies on a selection of
events based on the ZDC and CTB trigger signals,
Figure~\ref{fig:zdcvsctb}.  The most central collisions are
studied by selecting events below a ZDC threshold,
Figure~\ref{fig:centralityClasses}(a).
This leads to an ambiguity, since both central and peripheral
collisions can have a low ZDC signal, so we set a CTB cut only to
break this ambiguity (the cut was set at 8500 ADC counts, not
shown in the figure in the interest of clarity).  This approach
works up to a point, since the ZDC signal turns around and becomes
correlated with multiplicity instead of anti-correlated, and we
must therefore abandon the the simple ZDC threshold cuts in place
of cuts in both ZDC and CTB. Figure~\ref{fig:centralityClasses},
left panel, shows the cuts in ZDC \vs\ CTB signals; the right
panel shows the resulting event selection in the \hminus\
multiplicity distribution for reference.

\subsection{\pt\ and $\eta$ \vs\ Centrality}

\begin{figure}[htb]
  \centering
  \includegraphics*[width=.6\textwidth]{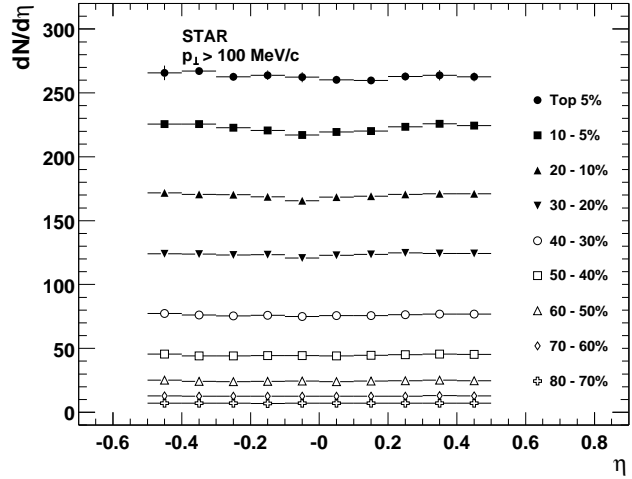}
  \caption[\hminus\ $\eta$ distribution for different centralities.]
  {The \hminus\ $\eta$ distribution for different centralities.}
  \label{fig:etacentrality}
\end{figure}

\begin{figure}[htb]
  \centering
  \includegraphics[width=.6\textwidth]{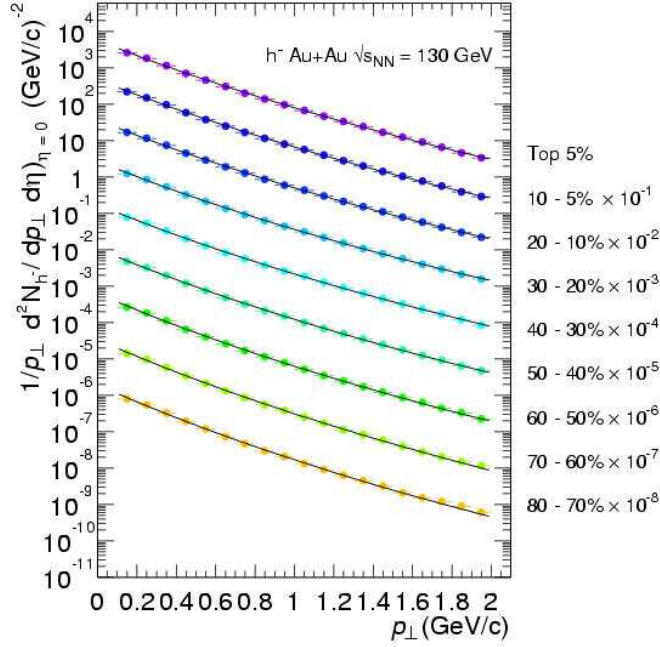}
  \caption[\hminus\ \pt\ distribution for different centralities.]
  {The \hminus\ \pt\ distribution for different centralities.  The
  data are scaled down by successive powers of 10 for display purpose.}
  \label{fig:ptcentrality}
\end{figure}

With the previous centrality selections, we obtain a $d^2N/d\pt
d\eta$ distribution for each event sample.  We again apply the
tracking corrections (Sec. \ref{sec:trackingCorrections}) for each
event and accumulate the corrected distributions for each
centrality class. The $\eta$ distribution for the different
centrality bins is shown in Fig.~\ref{fig:etacentrality}.
The shape of the distributions at mid-rapidity ($|\eta|<0.5$) is
very similar for all the centrality selections.
\begin{table}[htb]
\centering
\begin{tabular}{ccccccc}
\hline \hline
 $\sigma\ (\%)$ & \meanpt\ [\gevc] &
$dN/d\eta$ {\tiny(Meas.)}&
 $dN/d\eta$ {\tiny(Extrap.)} &
 $\langle \npart \rangle$ &
 $\langle \ncoll \rangle$ &
 $\langle b \rangle$ \fm \\
0 - 5\% & 0.511  &  266.5    &  289.8 & $345 \pm 7$ & $1050 \pm 19$ & $ 2.3 \pm 0.2$ \\
 5-10\% & 0.511  &  222.5    &  242.2 & $289 \pm 9$ & $ 826 \pm 25$ & $ 4.3 \pm 0.3$\\
10-20\% & 0.509  &  169.7    &  184.9 & $221 \pm 4$ & $ 566 \pm 10$ & $ 5.8 \pm 0.3$\\
20-30\% & 0.506  &  123.5    &  134.7 & $152 \pm 9$ & $ 336 \pm 25$ & $ 7.5 \pm 0.3$\\
30-40\% & 0.495  &   76.1    &   83.1 & $102 \pm 4$ & $ 209 \pm 15$ & $ 8.7 \pm 0.3$\\
40-50\% & 0.482  &   44.7    &   48.9 & $ 63 \pm 4$ & $ 105 \pm 4$  & $10.1 \pm 0.4$\\
50-60\% & 0.467  &   24.4    &   26.8 & $ 35 \pm 3$ & $  36 \pm 5$  & $11.2 \pm 0.6$\\
60-70\% & 0.453  &   12.7    &   13.9 & $ 20 \pm 2$ & $  22 \pm 4$  & $12.0 \pm 0.7$\\
70-80\% & 0.448  &    7.1    &    7.8 & $  9 \pm 4$ & $   9 \pm 4$  & $12.6 \pm 1.1$\\
\hline \hline
\end{tabular}
\caption[Mean \pt\ and $dN_{h^-}/d\eta$ for different fractions of
$\sigma_{\mathrm{AuAu}}$]{\meanpt\ and $dN_{h^-}/d\eta$ (in the
region $|\eta|<0.5$) for various centralities selected as
fractions of the cross section $\sigma_{\mathrm{AuAu}}$
(Fig.~\ref{fig:centralityClasses}).  The quantities statistically
related to the multiplicity using a Glauber model calculation are
also shown for reference.} \label{tab:meanptdndeta}
\end{table}
The  \pt\ distribution for the different centrality classes is
shown in Fig.~\ref{fig:ptcentrality}.  The $dN/d\eta$ and \meanpt\
values are collected in Table~\ref{tab:meanptdndeta} while the
\pt\ distributions for the various centralities are collected in
Table~\ref{tab:hminuspt}.
\begin{table}[htb]
  \centering
  \small
\begin{tabular}{c|r|r|r|r|r} 
    \pt & \multicolumn{5}{c}{Fraction of $\sigma_{\mathrm{hadronic}}$} \\
   {\tiny(\gevc)}\ & \multicolumn{1}{c}{0-5\%}  & \multicolumn{1}{c}{5-10\%} & \multicolumn{1}{c}{10-20\%} & \multicolumn{1}{c}{20-30\%} & \multicolumn{1}{c}{30-40\%} \\ \\ 
   0.15 & 2226.0 $\pm$   94.8 &  2195.9 $\pm$   84.0 & 1689.4 $\pm$   53.6  & 1244.6 $\pm$   42.5  & 787.3 $\pm$  21.6 \\
   0.25 & 1729.1 $\pm$   36.9 &  1514.3 $\pm$   52.5 & 1162.0 $\pm$   33.3  &  855.7 $\pm$   26.4  & 537.8 $\pm$  13.3 \\
   0.35 & 1117.1 $\pm$   26.1 &   965.8 $\pm$   34.5 &  734.8 $\pm$   21.8  &  532.6 $\pm$   17.0  & 331.2 $\pm$   8.6 \\
   0.45 &  732.2 $\pm$   13.7 &   588.7 $\pm$   22.6 &  450.1 $\pm$   14.3  &  326.3 $\pm$   11.2  & 201.4 $\pm$   5.6 \\
   0.55 &  478.8 $\pm$    7.3 &   381.8 $\pm$   15.9 &  291.2 $\pm$   10.1  &  211.2 $\pm$    7.9  & 130.1 $\pm$   4.0 \\
   0.65 &  304.3 $\pm$    6.0 &   251.2 $\pm$   11.5 &  190.0 $\pm$    7.2  &  137.4 $\pm$    5.7  &  83.8 $\pm$   2.8 \\
   0.75 &  204.7 $\pm$    4.0 &   169.6 $\pm$    8.5 &  128.6 $\pm$    5.3  &   92.7 $\pm$    4.2  &  56.4 $\pm$   2.1 \\
   0.85 &  139.5 $\pm$    4.6 &   115.0 $\pm$    6.3 &   86.9 $\pm$    4.0  &   63.6 $\pm$    3.1  &  38.0 $\pm$   1.5 \\
   0.95 &   97.6 $\pm$    2.0 &    80.7 $\pm$    4.8 &   60.4 $\pm$    3.0  &   43.8 $\pm$    2.4  &  26.5 $\pm$   1.2  \\
   1.05 &   68.4 $\pm$    1.9 &    56.3 $\pm$    3.7 &   43.1 $\pm$    2.3  &   31.2 $\pm$    1.8  &  18.47 $\pm$  0.89 \\
   1.15 &   47.0 $\pm$    1.5 &    39.8 $\pm$    2.8 &   30.3 $\pm$    1.8  &   21.7 $\pm$    1.4  &  12.71 $\pm$  0.67 \\
   1.25 &   34.2 $\pm$    1.2 &    28.0 $\pm$    2.2 &   21.4 $\pm$    1.4  &   15.33 $\pm$   1.07  &  9.03 $\pm$  0.51 \\
   1.35 &   23.8 $\pm$    1.0 &    20.1 $\pm$    1.7 &   15.24 $\pm$   1.06  &  10.95 $\pm$   0.83  &  6.33 $\pm$  0.39 \\
   1.45 &   16.6 $\pm$    0.9 &    14.20 $\pm$   1.29 &  10.99 $\pm$   0.82  &   7.69 $\pm$   0.63  &  4.50 $\pm$  0.30 \\
   1.55 &   11.72 $\pm$   0.74 &   10.46 $\pm$   1.02 &   7.91 $\pm$   0.64  &   5.70 $\pm$   0.50  &  3.25 $\pm$  0.23 \\
   1.65 &    8.69 $\pm$   0.46 &    7.63 $\pm$   0.79 &   5.76 $\pm$   0.50  &   4.02 $\pm$   0.38  &  2.32 $\pm$  0.18 \\
   1.75 &    6.20 $\pm$   0.43 &    5.61 $\pm$   0.62 &   4.18 $\pm$   0.39  &   2.96 $\pm$   0.30  &  1.68 $\pm$  0.14 \\
   1.85 &    4.44 $\pm$   0.39 &    3.93 $\pm$   0.47 &   2.99 $\pm$   0.30  &   2.14 $\pm$   0.24  &  1.20 $\pm$  0.11 \\
   1.95 &    3.38 $\pm$   0.32 &    2.88 $\pm$   0.37 &   2.20 $\pm$   0.23  &   1.57 $\pm$   0.18  &  0.851 $\pm$ 0.081 \\
\multicolumn{5}{c}{\ }\\ 
    \pt & \multicolumn{4}{c}{Fraction of $\sigma_{\mathrm{hadronic}}$} \\
   {\tiny(\gevc)}\ & \multicolumn{1}{c}{40-50\%} & \multicolumn{1}{c}{50-60\%} & \multicolumn{1}{c}{60-70\%} & \multicolumn{1}{c}{70-80\%} \\ \\ 
   0.15 & 472.0 $\pm$  14.9  & 268.5 $\pm$  10.7  & 143.5 $\pm$   6.4  &  80.6 $\pm$   4.8 \\
   0.25 & 325.3 $\pm$   9.3  & 183.1 $\pm$   6.6  &  97.1 $\pm$   3.9  &  54.7 $\pm$   2.9 \\
   0.35 & 198.8 $\pm$   5.9  & 110.1 $\pm$   4.1  &  57.8 $\pm$   2.4  &  32.6 $\pm$   1.8 \\
   0.45 & 119.1 $\pm$   3.8  &  65.3 $\pm$   2.7  &  34.5 $\pm$   1.6  &  19.3 $\pm$   1.2 \\
   0.55 &  76.0 $\pm$   2.7  &  41.2 $\pm$   1.8  &  21.2 $\pm$   1.0  &  11.91 $\pm$  0.78 \\
   0.65 &  48.8 $\pm$   1.9  &  26.3 $\pm$   1.3  &  13.37 $\pm$  0.72  &  7.26 $\pm$  0.53 \\
   0.75 &  32.3 $\pm$   1.4  &  16.90 $\pm$  0.92  &  8.68 $\pm$  0.50  &  4.67 $\pm$  0.37 \\
   0.85 &  21.5 $\pm$   1.0  &  11.42 $\pm$  0.68  &  5.56 $\pm$  0.36  &  2.97 $\pm$  0.26 \\
   0.95 &  15.12 $\pm$  0.76  &  7.78 $\pm$  0.51  &  3.84 $\pm$  0.26  &  2.08 $\pm$  0.20 \\
   1.05 &  10.03 $\pm$  0.56  &  5.18 $\pm$  0.37  &  2.52 $\pm$  0.19  &  1.35 $\pm$  0.137 \\
   1.15 &   6.89 $\pm$  0.41  &  3.50 $\pm$  0.27  &  1.73 $\pm$  0.14  &  0.910 $\pm$ 0.098 \\
   1.25 &   4.97 $\pm$  0.32  &  2.49 $\pm$  0.21  &  1.14 $\pm$  0.10  &  0.629 $\pm$ 0.072 \\
   1.35 &   3.50 $\pm$  0.24  &  1.71 $\pm$  0.15  &  0.828 $\pm$ 0.072  & 0.449 $\pm$ 0.053 \\
   1.45 &   2.47 $\pm$  0.18  &  1.23 $\pm$  0.12  &  0.592 $\pm$ 0.053  & 0.317 $\pm$ 0.039 \\
   1.55 &   1.76 $\pm$  0.14  &  0.866 $\pm$ 0.087  & 0.403 $\pm$ 0.037  & 0.210 $\pm$ 0.027 \\
   1.65 &   1.24 $\pm$  0.11  &  0.630 $\pm$ 0.066  & 0.279 $\pm$ 0.026  & 0.154 $\pm$ 0.020 \\
   1.75 &   0.940 $\pm$ 0.082  & 0.438 $\pm$ 0.049  & 0.205 $\pm$ 0.019  & 0.120 $\pm$ 0.014 \\
   1.85 &   0.652 $\pm$ 0.061  & 0.330 $\pm$ 0.038  & 0.151 $\pm$ 0.013  & 0.089 $\pm$ 0.010 \\
   1.95 &   0.474 $\pm$ 0.046  & 0.226 $\pm$ 0.027  & 0.115 $\pm$ 0.009  & 0.061 $\pm$ 0.006 \\
\end{tabular}
  \normalsize
  \caption[\hminus\ \pt\ distribution for various centralities.]
  {\hminus\ yield $d^2N/(\pt\ d\pt\ d\eta)$ at $\eta =0$
  for different centralities.}
  \label{tab:hminuspt}
\end{table}

It is instructive to do a comparison between peripheral and
central \pt\ distributions.  It is easier to study differences in
the distributions by plotting the ratios of the \pt\ spectra,
since slight changes in curvature are difficult to see in a log
scale such as in Fig.~\ref{fig:ptcentrality}. As a reference, we
take the shape of the measured \pt\ distribution for the most
peripheral bin ($70-80\%$ of \sigmahad).
\begin{figure}[htb]
  \centering
  \mbox{\subfigure[]{
  \includegraphics[width=.45\textwidth]{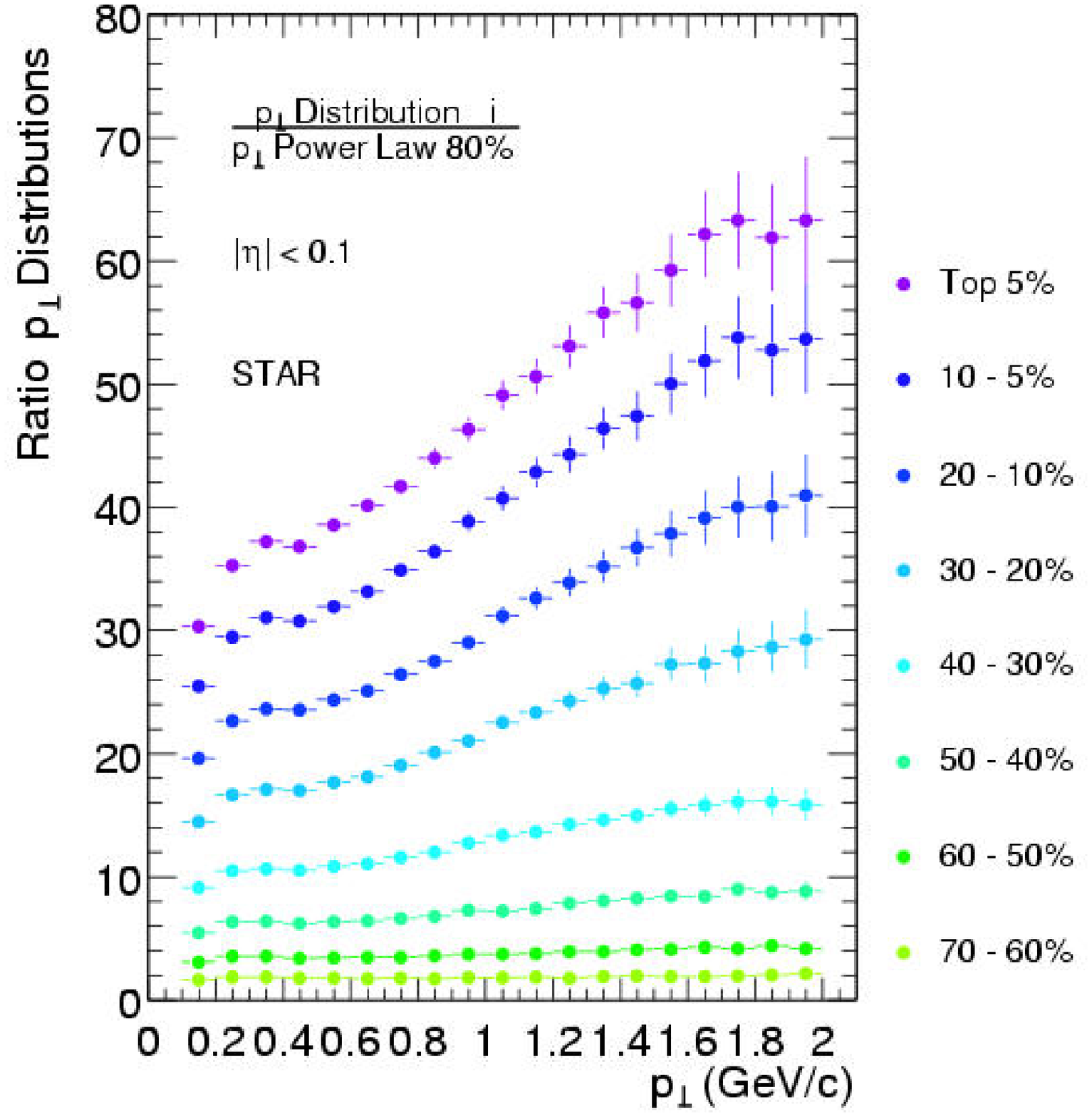}}\quad
  \subfigure[]{\includegraphics[width=.46\textwidth]{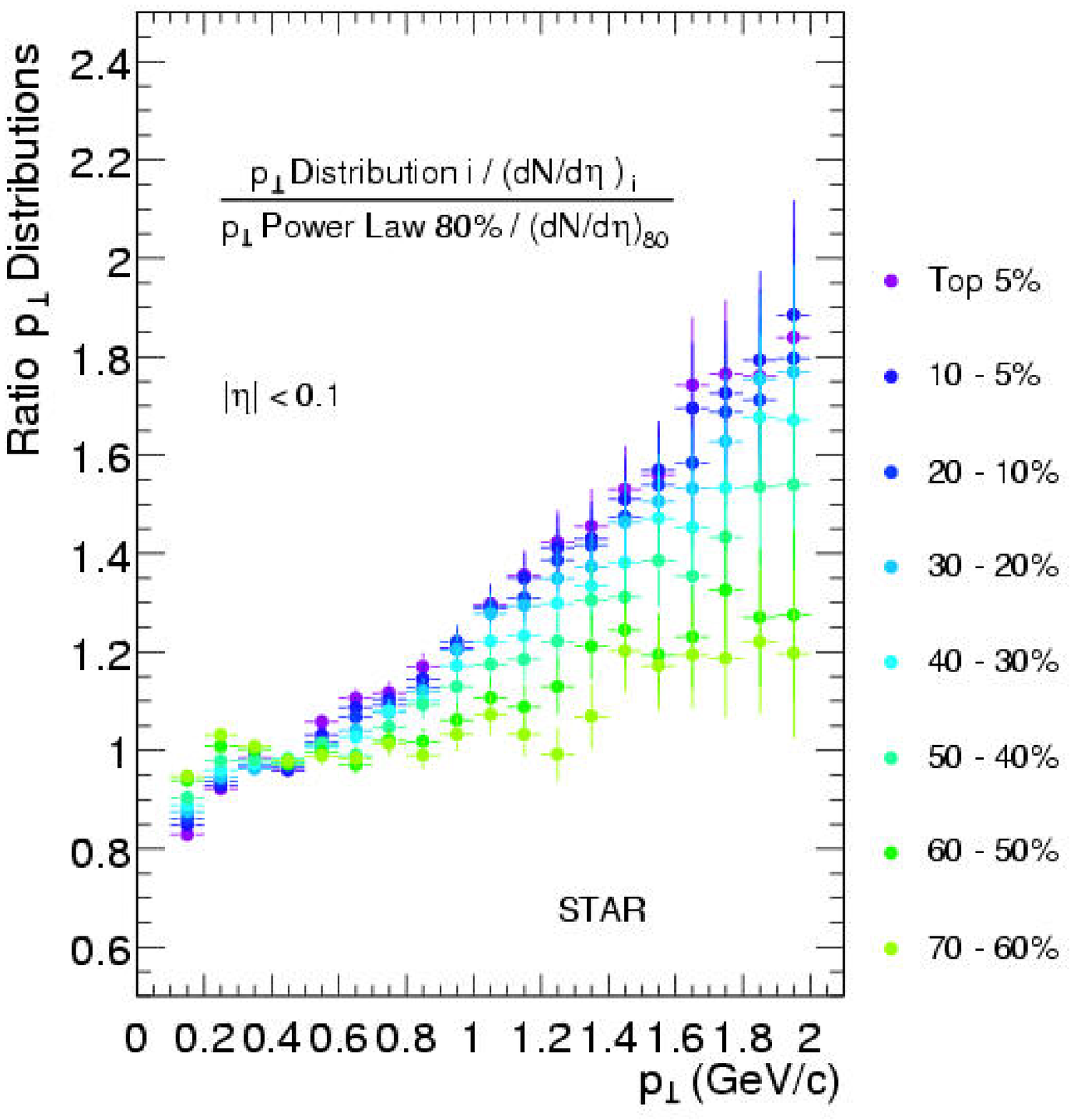}}}
  \caption[Ratio of \hminus\ \pt\ distributions to the most peripheral \pt\ spectrum.]{Ratio of the \pt\ distributions for different centralities to
  the \pt\ spectrum of the most peripheral bin (70-80\%).  Left panel: simple ratio of the data.
  Right panel: Ratio of data points to the power law fit to the most peripheral bin including
  a normalization of each distribution to the corresponding mid-rapidity density $dN/d\eta$ of
  each sample.}
  \label{fig:ptcurvatures}
\end{figure}
We then take the ratio of each of the distributions in
Fig.~\ref{fig:ptcentrality} to the most peripheral.  The simplest
exercise is to take the ratio without any further scaling. This is
shown in Fig.~\ref{fig:ptcurvatures}(a).  If we take the most
peripheral distribution as the one which most closely should
follow the spectrum from $pp$ or $\ppbar$ collisions, we see how
the shape of the \pt\ distributions change with centrality.
Similarly to the \ppbar\ comparison, we see an increase in the
\hminus\ yield with increasing \pt\ for a particular centrality
selection compared to the peripheral distribution.  We also see
that the excess at fixed \pt increases with centrality.  This is
more easily seen in a double ratio, where we normalize each \pt\
distribution by the corresponding $dN/d\eta$, \ie\ by the integral
with respect to \pt.  This double-ratio is shown in
Fig.~\ref{fig:ptcurvatures}.  The double-ratio curves then meet at
low-\pt\ independent of centrality, and the rise with increasing
\pt\ is evident.  We find that for the most central collisions,
the \hminus\ yield at 2 \gevc\ is almost a factor of 2 larger than
expected from a superposition of peripheral collisions that would
integrate to the same yield.  This change in behaviour is
consistent with the presence of collective effects such as
increased radial flow for central collisions. However, as
mentioned before, the \pt\ spectrum for even the most central
collisions still does not reach the expected scaling with \ncoll\
at 2 \gevc. The measurement of the \pt\ distributions at higher
\pt\ will help be an interesting extension to these studies, as
they will shed more light on the modification of the \pt\ spectra,
\eg\ whether the ``Cronin'' effect observed in \pA\ collisions and
lower energy $AA$ collisions should be seen at these moderate
momenta, and if so, whether the jet quenching hypothesis is able
to reproduce the centrality dependence of the \pt\ ratios such as
Fig.~\ref{fig:pt} and Fig.~\ref{fig:ptcurvatures}.

\begin{figure}[htb]
  \centering
  \includegraphics[width=.65\textwidth]{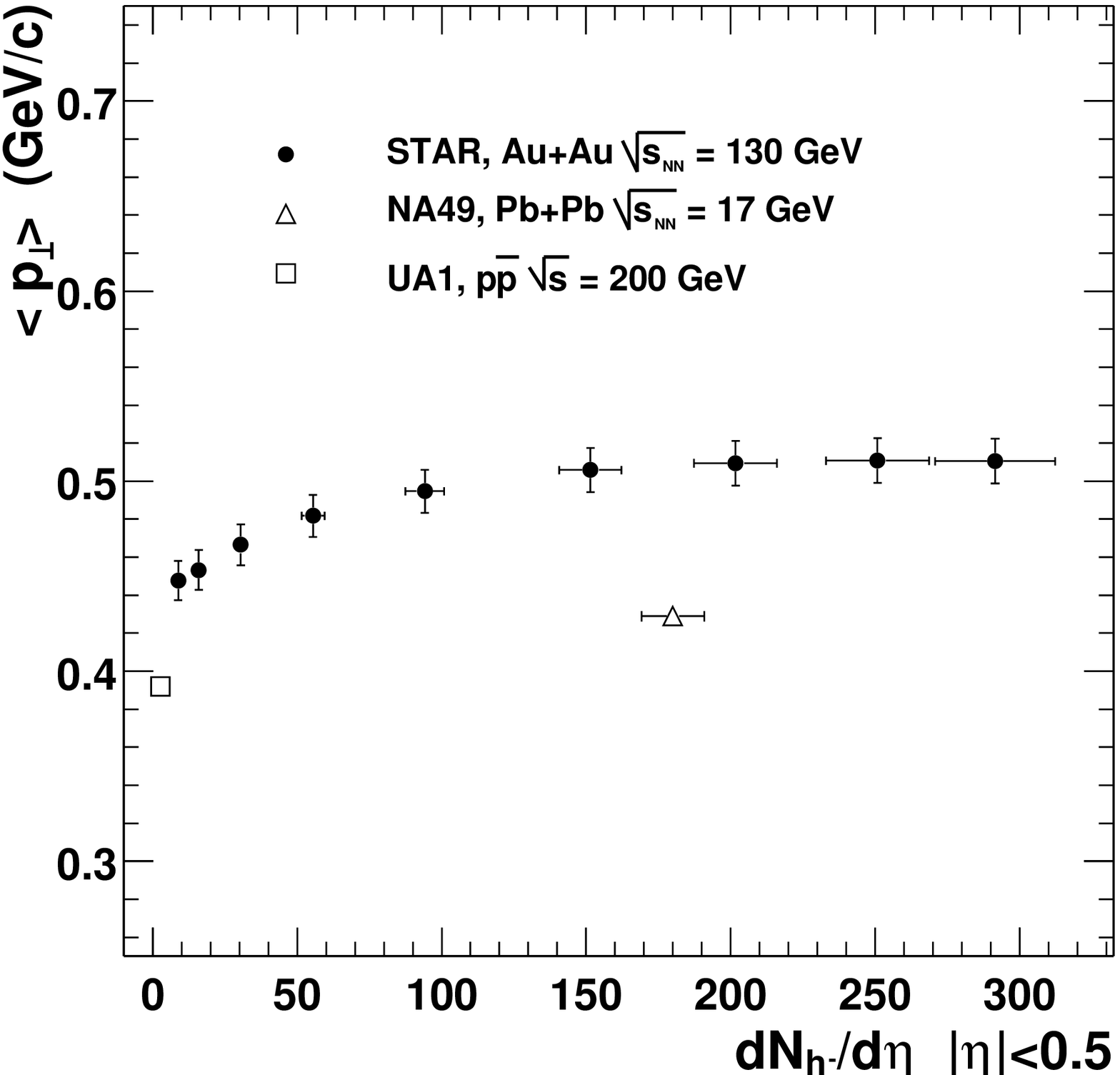}
  \caption[\hminus\ multiplicity dependence of mean \pt.]
  {The multiplicity dependence of \meanpt\ for \hminus.
  The values obtained from the UA1\cite{ua1:90} and NA49\cite{na49:99}
  \pt\ distributions in Fig.~\ref{fig:pt} are also shown.}
  \label{fig:meanptcentrality}
\end{figure}

Figure \ref{fig:meanptcentrality} shows the
multiplicity dependence of \meanpt\ for \hminus. We use the fitted
parameters to the power law in Fig.~\ref{fig:ptcentrality} to
obtain \meanpt, the corresponding multiplicity is obtained by
averaging the measured yield in the region $|\eta|<0.5$, $0.1 <
\pt < 2 \gevc$ (Fig.~\ref{fig:etacentrality}) and extrapolating to
all \pt\ using the power law. This has the effect that the average
multiplicity measured in the wider slice is higher (267, compared
to 261). The data are collected in

Table ~\ref{tab:meanptdndeta}, which shows the multiplicity and
\meanpt\ obtained from the spectra for the different fractions of
cross section. The error bars representing the systematic
uncertainty are 2\% for \meanpt, 6.4\% for measured and 7.1\% for
extrapolated $dN/d\eta$. By using the power-law fits to the \pt\
distributions in this range, we obtain the missing fraction in
order to obtain a total yield ($\sim7\%$ for $0< \pt < 0.1$ and
$1\%$ for $\pt > 2\ \gevc$). The centrality selection is based on
cuts in ZDC and CTB trigger signals as shown in
Fig.\ref{fig:centralityClasses}.

We see an increase in \meanpt\ in the first peripheral bins and
then a saturation.  The value of \meanpt\ increases by 18\%
compared to SPS energies, when comparing to central collisions. To
put into perspective the stiffness of the observed \AuAu\ \pt\
spectra, the measured \meanpt\ at \sqrts\ = 900 \gev\ reported in
\cite{ua1:90} is \meanpt\ = $0.447 \pm 0.003$ \gevc. At the
highest Tevatron energies, CDF reports \cite{Abe:1988yu} \meanpt\
= $0.495 \pm 0.014$.  The saturation of \meanpt\ contrasts the
structure proposed in Ref.~\cite{vanHove:82}, where the
qualitative dependence for a first order transition (under the
assumption that the central multiplicity is approximately
proportional to the $sV$, where $V$ is the volume of the system
and $s = \epsilon + p$ is the entropy density in Landau
hydrodynamics, and that \meanpt\ reflects the temperature of the
system) should be an initial correlation between $dN/d\eta$ and
\meanpt, followed by a saturation, and then a subsequent rise.  We
find no such structure.  This behaviour helps to strengthen the
growing evidence against a strong first order transition.  In
addition, it is also consistent with current expectations of a
second order phase transition region at higher baryo-chemical
potentials with a critical point followed by a region of a smooth
cross over between the QGP and hadron gas phase below some as yet
undetermined value for $\mu_B$.  Initial estimates from particle
ratios at RHIC yield values of $\mu_B \simeq\ 45\ \mev$, which
would then most probably place RHIC collisions at this energy in
the cross-over region.  This will of course need to be studied
with more sophisticated analyses, in particular, an interesting
future direction could be the measurement of the \meanpt\ \vs\
$dN/d\eta$, along with \meanpt\ fluctuations event-by-event,
studied for different centre-of-mass energies or collision systems
to search for the possible critical point.  Theoretical guidance
in this direction would also be fruitful in this search.

%
%
\chapter{Results and Discussion: Identified Pions}
\label{ch:PiResults} The next step in the analysis of charged
particles is to use the particle identification capabilities of
the TPC.  This allows us to select a specific particle for further
study. The first natural candidate is the $\pi$ meson, the
lightest of the hadrons and the most copiously produced particle
in a high energy heavy ion collision.

\section{Raw Yields}
The starting point to obtain the raw yields for the identified
particle analyses is Fig.~\ref{fig:dedxvsp}.  One can obtain the
raw yields of pions, kaons and protons (and their antiparticles)
selecting events according to their centrality and fitting the
\dedx\ distribution in a given y-\pt\ phase space cell.  The
procedure adopted here is a variation from this which offers a few
advantages.   Tracks were selected according to quality criteria
based on number of points on the track and on the pointing
accuracy to the primary event vertex.  The cuts used in the pion
analysis were more stringent than the ones used for the \hminus\
analysis.  This was done mainly to select tracks with a good
\dedx\ resolution as it depends mainly on the number of valid
ionization samples, see Fig.\ref{fig:dedxResolution}.

For this analysis, we followed a procedure similar to the one
outlined in Ref.~\cite{aguilar-benitez:91} for ionization
measurements.  We use the truncated mean to estimate the
ionization of a given track ($I_m$).  The expected ionization for
a given momentum and a given mass hypothesis ($I_h$)is known to an
accuracy of better than 0.1\%, as represented by the curves in
Fig.\ref{fig:dedxvsp}.  We then construct the $z$ variable defined
as
\begin{equation}\label{eq:zvariable}
  z = \ln\left(\frac{I_m}{I_h}\right)
\end{equation}
The $z$ variable follows a Gaussian distribution with mean zero
for a given particle population and with standard deviation given
by the measured \dedx\ resolution as a function of the number of
\dedx\ samples, Fig.\ref{fig:dedxResolution}.  The mean will be
zero independent of momentum for the given particle species under
consideration, since the momentum dependence is contained in
$I_h$. By incorporating $I_h$ and the measured resolution
$\sigma_{dE/dx}$, we can obtain a new distribution, $Z_\pi$, that
has unit width and zero mean for the particle of interest:
$Z_{\pi}=z_{\pi}/\sigma_{dE/dx}$.

\begin{figure}[htb]
\begin{center}
\includegraphics[width=.6\textwidth]{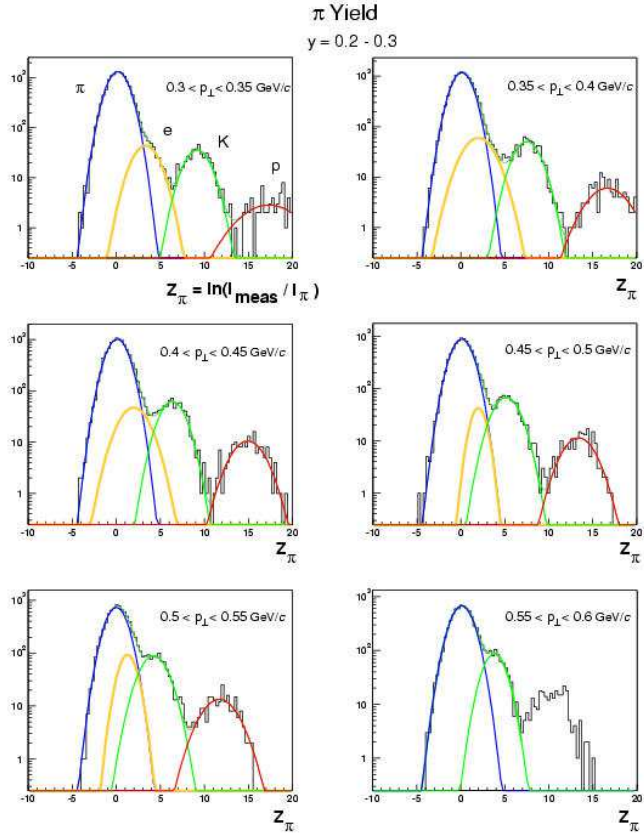}
\caption[Fits to $Z_{\pi}$ distributions for raw \piminus\
yields.]{Fits to the $Z_{\pi}$ distribution to extract the yield
of $\pi^-$. Shown is the region $0.2 < y < 0.3$ and $0.3 < \pt <
0.6\ \gevc$.} \label{fig:pionzfits}
\end{center}
\end{figure}

Fig.~\ref{fig:pionzfits} shows the result of fitting to the $Z$
distributions in different phase-space bins.  Shown is the region
0.2 < y < 0.3 for 6 different \pt\ bins in steps of 50 \mevc\
starting from \pt = 0.3 \gevc.
The abscissa is in log scale, and the ordinate is in linear scale
(note that $Z_\pi$ is already the log of the ionization).
Strictly, only the pion population should follow a Gaussian in the
$z_{\pi}$ variable. One can of course construct a $z$ variable for
the other species ($z_{K}$, $z_{p}$ and $z_{e}$) in order to
extract their yields. The analysis of the other particle species
is being carried out in STAR also. Here we focus on the pions
which are the bulk of the produced particles as can be seen from
the relative heights of the different Gaussians in the plot. The
figure shows the result of the different fits using a Gaussian for
each of the particle species. The pion yield is extracted from the
Gaussian centered at zero. In the lower right panel, the electron
yield is small enough and close enough to the kaon population that
one can use a single Gaussian function.  The fit also stops at
$Z_{\pi}=5$ in the bottom right panel, thus the anti-proton
Gaussian is not shown.

\section{Corrections}
\begin{figure}[htb]
  \centering
  \includegraphics[width=.7\textwidth]{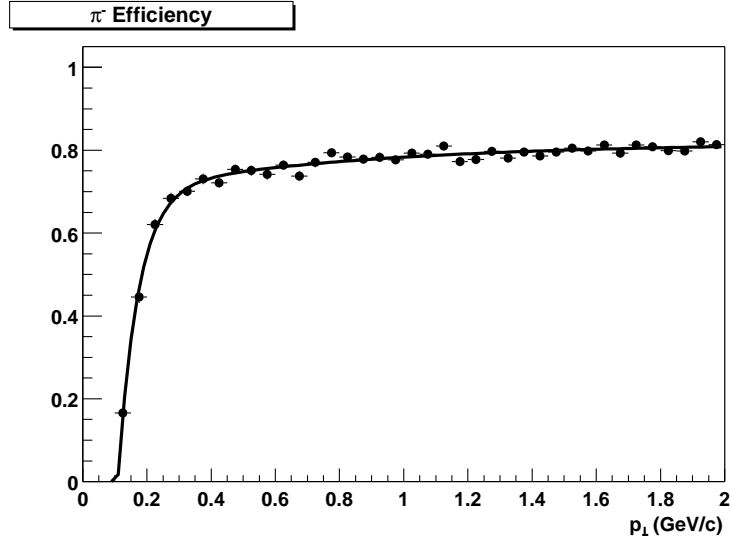}
  \caption[\piminus\ reconstruction efficiency \vs\ \pt.]
  {The track reconstruction efficiency
  as a function of \pt\  for \piminus\ mesons.  The efficiency
  is lower than in Fig.~\ref{fig:efficiency} as the
  requirements for tracks in the \dedx\ analysis are more
  stringent than for the \hminus\ analysis.}
  \label{fig:piEfficiency}
\end{figure}

The corrections applied to the raw data follow the same procedure
as for the \hminus\ distribution with a few important differences.
First of all, we chose to do them in cells of ($y,\pt$) instead of
($\eta,\pt$).  The acceptance and efficiency use the embedding
procedure as in the \hminus\ analysis.  Embedding was done in the
region .1 < \pt\ < 2 \gevc, and |$y$|<1.  For the lowest \pt\ bin,
50 < \pt\ < 100 \mevc, we used a full \Hijing\ simulation. The
efficiency obtained from embedding is shown in
Fig.\ref{fig:piEfficiency}.
The stringent requirement on 24 or more fit points on the track
effectively removed the need to undertake a split track
correction. For the momentum range $\pt<0.6\ \gevc$ the distortion
of the spectrum due to resolution effects was found to be
negligible. This fact and the good statistics of the sample
allowed for a much finer bin size in the \pt\ spectrum.  This was
also desirable since this narrowed the phase space where the fits
were made to obtain the yield. Event though the use of the $z$
variable fixes the region where the pions are found, the other
particles will still vary with momentum.  Therefore, a narrower
\pt\ bin will reduce this effect making the fits more stable.  The
final spectrum was also corrected for the products of weak decays
and for the expected contribution of pions resulting from
secondary interactions. Since for this analysis the requirement on
the distance of closest approach to the primary interaction vertex
was set at 1 cm, the background correction was smaller than for
the \hminus\ analysis, and was found to be $\lesssim 5\%$. We
present now the resulting distributions for identified \piminus\
mesons.

\section{\pt\ and \mt\ Distributions}

\subsection{Results}
\begin{figure}[htb]
\begin{center}
\includegraphics[width=0.5\textwidth]{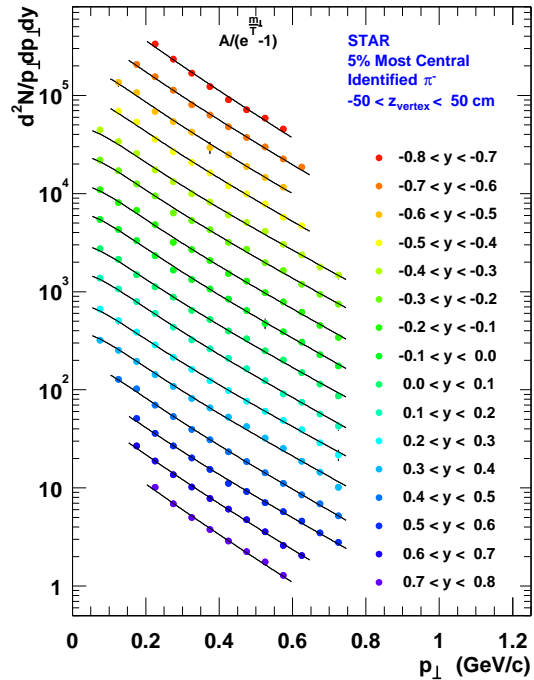}
\caption[\piminus\ \pt\ spectrum for central collisions]{\piminus\
 \pt\ distribution for central collisions.  The different bins in rapidity
 are scaled by successive powers of 2 for display purpose.} \label{fig:pionspectraptcentral}
\end{center}
\end{figure}

We first look at the spectra for central events.
Figure~\ref{fig:pionspectraptcentral} shows the \pt\ distribution
for \piminus\ for the 5\% most central collisions (selected by a
cut on raw \hminus\ multiplicity).  The data are grouped for
different bins in rapidity, and scaled by successive powers of 2
for the figure.

\begin{figure}[htb]
\begin{center}
\includegraphics[width=0.6\textwidth]{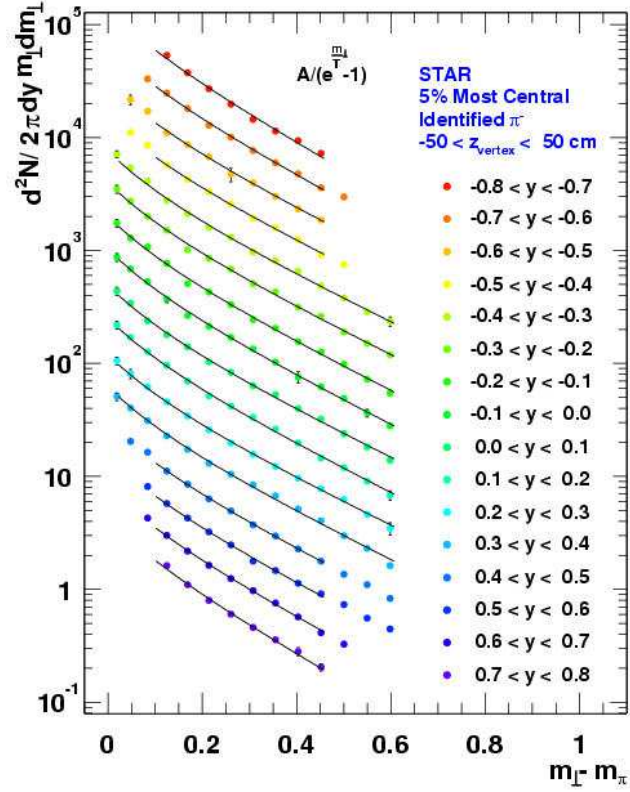}
\caption[\piminus\ \mt\ spectrum for central collisions]{\piminus\
$d^2N/(2\pi\ dy\ \mt d\mt)$ distribution for central collisions.
The rapidity bins are scaled by successive powers of 2 for display
purpose.} \label{fig:pionspectracentral}
\end{center}
\end{figure}

The more common way to show such a distribution is in transverse
mass, \mt, and this is shown in Figure
~\ref{fig:pionspectracentral}.
The main corrections applied to the raw data are acceptance,
efficiency and contributions from weak decay background and
secondary interactions.  The effect of energy loss in the
determination of the \piminus\ momenta was found to be less than
2\% at 100 \mevc, decreasing rapidly for larger momenta.  We also
studied the effect of momentum resolution as for the \hminus\ data
and found that the main effect was the feeding of the lowest \pt\
bin from its nearest neighbour. The change in the slope for $\pt\
> 100\ \mevc$ was found to be $\ll 1\%$.  The data for the central
rapidity bins, $|y|<0.4$, was obtained in the range $0.05 < \pt <
0.75\ \gevc$, which accounts for $\sim 85\%$ of the total yield.
As we go towards the forward and backward regions, the \pt\ range
where the identification via \dedx\ in the TPC can be done
shrinks, and we can only measure in the range 0.2 < \pt\ < 0.6
\gevc.

\subsection{Discussion}
Pions are the bulk of the produced particles, they are the
lightest of the hadrons and have large interaction cross sections
in nuclear matter; they are therefore expected to thermalize
easily.  In addition, their spectra are the least affected by a
given collective transverse flow velocity, and are thus good
probes to study the kinetic freeze-out properties of the system.
The pion data are fit with a Bose-Einstein distribution of the
form $A/(\exp(\mt/T) - 1)$. Typically, these distributions are fit
with a Boltzmann distribution \cite{schukraft:91}.  The data at
low-\pt\ fit with a Boltzmann distribution always seemed to show
an enhancement.  This partly was attributed to contributions
coming from resonance decays. There are also enhancements in the
high-\pt\ region which can arise from contributions of pions
coming from high momentum transfer of partons early in the
collision.  A consequence of all these structures was that simple
Boltzmann fits to the same spectrum would lead to different slope
parameters depending upon the \pt\ region included in the fit.
Since we focus here on the low \pt\ part of the spectrum, and
since pions are bosons, we chose to fit the data with a
Bose-Einstein distribution.  We find that this naturally gives an
enhancement in the very low \pt\ region of the spectrum.  Also, we
have already estimated the contribution from weak decays to the
distribution and corrected for this effect.  We have not estimated
the contribution from resonance decays however. We fit the data
over the measured range, instead of excluding the low momentum
data from the fit, and find no significant enhancement at low \pt.
The absence of a noticeable contribution from resonances might be
due to reinteraction of the daughter decay particles. A fit using
a Boltzmann distribution for the range above $\pt = 0.3\ \gevc$
causes a variation in the slope parameters of $\sim 10\ \mev$.
Since the data at forward and backward rapidity have a smaller
range in \pt\, as a systematic check we performed a fit restricted
to the \pt\ range which was measured for all the rapidity bins.
This resulted in a change in the slope parameters of 5-10 \mev\
for the $|y|<0.4$ spectra.  We find the slope parameter for the
central collisions to be $\Teff = 210\ \pm 20\ \mev$, where the
systematic uncertainty comes from studying the variation in the
slope parameter with several changes, including variations in the
fit range, choice of fit function, centrality selection, and
analyzing the pions in $\eta$ instead of $y$.  The error on
$\Teff$ from any particular fit is on the order of $5\ \mev$. The
rapidity dependence of the \piminus\ yield and of the slope
parameters is discussed in the following section.

The shape of the low \pt\ pion spectra, in contrast to the
\hminus\ distribution, shows very little change when going from
SPS energies to RHIC.  The slope parameters reported by NA49
\cite{Jones:1996xc} for \piminus\ near midrapidity are $\Teff =
188 \pm 6\ \mev$ for \piplus\ and $\Teff = 192 \pm 3\ \mev$ using
an exponential fit.  WA98 data for $\pi^0$ spectra for central
collisions yields also similar values of $\Teff \simeq 204$
(although fitting the data with a Bessel function, $\mt
K_1(\mt/\Teff)$ yields a smaller value of $\Teff=155$, so one must
be careful when comparing). The pion spectra from NA45
\cite{ceretto:98} were analyzed in terms of a local inverse slope
changing with \mt\ to take into account the difference at low and
high \pt.  For the low \mt\ range ($\mt-m_{\pi} < 0.8\ \gevcc$)
the local inverse slope is found to be in the range $180-200\
\mev$. The slopes at the AGS have smaller values, E895 finds
\cite{Liu:1998rt} $\Teff \simeq 110\ \mev$.  While the initial
temperature reached at RHIC is expected to be larger than at the
SPS and the AGS, the \piminus\ spectrum probes the final state.
The saturation of the \piminus\ slope parameter would indicate
that the freeze-out temperature, assuming that thermal equilibrium
is reached, is the same at SPS and RHIC. This does not mean that
the final state is the same, for there is evidence of an increased
collective radial flow velocity at RHIC based on the preliminary
\pt\ distributions of heavier particles from STAR
\cite{harris:01,calderon:01}. Since the collective velocity
affects the heavier particles the most, the pion inverse slope is
the one numerically closer to the actual freeze-out temperature,
\Tfo.  The slopes of the \kminus\ and \pbar\ will be much larger
than for the \piminus\ even if they all freeze out at the same
temperature in the presence of large radial flow.

In order to gain further insight into the dynamics of the system,
a study of the heavier particle spectra can be made to address the
question of radial flow.  In a different approach, the combination
of the 2-particle correlation results\cite{Adler:2001zd} along
with the pion spectra is also useful: they determine the
6-dimensional pion \textit{phase space density}, \ie\ the
dimensionless average number of pions per 6-dimensional phase
space cell $\hbar^3$ (see \eg\ \cite{Ferenc:1999ku}).  Such
analyses can help to disentangle the intrinsic freeze-out
temperature and the transverse flow velocity.  For example, for
the 5\% most central data, preliminary analysis of the phase space
density yield values of $\Tfo \sim 120\ \mev$ and $\langle
\beta_\perp \rangle \sim 0.53$.  This is a rather large flow
velocity, yet since the pions are so light, the slope parameter
$\Teff$ is only increased to $\sim 200$ \mev, while the slope
parameter for the heavier \kminus\ increases to $\sim 300\ \mev$
and is even larger for the \pbar\ \cite{harris:01}. The
systematics of the centrality dependence of the pion phase space
density is a currently active area of study in STAR which will
certainly help to shed light on the determination of the kinetic
freeze-out conditions of the system.

\section{Rapidity Distribution}
\subsection{Results}
\begin{figure}[htb]
\begin{center}
\includegraphics[width=.6\textwidth]{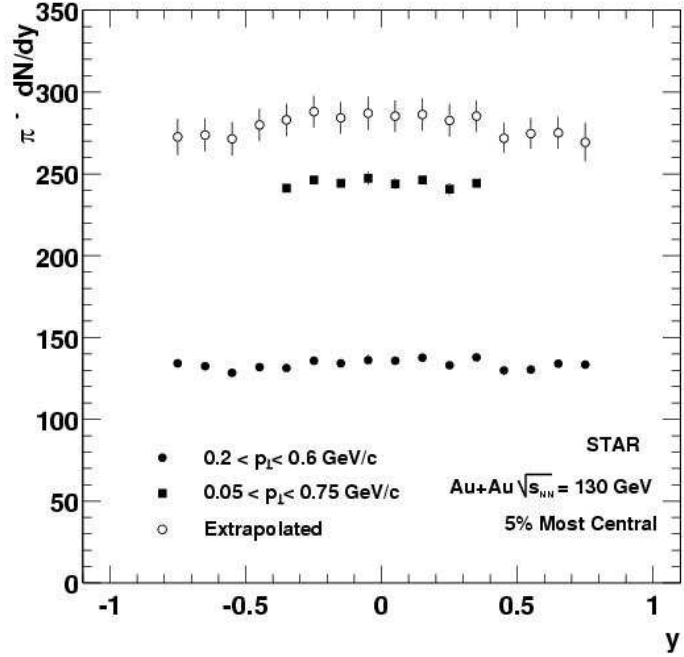}
\caption[\piminus\ rapidity distribution, $dN/dy$]{\piminus\
rapidity distribution. The black circles (squares) are obtained
summing the yield between \pt\ 0.2 - 0.6 \gevc\ (0.05 - 0.75) from
Fig.~\ref{fig:pionspectraptcentral}. The hollow data points are
the yields obtained using the fit function extrapolated to all
\pt.} \label{fig:piondndy}
\end{center}
\end{figure}

\begin{figure}[htb]
\begin{center}
\includegraphics[width=.7\textwidth]{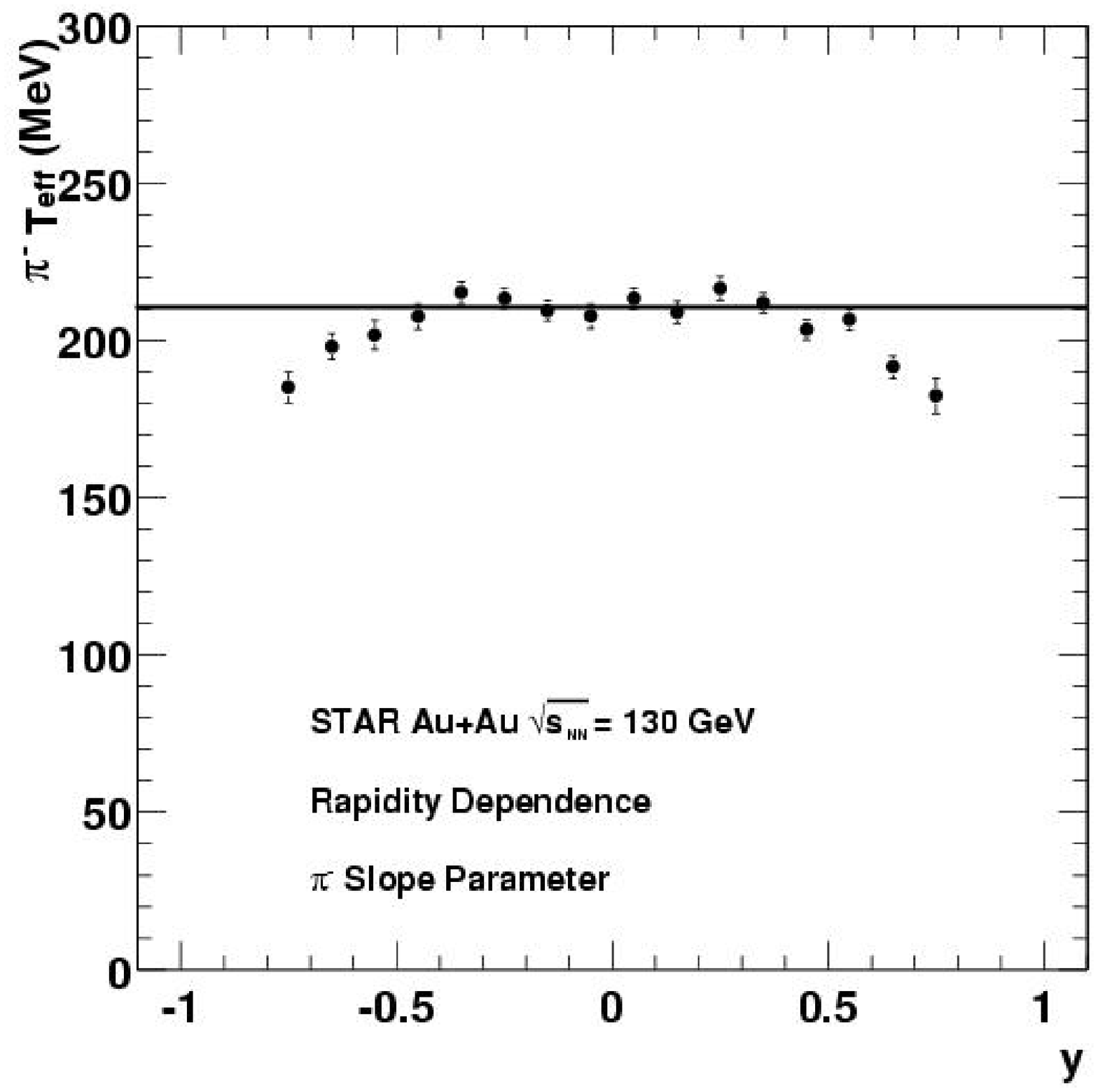}
\caption[\piminus\ slope parameter $T_{\mathrm{eff}}$ \vs\ $y$.
]{\piminus\ slope parameter for the different rapidity bins from
Fig.~\ref{fig:pionspectracentral}. The flat line at $\Teff = 210$
is the average for $|y|<0.1$} \label{fig:pionTeffY}
\end{center}
\end{figure}

By integrating the \pt\ spectra in
Figure~\ref{fig:pionspectraptcentral} for each of the rapidity
bins, we are able to obtain the $dN/dy$ \vs\ $y$ distribution for
\piminus.  The result is shown in Figure~\ref{fig:piondndy}.
Several data points are shown.  The black circles are obtained by
integrating the \pt\ spectrum in the limited \pt\ region $0.2 -
0.6\ \gevc$ where we have data for $|y|<0.8$.  This shows that the
measured yield in this region is relatively independent of
rapidity. The measured yield in this region is, however, only
about half of the total yield.  To illustrate the behaviour of the
yield in the full available $\pt$ region ($0.05 - 0.75\ \gevc$)
for a narrower $y$ region, we plot the data shown as the black
squares.  Finally, the open circles are the extrapolated yield to
all $\pt$ based on the fits to the data using the Bose-Einstein
distribution.  We find the pion yield to be $\dndy|_{|y|<0.1} =
286 \pm 10$ for the 5\% most central events.
Not only is it important to study the yields as a function of
rapidity, but it is equally important to examine the rapidity
dependence of the slope parameters.  At the SPS and at the AGS,
they have been found to decrease with increasing $|y|$.  The
$\Teff$ slope parameter extracted from the Bose-Einstein fit to
the data is shown in Figure~\ref{fig:pionTeffY}.
The error bars in the figure are the uncorrelated point-to-point
systematic uncertainties on $\Teff$, the overall correlated
systematic uncertainty is $\pm 20\ \mev$.  The rapidity dependence
of the yield, the slope parameter \Teff\ and \meanpt\ are
collected in
\begin{table}[htb]
    \centering
    \begin{tabular}{cccc}
        \hline \hline
        $y$   &      \dndy\            &  \Teff\ [\mev]   &  \meanpt\ [\mevc] \\
        -0.75 & 273 $\pm$ 13  &  185.0 $\pm$ 5.0 &  356 $\pm$ 9.9  \\
        -0.65 & 274 $\pm$ 11  &  198.1 $\pm$ 4.2 &  376 $\pm$ 8.4  \\
        -0.55 & 273 $\pm$ 10  &  201.7 $\pm$ 4.6 &  385 $\pm$ 9.2  \\
        -0.45 & 280 $\pm$ 10  &  207.5 $\pm$ 4.0 &  391 $\pm$ 7.9  \\
        -0.35 & 283 $\pm$ 10  &  215.2 $\pm$ 3.6 &  397 $\pm$ 7.2  \\
        -0.25 & 288 $\pm$ 10  &  213.4 $\pm$ 3.1 &  402 $\pm$ 6.2  \\
        -0.15 & 284 $\pm$ 10  &  209.5 $\pm$ 3.2 &  396 $\pm$ 6.4  \\
        -0.05 & 287 $\pm$ 10  &  207.7 $\pm$ 3.9 &  394 $\pm$ 7.8  \\
         0.05 & 285 $\pm$ 10  &  213.3 $\pm$ 3.4 &  400 $\pm$ 6.8  \\
         0.15 & 286 $\pm$ 10  &  208.9 $\pm$ 3.4 &  392 $\pm$ 6.9  \\
         0.25 & 284 $\pm$ 10  &  216.5 $\pm$ 3.8 &  404 $\pm$ 7.6  \\
         0.35 & 285 $\pm$ 10  &  211.9 $\pm$ 3.1 &  400 $\pm$ 6.2  \\
         0.45 & 272 $\pm$ 10  &  203.4 $\pm$ 3.2 &  383 $\pm$ 6.4  \\
         0.55 & 275 $\pm$ 10  &  206.5 $\pm$ 3.4 &  388 $\pm$ 6.8  \\
         0.65 & 275 $\pm$ 11  &  191.5 $\pm$ 4.0 &  367 $\pm$ 8.0  \\
         0.75 & 269 $\pm$ 14  &  182.4 $\pm$ 5.6 &  350 $\pm$ 11.2 \\
         \hline \hline
\end{tabular}
\caption[\piminus\ \dndy, \Teff\ and mean \pt\ \vs\ $y$.]{\piminus\ \dndy, \Teff\ and \meanpt\ \vs\ $y$.}
\label{tab:piRapidity}
\end{table}
Table~\ref{tab:piRapidity}. The \pt\ distributions for the most
central data for the various $y$ bins are collected in
Table~\ref{tab:piPtRapidity}.

\subsection{Discussion}
The rapidity distribution has nearly a plateau shape at
mid-rapidity. However, there remains a slight rapidity dependence
of the slope parameter $\Teff$.  This behaviour is indicative that
the idealized boost-invariant mid-rapidity region is not yet
reached in \AuAu\ collisions at \sqrtsNN = 130 \gev.  A possible
cause for the observed effect can be a change in the baryon
content of the system with rapidity. In the presence of radial
flow driven by the dominant \piminus\ mesons, a larger fraction of
baryons would take away some of the pion's kinetic energy,
reducing the observed slope parameter.  The anti-proton to proton
ratios reported by STAR~\cite{Adler:2001bp} and
BRAHMS~\cite{Bearden:2001kt} show a drop in the $\pbar/p$ ratio
from 0.65 at mid-rapidity to 0.41 for the region $y\approx 2$,
which already hints at such effects, but a better understanding of
the dynamics must wait for the measured baryon yields and the
baryon rapidity distributions.

\section{Centrality Dependence}
\subsection{Centrality Selection}
For the \piminus\ centrality dependence, we chose to do a
selection based on the event multiplicity.  This is in contrast to
the centrality selection for the \hminus\ analysis which was based
on ZDC and CTB trigger signals.  The reason for this choice was
simply to use a common set of cuts which would be useful not just
for this analysis, but also for other studies such as particle
correlations (Hanbury-Brown Twiss) and in particular for the
studies of the pion phase space density. Having a standard set of
values based on the (uncorrected) \hminus\ yield facilitates
comparison and combination of different observables.  There was a
worry that since the pions are the main component of the charged
hadron spectra, a centrality selection based on charged
multiplicity would introduce a sizable auto-correlation.
Certainly, the shape of the \hminus\ multiplicity distribution for
5\% most central events (the shaded region in
Fig.~\ref{fig:hminus}) would look different by selecting events
with a straight cut on uncorrected multiplicity.  The mean of this
distribution, \ie\ the value of $dN/d\eta$ for the region
$|\eta|<0.5$, will also shift to higher values by doing such a
selection.  The question is how significant is this shift.  We
studied this by selecting the 5\% most central events based on
uncorrected multiplicity and comparing the mean of the resulting
distribution to the mean of the histogram in the shaded region of
Figure ~\ref{fig:hminus}.  The mean of the histogram using the
centrality definition based on multiplicity was 4.5\% higher than
than the one using the ZDC as centrality definition.  The
difference in the mean of the histograms using these two
centrality definitions will also decrease for more peripheral
collisions.  We therefore conclude that the differences are not
problematic, and when encountered, they can be reconciled by at
most a 4.5\% effect.

\subsection{\mt\ Distributions \vs\ Multiplicity}
\begin{figure}[htb]
\begin{center}
\includegraphics[width=.6\textwidth]{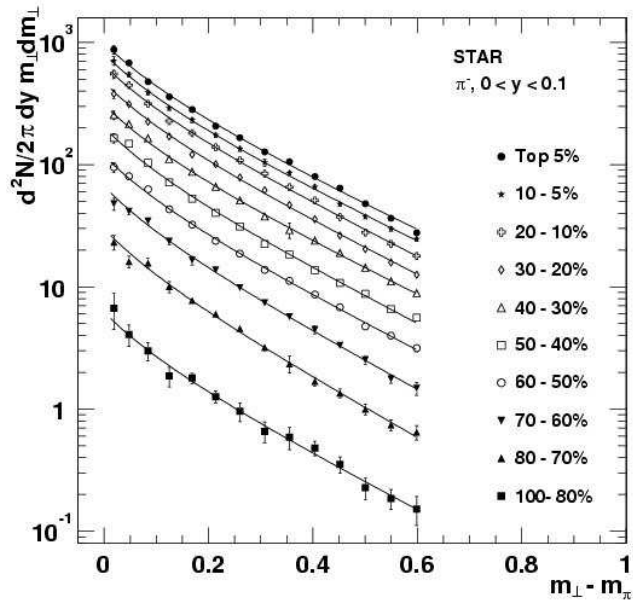}
\caption[\piminus\ \mt\ distribution for different
centralities.]{\piminus\ \mt\ distribution for different
multiplicity selections.  The distributions are plotted in the
rapidity slice 0 < $y$ < 0.1 units. The data are fit to
Bose-Einstein distributions (curves) from which we extract the
slope parameter \Teff.} \label{fig:pionPtCent}
\end{center}
\end{figure}

\begin{figure}[htb]
\begin{center}
\includegraphics[width=.6\textwidth]{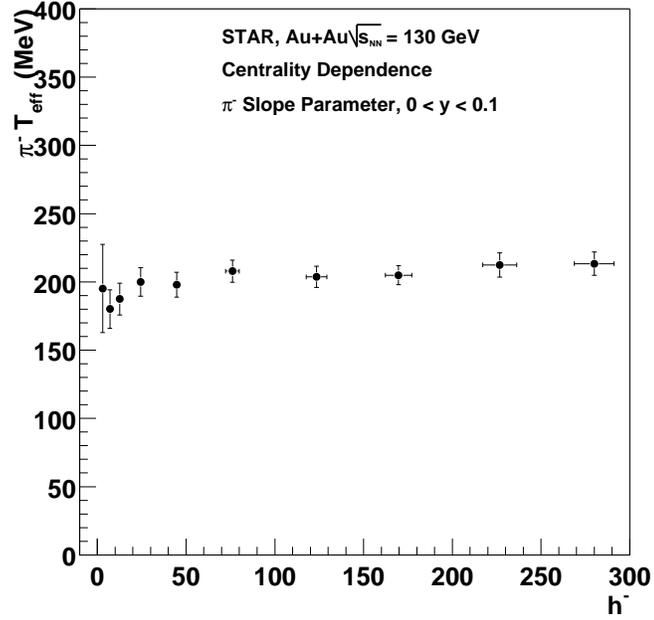}
\caption[Centrality dependence of the \piminus\ slope
parameter.]{Centrality dependence of the slope parameter obtained
by a fit to the \piminus\ \mt\ distribution.}
\label{fig:pionTeffCent}
\end{center}
\end{figure}

Figure~\ref{fig:pionPtCent} shows the \mt\ distribution for 10
different centrality selections.  The data are taken in the
rapidity interval $0 < y < 0.1$.  Again, the data are fit by a
Bose-Einstein distribution which agrees well with the spectra even
at the lowest \mt.  The slope parameters are very similar for the
different centralities.
They are collected and plotted \vs\ the \hminus\ mean multiplicity
of the corresponding centrality bin in
Figure~\ref{fig:pionTeffCent}.
The horizontal error bars are the systematic uncertainty in the
determination of the mean multiplicity of each centrality bin. The
vertical error bars are the uncorrelated systematic uncertainty in
the measurement of the slope parameter.  For the most peripheral
bins, the statistical error is also important, and the error bar
is therefore larger.

We see that there is a only a slight dependence of the slope
parameter with centrality.  The main difference happens for the
2nd and 3d most peripheral bins (the most peripheral bin has a
larger error, so it is hard to see a systematic effect).  The 2nd
most peripheral bin has a slope $\Teff = 177 \pm 8\ \mev$, while
the most central bin has a slope $\Teff = 210 \pm 4\ \mev$ (where
the errors are the uncorrelated systematic uncertainties for the
comparison). This yields an increase of $19 \pm 5 \%$. The
increase occurs rapidly, as the 60\% of \sigmahad\ bin already has
a slope parameter of $\Teff = 196 \pm 9 \mev$, and from then on
the data are consistent with having no further centrality
dependence.  For comparison, the increase in \hminus\ \meanpt\
from the $70-80\%$ bin to the most central bin is $14 \pm 3\%$. In
addition, the \hminus\ distribution includes a \pbar\ component,
and the \pbar\ slope has a stronger centrality dependence than the
\piminus\ slope. At SPS energies, the centrality dependence of the
pion slopes was also found to be rather constant. This finding is
also in agreement with the argument that the slope of the
\piminus\ spectrum is not changed significantly by the collective
expansion.

\begin{table}[htb]
  \centering
  \begin{tabular}{c|r|r|r|r|r} 
    $\mt - m_\pi$ {\tiny(\gevcc)}\ & \multicolumn{1}{c}{0-5\%}  & \multicolumn{1}{c}{5-10\%} & \multicolumn{1}{c}{10-20\%} & \multicolumn{1}{c}{20-30\%} & \multicolumn{1}{c}{30-40\%} \\ \hline
 0.0189 & 871  $\pm$ 70  & 707  $\pm$ 57   & 550  $\pm$ 43   &  377  $\pm$ 31   & 254  $\pm$ 21    \\
 0.0478 & 679  $\pm$ 30  & 544  $\pm$ 25   & 446  $\pm$ 19   &  312  $\pm$ 14   & 214  $\pm$ 9.5   \\
 0.0842 & 473  $\pm$ 25  & 388  $\pm$ 20   & 312  $\pm$ 13   &  224  $\pm$ 9.7  & 164  $\pm$ 7.0   \\
 0.125  & 357  $\pm$ 12  & 289  $\pm$ 11   & 226  $\pm$ 6.9  &  169  $\pm$ 5.9  & 111  $\pm$ 3.9   \\
 0.169  & 281  $\pm$ 11  & 231  $\pm$ 11   & 180  $\pm$ 6.4  &  120  $\pm$ 5.2  & 87.1 $\pm$ 2.8   \\
 0.214  & 206  $\pm$ 7.2 & 173  $\pm$ 6.7  & 138  $\pm$ 4.1  &  100  $\pm$ 2.8  & 65.7 $\pm$ 2.4   \\
 0.261  & 165  $\pm$ 5.8 & 134  $\pm$ 5.8  & 108  $\pm$ 3.2  &  78.1 $\pm$ 2.1  & 50.8 $\pm$ 1.5   \\
 0.308  & 127  $\pm$ 5.6 & 104  $\pm$ 6.9  & 84.3 $\pm$ 2.9  &  61.6 $\pm$ 1.8  & 37.6 $\pm$ 1.0   \\
 0.356  & 105  $\pm$ 3.2 & 85.4 $\pm$ 3.1  & 65.1 $\pm$ 2.1  &  46.6 $\pm$ 1.6  & 29.1 $\pm$ 1.2   \\
 0.404  & 79.6 $\pm$ 2.9 & 65.6 $\pm$ 2.6  & 50.9 $\pm$ 1.5  &  35.6 $\pm$ 1.6  & 23.8 $\pm$ 0.74  \\
 0.452  & 64.0 $\pm$ 1.9 & 47.5 $\pm$ 1.7  & 37.2 $\pm$ 1.2  &  26.2 $\pm$ 0.90 & 18.8 $\pm$ 0.60  \\
 0.501  & 47.6 $\pm$ 1.6 & 37.4 $\pm$ 1.4  & 27.7 $\pm$ 0.95 &  20.3 $\pm$ 0.77 & 14.2 $\pm$ 0.50  \\
 0.550  & 36.4 $\pm$ 1.3 & 29.7 $\pm$ 1.1  & 22.4 $\pm$ 0.77 &  15.9 $\pm$ 0.64 & 11.1 $\pm$ 0.42  \\
 0.599  & 27.7 $\pm$ 1.3 & 24.5 $\pm$ 0.98 & 17.8 $\pm$ 0.63 &  12.5 $\pm$ 0.56 & 8.82 $\pm$ 0.35  \\
\multicolumn{6}{c}{\ }\\ 
$\mt - m_\pi$ {\tiny(\gevcc)}\ & \multicolumn{1}{c}{40-50\%} & \multicolumn{1}{c}{50-60\%} & \multicolumn{1}{c}{60-70\%} & \multicolumn{1}{c}{70-80\%} & \multicolumn{1}{c}{80-100\%} \\
\hline
 0.0189 & 165  $\pm$ 14.4 & 95.1 $\pm$ 9.0  & 47.7 $\pm$ 5.3  & 23.2  $\pm$ 3.2   & 6.67  $\pm$ 2.2   \\
 0.0478 & 148  $\pm$ 7.0  & 80.2 $\pm$ 4.6  & 41.4 $\pm$ 2.7  & 16.0  $\pm$ 1.8   & 4.07  $\pm$ 0.81  \\
 0.0842 & 103  $\pm$ 4.5  & 62.8 $\pm$ 3.7  & 34.4 $\pm$ 2.2  & 15.7  $\pm$ 1.5   & 3.00  $\pm$ 0.52  \\
 0.125  & 71.3 $\pm$ 2.8  & 43.0 $\pm$ 2.3  & 23.5 $\pm$ 1.5  & 10.0  $\pm$ 1.0   & 1.86  $\pm$ 0.34  \\
 0.169  & 52.6 $\pm$ 3.0  & 32.2 $\pm$ 1.2  & 16.4 $\pm$ 1.3  & 7.71  $\pm$ 0.41  & 1.79  $\pm$ 0.17  \\
 0.214  & 40.3 $\pm$ 2.2  & 23.7 $\pm$ 1.2  & 13.7 $\pm$ 0.56 & 5.97  $\pm$ 0.34  & 1.26  $\pm$ 0.14  \\
 0.261  & 31.1 $\pm$ 1.2  & 18.7 $\pm$ 0.71 & 9.79 $\pm$ 0.48 & 4.55  $\pm$ 0.27  & 0.956 $\pm$ 0.16  \\
 0.308  & 22.3 $\pm$ 1.1  & 13.7 $\pm$ 0.55 & 7.38 $\pm$ 0.35 & 3.21  $\pm$ 0.21  & 0.656 $\pm$ 0.13  \\
 0.356  & 18.4 $\pm$ 0.66 & 11.2 $\pm$ 0.45 & 5.67 $\pm$ 0.28 & 2.34  $\pm$ 0.37  & 0.588 $\pm$ 0.12  \\
 0.404  & 13.6 $\pm$ 0.50 & 8.65 $\pm$ 0.37 & 4.46 $\pm$ 0.33 & 1.68  $\pm$ 0.13  & 0.480 $\pm$ 0.066 \\
 0.452  & 10.8 $\pm$ 0.42 & 6.78 $\pm$ 0.30 & 3.31 $\pm$ 0.19 & 1.35  $\pm$ 0.11  & 0.351 $\pm$ 0.053 \\
 0.501  & 8.76 $\pm$ 0.40 & 4.75 $\pm$ 0.27 & 2.51 $\pm$ 0.18 & 0.996 $\pm$ 0.10  & 0.228 $\pm$ 0.045 \\
 0.550  & 6.54 $\pm$ 0.31 & 3.98 $\pm$ 0.23 & 1.76 $\pm$ 0.14 & 0.740 $\pm$ 0.076 & 0.185 $\pm$ 0.034 \\
 0.599  & 5.59 $\pm$ 0.30 & 3.14 $\pm$ 0.20 & 1.47 $\pm$ 0.18 & 0.645 $\pm$ 0.087 & 0.152 $\pm$ 0.040 \\
  \end{tabular}
  \caption[\piminus\ $d^2N/(2\pi \mt d\mt dy)$ \vs\ multiplicity.]
  {\piminus\ $d^2N/(2\pi \mt d\mt dy)$ at $y=0.05$ for various \hminus\ multiplicity bins, Fig.~\ref{fig:pionPtCent}}
  \label{tab:piMtMultiplicity}
\end{table}

\begin{sidewaystable}[htb]
    \centering
    \begin{tabular}{c|c|c|c|c|c|c|c|c}

\pt\ [\gevc] & $y = -0.75$& $y = -0.65$    & $y = -0.55$    &  $y = -0.45$   & $y = -0.35$     &   $y = -0.25$  &  $y = -0.15$   &  $y = -0.05$  \\
0.075     &               &                &                &                & 5583 $\pm$ 415  & 5480 $\pm$ 437 & 5475 $\pm$ 437 & 5453 $\pm$ 435\\
0.125     &               &                & 4242 $\pm$ 428 & 4332 $\pm$ 190 & 4225 $\pm$ 205  & 4265 $\pm$ 195 & 4039 $\pm$ 214 & 4297 $\pm$ 181\\
0.175     &               & 3238 $\pm$ 108 & 3332 $\pm$  95 & 3347 $\pm$  99 & 3213 $\pm$ 102  & 3145 $\pm$ 112 & 3355 $\pm$ 118 & 3320 $\pm$ 115\\
0.225     & 2613 $\pm$ 77 & 2428 $\pm$  60 & 2139 $\pm$  85 & 2244 $\pm$  64 & 2183 $\pm$  60  & 2361 $\pm$  71 & 2401 $\pm$  77 & 2329 $\pm$ 167\\
0.275     & 1829 $\pm$ 47 & 1767 $\pm$  41 & 1699 $\pm$  45 & 1667 $\pm$  59 & 1652 $\pm$  47  & 1587 $\pm$  85 & 1590 $\pm$  71 & 1668 $\pm$  77\\
0.325     & 1320 $\pm$ 32 & 1253 $\pm$  39 & 1317 $\pm$  33 & 1300 $\pm$  44 & 1244 $\pm$  40  & 1329 $\pm$  52 & 1340 $\pm$  47 & 1327 $\pm$  53\\
0.375     &  963 $\pm$ 23 &  988 $\pm$  22 &  924 $\pm$ 128 & 1016 $\pm$  44 & 1028 $\pm$  29  & 1067 $\pm$  31 & 1038 $\pm$  30 & 1065 $\pm$  30\\
0.425     &  705 $\pm$ 30 &  750 $\pm$  21 &  774 $\pm$  24 &  748 $\pm$  27 &  758 $\pm$  25  &  796 $\pm$  28 &  762 $\pm$  25 &  839 $\pm$  53\\
0.475     &  557 $\pm$ 27 &  582 $\pm$  18 &  590 $\pm$  22 &  621 $\pm$  24 &  633 $\pm$  25  &  674 $\pm$  21 &  639 $\pm$  19 &  642 $\pm$  32\\
0.525     &  457 $\pm$ 12 &  466 $\pm$  13 &  456 $\pm$  15 &  490 $\pm$  15 &  515 $\pm$  18  &  493 $\pm$  15 &  490 $\pm$  13 &  476 $\pm$  56\\
0.575     &  354 $\pm$ 21 &  352 $\pm$  21 &  362 $\pm$  12 &  361 $\pm$  14 &  381 $\pm$  12  &  409 $\pm$  14 &  393 $\pm$  12 &  390 $\pm$  12\\
0.625     &               &  290 $\pm$  14 &                &  293 $\pm$   9 &  296 $\pm$  10  &  297 $\pm$  10 &  310 $\pm$  10 &  306 $\pm$  15\\
0.675     &               &                &                &                &  223 $\pm$   9  &  236 $\pm$   7 &  227 $\pm$  11 &  228 $\pm$  18\\
0.725     &               &                &                &                &  185 $\pm$  18  &  186 $\pm$   7 &  170 $\pm$   7 &  175 $\pm$   9\\ \\

\pt\      &  $y = 0.05$    & $y = 0.15$     & $y = 0.25$     &  $y = 0.35$    &  $y = 0.45$    &  $y = 0.55$   &   $y = 0.65$   &   $y = 0.75$  \\
0.075     & 5472 $\pm$ 437 & 5490 $\pm$ 438 & 5295 $\pm$ 424 & 5112 $\pm$ 410 &                &               &                &               \\
0.125     & 4266 $\pm$ 186 & 4260 $\pm$ 215 & 4060 $\pm$ 395 & 4047 $\pm$ 219 & 4093 $\pm$ 190 &               &                &               \\
0.175     & 2975 $\pm$ 159 & 3160 $\pm$ 126 & 3103 $\pm$  97 & 3109 $\pm$ 146 & 3274 $\pm$ 107 & 3262 $\pm$ 95 & 3444 $\pm$ 102 &               \\
0.225     & 2244 $\pm$  75 & 2401 $\pm$  75 & 2299 $\pm$  78 & 2289 $\pm$  68 & 2237 $\pm$  59 & 2310 $\pm$ 61 & 2415 $\pm$  62 &  2597 $\pm$ 83\\
0.275     & 1765 $\pm$  71 & 1739 $\pm$  69 & 1723 $\pm$  53 & 1738 $\pm$  48 & 1712 $\pm$  44 & 1721 $\pm$ 43 & 1741 $\pm$  43 &  1768 $\pm$ 52\\
0.325     & 1291 $\pm$  45 & 1286 $\pm$  36 & 1299 $\pm$  40 & 1307 $\pm$  40 & 1268 $\pm$  34 & 1298 $\pm$ 31 & 1312 $\pm$  34 &  1276 $\pm$ 36\\
0.375     & 1040 $\pm$  36 & 1045 $\pm$  30 &  991 $\pm$  31 & 1047 $\pm$  29 &  990 $\pm$  26 &  990 $\pm$ 24 & 1002 $\pm$  31 &   964 $\pm$ 26\\
0.425     &  798 $\pm$  35 &  839 $\pm$  31 &  784 $\pm$  29 &  842 $\pm$  24 &  749 $\pm$  22 &  714 $\pm$ 21 &  777 $\pm$  19 &   737 $\pm$ 18\\
0.475     &  660 $\pm$  20 &  657 $\pm$  20 &  613 $\pm$  21 &  677 $\pm$  21 &  594 $\pm$  18 &  589 $\pm$ 16 &  605 $\pm$  15 &   574 $\pm$ 15\\
0.525     &  500 $\pm$  18 &  494 $\pm$  14 &  484 $\pm$  15 &  517 $\pm$  16 &  457 $\pm$  16 &  454 $\pm$ 22 &  457 $\pm$  14 &   452 $\pm$ 37\\
0.575     &  402 $\pm$  12 &  364 $\pm$  13 &  389 $\pm$  14 &  404 $\pm$  13 &  357 $\pm$  22 &  368 $\pm$ 13 &  331 $\pm$  14 &   328 $\pm$ 24\\
0.625     &  299 $\pm$  10 &  299 $\pm$  10 &  315 $\pm$  10 &  299 $\pm$  10 &  273 $\pm$  11 &  294 $\pm$ 11 &  262 $\pm$  14 &               \\
0.675     &  229 $\pm$   8 &  227 $\pm$   9 &  231 $\pm$   9 &  231 $\pm$  10 &  221 $\pm$  11 &  224 $\pm$ 10 &                &               \\
0.725     &  174 $\pm$   8 &  171 $\pm$  16 &  174 $\pm$  22 &  163 $\pm$   7 &  166 $\pm$   7 &  179 $\pm$  9 &                &               \\

\end{tabular}
\caption{\piminus\ $d^2N/(\pt d\pt dy)$ for the 5\% most central
events.}
\label{tab:piPtRapidity}
\end{sidewaystable}

%
%
\chapter{Conclusions and Summary}
\label{ch:Conclusions} We have studied the general characteristics
of charged particle production in \AuAu\ collisions at \sqrts\ =
130 \gev. The gross features of the cross section indicate that
the collision is dominated by geometry, as evidenced by the shape
of the multiplicity distribution.  Glauber model studies are able
to reproduce the measured $d\sigma/dN_{h^-}$ distributions and are
useful to obtain a statistical determination of the number of
participants in the collision for a given event sample, a quantity
related to the impact parameter of the collision.  This
determination of the collision centrality will be used in future
studies of the centrality dependence of many experimental
observables, such as the production of strange mesons and baryons,
charge fluctuations and identical particle correlations to name a
few.

From the yield of charged hadrons, we find that particle
production per participant in central collisions increases by 38\%
relative to \ppbar\ and 52\% compared to nuclear collisions at
\sqrtsNN\,= 17 \gev. The \hminus\ \pt\ spectrum in \AuAu\
distribution is harder than that of the \ppbar\ reference system
at similar centre-of-mass  energies for the \pt\ region up to 2
\gevc. Scaling of produced particle yield with number of
participants shows a strong dependence on \pt, with Wounded
Nucleon scaling achieved only at the lowest measured \pt. The
\hminus\ pseudorapidity distribution is almost constant within
$|\eta|<1$, indicating that at these energies we are approaching a
boost invariant mid-rapidity region. This finding contrasts the
results from lower energies where the $\eta$ distribution was
found to peak at mid-rapidity. The shape of the $\eta$
distribution in the mid-rapidity region is found to be similar for
all centralities. Comparing the \pt\ distributions for different
centralities to the distribution measured in peripheral
collisions, we find a rise in the number of produced particles
with increasing \pt\ up to $\pt < 2\ \gevc$. This is also
consistent with the observed rise when comparing to $\ppbar$
collisions. However, the ratio of \AuAu\ to \ppbar\ indicates that
the limit of scaling with the number of binary collisions \ncoll\
is not reached. As for future directions with \pt\ spectra, an
interesting indication coming from preliminary data at high \pt\
is that there a turnover in the shape of this ratio beyond 2
\gevc, which has been intensely debated in the context of being a
scenario consistent with partonic energy loss in a QGP.

The measured \hminus\ \meanpt\ increases by 14\% from peripheral
($70-80\%$) to central collisions.  The shape is found to be
relatively smooth.  If deconfinement is reached in these
collisions, this behaviour is consistent with a cross over region
between the deconfined and hadronic phase, instead of a strong
first order phase transition.  From the combination of the
pseudorapidity and \pt\ distribution results, we discussed the
applicability of the Bjorken estimate of the initial energy
density. The lower limit of $\epsilon \geq 3.9\ \gev/\fm^3$, taken
at face value and comparing to $\mu=0$ lattice results, leads to
the conclusion that necessary conditions for QGP formation are in
fact reached. Stated another way, to the question of whether the
QGP can be formed in the high energy RHIC collisions, the answer
is affirmative. Of course, this does not mean that one has a proof
of deconfinement, but it helps to set the stage for the
measurement of the additional probes of QGP formation, which are
needed to provide the detailed evidence.

The rapidity distribution of negative pions is found to be
relatively flat in the range $|y| <\ 0.8$. Although this is still
a small range compared to the rapidity gap of $\sim$10 units
between the colliding nuclei, this again signals the approach to
the limit of a boost-invariant baryon-free mid-rapidity region.
The pion transverse mass spectra below 0.6 \gevc\ are well
parameterized by a Bose-Einstein distribution with inverse slope
parameter $T \simeq 190\ \mev$.  The slope parameter is relatively
independent of centrality, perhaps growing with increasing
centrality up to a centrality of 60\% of \sigmahad\ but then
saturating. For the most central collision, the slope parameter
shows a small but systematic drop with rapidity. This is an
indication that boost invariance is not fully achieved. A possible
cause for this is increase in the baryon content with rapidity.
The rapidity and transverse momentum distributions of protons and
anti-protons will help shed more light on this subject.   The
measurement of the \mt\ spectra of more massive hadrons (kaons,
protons, $\lambda$s, etc.), as well as a combination of identical
particle correlations and spectra, are natural extensions to the
studies presented here in order to determine the conditions at
kinetic freeze-out such as transverse flow velocity \betat\ and
the kinetic freeze-out temperature $\Tfo$.

Identifying deconfinement and a complete characterization of the
QGP will require data from a wide spectrum of experimental
observables, systematic variations of centre-of-mass energies and
nuclear species, \pp\ and \pA\ reference data and candid guidance
from theory. We hope that the results presented here provide
useful baseline information to both the experimental and
theoretical efforts in this search.

{
\appendix
}

\pagebreak
\addcontentsline{toc}{chapter}{Bibliography}
\bibliography{mrabbrev,Bibliography}
\end{document}